%% file: main.tex
\documentclass[
12pt, 
english, 
onehalfspacing, 
nolistspacing, 
liststotoc, 
headsepline, 
]{MastersDoctoralThesis} 
\usepackage{color}
\usepackage{tikz}
\usepackage{amsmath}
\usepackage{amssymb}
\usepackage{bm}
\usepackage{tikz} 
\usepackage{doi}
\usepackage{url}
\usepackage{xcolor}
\usepackage{enumitem}
\usepackage{lscape}
\usepackage{float}
\usepackage{pdfpages}
\usepackage{epstopdf}
\newcounter{numln}
\newlist{numitemise}{itemize}{2}
\setlist[numitemise]{wide}%
\setlist[numitemise, 1]{labelindent=0pt,labelwidth=2em, label=\stepcounter{numln}\makebox[2em]{\thenumln.\hfill}, leftmargin=\dimexpr\labelwidth+\labelsep\relax}%
\setlist[numitemise, 2]{labelindent=\dimexpr -2em-\labelsep\relax, labelwidth=\dimexpr 2em+\labelsep\relax, label=\stepcounter{numln}\makebox[\dimexpr\labelwidth + \labelsep\relax]{\thenumln.\hfill\textbullet}, leftmargin=\dimexpr\leftmargin+2\labelsep\relax}%

\usepackage[utf8]{inputenc} 
\usepackage[T1]{fontenc} 

\usepackage{mathpazo} 

\usepackage[backend=bibtex,style=nature,natbib=true]{biblatex}
\usepackage[autostyle=true]{csquotes} 
\usepackage{setspace}
\usepackage{graphicx}
\usepackage{amsfonts}
\usepackage{amssymb}
\usepackage{amsmath}
\usepackage{amsbsy}
\usepackage{mathrsfs}
\usepackage{latexsym}
\usepackage{subfigure}
\usepackage{wasysym}
\usepackage{bm}
\usepackage{subfigure}
\usepackage{color}
\usepackage{comment}
\usepackage{textgreek}
\usepackage{hyperref}
\usepackage[T1]{fontenc}
\usepackage{pbsi}
\usepackage{calligra}
\usepackage{yfonts}
\usepackage{multicol}
\usepackage{caption}

\DeclareMathOperator{\csch}{csch}
\DeclareMathOperator{\arcsinh}{arcsinh}

\usepackage{array}
\newcolumntype{M}[1]{>{\centering\arraybackslash}m{#1}}
\newcolumntype{N}{@{}m{0pt}@{}}

\thesistitle{Thermodynamical Aspects\\ of Some Cosmological Models} 

\supervisor{NB} 

\examiner{} 

\degree{Doctor of Philosophy} 

\author{Tanima Duary} 
\addresses{} 

\subject{Physical Sciences} 
\keywords{} 
\university{\href{https://www.iiserkol.ac.in/}{IISER Kolkata}} 
\department{\href{https://physics.iiserkol.ac.in/}{Department of Physical Sciences}} 
\def\DateSub{April, 2024}

\AtBeginDocument{
\hypersetup{pdftitle=\ttitle} 
\hypersetup{pdfauthor=\authorname} 
\hypersetup{pdfkeywords=\keywordnames} 
}

\begin{document}
\sloppy
\frontmatter 

\pagestyle{plain} 


\begin{titlepage}
\begin{center}

\textsc{\Large Doctoral Thesis}\\[0.5cm] 
\HRule \\[0.4cm] 
{\huge \bfseries \ttitle\par}\vspace{0.4cm} 
\HRule \\[1.5cm] 

\begin{center}
	\large{By\\
     {{Tanima Duary}} \\
	Roll No.: 14IP021\\[1ex]
	\emph{Supervisor:} Prof. Narayan Banerjee\\[1ex]
    Department of Physical Sciences\\\vspace{0.0cm}
    Indian Institute of Science Education and Research Kolkata\vspace{0.5 em}}
\end{center}

\vspace{1.5cm}

\centering
	\includegraphics[width=0.3\textwidth]{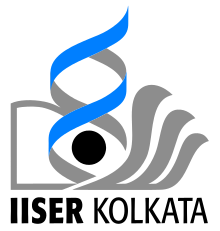}
	
\vspace{1.2cm}

\textit{ A thesis submitted in fulfillment of the requirements for 
the degree of \degreename \ 
in the \deptname \ at \ {\href{https://www.iiserkol.ac.in/}{Indian Institute of 
Science Education and Research Kolkata}} }

\vspace{0.7cm}

{\large \DateSub}\\[4cm] 

\vfill
\end{center}
\end{titlepage}


{\begin{declaration}

\noindent
\par I, Ms. \textbf{\emph{Tanima Duary}}, Registration No. \textbf{\emph{14IP021}} dated 
\emph{24th July 2014}, 
a student of the {Department of Physical Sciences} of the
Integrated PhD Programme of Indian Institute of Science Education and Research 
Kolkata (IISER Kolkata), hereby declare that this thesis is my own work and, to 
the best of my knowledge, it neither contains materials previously published or 
written by any other person, nor has it been submitted for any degree/diploma or 
any other academic award anywhere before. I have used the originality checking service to prevent inappropriate copying.\\
 
I also declare that all copyrighted material incorporated into this thesis is in 
compliance with the Indian Copyright
Act, 1957 (amended in 2012) and that I have 
received written permission from the copyright owners for my use of their work.\\
 
I hereby grant permission to IISER Kolkata to store the thesis in a database 
which can be accessed by others.

\vspace{2.cm}
\begin{flushleft}
Date: 
\end{flushleft}
\begin{flushright}
	\noindent \rule{4cm}{0.4pt}\\
\vspace{0.4cm}
	\textbf{\emph{Tanima Duary}}\\
	Department of Physical Sciences\\
	Indian Institute of Science Education and Research Kolkata \\
	Mohanpur 741246, West Bengal, India
\end{flushright}

\vspace{3.0cm}

\end{declaration}

\cleardoublepage

\include{Others/certificate}

\cleardoublepage
\include{Others/acknowledgements}

\cleardoublepage
\dedicatory{\bsifamily{\calligra{\huge{\textbf{To Moni and my adoring parents...}}}}}
\cleardoublepage
\include{Others/abstract}
\cleardoublepage


\doublespacing

\include{Others/preface}
\cleardoublepage


\tableofcontents 
\cleardoublepage
\include{Others/abbreviations}

\cleardoublepage
\listoffigures 
\cleardoublepage
\include{Others/publication}

\cleardoublepage






%


%
%
%



\mainmatter 

\pagestyle{thesis} 


\def\rs{r_{s}}
\def\rh{r_H}
\def\rstar{r_{\star}}
\def\rsh{r_{\square}}
\def\roplus{r_{\oplus}}
\def\scriplus{\mathscr{I}^{+}}
\def\scriminus{\mathscr{I}^{-}}

\def\observerminus{\mathbb{O}^{-}}
\def\observerplus{\mathbb{O}^{+}}
\def\kr{\kappa}
\def\ksg{\mathrm{\varkappa}}
\def\rhash{r_{\sharp}}

\def\Phit{\tilde{\Phi}}
\def\omegat{\tilde{\omega}}
\def\mstar{m_{\star}}
\def\mk{\mathbb{K}}

\def\Tr{\mathrm{Tr}\hspace{1pt}}
\def\varphit{\tilde{\varphi}}
\def\pit{\tilde{\pi}}

\def\lstar{l_{\star}}
\def\sstar{s_{\star}}

\include{Chapters/Chapter1}

\cleardoublepage
\include{Chapters/Chapterthermo}

\cleardoublepage
\include{Chapters/Chapter2}

\cleardoublepage
\include{Chapters/Chapter3}

\cleardoublepage
\include{Chapters/Chapter4}

\cleardoublepage
\include{Chapters/Chapter5}

\cleardoublepage
\include{Chapters/Chapter6}

\cleardoublepage
\include{Chapters/Chapter7}







\printbibliography[heading=bibintoc]

\end{document}

%% file: Others/certificate.tex
\addcontentsline{}{chapter}{Certificate from the Supervisor}
\thispagestyle{plain}
\null\vfil
{\noindent\huge\bfseries Certificate from the Supervisor \par\vspace{10pt}}

\noindent
\par
{\doublespacing This is to certify that the thesis entitled \textbf{\emph{``Thermodynamical Aspects of Some Cosmological Models''}} submitted by Ms. \textbf{\emph{\authorname}}, Registration No. \textbf{\emph{14IP021}} dated \textbf{\emph{24th July 2014}}, a
student of the Department of Physical Sciences of the Integrated PhD Programme of IISER Kolkata, is based upon her own research work under my supervision. I also certify, to the best of my knowledge, that neither the thesis nor any part of it has been submitted for any degree/diploma or any other academic award
anywhere before. In my opinion, the thesis fulfils the requirement for the award of the degree of Doctor
of Philosophy.} \\

\vspace{2.5cm}
\begin{flushleft}
Date: 
\end{flushleft}
\begin{flushright}
\noindent \rule{6cm}{0.4pt}\\
\vspace{0.4cm}	\textbf{\emph{Prof. Narayan Banerjee}} \\
	 Professor\\
	Department of Physical Sciences \\
	Indian Institute of Science Education and Research Kolkata\\
    Mohanpur 741246, West Bengal, India
\end{flushright}

%% file: Others/acknowledgements.tex
\addcontentsline{}{chapter}{Acknowledgements}
\thispagestyle{plain}
{\noindent\huge\bfseries Acknowledgements \par\vspace{6pt}}

 I extend my heartfelt gratitude to all those who played important roles in making this journey possible. Without their unwavering support and encouragement, the completion of this thesis would not have been possible. I want to take this moment to express my deep appreciation for their contributions.
 
 First and foremost, I wish to express my sincere and deepest gratitude to my supervisor, Prof. Narayan Banerjee, for being a guiding light throughout my doctoral journey.  He charted the course of my research with his visionary guidance and profound wisdom. I am truly indebted to him for his boundless patience and efforts, which have not only made this thesis possible but also enriched my understanding of the subject. I express my gratitude to him for maintaining faith in my abilities even during times when I doubted myself. Having the privilege of working under his guidance has been a rare and invaluable opportunity. I am grateful for the incredible teachings, enlightening discussions, and invaluable advice that have consistently motivated me on this scientific expedition. I am thankful for the time and effort he dedicated to reviewing drafts, providing constructive feedback, and offering valuable suggestions that significantly enhanced the overall quality of this thesis. I extend my heartfelt thanks to him for providing guidance and support when I felt directionless, helping me discover the strength to overcome challenges.\vspace{0.5em}

  I am profoundly grateful to Dr. Ananda Dasgupta, whose exceptional teaching prowess and constant support have left an indelible mark on my academic journey. His teaching style, characterized by clarity, enthusiasm, and a genuine passion for the subject matter, has made learning not just a scholarly pursuit but also an enriching experience. His wisdom and insight have been a source of inspiration, instilling in me a resilience that extends beyond academic pursuits. His impact on my education and personal growth is immeasurable, and I am privileged to have had the opportunity to learn under his tutelage. \vspace{0.5em}

  I extend my gratitude to the members of my research progress committee, Dr. Golam Mortuza Hossain and Prof. Rajesh Kumble Nayak, for their valuable insights and recommendations. I express my gratefulness to Dr. Koushik Dutta for his encouragement and I am thankful for providing me with the opportunity to serve as a volunteer at the 32nd IAGRG conference in 2022. This experience has been enriching. I am thankful to all the faculty members of Department of Physical Sciences, IISER Kolkata. \vspace{0.5em}

  I would like to express my heartfelt thanks to Sangita di, Munna da, and Ipsita di from the departmental office for their invaluable assistance with all official matters. My sincere appreciation goes to the librarian and Assistant Librarian at IISER-K Library for their cooperative support. Gratitude also to the individuals from DOAA, DOSA, DORD, CCC, SAC, and the Medical unit for their assistance. I would like to specially thank Ms. Saberi Roy Choudhury and Mr. Arun Dutta for helping us in CSIR fellowship matters. \vspace{0.5em}

  In the close-knit community of IISER Kolkata, Shibendu, Prashanti, Purba, Shreya, Basabendra, and Sourav have been more than friends; they are cherished companions who have shared in the joys and difficulties of academic pursuits. Their friendship has added richness to my experience at IISER Kolkata, and I am truly fortunate to have them as the closest allies. With deep reverence, I extend my regards and profound gratitude to my beloved friend, the late Subhadip Roy. Each passing day is a poignant reminder of his absence, and not a moment goes by when I do not miss the warmth of his friendship. \vspace{0.5em}

  I would like to thank my seniors- Shantanu da, Subhajit da, Chiranjeeb da, Ankan da, Soumya da, Sachin da, Avijit da, Srijita di and Anushree di  for the precious pieces of advice they generously shared. I would also like to thank Sampurna, Debraj, Medha, Siddhartha, Abhirup, Toushik da, Priyanka di, Diganta, Budhaditya, Poulomi, Saikat, Arkayan, Amulya, Chiranjit, Swarup, Brotoraj, Soumya, Debajyoti, Narayan, Arnab, Samit, Roshan, Kakali, Madhura, Soumi, Ananya, Fareeha, Debanajana and Lucky for making this journey memorable. I wish to thank Branali, Rajrupa, Poulami, Beetihotra, Navonil, Sayantan, and Siddharth. It was fun working with you side by side at the IAGRG conference. I am thankful to all my friends from Scottish Church College. \vspace{0.5em}

  I am grateful to my grandparents (dadu, dida), maternal uncle (mama), aunty (mami), sister Sushree, and little brother Sashreek,  for the affection they have bestowed upon me. \vspace{0.5em}

  I express my deepest gratitude to my parents, whose love, encouragement, and support which has been a steadfast anchor during both the peaks and valleys of my journey. Their encouragement and presence have been invaluable, providing solace in challenging times and magnifying the joy in moments of success. Their sacrifices, guidance, and belief in my potential have shaped the person I am today. I am truly blessed to have such caring and supportive parents by my side. I want to extend my heartfelt acknowledgment to my little brother, Moni, whose presence and enthusiasm have brought joy and inspiration to my life. His infectious energy, and genuine camaraderie have been a constant source of encouragement throughout my journey. I am grateful for the special bond we share, and I appreciate the positive impact he has had on my life. I express my gratitude to my dearest friend, Aritra, for being by my side through thick and thin. In the symphony of life, grateful for the serendipitous notes that brought him into the melody of my journey. Across the tapestry of time and the vastness of space, we share an epoch and a planet, a harmonious rhyme.\vspace{0.5em}

  I extend my sincere gratitude to the Council of Scientific and Industrial Research, India, for granting financial assistance through the CSIR-NET fellowship (Award No. 09/921(0171)/2017-EMR-I).\vspace{0.5em}

 I would like to express my appreciation for the enchanting campus of IISER Kolkata, characterized by its captivating beauty, vibrant sunset skies, and picturesque walking paths.

%% file: Others/abstract.tex
\addcontentsline{}{chapter}{Abstract}
\thispagestyle{plain}
\null\vfil
{\noindent\huge\bfseries Abstract \par\vspace{12pt}}

This thesis is focused on the thermodynamic analysis of cosmological models, specially the models that explain late-time cosmic acceleration. The cosmological principle says that the universe exhibits spatial homogeneity and isotropy. To describe it we consider the Friedmann-Lemaître-Robertson-Walker (FLRW) metric. A thorough evaluation of the feasibility of the models was conducted through the application of the Generalized Second Law (GSL). This law says that the overall entropy, i.e., the sum of the entropy of the horizon and the fluid enclosed within the horizon, should never decrease. Considering the dynamic nature of the universe, our methodology focused on the apparent horizon, instead of the event horizon. Within this framework, we have considered a condition of thermodynamic equilibrium between the apparent horizon and the fluid contained within it. In this state of equilibrium, we have considered the Hayward-Kodama temperature as the temperature associated with the apparent horizon.\\

The first chapter contains concise overview of cosmology. Chapter \ref{Chapterthermo} goes deeper into the thermodynamics applied to cosmology. It focuses more on the Generalized Second Law of Thermodynamics and explains it in more detail. Furthermore, we go into extensive details regarding Hayward-Kodama temperature. This chapter also contains a detail discussion about apparent horizon. The conditions required for thermodynamic stability has been discussed in this chapter.\\

In chapter \ref{Chapter2}, we conduct a thermodynamic comparison between quintessence models involving thawing and freezing scenarios. We have considered an ansatz on the energy density of the scalar field, which is picked up from the literature. The motivation for picking the ansatz was that, by choosing values of just one parameter, we can get either thawing or freezing behaviour. Both of these models are observed to violate the Generalized Second Law of Thermodynamics. Nevertheless, in the case of freezing models, there is still a possible way to resolve this, as this violation occurs in the distant past, deep within the radiation-dominated era, a period where a conventional scalar field model combined with pressureless matter is not an accurate representation of the matter content. In contrast, the thawing model exhibits a violation of GSL, manifesting as a finite future breakdown. Therefore, we conclude that the freezing models are favoured compared to the thawing ones on the considerations of GSL viability.\\  

In chapter \ref{Chapter3}, we scrutinize Brans-Dicke cosmological models within the context of a spatially isotropic and homogeneous universe, evaluating their compatibility with the GSL. Our investigation is carried out within the Einstein frame. We find that in dust era, these models exhibit thermodynamic feasibility when the Brans-Dicke parameter $\omega$ assumes negative values. This range has strong alignment with the range that is required for the recent observations of the cosmic acceleration.\\

In chapter \ref{Chapter4}, we explore the thermodynamic viability of a selection of dark energy models, which have been reconstructed using the cosmological jerk parameter. Our investigation involves the adoption of models previously documented in the literature. These models are categorized into two groups based on the presence or absence of interactions in the dark sector. We employ the GSL as a diagnostic tool for our analysis. In an attempt to capture the dynamic nature of spacetime, we replace the Hawking temperature with the Hayward-Kodama temperature. Our results indicate that, dependent on the chosen parametrization ansatz for jerk, the total entropy exhibits a time-increasing trend. This suggests the potential existence of viable models within this framework. This trend persists even when there is interaction in the dark sector.\\

In chapter \ref{Chapter5}, we have considered a model in spatially flat FRW spacetime, that mimics the characteristics of $\Lambda$CDM model and checked the thermodynamic stability. In this chapter also we have  utilized the Hayward-Kodama temperature as the temperature of the apparent horizon. Assuming the thermal equilibrium between the apparent horizon and the fluid inside the horizon, we investigated the thermodynamic stability of the matter composition within the universe and found out that it lacks the thermodynamic stability. We found out an interesting result while calculating the heat capacity at constant volume ($C_V$). It is shown that the transition from the decelerated to the accelerated cosmic expansion is a second-order thermodynamic phase transition, while the deceleration parameter $q$ serves as the order parameter. 
\\

In chapter \ref{Chapter6}, we reach the epilogue of this thesis, where we not only provide our final conclusions but also briefly discussed the aspects of the work presented in this dissertation and future prospects.

%% file: Others/Preface.tex
\addchaptertocentry{Preface}
\thispagestyle{plain}
\null\vfil
{\noindent\huge\bfseries Preface \par \vspace{12pt}}

\onehalfspacing

The research presented in this dissertation was conducted at the Department of Physical Sciences, Indian Institute of Science Education and Research Kolkata, India. The initial chapter encapsulates a preamble to the realm of cosmology,  centering upon the different models to explain cosmic acceleration. The next chapter (chapter \ref{Chapterthermo}) focuses on the intricate interrelation linking cosmology and the principles of thermodynamics. The succeeding chapters are grounded on the subsequent scholarly articles:

\begin{itemize}
\item Chapter \ref{Chapter2} : Thermodynamics of Thawing and
Freezing Quintessence Models\\
  {\textbf{Tanima Duary}, Ananda Dasgupta and Narayan Banerjee , {\it ``Thawing and Freezing Quintessence Models: A thermodynamic Consideration"}, \href{https://doi.org/10.1140/epjc/s10052-019-7406-z}{Eur.Phys.J.C {\bf 79} (2019) 11, 888}, \href{https://doi.org/10.48550/arXiv.1906.10408}{arXiv:1906.10408}.}
  
\item Chapter \ref{Chapter3} : Thermodynamics of Brans-Dicke Cosmology\\
{\textbf{Tanima Duary}, Narayan Banerjee, {\it Brans–Dicke cosmology: thermodynamic viability}, \href{https://doi.org/10.1140/epjp/s13360-019-00078-z}{Eur.Phys.J.Plus {\bf 135} (2020) 1, 4}, \href{https://doi.org/10.48550/arXiv.1910.10931}{arXiv:1910.10931}.}

\item Chapter \ref{Chapter4} : Thermodynamic Analysis of
Cosmological models reconstructed
from jerk parameter\\
 {\textbf{Tanima Duary}, Narayan Banerjee,  {\it Cosmological models reconstructed from jerk: A thermodynamic analysis}, \href{https://doi.org/10.1016/j.newast.2021.101726}{New Astron. {\bf 92} (2022) 101726}.}

\item Chapter \ref{Chapter5} : A Possible Thermodynamic Phase
Transition: Signature flip of the Deceleration Parameter\\
{\textbf{Tanima Duary}, Narayan Banerjee and Ananda Dasgupta {\it Signature flip in deceleration parameter: A thermodynamic phase transition?}, \href{https://doi.org/10.1140/epjc/s10052-023-11995-w}{Eur.Phys.J.C {\bf 83} (2023) 9, 815}, \href{
https://doi.org/10.48550/arXiv.2303.14031}{arXiv:2303.14031v2}.}
\end{itemize}

%% file: Others/abbreviations.tex

\begin{abbreviations}{ll} 
\textbf{$\Lambda$CDM} &  {\bf L}ambda {\bf C}old {\bf D}ark {\bf M}atter\\[1ex]

\textbf{BD} & {\bf B}rans--{\bf D}icke \\[1ex]
\textbf{CMBR} & {\bf C}osmic {\bf M}icrowave {\bf B}ackground {\bf R}adiation \\[1ex]

\textbf{FRW} & {\bf F}riedmann--{\bf R}obertson--{\bf W}alker\\[1ex]
 
\textbf{GTR} & \textbf{G}eneral \textbf{T}heory of \textbf{R}elativity\\[1ex]
\textbf{GSLT} & {\bf G}eneralised {\bf S}econd {\bf L}aw of {\bf T}hermodynamics
\\[1ex]
\textbf{KG equation} & {\bf K}lein {\bf G}ordon equation
\\[1ex]

\textbf{EoS} parameter & {\bf E}quation {\bf o}f {\bf S}tate parameter \\[1ex]

\textbf{BDT} & {\bf B}rans--{\bf D}icke {\bf T}heory \\[1ex]

\textbf{NMCSTT} & {\bf N}on-{\bf M}inimally {\bf C}oupled {\bf S}calar-{\bf T}ensor {\bf T}heories  \\[1ex]

\textbf{DDE} & {\bf Dy}namical {\bf D}ark {\bf E}nergy \\[1ex]

\textbf{CPL model} & {\bf C}hevallier --{\bf P}olarski--{\bf L}inder model. \\[1ex]
\textbf{CMB} & {\bf C}osmic {\bf M}icrowave {\bf B}ackground. \\[1ex]
\textbf{HDE} & {\bf H}olographic {\bf D}ark {\bf E}nergy. \\[1ex]

\end{abbreviations}

%% file: Others/publication.tex
\addchaptertocentry{Dissemination}
\thispagestyle{plain}
\null\vfil
{\noindent\huge\bfseries List of Publications \par \vspace{12pt}}

\onehalfspacing

\paragraph*{This thesis is based on the following works:}
  \begin{numitemise}
  
        \item {\textbf{Tanima Duary}, Ananda Dasgupta and Narayan Banerjee , {\it Thawing and Freezing Quintessence Models: A thermodynamic Consideration}, \href{https://doi.org/10.1140/epjc/s10052-019-7406-z}{Eur.Phys.J.C {\bf 79} (2019) 11, 888}, \href{https://doi.org/10.48550/arXiv.1906.10408}{arXiv:1906.10408}.}
        
        \item {\textbf{Tanima Duary}, Narayan Banerjee, {\it Brans–Dicke cosmology: thermodynamic viability}, \href{https://doi.org/10.1140/epjp/s13360-019-00078-z}{Eur.Phys.J.Plus {\bf 135} (2020)  1, 4}, \href{https://doi.org/10.48550/arXiv.1910.10931}{	arXiv:1910.10931}.}

     	\item {\textbf{Tanima Duary}, Narayan Banerjee,  {\it Cosmological models reconstructed from jerk: A thermodynamic analysis}, \href{https://doi.org/10.1016/j.newast.2021.101726}{New Astron. {\bf 92} (2022) 101726}.}

        \item {\textbf{Tanima Duary}, Narayan Banerjee and Ananda Dasgupta {\it Signature flip in deceleration parameter: A thermodynamic phase transition?}, \href{https://doi.org/10.1140/epjc/s10052-023-11995-w}{Eur.Phys.J.C {\bf 83} (2023) 9, 815}, \href{
https://doi.org/10.48550/arXiv.2303.14031}{arXiv:2303.14031v2}.}
  	
\end{numitemise}

%% file: Chapters/Chapter1.tex
\chapter{Introduction} 
\label{Chapter1}

\section{Concise Overview:}
\subsection{Brief chronological background:}

    Einstein's General Theory of Relativity is a revolutionary framework in physics that transformed our understanding of gravity and hence, the evolution of the universe. Proposed by Albert Einstein in 1915 \cite{Einstein:1915ca}, it represents one of the most significant intellectual achievements in human history. At its core, the theory suggests that gravity is not a force as traditionally understood but rather a curvature in the fabric of spacetime caused by the presence of mass and energy. According to General Relativity, massive objects such as stars and planets distort the geometry of spacetime, causing objects to follow curved paths. The theory provides a new mathematical description of gravity, utilizing a set of equations that relate the distribution of matter and energy to the curvature of spacetime. \vspace{0.5em}
      
     General Theory of Relativity provides a theoretical framework for examining the structure and dynamics of the entire universe, and Einstein devised this framework in 1917 \cite{Einstein:1917ce}. At that time, physicists believed the cosmos was stationary and not evolving, a celestial clockwork mechanism that would run forever. However, this prevailing view was refuted by Einstein's equations in general relativity, which indicated that the universe could either expand or contract. To account for this, Einstein introduced a modification to his equations by including a cosmological constant ($\Lambda$), which acted as a repulsive force to counterbalance gravity and maintain a stable universe. Willem de Sitter proposed the maximally symmetric vacuum solution of Einstein's field equations with a positive cosmological constant $\Lambda$ \cite{deSitter2017-DESOTR,deSitter:1917zz}. Karl Schwarzschild derived a solution to Einstein's equations that described the spacetime curvature around a massive object, such as a star or a black hole \cite{Schwarzschild:1916uq}. This solution laid the foundation for our understanding of gravitational collapse and singularities. In 1919, Arthur Eddington led an expedition to observe a solar eclipse and confirmed Einstein's prediction that light rays would bend in the presence of massive objects \cite{Dyson:1920cwa}.\vspace{0.5em}
     
     In 1922, Alexander Friedmann introduced the idea of an expanding universe in his paper "On the Curvature of Space" \cite{Friedman:1922kd}. He derived a set of mathematical solutions to Einstein's equations that described a dynamic cosmos, where the matter was in motion and the universe itself was evolving over time. Friedmann's work laid the groundwork for modern cosmology by challenging the prevailing belief in a static universe and introducing the concept of cosmic expansion. It provided a theoretical basis for the subsequent development of the Big Bang model and set the stage for further investigations into the nature, origin, and evolution of the universe. \vspace{0.5em}
     
      In 1929, Edwin Hubble made a significant discovery that revolutionized our understanding of the universe. He studied distant galaxies and their light spectra, which contained characteristic features called spectral lines \cite{Hubble:1929ig}. These lines could be used to determine the motion of an object. By examining the spectra of galaxies, Hubble noticed a peculiar pattern: the spectral lines were consistently shifted towards longer wavelengths, known as a redshift. Based on the known Doppler effect, i.e., the redshift of light, Hubble concluded that the galaxies were moving away from each other. Moreover, he observed that the more distant a galaxy was, the greater its redshift. This finding implied that if galaxies were moving away from us, then at some point in the past, they must have been closer together.\vspace{0.5em}
      
       Based on Hubble's discovery, Georges Lemaître suggested that the universe began from an extremely hot and dense state, which he referred to as the \textit{primeval atom} or the \textit{cosmic egg}. He theorized that this primordial atom contained all the matter and energy in the universe and exploded in an event he called the \textit{explosion of the primeval atom}. This explosion marked the beginning of the universe, a moment commonly referred to as the \textit{Big Bang}. Though he proposed the model in 1931 \cite{Lemaitre:1931zza}, this gained wider recognition in the scientific community later, with additional evidence supporting the Big Bang model emerged, such as the discovery of the cosmic microwave background radiation, which is considered a remnant of the early stages of the universe. In 1948, Alpher, Bethe, Gamow, and Herman developed the first detailed model of Big Bang nucleosynthesis, which provided an explanation for the formation of light elements, such as hydrogen and helium, in the early universe \cite{Alpher:1948ve,Alpher:1948srz,Alpher:1948gsu}. Alpher, Gamow, and Herman's model proposed that in the first few minutes after the Big Bang, the universe was extremely hot and dense, with temperatures of billions of degrees. During this time, the conditions were favorable for nuclear reactions to take place. According to their calculations, the extreme temperatures and densities allowed for the fusion of protons and neutrons to form light elements, primarily hydrogen and helium.\vspace{0.5em}

        In 1965, Arno Penzias and Robert Wilson made a groundbreaking discovery known as the cosmic microwave background radiation (CMB) or the Penzias-Wilson radiation \cite{Penzias:1965wn}. Penzias and Wilson were astronomers working at Bell Labs in New Jersey, conducting experiments using a large horn antenna originally built for satellite communication. However, no matter what they did, they encountered a persistent noise signal in their measurements. They could not eliminate this noise, which seemed to be coming from every direction in the sky. It was unknown to Penzias and Wilson that their observations were coinciding with the work of  Robert Dicke, Jim Peebles, and others at Princeton University \cite{Dicke:1946glx,Peebles:1962fbw}. Theoretical work at Princeton had predicted the existence of a faint radiation left over from the early stages of the universe, known as the cosmic microwave background. When Penzias and Wilson learned about the Princeton team's prediction, they realized that the noise they were observing matched the characteristics of the cosmic microwave background radiation (CMB). This radiation is the remnant of the intense heat and energy that filled the universe shortly after the Big Bang. The discovery of the CMB \cite{Dicke:1965zz} provided strong evidence in support of the Big Bang model.\vspace{0.5em}
        
         According to the Big Bang model, the universe expanded rapidly from an extremely hot and dense state. The observations indicated that different regions of the universe that were now far apart appeared to have the same temperature, properties and flat. To address this issue, Alan Guth(1981) proposed the concept of cosmic inflation \cite{Guth:1980zm} and suggested that a brief period of exponential expansion occurred in the very early universe, causing it to grow exponentially larger in size. This rapid expansion would have smoothed out the initial irregularities and made the universe homogeneous and isotropic, providing an explanation for the observed uniformity across vast regions of space. The cosmic inflation  also indicates that quantum fluctuations during inflation could lead to the formation of tiny density perturbations in the early universe. These fluctuations served as seeds for the later formation of galaxies, clusters of galaxies, and other cosmic structures. Guth's work had a profound impact on our understanding of the universe. Subsequent observations, such as the measurement of the cosmic microwave background radiation by the Cosmic Background Explore(COBE) \cite{PhysRevLett.69.3602} and  The Wilkinson Microwave Anisotropy Probe(WMAP) satellites \cite{WMAP:2003syu}, provided strong evidence in support of cosmic inflation and moreover the Big Bang model.\vspace{0.5em}

       The discovery of cosmic acceleration in late 1990s posed a threat to our understanding of the universe. Two independent teams of astronomers, the High-Z Supernova Search Team (Schmidt \textit{et al}., 1998) \cite{SupernovaSearchTeam:1998bnz} and the Supernova Cosmology Project (Perlmutter \textit{et al}., 1999) \cite{SupernovaCosmologyProject:1998vns} provided supporting data that Hubble expansion is accelerating over time, while studying distant supernovae, specifically Type Ia supernovae(SNIa) as standard candles to measure the correlation between distance and redshift. This discovery was contrary to the prevailing belief that the gravitational pull of matter should be causing the expansion to decelerate. Adam G. Riess and his colleagues analysed the data provided by the previously mentioned groups in the article titled "Observational Evidence from Supernovae for an Accelerating universe and a Cosmological Constant" \cite{SupernovaSearchTeam:1998fmf} and concluded that  the expansion of the universe was accelerating rather than slowing down and proposed the presence of a cosmological constant (i.e., the vacuum energy density) in the equations of general relativity. The cosmological constant acts as a repulsive force, causing the expansion of the universe to accelerate over time. This discovery of an accelerating universe and the need for dark energy revolutionized cosmology and deeply influenced our comprehension of the basic nature of the universe.\vspace{0.5em}

        Later, WMAP team discovered that approximately 71.4\% of the universe consists of dark energy \cite{WMAP:2003gmp}. Planck space observatory, designed to map the anisotropies of the cosmic microwave background (CMB) at infrared and microwave frequencies, revealed that the ordinary matter makes up only about 4-5\% only, with dark matter(25-26\%) and dark energy(69-70\%) accounting for the remaining 95\% \cite{Planck:2015fie}. This dominating dark energy is thought to exert a repulsive gravity that counteracts the attractive force of gravity, causing galaxies and other cosmic structures to move away from each other at an accelerating rate. There are several theories and hypotheses regarding the source of this exotic dark energy, but no definitive explanation has been established. Such theories include two most talked about categories: (i) cosmological constant, and (ii) quintessence field.\vspace{0.5em}

        The other way of explaining this accelerated expansion is to modify the theory of gravity. Some eminent researchers of this era have proposed that our understanding of gravity on cosmological scales may need modification. Modified theories of gravity, such as the f(R) theory, Brans-Dicke theory attempt to explain the observed cosmic acceleration without invoking dark energy. These theories suggest that gravity behaves differently on large scales and can account for the accelerated expansion. But there is no certain observational evidence to choose one theoretical model. There are reverse engineering techniques, known as the Reconstruction of dark energy models from observational datas to understand the nature of dark energy. \vspace{0.5em}
        
        However, theorists use several methods to compare the available theories of dark energy in literature and try to find out which one is preferable. In this thesis, we have used the \textit{Generalised Second Law of Thermodynamics} to study different cosmological models. Before delving into the specifics of that topic,  initially,  a concise aspect related to cosmology is explored here.

\subsection{Geometry and Dynamics of the Spacetime:}
 In 1915, Einstein's work on a new theory of gravity did not just result in a different way to think about forces or gravitational fields. It led to a big change in how we see space and time, shaking up our understanding in a major way. Einstein recognized that the empirical observation of all objects falling with the same acceleration in a gravitational field naturally pointed towards an interpretation of gravity based on the curvature of spacetime.  According to the General theory of relativity, gravity is manifested by the interaction of spacetime curvature and matter \cite{Misner:1974qy,Wald:1984rg,Carroll:1997ar,Carroll:2004st,JayantVNarlikar:2010vvf,d1992introducing,hooft2001introduction,book:Schutz, hartle2003gravity}. Einstein's field equations give the mathematical manifestation of this theory. By varying Einstein-Hilbert action with respect to $g_{\mu\nu}$, 
\begin{equation}\label{EHaction}
\mathcal{S}_{EH}=\int{\sqrt{-g}\left(\frac{R}{16\pi G}+\mathcal{L}_m\right)d^4x},
\end{equation}
we get the Einstein's field equation, 
\begin{equation} \label{einfld}
G_{\mu\nu}\equiv R_{\mu\nu}-\frac{1}{2}g_{\mu\nu}R=8\pi G T_{\mu\nu}.
\end{equation}
In the above equation,  $g^{\mu\nu}$ denotes the metric tensor and $g\equiv \det{g_{\mu\nu}}$; $g_{\mu\nu}$ is symmetric and throughout the thesis we follow the signature convention (-,+,+,+). Here, $R$ is the scalar curvature, known as Ricci scalar and defined as, $R=g^{\mu\nu} R_{\mu\nu}$. $R_{\mu\nu}$ is the Ricci tensor that captures the local curvature of spacetime at each point. $G\approx 6.7\times 10^{-11} \text{m}^3\text{kg}^{-1}\text{s}^{-2}$ 
represents the Newtonian constant of gravitation. In the action  \eqref{EHaction}, the Lagrangian density is divided in two parts: (i) gravitational part: $R$ and (ii) matter part: $\mathcal{L}_m$. The volume element is $\sqrt{-g}d^4x$. Equation \eqref{einfld} signifies the connection between geometry, contributed by $G_{\mu\nu}$ and matter, contributed by $T_{\mu\nu}$. Energy-momentum tensor or stress-energy tensor $T_{\mu\nu}$ is defined as, 
\begin{equation}\label{energymomten}
T_{\mu\nu} = -2 \frac{\partial\mathcal{L}_m}{\partial g^{\mu\nu}}+g_{\mu\nu}\mathcal{L}_m, 
\end{equation}
describes the distribution of mass, energy, and momentum in spacetime and  depends on the specific physical system under consideration.
\subsection{Cosmology:}
  The foundation of the contemporary cosmological model lies in the Cosmological Principle, asserting that the universe exhibits spatial homogeneity i.e., translationally invariant (Copernican principle) and isotropy i.e., rotationally invariant at scales much larger than the galaxy cluster \cite{blau2011lecture,Carroll:2004st}. This suggests that the space is maximally symmetric. Thus it becomes feasible to depict the universe using a lucid geometric framework by taking 1+3 foliation of the spacetime. The invariant interval \cite{LANDAU1975259},
  \begin{equation}
  ds^2= g_{\mu\nu}dx^\mu dx^\nu,
\end{equation}            
does not depend on the choice of co-ordinates $x^\mu$s ($x^0$ denotes cosmic time $t$ and $x^1$,$x^2$,$x^3$ are space co-ordinates).  The curvature of the spatial part can be of three forms: (i) positive curvature, (ii) zero curvature or (iii) negative curvature. A possible way to represent the metric in 1+3 foliation is, 
\begin{equation}
ds^2= -dt^2+a^2(t)h_{ij}x^ix^j,
\end{equation}
where $a(t)$ is known as the cosmic scale factor.  The 3-metric $h_{ij}$ is, 
\begin{align} \label{hij}
h_{ij}= \delta_{ij}+k\frac{x_ix_j}{1-k|{x}|^2}, \hspace{3em} k=
\begin{cases}
0 \hspace{1em} &\text{flat} \\ +1\hspace{1em} &\text{positive curvature} \\ -1\hspace{1em} &\text{negative curvature}
\end{cases}.
\end{align}
In the above equation $k$ is the curvature parameter. 
\begin{itemize}
\item \textbf{Friedmann-Robertson-Walker metric (FRW metric):}
\end{itemize}
From the above-written metric in the maximally symmetric spacetime, we can write the FRW metric \cite{Friedman1922berDK,1924ZPhy...21..326F,Lemaitre:1931zza,1927ASSB...47...49L,1933ASSB...53...51L,Robertson:1935zz,1937PLMS...42...90W}, 
\begin{equation}\label{FRWmetric}
ds^2= -dt^2+a^2(t)\left(\frac{dr^2}{1-kr^2}+r^2d\theta^2+r^2 sin^2\theta d\phi^2\right), 
\end{equation}
 in the comoving spatial coordinates ($r$,$\theta$,$\phi$).   
The scale factor $a(t)$ is the length scale of the universe serves as a gauge for understanding how physical distances evolve over time, where the coordinate distances, denoted as $r$, are considered fixed by definition. The present value of the cosmic scale factor $a(t)$ is often considered as, $a_0=a(t_0)=1$.
\begin{itemize}
\item \textbf{Cosmological Redshift:}
\end{itemize}  
  The primary and critical information about the cosmic scale factor, $a(t)$, is obtained from the observation of redshifts in the frequency of light emitted by distant celestial sources. As the universe expands, the wavelengths ($\lambda$) of photons traveling through space also stretch, leading to a redshift in their observed frequency. This is a key observational evidence of the expanding universe. Astronomer Edwin Hubble interpreted this redshift as due to a Doppler effect and therefore, ascribed a recessional velocity $v$ to the galaxy, which is related to the shift in wavelength by the Doppler formula,
\begin{equation}
\frac{v}{c}=\frac{\Delta\lambda}{\lambda}\equiv z,
\end{equation} 
where $z$ is known as the redshift. 

 The cosmological redshift behaves symmetrically between the receiver (observer) and the emitter (distant galaxy). In other words, if we observe a redshift in the light coming from a distant galaxy, the light sent from the Earth to that galaxy would also experience a redshift when observed from the distant galaxy's perspective.
 According to \textbf{\textit{Hubble's Law}}, the recessional velocity increases proportionally with the distance $d$ of the galaxy \cite{Hubble:1931zz}, i.e., 
\begin{align}
v \propto d, \nonumber \\
\therefore v = H d,
\end{align}
     here the proportionality constant $H$ is known as \textit{Hubble constant}. To provide a more accurate historical account, credit for this fundamental discovery, known as Hubble's law, should potentially be shared with G. Lema{\^i}tre as well. In many cosmological models, the Hubble constant $H$, is a time-dependent function, and therefore, nomenclatured as Hubble parameter. However, in the given equation, $H$ represents the present-day value of the Hubble constant, often denoted as $H_0$. Mega-parsecs (Mpc) are a common unit of measurement for galactic distances. The term \textit{parsec} is derived from \textit{parallax of one arcsecond} and is based on the phenomenon of parallax, which is the apparent shift in the position of a nearby object when observed from different vantage points. One parsec is defined as the distance at which an object shows a parallax angle of one arcsecond, or approximately $3.26$ light-years (about $3.086 \times 10^{13}$ kilometers or $1.917 \times 10^{13}$ miles). 
     Due to the old trigonometric technique of determining star distances, this unit was created. The unit in which $H$ is measured, is $\text{km} \text{Mpc}^{-1}\text{s}^{-1}$. At current epoch its value is, $H_0\approx 70 \text{km} \text{Mpc}^{-1}\text{s}^{-1}$.
     
     In view of FRW cosmology, redshift parameter $z$ can be expressed as, 
 \begin{equation}
 z=\frac{a_0}{a(t)}-1,
 \end{equation}
     and the Hubble parameter is expressed as,
     \begin{equation}
     H=\frac{\dot{a}(t)}{a(t)}.
     \end{equation}
     An overdot signifies a derivative of the variable with respect to cosmic time ($t$). 
   
 \begin{itemize}
 \item \textbf{\textit{A Perfect Fluid:}}
\end{itemize}   
 We consider that the universe is composed of a fluid. By definition, a perfect fluid is characterized by the property that a comoving observer perceives the fluid surrounding them as isotropic. In case of a perfect fluid, the heat conduction, viscosity or other transport or dissipative processes are considered negligible. One can write the energy-momentum tensor for perfect fluid in comoving coordinates as, 
 \begin{equation}\label{stresstensor}
 T^{\mu\nu}=(\rho+p)u^\mu u^\nu+pg^{\mu\nu},
 \end{equation}
 which acts as a source of spacetime curvature in the Einstein equation. 
 In above equation $u_\mu$ represents the 4-velocity, in comoving frame its componennts are, 
 \begin{equation}
 u^\mu =(1,0,0,0).
 \end{equation}
 In equation \eqref{stresstensor}, $\rho$ and $p$ are the energy density and pressure of the perfect fluid respectively.
The continuity equation is then expressed as, 
\begin{equation}
\nabla_\nu T^{\mu\nu}=0,
\end{equation} 
where $\nabla_\nu$ is the covariant derivative and is often indicated by the symbol ;.
The energy density ($\rho$) and the pressure are related by the equation of state (EoS), 
\begin{equation}
p=w\rho,
\end{equation}
where $\omega$ is called the equation of state parameter.
Here are some examples of perfect fluid:
\begin{enumerate}
\item \textit{Dust:}
   In case of dust particles, the pressure is zero, as in this case the kinetic energy is negligible compared to the rest energy. Therefore, $p=0$ and thus $w=0$. Dust is widely accepted as an accurate representation of baryonic matter (and cold dark matter) in the present epoch.

 \item \textit{Radiation:}
 For electro-magnetic radiation, the equation of state is, $p=\frac{\rho}{3}$. Therefore, the EoS parameter is $w=1/3$.
 
 \item \textit{Cosmological constant:}
 For $\Lambda$, EoS is, $p=-\rho$. Therefore, EoS parameter is $w=-1$.
 \end{enumerate}
 For the fluids mentioned above, we see that EoS parameter is constant. 
 
 \begin{itemize}
 \item \textbf{\textit{The Einstein Field Equations:}}
 \end{itemize}
 In FRW cosmology, the Einstein equations give two independent equations,
 \begin{align}
 \frac{\dot{a}^2}{a^2}+\frac{k}{a^2}&=\frac{8\pi G}{3}\rho,\label{F1}\\
 2\left(\frac{\ddot{a}}{a}\right)+\left(\frac{k+\dot{a}^2}{a^2}\right)&=-8\pi Gp.\label{F2}
 \end{align}

The fluid conservation equation is, 
\begin{equation}\label{conservation}
\dot{\rho}+3H(\rho+p)=0.
\end{equation}
Due to the Bianchi identities ($G^{\mu\nu};_{\nu}=0$), the conservation equation is not independent of Einstein equations and can be derived from them. 

\begin{itemize}
\item \textbf{\textit{Some Important Kinematic Quantities:}}
\end{itemize}

\textbf{1.} \textit{Deceleration Parameter:}
The deceleration parameter, denoted by $q$, measures the rate of change of expansion rate and defined as, 
\begin{equation}\label{decel}
q\equiv-\frac{\ddot{a}/a}{\left(\dot{a}/a\right)^2} =-\frac{1}{aH^2}\frac{d^2a}{dt^2}.
\end{equation}
(i) If q > 0: The universe is decelerating. The expansion rate is slowing down over time. Normal baryonic matter shows $q>0$. 
(ii) If q < 0: The universe is accelerating. The expansion rate is increasing over time. Observational evidence suggests that our universe is currently in an accelerated expansion phase. This discovery led to the proposal of the concept of dark energy which gives $q<0$.
(iii) If q = 0: The universe is coasting. The expansion rate remains constant over time.

\textbf{2.} \textit{Jerk Parameter:}
Jerk parameter is related to the third derivative of the scale factor with respect to cosmic time and defined as, 
\begin{equation}
j\equiv -\frac{1}{aH^3}\frac{d^3a}{dt^3}.
\end{equation}
The behaviour of the jerk parameter can give insights into the presence of dark energy, as it plays a crucial role in the accelerated expansion of the universe. The expansion of the universe is currently accelerating, and this acceleration is attributed to dark energy, represented by the cosmological constant ($\Lambda$). The value of the jerk parameter for $\Lambda$CDM model is $j_{\scriptsize{\Lambda CDM}} =-1$.  A value of 
$j$ not equal to $-1$ could suggest the presence of additional forces or components in the universe that contribute to its acceleration and whose effects are not fully captured by the cosmological constant.
It is to be noted that there is another parameter, known as the snap parameter, $s=\frac{1}{aH^4}\frac{d^4a}{dt^4}$, related to fourth order derivative of the scale factor. However, in the aspects of observational cosmology, the evolution of $q$ holds physical significance. Therefore, we only narrow our analysis to the jerk parameter.

\textbf{3.} \textit{Critical Density \& Density Parameter:}
The critical density ($\rho_c$) is a critical value of density that determines the geometry of the universe. It is defined as,
\begin{equation}\label{critden}
\rho_{c}(t)\equiv\frac{3H^2}{8\pi G}.
\end{equation}

The critical density represents the dividing line between an open, flat, or closed universe. $\Omega$ represents the ratio of the average density of the universe to the critical density. It is defined as,
\begin{equation}\label{denpara}
\Omega=\frac{\rho}{\rho_{c}}.
\end{equation}

The spatial shape and curvature of the cosmos are determined by the value of the density parameter in the following way:\\
(i)$\Omega<1 (\Rightarrow k=-1)$ : implies the universe has an open geometry.\\
(ii)$\Omega=1 (\Rightarrow k=0)$: the universe has a flat geometry.\\
(iii)$\Omega>1 (\Rightarrow k=+1)$:the universe has a closed geometry.
To gain a deeper understanding of the fundamental principles underlying modern cosmology, readers may refer to \cite{Weinberg:1972kfs,Hawking:1973uf,Raychaudhuri:1992qz,Kolb:1990vq,padmanabhan2000theoretical,rich2009fundamentals,liddle2003introduction,}.

\section{Thermal Evolution of the Universe in Brief:}
Understanding the thermal evolution history is crucial in cosmology as it provides insights into the behaviour of different constituents of the universe and the processes that shaped its current state. Here is an overview of the key stages in the thermal evolution of the universe:
\begin{enumerate}
\item \textit{Planck Era}: (Time: $0$ to $10^{-43}$ seconds after the Big Bang):
Temperature Range: $> 1.416 \times 10^{32}$ K. During this incredibly brief epoch, the universe was in a state of extremely high energy and temperature. 
In the Planck epoch, it is presumed that the characteristics of cosmology and physics were primarily governed by the quantum effects of gravity.

\item \textit{Grand Unification Era}: (Time: $10^{-43}$ seconds to $10^{-36}$ seconds):
Temperature Range: $1.416 \times 10^{32}$ K to $10^{27}$ K. The three forces of the Standard Model remained unified (gravity excluded).

\item \textit{Inflationary epoch and Electroweak Era}: (Time:$ 10^{-36}$ seconds to $10^{-12}$ seconds):
Temperature Range: $10^{27}$ K to $10^{15}$ K. Cosmic inflation stretched the fabric of space by approximately a factor of $10^{26}$ within a time span ranging from $10^{-36}$ to $10^{-32}$ seconds. The electroweak force (combining electromagnetism and weak nuclear force) separated from the strong force, and elementary particles gained mass through interactions with the Higgs field.

\item \textit{Quark-Gluon Plasma Era}: (Time: $10^{-12}$ seconds to $10^{-6}$ seconds):
Temperature Range: $10^{15}$ K to $10^{12}$ K. The temperature was still extremely high, causing quarks and gluons to roam in what is known as a quark-gluon plasma. As the universe expands and cools, this state transitioned to confinement, forming protons and neutrons.

\item \textit{Nucleosynthesis Era}: (Time: $10$ seconds to $10^3$ seconds):
Temperature Range: $10^{10}$ K to $10^9$ K. The universe had cooled enough for nuclear reactions to take place. These reactions, known as nucleosynthesis, created the primordial nuclei of hydrogen, helium, and trace amounts of lithium and beryllium. Most of the helium in the universe was formed during this phase.

\item \textit{Recombination Era}: (Time: $\approx 380,000$ years):
Temperature Range: $\approx 3000$ K. At this point, the temperature was low enough for protons and electrons to combine and form neutral hydrogen atoms through a process called \textit{recombination.} The positively charged protons captured negatively charged electrons to create stable neutral atoms. This transition led to a significant decrease in the scattering of photons by charged particles, making the universe transparent to light. As a result, the photons that were once tightly coupled to matter were released, filling space and creating the Cosmic Microwave Background (CMB) radiation.

\item\textit{Dark Ages}: (Time: $\approx 380,000$ years to $\approx 150$ million years):
Temperature Range: Cooling from $\approx 3000$ K to a few K.
 Following recombination and decoupling, the universe entered a transparent state. However, the process of hydrogen clouds collapsing to form stars and galaxies was extremely slow, leading to a lack of new sources of light. Consequently, the only remaining photons (electromagnetic radiation or \textit{light}) in the universe were those released during decoupling, which are now observable as the cosmic microwave background. Additionally, occasional 21 cm radio emissions emitted by hydrogen atoms contributed to the available light. Initially, the decoupled photons filled the cosmos with a brilliant pale orange glow, gradually redshifting to wavelengths that are not visible after approximately 3 million years. As a result, the universe ended up without visible light. During this era, matter started dominating over radiation.

\item \textit{Formation of the First Stars and Galaxies}: (Time: $\approx 150$ million years and onward):
Temperature Range: Cooling from a few K to thousands of K.
 The first stars and galaxies began to form, releasing energy and ionizing the surrounding gas.

\item \textit{Present-Day universe}: (Time: $\approx 13.8$ billion years):
Temperature Range: $\approx 2.73$ K (Temperature of the Cosmic Microwave Background Radiation).
 The universe continues to expand and cool. Its average temperature is now very low, and most of the thermal radiation is in the form of the CMB. Now the universe is dominated by the dark energy and causes acceleration in the expansion. 
\end{enumerate}

\section{Possible ways to Elucidate the Cosmic Acceleration:}
The discovery of the late-time accelerated expansion of the universe occurred in 1998 when two separate groups, the Supernova Cosmology Project \cite{SupernovaCosmologyProject:1998vns} and the High-Z Supernova Search Team \cite{SupernovaSearchTeam:1998fmf,SupernovaSearchTeam:1998bnz,Riess:1998dv}, independently observed supernovae with redshifts $z < 1$. Type Ia supernovae, known for their consistent intrinsic brightness, serve as reliable standard candles, easily distinguishable over varying distances.  These supernovae are thought to occur when a white dwarf star in a binary system accumulates enough mass from its companion, triggering a thermonuclear explosion. As the universe expands, the separation between the observer and celestial objects increases. Consequently, the emitted photons experience redshift. By analyzing the observed brightness of these objects and the redshift of the photons, researchers can gauge the rate of the expansion of the universe. \vspace{0.5em}

The brightness of a supernova is represented by its absolute magnitude. It can be utilized in calculating the cosmic luminosity distance ($d_L$). The connection between the apparent luminosity ($f$) and the intrinsic luminosity ($L$) is expressed as, $f=\frac{L}{4\pi d_L^2}$.  Measurements of the luminosity distance ($d_L$) for supernovae are recorded at various redshifts in the format of distance modulus ($\mu_B$). This modulus is defined as the disparity between the apparent magnitude ($m_B$) and the absolute magnitude ($M_B$) within the  wavelength of the blue line of the observed spectrum as, $\mu_B=m_B-M_B = 5\log_{10}\frac{d_L}{1 \text{Mpc}}+25.$ In terms of $H_0$ and $q_0$, i.e., the e present values of the Hubble parameter and the deceleration parameter respectively, we can write, $d_L=\frac{cz}{H_0}\left(1+\frac{z}{2}[1-q_0]+\mathcal{O}(z^2) \right)$. \vspace{0.5em}

Both supernova groups gauged the luminosity distances and witnessed the fading of supernovae. The determined luminosity distances surpassed their anticipated values, signifying a negative $q_0$. This affirmation supported the conclusion that light sources are moving away from each other at an accelerated pace. \vspace{0.5em}

This observed accelerated expansion of the universe is a perplexing phenomenon that cannot be fully accounted for by the known components of the universe. All known forms of matter and energy that we encounter in the universe, such as ordinary matter, radiation (including light), and the dark matter, contribute to slowing down the expansion of the universe. These fluids follow the strong energy condition, $\rho+3p>0$. 
Now if we combine the equations, \eqref{F1} and \eqref{F2}, we can write, 
\begin{equation}\label{rceqn}
\frac{\ddot{a}}{a}=-\frac{4\pi G}{3}(\rho+3p).
\end{equation}
From the definition of deceleration parameter \eqref{decel}, $q =-\frac{1}{aH^2}\frac{d^2a}{dt^2}$ and the equation \eqref{rceqn}, we see that for these fluids $q>0$, i.e., the decelerating cosmic expansion . However, rather surprisingly, the universe has undergone two separate periods of accelerated expansion: one marked by early exponential inflation and the other by late-time cosmic acceleration. In the interim, these two phases of rapid expansion were separated by a period characterized by a decelerated expansion. \vspace{0.5em}

 As a result, scientists have put forward various theoretical models to try to elucidate the acceleration in the expansion history in two distinct times, (i)inflation and (ii)late-time cosmic acceleration. Nevertheless, none of these models has achieved widespread acceptance as flawless or has been strongly supported by irrefutable observational evidence. Additionally, these models have not yet provided a direct and conclusive method for detecting the underlying cause of cosmic acceleration. \\

\subsection{Inflation:}
Inflation is a suggested paradigm in cosmology that proposes a rapid exponential expansion of the universe during its early epoch ($\sim 10^{-36} - 10^{-32} $ seconds). This model was developed to address several outstanding problems and observed phenomena in the standard Big Bang model. \vspace{0.5em}

The concept of inflation was first proposed in the early 1980s by Alan Guth \cite{Guth:1980zm} and later refined by other physicists, (see \cite{Barrow:1981pa,Linde:1981mu, Turner:1982ah,Albrecht:1982mp,Guth:1982ec,Starobinsky:1982ee,Kofman:1985aw}). The key idea behind inflation is that in the tiny fraction of a second after the Big Bang, the universe underwent an exponential expansion, increasing its size by an enormous factor. This rapid expansion allowed the universe to smooth out irregularities and achieve a high degree of homogeneity and isotropy, which explains why the cosmic microwave background radiation appears so uniform in all directions. (For introductions to inflation, see \cite{Ryden:1970vsj}). \vspace{0.5em}

Inflation addresses several important cosmological puzzles:

1. \textit{The Horizon Problem:} The universe appears to have a very uniform temperature on large scales, even regions that are too distant to have ever been in causal contact. Without inflation, there would not have been enough time for these regions to reach thermal equilibrium and have the same temperature. Inflation solves this by stretching these regions out of causal contact, so they can reach the same temperature before inflation ends. (See \cite{Plebanski:2006sd}, for a comprehensive discussion on to solve the horizon problem without invoking inflation. )

2. \textit{The Flatness Problem:} The universe appears to be spatially flat on cosmological scales. Inflation can flatten and stretch the geometry of space, making it very close to flat, $k=0$ in equation \eqref{hij}.

3. \textit{The Monopole Problem:} Some particle physics theories predict the existence of magnetic monopoles, which are highly magnetically charged particles. However, these monopoles are not observed in the universe in the abundance predicted. Inflation dilutes their numbers, making them incredibly rare and explaining their absence. \vspace{0.5em}
  
Inflationary theory suggests that the exponential expansion was driven by a scalar field, often called the \textit{inflaton field}. As the universe expanded and cooled, this field settled into its lowest energy state, ending the inflationary phase. The energy released during the decay of the inflaton field reheated the universe, initiating the subsequent hot  phase and the standard cosmological evolution that followed. \vspace{0.5em}

While inflation has been highly successful in explaining various observed features of the universe, it remains a theoretical framework.  Direct evidence for inflation is still a subject of intense research and observational efforts, such as studying the cosmic microwave background radiation and the large-scale structure of the universe. Confirming inflation would be a profound confirmation of our understanding of the early history of the universe and the processes that shaped its current state.

\subsection{Late-time Cosmic Acceleration:}

Late-time cosmic acceleration was discovered in the late 1990s from observational data of type IA supernova. In addition to using type-Ia supernovae \cite{SupernovaSearchTeam:2003cyd,SupernovaCosmologyProject:2011ycw,SDSS:2014iwm}, Baryon Acoustic Oscillations (BAO) \cite{SDSS:2005xqv,Eisenstein:2005sbt}, the WMAP \cite{Barris:2003dq}, Planck \cite{Planck:2015fie,Planck:2018vyg,Allen:2007ue,Hicken:2009dk} satellite missions, and the Dark Energy Survey (DES) \cite{DES:2018paw} data also provide substantial support for the existence of a seamless transition. This transition involves the universe transitioning from a prior phase of decelerated expansion to its current state of accelerated expansion, occurring at an intermediary redshift of approximately $z \approx 0.5$ \cite{SupernovaSearchTeam:2004lze,Riess:2006fw,Turner_2002,Cunha:2008ja,Cunha:2008mt,A.C.C.Guimaraes_2009,Lu:2011ue,Cattoen:2007sk}. \vspace{0.5em}

There are two potential explanations to find out a way to unravel the mystery. One of those is that gravity acts in a manner unlike from our current understanding, hence the theory of gravity needs to be modified. The other one is that the cosmos contains an exotic element with unconventional gravitational attributes, leading to an apparent negative pressure effect. These two main broad categories are discussed below.

\subsubsection{Modified Theories of Gravity:}\label{BDTheory}
  Modified gravity theory refers to a class of theoretical frameworks that propose modifications or extensions to Einstein's general theory of relativity (GR). The idea is to account for the cosmic acceleration without invoking any exotic fluid component, such as dark energy. Not only cosmic acceleration, but the theories also attempt to give insight into some other problems such as addressing the nature of dark matter, resolving singularities etc. Additionally, they aim to provide a semi-classical description of gravitational interactions, where quantum effects are taken into account through an effective action. There are many different proposals for modified gravity, each with its own set of mathematical equations and predictions \cite{Tsujikawa2010,Sotiriou:2008rp,DeFelice:2010aj,Maartens:2010ar,Clifton:2011jh,Joyce:2014kja,Koyama_2016}. Some of the well-known examples of modified gravity theories include: f(R) Gravity \cite{Capozziello:2002rd,Capozziello:2002rd,Carroll:2003wy,Carroll:2004de,Nojiri:2003ft,Nojiri:2006gh,Nojiri:2007jr,Nojiri:2007as,Vollick:2003aw,Mena:2005ta}, scalar-tensor theories \cite{Banerjee:1996iy,Amendola:1999qq,Uzan:1999ch,Chiba:1999wt,Amendola:1999er,Bartolo:1999sq,Perrotta:1999am,Riazuelo:2001mg,Banerjee:2000mj,Banerjee:2000gt,Sen:2000zk,Mota:2003tm,Mota:2003tc,Mukherjee:2019see,Bergmann:1968ve}, Scalar-Einstein-Gauss-Bonnet Gravity \cite{Nojiri:2005vv,Nojiri:2005jg,Nojiri:2006je}, Galileon Gravity \cite{Nicolis:2008in,Babichev:2012qs,Gao:2011qe}, TeVeS (Tensor-Vector-Scalar) theory \cite{Park:2000an,Cho:1987yc,Park:1997zf}, f(T) Gravity \cite{Cai_2016}, Horndeski Gravity \cite{Horndeski:1974wa,Barrow:2012ay,Koyama:2013paa}, Dilaton Gravity \cite{Fujii:1971vv,Sugimoto:1972zx}, Dvali-Gabadadze-Porrati (DGP) Model(braneworld scenario) \cite{Dvali:2000hr,Deffayet:2001pu}, Chameleon Gravity \cite{Khoury:2003rn,Brax:2004qh,Waterhouse:2006wv,Brax:2008hh} and some extended theories \cite{Bamba:2013aca,Hossain:2014xha}. Among these, f(R) gravity and scalar-tensor theories will be discussed here.
  
\textbf{1.}  \textbf{\textit{$f(R)$ Theories:}}
In the realm of gravitational theories, a straightforward modification to General Relativity (GR) is presented through $f(R)$ gravity. This modification centers on the spacetime action and its transformation. Rather than adhering to the conventional Ricci scalar $R$, $f(R)$ gravity introduces a more flexible approach, incorporating an analytical function, $f = f(R)$. As a result, the action governing $f(R)$ gravity can be succinctly represented as,
\begin{equation}\label{fr-action}
\mathcal{S}= \int{\sqrt{-g}\left(\frac{f(R)}{16\pi G}+\mathcal{L}_m\right)d^4x}.
\end{equation}

By taking variation of the action given in equation \eqref{fr-action} with respect to the metric tensor $g_{\mu\nu}$, we arrive at the field equations, which are expressed as,
\begin{align}
f^\prime R_{\mu\nu}-\frac{f}{2}g_{\mu\nu}-\left(\nabla_\mu\nabla_\nu-g_{\mu\nu}\square\right)f^\prime=T_{\mu\nu}.
\end{align}
Here superscript $\prime$ implies the derivative with respect to $R$ and the symbol $\square$ represents the d'Alembertian operator and given by, $\square=g^{\mu\nu}\nabla_\mu\nabla_\nu$.
In terms of Einstein tensor this equation can be written as,
\begin{align}\label{geff}
G_{ab}=\frac{1}{f^\prime(R)}\left(T_{\mu\nu}+T^{\mbox{eff}}_{\mu\nu}\right),
\end{align}
where the effective stress-energy tensor is $T^{\mbox{eff}}_{\mu\nu}=\left[\frac{f-Rf^\prime}{2}g_{\mu\nu} +\left(\nabla_\mu\nabla_\nu-g_{\mu\nu}\square\right)f^\prime\right]$. The source of this term is entirely geometric in nature.

The inclusion of $f^\prime$ in the denominator within Equation \eqref{geff} signifies the existence of a non-minimal coupling, which in turn renders the effective gravitational constant a variable in these theories.

These $f(R)$ gravity theories have the capability to elucidate phases of cosmic acceleration without the requirement for exotic or unconventional matter constituents. This is achievable through the careful selection of the function $f$. These $f(R)$ models, characterized by $f(R) \sim R^2$ have proven their capability in generating scenarios of early inflation. The models featuring $f(R) \sim 1/R^n$, where n is positive, are put forward to account for the  accelerated expansion of the universe during its later phases. A thought-provoking discourse of how this theory harmonizes with the established cosmological model to explicate cosmic acceleration can be encountered in the extensive review \cite{Capozziello:2009nq}. To know more detail about $f(R)$ theories of gravity, we refer to the articles \cite{Carroll:2003wy,Vollick:2003aw,Starobinsky:1980te,Kerner:1982yg,Duruisseau:1986ga,Capozziello:2003gx,Nojiri:2003ni,Das:2005bn,Bean:2006up,Pogosian:2007sw,Evans:2007ch,Mukherjee:2014fna,Amendola:2006kh,Amendola:2006we,DeFelice:2007ez,Capozziello:2009ka}

\textbf{2.} \textbf{\textit{Scalar-tensor theories:}}

  In gravitation and cosmology, scalar-tensor theories are rooted in the concept of a non-minimal coupling between the scalar field and the spacetime geometry. The origins of these theories can be traced back to the pioneering work of Jordan \cite{Jordan:1949zz} and were subsequently advanced by Brans and Dicke \cite{Brans:1961sx}. The generality of their findings was later augmented by the insights of Nordvedt \cite{Nordtvedt:1970uv} and Wagoner \cite{Wagoner:1970vr}. The action to represent a broad category of Non-Minimally Coupled Scalar-Tensor Theories (NMCSTT) is,
 \begin{equation}\label{NMCSTT-action}
 \mathcal{S}=\frac{1}{2}\int \sqrt{-g}\left[\left( f(\phi)R-\frac{\omega(\phi)}{\phi}\partial_\mu\phi\partial^\mu\phi\right)+\mathcal{L}_m\right]d^4x.
 \end{equation}

Within this framework, the scalar field $\phi$ is coupled to the Ricci scalar $R$ in a non-minimal manner, leading to a consequential adjustment in the gravitational coupling strength.
The field equations governing the NMCSTT encompassed within the action \eqref{NMCSTT-action} are derived through variations in both the metric $g^{\mu\nu}$ and the scalar field $\phi$. These equations can be expressed as follows:
\begin{equation}\label{fsce1}
f\left(R_{\mu\nu}-\frac{1}{2}g_{\mu\nu}R\right)+g_{\mu\nu}\square f-\nabla_\nu \nabla_\mu f-\frac{\omega}{\phi}\partial_\mu\phi \partial_\nu\phi+\frac{1}{2}g_{\mu\nu}\frac{\omega}{\phi}\partial_\mu\phi \partial^\mu\phi=T_{\mu\nu},
\end{equation}
and
\begin{equation}\label{fsce2}
 R \frac{df}{d\phi}+\frac{2\omega}{\phi}\square\phi+\left(\frac{1}{\phi}\frac{d\omega}{d\phi}-\frac{\omega}{\phi^2}\right)\partial_a\phi \partial^a\phi=0.
\end{equation}
Here the effective gravitational coupling is $G_{\mbox{eff}}=f(\phi)^{-1}$. As a consequence, it is presumed that $f(\phi)$ is positive to guarantee a positive coupling.

 Brans-Dicke (BD) theory serves as the quintessential example within this classification. Diverse scalar-tensor theories are born from the overarching action presented in equation \eqref{NMCSTT-action} each distinguished by their unique choices of the functions $f(\phi)$ and $\omega(\phi)$. For comprehensive discussions on scalar-tensor theories, one may follow the articles \cite{Fujii:2003pa,Faraoni:2004pi,Quiros:2019ktw}. Here we shall discuss about the Brans-Dicke theory.
 \begin{itemize}
 \item \textbf{Brans-Dicke theory}
 \end{itemize}
 
  Brans-Dicke theory  stands as one of the most widely discussed  theoretical framework in the field of theoretical physics and cosmology that extends Einstein's theory of general relativity. Proposed by Carl H. Brans and Robert H. Dicke in 1961 \cite{Brans:1961sx}, this theory introduces a scalar field, known as the Brans-Dicke field, alongside the familiar metric tensor used in general relativity. This addition was motivated by the desire to explore a broader range of gravitational theories that could accommodate variations in the strength of gravity and address certain shortcomings of general relativity.
 
Brans-Dicke action follows from equation \eqref{NMCSTT-action} by the choice $f(\phi)=\phi$ and $\omega=a$ constant (dimensionless). Therefore, now the action looks like, 
 \begin{equation}\label{BDgenaction}
 \mathcal{S}=\frac{1}{16\pi G_0}\int \sqrt{-g}\left[\left(\phi R-\frac{\omega}{\phi}\partial_\mu\phi\partial^\mu\phi\right)+\mathcal{L}_m\right]d^4x.
 \end{equation}

 Upon taking the conformal transformation \cite{Dicke:1961gz}, 
 \begin{equation}
 \tilde{g}_{\mu\nu}=\phi g_{\mu\nu},
 \end{equation}
 the action becomes, 
 \begin{equation}\label{BDaction}
 \bar{\mathcal{S}} =\frac{1}{16\pi G_0} \int \sqrt{-\bar{g}}\left[\phi_0\left(\bar{R}-\frac{2\omega+3}{2}\psi_{,\alpha}\psi_{,\beta}\bar{g}^{\alpha\beta}\right)+\bar{\mathcal{L}_M}\right]d^4x ,
\end{equation} 
where $\psi = \ln (\frac{\phi}{\phi_0})$. $\phi_0$ is a constant.

This version is popularly known as \textit{BD theory in Einstein frame}. Then the field equations assume a significantly simplified form, which can be expressed as follows:

\begin{align}\label{B1}
G_{\alpha\beta} = T_{\alpha\beta}+\frac{2\omega+3}{2}(\psi_{,\alpha}\psi_{,\beta}-\frac{1}{2}g_{\alpha\beta}\psi^{,\mu}\psi_{,\mu}).
\end{align}
This is written in the unit of $8\pi G_0=1$. The equation of motion for the scalar field $\psi$, derived by varying the action (\ref{BDaction}) with respect to $\psi$, results in the expression,
\begin{align}\label{B2}
\Box \psi =\frac{T}{2\omega+3},
\end{align} 
 where $T$ represents the trace of the energy-momentum tensor for the matter sector.
 
  The Brans-Dicke theory has been subjected to various experimental tests to determine its validity compared to General Relativity. These tests include measurements of the deflection of light by massive objects, the Nordtvedt effect, precision tests in the solar system, and cosmological observations. Observational constraints limit $\omega$ to a very large value \cite{Will:2018bme,Reasenberg:1979ey,Bertotti:2003rm}. The Brans-Dicke (BD) theory was initially suggested to converge into the General Theory of Relativity (GTR) in the limit as ω approaches infinity, a notion initially posited in \cite{Weinberg:1972kfs}. Subsequently, it has been elucidated in the literature that the inclusion of the trace of the energy-momentum tensor of matter distribution imposes certain constraints on this proposition, as expounded in studies \cite{Banerjee:1996iy,Faraoni:1999yp}.  The static spherically symmetric vacuum solutions of this theory can be found in articles \cite{Brans:1962zz}. To know more aspects of these solutions, readers may refer to \cite{Bhadra:2001my,Bhadra:2005mc,Campanelli:1993sm,Ruffini:1971bza,Faraoni:2016ozb,Bhadra:2001fx,Agnese:1995kd,Nandi:1997mx,Anchordoqui:1997yb,PhysRevD.65.084022,Bhadra:2004hu,Nandi:2004ha,Bronnikov:1998hm,Nandi:2000gt}. Numerous spatially homogeneous and isotropic cosmological solutions within the framework of Brans-Dicke (BD) theory can be found in the literature. A few such examples are listed in Table \ref{BDCosSol}.
  \begin{table}[t!]
\begin{center}
\begin{tabular}{  m{6cm}  m{3.5cm} m{3.5cm}  } 
  \hline
  \textbf{Solution}& \textbf{Constraint on ${\omega}$} & \textbf{References} \\ 
  \hline
   \vspace{1em} O'Hanlon and Tupper solution & \vspace{1em}$\omega>-\frac{3}{2}$, $\omega\neq 0,-\frac{4}{3}$ & \vspace{1em} \cite{OHanlon:1972ysn} \\ 
 \vspace{1em} BD dust solution & \vspace{1em} $\omega\neq-\frac{4}{3}$ &\vspace{1em}  \cite{Brans:1961sx} \\ 
  \vspace{1em} Nariai solution & \vspace{1em} ... &\vspace{1em} \cite{10.1143/PTP.40.49,gurevich1973problem} \\
 \vspace{1em}  Other perfect fluid solutions & \vspace{1em} ... & \vspace{1em} \cite{lorenz1984exact,Morganstern:1971dz}\\
  \vspace{1em} Only matter source: cosmological constant
   & \vspace{1em} ... & \vspace{1em}  \cite{Mathiazhagan:1984vi,La:1989za,Romero:1992xx}\\
\vspace{1em} FRW universes with closed and open
spatial sections & \vspace{1em} ... & \vspace{1em} \cite{lorenz1984exact,Morganstern:1971dz,Liddle:1993fq,PhysRevD.47.5329,Dehnen:1971zz,Levin:1993wr,Mimoso:1994wn} \\
  \hline
\end{tabular}
\end{center}
\caption{Exact solutions of BD theories in cosmology that have been studied
in the existing literature.}\label{BDCosSol}
\end{table}
BD theory provides solutions to a variety of cosmological quandaries.  It offers solutions to issues like the graceful exit problem during inflationary phases \cite{Mathiazhagan:1984vi,La:1989za} and the late-stage accelerated expansion of the universe, all without requiring the introduction of dark energy \cite{Banerjee:2000mj}. The monograph \cite{Faraoni:2004pi} by Faraoni provides  an extensive exploration of exact cosmological solutions within BD theory.

\subsubsection{Dark Energy Models}\label{Darkenergy}

Dark energy models seek to elucidate the cosmic acceleration within the framework of General Relativity (GR) \cite{Sahni:1999gb,Sahni:2004ai,Padmanabhan:2002ji,Padmanabhan:2005ur,Copeland:2006wr,Ruiz-Lapuente:2007ihz,Durrer:2007re}. Equation \eqref{rceqn} illustrates that an accelerated expansion can arise when a constituent within the energy sector exerts substantial negative pressure. This exotic component, commonly referred to as dark energy, instigates acceleration due to its negative pressure, distinct from the fluid pressure attributed to particle motion. To facilitate analysis, we introduce dimensionless representations of energy densities by scaling them with the critical density ($\rho_{c}$), defined in equation \eqref{critden} and density parameter ($\Omega_i$), defined in equation, \eqref{denpara}. Subscript $'i'$ implies either dust ($i = m$), radiation ($i = r$) or dark energy ($i = d$).
Now, the Hubble parameter $H$, scaled by its present value $H_0$, can be written in terms of density parameters as (for spatially flat models),
\begin{equation}
h^2\equiv \frac{H^2}{H^2_0}=\Omega_m+\Omega_r+\Omega_d.
\end{equation}
 Dust matter, a significant component in the energy budget, is pressureless ($p_m = 0$), thus yielding $w_m \equiv \frac{p_m}{\rho_m} = 0$ as its equation of state parameter. The dark energy equation of state parameter ($w_d$) is defined as, $w_d \equiv p_d/\rho_d$.

We denote the density parameters at the present epoch, as $\Omega_{m0}$, $\Omega_{r0}$, $\Omega_{d0}$. For non-pressured matter, $\Omega_m =\Omega_{m0}(1 + z)^3$; for radiation, $\Omega_r =\Omega_{r0}(1 + z)^4$, using the standard scaling of the scale factor with $a_0 = 1$. 
The composite model's effective equation of state (EoS) is expressed as:
\begin{equation}
w_{eff} = (p_r + p_d) / (\rho_m + \rho_r + \rho_d).
\end{equation}

To achieve an accelerated universe, $w_{eff}$ must be less than $-1/3$, as indicated by equation \eqref{rceqn}.

Given the negligible contribution of radiation in the late-time universe compared to other components, we can estimate q for a spatially flat universe composed of pressureless matter and dark energy as follows,
\begin{equation}
q \approx \frac{1}{2}(1 + 3w_d \Omega_d).
\end{equation}

Hence, the condition for late-time cosmic acceleration is $w_d < -\frac{1}{3}\Omega^{-1}_d$. Recent cosmological observations  data  suggests $\Omega_{d0} =\Omega_d|_{z=0}\approx 0.7$. Consequently, the value of the dark energy equation of state (DE EoS) parameter at the present epoch should be $w_{d0} = w_d|_{z=0} \lesssim -0.5$.

Various theoretical frameworks exist for explaining dark energy, yet none have gained universal acceptance, each exhibiting its own drawbacks and limitations. The most straightforward model of dark energy is the cosmological constant $\Lambda$, characterized by an equation of state $w = -1$.

Below, we delve into an array of diverse models that seek to understand the nature of the exotic dark energy. These models offer distinct perspectives and approaches, each contributing to our ongoing quest to unravel the mysteries of cosmic acceleration.

\begin{itemize}

\item \textbf{\textit{Cosmological constant}}
\end{itemize}

 In 1917, Einstein introduced the cosmological constant, denoted by $\Lambda$, into his equations of general relativity. Einstein added $\Lambda$ to his equations to achieve a static solution for the universe. At the time, the prevailing view was that the universe was static and unchanging. However, when Edwin Hubble discovered the expansion of the universe in the late 1920s, Einstein famously referred to the inclusion of the cosmological constant as his \textit{greatest blunder} \cite{Einstein:1917ce}. He removed the cosmological constant from his equations because it was no longer needed to describe a static universe.

The cosmological constant made a comeback as vacuum energy in the context of inflation \cite{Guth:1980zm}. In the late 20th century, the cosmological constant again resurged in cosmology to explain the observed accelerated expansion of the universe \cite{SupernovaSearchTeam:1998fmf,SupernovaCosmologyProject:1998vns,SupernovaSearchTeam:1998bnz}. This acceleration was discovered in the late 1990s through observations of distant supernovae and is attributed to a mysterious dark energy that permeates the universe. One may refer to \cite{Padmanabhan:2002ji,Sahni:1999gb,Carroll:2000fy,Peebles:2002gy,Frieman:2008sn,Amendola:2010ub,Mehrabi:2018dru}, to learn more about $\Lambda$ as a representative of this dark energy.

In a universe where the cosmological constant dominates, the solution corresponds to an exponential rate of expansion, where the scale factor $a(t)$ grows exponentially with time (see reference \cite{Liddle:2000cg}), 
\begin{equation}\label{alambda}
a(t)\propto \exp\left(\sqrt{\frac{\Lambda}{3}}t\right) = \exp(Ht).
\end{equation}

The $\Lambda$CDM model is the prevailing cosmological framework, consisting of cold dark matter (CDM) with no pressure and the cosmological constant serving as dark energy. It is commonly referred to as the standard cosmological model. The energy density tied to $\Lambda$ is defined as,
\begin{equation}
\rho_\Lambda=\frac{\Lambda}{8\pi G}.
\end{equation} 

To maintain $\rho_\Lambda$ as a constant by definition, it necessitates that the pressure associated with $\Lambda$, must be, $p_\Lambda=-\rho_\Lambda$. Therefore, the cosmological constant exhibits an effective negative pressure, with the equation of state parameter $w=-1$. 

In a universe exhibiting matter dominance, the scale factor evolves according to the relationship $a(t)\propto t^{2/3}$. As the universe enters a phase where cosmological constant holds sway,  with EoS parameter $w=-1$, the scale factor approaches an asymptotic form expressed in equation \eqref{alambda}. When we consider a universe with a spatially flat geometry, accommodating both matter and cosmological constant, the solution seamlessly integrates the properties of these two components across different cosmic eras and as given by \cite{Frieman:2008sn}, 
\begin{align}
 a(t)&=\left(\frac{\Omega_m}{\Omega_{vac}}\right)^{1/3} \left(\sinh[3\sqrt{\Omega_{vac}}H_0 t/2]\right)^{2/3}\nonumber \\&=\left(\frac{\Omega_m}{\Omega_{vac}}\right)^{1/3}\sinh^{2/3}(t/t_0).
\end{align}
This solution adeptly mirrors the evolving characteristics of matter and vacuum energy as we traverse from earlier to later cosmic eras. It essentially embodies the principles of the $\Lambda$CDM model. Various facets of the cosmological constant model have been thoroughly examined and elaborated upon by Carroll \cite{Carroll:2000fy} and extensively discussed by Padmanabhan \cite{Padmanabhan:2002ji,Padmanabhan:2004av}.

The cosmological constant model stands as the favoured choice in light of compelling observational support. Nevertheless, it is essential to acknowledge the inherent complexities of this model. Notably, the only plausible candidate to account for the cosmological constant is the energy density inherent in the vacuum. The notable hurdle lies in the staggering disparity between the energy density's observed value ($\rho^{\mbox{obs}}_\Lambda$) and its theoretically calculated counterpart ($\rho^{\mbox{theory}}_\Lambda$). This incongruity is striking, with the ratio $\frac{\rho^{\mbox{obs}}_\Lambda}{\rho^{\mbox{theory}}_\Lambda}$ hovering around the minuscule figure of $10^{-120}$. This enigma is fundamentally recognized as the \textit{fine-tuning} dilemma inherent to the cosmological constant model. Recent efforts have been directed towards achieving a reduced cosmological constant in the current cosmic era. A concise overview of various models featuring a cosmological constant, diminishing with time, can be found in table \ref{lambda}, which is adapted from \cite{Overduin:1998zv} (see also \cite{Sahni:1999gb}).

\begin{table}[t!]
\begin{center}
\begin{tabular}{  m{6cm}  m{4cm}  } 
  \hline
  \vspace{1em}\textbf{Evolutionary relation for $\Lambda(t)$} & \vspace{1em}\textbf{References} \\ 
  \hline
\vspace{1em} $\Lambda \propto t^{-2}$   &\vspace{1em}  \cite{endo1977cosmological,Canuto:1977zz,Bertolami:1986bg,berman1990brans,Beesham:1994ni,Lopez:1995eb,Overduin:1998zv} \\ 
\vspace{1em} $\Lambda \propto t^{-\alpha}$  &\vspace{1em} \cite{kalligas1992flat,Kalligas:1995qh,Beesham:1993bh}  \\
\vspace{1em} $\Lambda \propto A+B\exp(-\alpha t)$  &\vspace{1em} \cite{Beesham:1993bh,Spindel:1993bz}  \\
\vspace{1em} $\Lambda \propto a^{-2}$  &\vspace{1em} \cite{Lopez:1995eb,OZER1986363,OZER1987776,Abdel-Rahman:1992msa,Chen:1990jw,Gott227,Abdussattar:1997aa}  \\
\vspace{1em} $\Lambda \propto a^{-\alpha} $  &\vspace{1em}  \cite{Olson:1987gy,Pavon:1991uc,Sahni:1991ks,Maia:1994vz,Matyjasek:1994vp,Silveira:1994yq,Silveira:1997fp,John:1997ery,Overduin:1998zv,Turner:1997npq,Wang:1998gt,Caldwell:1997mh,Caldwell:1997ii,hoyle1997hubble,Hu:1998tk} \\
\vspace{1em} $\Lambda \propto \exp(-\alpha a)$  &\vspace{1em} \cite{Rajeev:1983pr}  \\
\vspace{1em} $\Lambda \propto T^\alpha$  &\vspace{1em} \cite{Canuto:1977zz,Kazanas:1980tx}  \\
\vspace{1em} $\Lambda \propto H^2$  &\vspace{1em} \cite{Copeland:1997et,Ferreira:1997au,Lima:1994ni,Wetterich:1994bg,Wetterich:1987fm}  \\
\vspace{1em} $\Lambda \propto H^2+Aa^{-\alpha}$  &\vspace{1em} \cite{Waga:1992hj,Salim:1992mx,Carvalho:1991ut,Arbab:1994db}  \\
\vspace{1em} $\Lambda \propto f(H)$  &\vspace{1em}  \cite{Lima:1994gi,Lima:1995ea} \\
\vspace{1em} $\Lambda \propto g(\Lambda,H)$  &\vspace{1em} \cite{Hiscock:1986ec,Reuter:1986wm}  \\
  \hline
\end{tabular}
\end{center}
\caption{List of some phenomenological $\Lambda$ models.}\label{lambda}
\end{table}

\begin{itemize}

\item \textbf{\textit{Constant dark energy equation of state model ($w_{d}\neq -1$)}}
\end{itemize}

In the pursuit of understanding dark energy through phenomenological analysis, cosmologists often opt for a constant value when characterizing the dark energy Equation of State (EoS). Importantly, this constant is not limited to the value of $-1$, providing a degree of flexibility in the modeling process. Dark energy models characterized by this constant EoS are referred to as \textit{Quiessence models} \cite{Sahni:2006pa}.

To illustrate, consider a scenario where the universe consists of Cold Dark Matter (CDM) and Dark Energy (DE), with the latter being defined by a constant EoS, $w_d$. This framework is commonly referred to as the $w$CDM model. Its purpose is to scrutinize observational data for any potential deviations from the established $\Lambda$CDM model. Within the context of the $w$CDM model, it is noteworthy that the dark energy density is no longer fixed but instead exhibits an evolution with changing redshift. This dynamic characteristic allows for a more comprehensive examination of the behaviour of dark energy over cosmic history.

\begin{itemize}

\item \textbf{\textit{Variable dark energy equation of state model}}
\end{itemize}

To tackle the cosmic coincidence problem, i.e.,  the mystery of why the energy densities of dark matter and dark energy in the universe are of the same order of magnitude in the present cosmic epoch and gain insights into the evolving nature of dark energy, researchers have turned their attention to models that exhibit dynamic behaviour over cosmic history. These models operate on the assumption that the equation of state for dark energy ($w_d$) has undergone temporal changes throughout the evolution of the universe. This paradigm shift has given rise to a multitude of dynamical dark energy (DDE) models, each characterized by a time-dependent EoS parameter. Among these models, the Chevallier-Polarski-Linder (CPL) model,  \cite{Chevallier:2000qy,Linder:2002et}, has gained significant prominence. It is defined by the functional form,
\begin{equation}
w^{CPL}(z) = w_0 + w_a \frac{z}{1 + z},
\end{equation} 
 where $w_0$ and $w_a$ are real numbers. This model is widely regarded as the most popular DDE model.  The conventional approach often involves parameterizing our limited understanding of how dark energy behaves. This methodology found extensive application within the Dark Energy Task Force \cite{Albrecht:2006um}. It has served as a practical benchmark for evaluating and contrasting the performance of various techniques aimed at investigating dark energy dynamics (see, for instance, \cite{Virey:2006em}). The simplicity of the CPL parameterization belies its rich characteristics, as thoroughly explored by Linder \cite{Linder:2007wa}. Of particular note are the two parameters it introduces: $w_0$, serving as a representation of the equation of state's present condition, and $w_a$, which encapsulates its broader temporal dynamics. 

An interesting proposition, detailed in references \cite{Linder:2002et,Linder:2007wa}, posits that the optimal way to characterize $w_a$ in relation to the derivative of $w$ is through the following expression: 
\begin{equation}
w_a=-2w^\prime|_{z=1}.
\end{equation}

  \begin{table}
\begin{center}
\begin{tabular}[t!]{  m{4cm}  m{6.5cm} m{3cm}  } 
  \hline
  \textbf{Model}& \textbf{$w(z)$} & \textbf{References} \\ 
  \hline
  \vspace{1em} Jassal-Bagla-Padmanabhan (JBP) & \vspace{1em}$w(z) = w_0 + w_1 \frac{z}{(1 + z)^2}$ & \vspace{1em}\cite{Jassal:2004ej,Jassal:2006gf} \\ 
 \vspace{1em} Barboza-Alcaniz parametrization & \vspace{1em} $w(z) = w_0 + w_1 \frac{z(1+z)}{1 + z^2}$ &\vspace{1em} \cite{Barboza:2008rh}  \\ 
  \vspace{1em} Wetterich parametrization & \vspace{1em} $w(z) = \frac{w_0}{1+w_2\ln(1+z)}$ &\vspace{1em} \cite{Wetterich:2004pv}  \\
 \vspace{1em}  Ma-Zhang parametrization & \vspace{1em} $w(z) = w_0 + w_1\left(\frac{\ln(2+z)}{1+z}-\ln 2\right)$ & \vspace{1em} \cite{Ma:2011nc} \\
  \hline
\end{tabular}
\end{center}
\caption{Various dark energy models with evolving equations of state, $w(z)$, investigated in the literature other than the CPL parametrization.}\label{wz}
\end{table}

 In this equation, $w^\prime$ represents the derivative of $w$, defined as $w^\prime \equiv\frac{dw}{d\ln a}$, with "$a$" denoting the scale factor. While it may not encompass the full spectrum of potential dynamics \cite{Linder:2007wa,Wang:2003gz,Wang:2004py,Wang:2004ru}, the CPL parameterization appears to strike a favorable balance for conducting a model-independent analysis. Other models with EoS dependent on $z$ have been listed in table \ref{wz}.

\begin{itemize}

\item \textbf{\textit{Quintessence scalar field model}}
\end{itemize}

The quintessence scalar field model is a captivating and influential theoretical framework in the field of cosmology. The concept of a quintessence scalar field was initially presented by Ratra and Peebles \cite{Ratra:1987rm} and independently by Wetterich \cite{Wetterich:1987fm} in order to facilitate the inflationary paradigm. In trying to grasp how the universe works on a big scale, the quintessence scalar field model is different from the usual cosmological constant model. It gives a dynamic and evolving explanation for what is causing the universe to expand. At its core, the quintessence scalar field model introduces a new dynamic component to the cosmic energy budget in the form of a scalar field, often denoted as $\Phi$, that permeates the universe. This scalar field possesses unique properties, including a potential energy function $V(\Phi)$ and negative pressure, allowing it to act as a driving force for the accelerated expansion of the cosmos. $\Phi$ slowly rolls down to its potential $V(\Phi)$, which leads to the dominant potential term over the kinetic term. Unlike the cosmological constant  associated with the $\Lambda$CDM model, which represents a static form of dark energy, quintessence offers a dynamic alternative that evolves over time.
The relevant action for quintessence field $\Phi$ is,
\begin{equation} \label{Qaction1}
S^{field} = \int \mathrm{d}^4 x \sqrt{-g}\Big[ \frac{R}{16\pi G}-\frac{1}{2} g^{\mu\nu}\mathrm{\partial} _\mu \Phi \mathrm{\partial} _\nu \Phi-V(\Phi)\Big].
\end{equation} 
Stress-energy tensor of the quintessence field is given by the equation,
\begin{equation}\label{em-q1}
T_{\mu\nu}^{\Phi} = \Big[\mathrm{\partial} _\mu \Phi \mathrm{\partial} _\nu \Phi-\frac{1}{2}g_{\mu\nu}g^{\alpha\beta} \mathrm{\partial} _\alpha \Phi \mathrm{\partial} _\beta \Phi - 2g_{\mu\nu} V(\Phi)\Big].
\end{equation} 
From the above equation, we can write the quintessence field energy density ($\rho_\Phi$) and pressure ($p_\Phi$) and respectively given by, 

\begin{equation}\label{q-den}
\rho_\Phi \equiv T^{0(q)}_0= \frac{1}{2}(\dot{\Phi})^2+V(\Phi),
\end{equation} 
and
\begin{equation}\label{q-press}
p_\Phi \equiv T^{1(q)}_1= T^{2(q)}_2=T^{3(q)}_3=\frac{1}{2}(\dot{\Phi})^2-V(\Phi).
\end{equation}

 The Klein-Gordon(KG) equation for $\Phi$ is,
\begin{equation}\label{kg-phi}
\ddot{\Phi}+3H\dot{\Phi}+V^\prime(\Phi)=0.
\end{equation}

The EoS parameter is therefore, 
\begin{equation}
w_\Phi=\frac{p_\Phi}{\rho_\Phi}=\frac{\dot{\Phi}^2-2V(\Phi)}{\dot{\Phi}^2+2V(\Phi)}.
\end{equation}

Therefore, the range of evolution of $w_\Phi$ is, $-1<w_\Phi<1$. Quintessence models are categorized into three distinct classes based on the characteristics of the potential energy function $V(\Phi)$. This classification is determined by the specific nature of the potential of the scalar field.

\begin{itemize}
\item $V(\Phi)<<\dot{\Phi}^2, w_\Phi \approx 1 \implies \rho_\Phi \sim a^{-6}$, resembling stiff matter, and it does not play a role in contributing to dark energy.
\item $V(\Phi)>>\dot{\Phi}^2 , w_\Phi \approx -1 \implies \rho_\Phi \sim  $ constant, resembling cosmological constant.
\item $-1< w_\Phi <1 \implies \rho_\Phi \sim a^{-m} $, leads to accelerated cosmic expansion for the range $0<m<2$ \cite{Copeland:2006wr}.
\end{itemize}
 
\begin{table}[t]
\begin{center}
\begin{tabular}{  m{6cm}  m{4cm}  } 
  \hline
  \vspace{1em}\textbf{$V(\Phi)$} & \vspace{1em}\textbf{References} \\ 
  \hline
\vspace{1em} $V_0e^{-\lambda\Phi}$   &\vspace{1em}  \cite{Ratra:1987rm,Wetterich:1987fm} \\ 
\vspace{1em} $V_0\Phi^{-\alpha}, \alpha>0$   &\vspace{1em}  \cite{Ratra:1987rm} \\ 

\vspace{1em} $V_0\left[\sinh^{\alpha\sqrt{k_0}\Delta_\Phi}\right]^\beta$ & \vspace{1em}  \cite{Urena-Lopez:2000ewq} \\
\vspace{1em} $V_0\left(e^{\alpha\Phi}+e^{\beta\Phi}\right)$   &\vspace{1em}  \cite{Sen:2001xu,Barreiro:1999zs} \\
\vspace{1em} $M^{4+\alpha}\Phi^{-\alpha}$   &\vspace{1em}  \cite{Zlatev:1998tr} \\
\vspace{1em} $V_0\left(e^{M_p/\Phi}-1\right)$   &\vspace{1em}  \cite{Zlatev:1998tr} \\
\vspace{1em} $V_0\left[\cosh(\lambda\Phi/M_P)-1\right]^p$   &\vspace{1em}  \cite{Sahni:1999qe} \\
\vspace{1em} $V_0\left[1+\cosh\left(\Phi/f\right)\right]$   &\vspace{1em}  \cite{Kim:1998kx,Frieman:1995pm} \\
\vspace{1em} $V_0 e^{\lambda\phi^2}/\Phi^\alpha, \alpha>0$   &\vspace{1em}  \cite{Brax:1999yv,Brax:1999gp} \\
\vspace{1em} $V_0e^{\lambda\Phi\left[(\Phi-B)^\alpha+A\right]}$   &\vspace{1em}  \cite{Albrecht:1999rm} \\
\vspace{1em} $V_0e^{-\lambda\Phi}\left(1+A\sin v\Phi\right)$   &\vspace{1em}  \cite{Dodelson:2000jtt} \\
\vspace{1em} $\frac{9A^2}{2}\cot^2\frac{(\sqrt{3}\Phi)}{4}+3A^2$ & \cite{Banerjee:2005ef}\\
  \hline
\end{tabular}
\end{center}
\caption{Various quintessence scalar field potentials $V(\Phi)$ explored in prior research.}\label{qfieldlist}
\end{table}

The classification of quintessence fields into thawing and freezing models is a significant concept in the study of dark energy and its role in the evolution of the universe. These models are categorized based on the behaviour of the effective EoS parameter, $w$ of the quintessence field as the universe expands over time \cite{Scherrer:2007pu}.
\begin{itemize}
\item \textbf{Thawing Models:}
\end{itemize}

Thawing models are characterized by a quintessence field with an effective EoS parameter, $w$ that starts out as almost a constant close to $-1$ (similar to the cosmological constant), which corresponds to a vacuum energy with $w = -1$. Over time, the equation of state parameter \textit{thaws} or gradually evolves into a more dynamic, time-dependent value that deviates from $-1$. Thawing models are particularly interesting because they represent quintessence fields that initially behave like a cosmological constant but then change their behaviour, which can have implications for the evolution of the universe.
\begin{itemize}
\item \textbf{Freezing Models:}
\end{itemize}

In contrast, freezing models are characterized by a quintessence field for which the effective EoS parameter, $w$ settles down to a constant value close to $-1$, but this happens relatively late in the evolution of the universe.
Initially, the quintessence field may have a dynamic behaviour, but it eventually reaches a stage where its equation of state parameter becomes nearly indistinguishable from that of a cosmological constant.
One important subclass of freezing models is known as \textbf{\textit{trackers}}, which are of particular interest \cite{Zlatev:1998tr,Steinhardt:1999nw,Johri:2000yx}. 

In tracker models, the energy density of the scalar field evolves almost parallel to the energy density of dark matter for most of cosmic history without dominating  dark matter. However, it eventually freezes to a value greater than the dark matter density at a later stage. Thakur, Nautiyal, Sen, and Seshadri \cite{Thakur:2012rp} have undertaken a comparison between a thawing model and freezing models that exhibit tracking behaviour. Their investigation revolves around the compatibility of these models with empirical data and offers valuable insights into their observational viability.

It is worth noting that not all quintessence fields neatly fit into either of these two categories, and there can be variations and more complex behaviours. The classification into thawing and freezing models is primarily a way to categorize and study the different possible behaviours of quintessence fields in the context of cosmic expansion and their ability to address issues such as the coincidence problem \cite{Wang:1999fa}, which relates to why dark energy and dark matter densities are of similar magnitude in the present universe. Both thawing and freezing models have been extensively studied in cosmology to understand their implications for the evolution of the universe.

Table \ref{qfieldlist} provides an exhaustive compilation of the various scalar field potentials that have been explored within the context of quintessence models. To explore further into the vast body of research about late-time cosmic acceleration, which includes various versions of quintessence potentials, one may refer to \cite{Sen:2001xu,Zlatev:1998tr,Zlatev:1998yg,Steinhardt:1999nw,Wang:1999fa,Urena-Lopez:2000ewq,Sahlen:2006dn,Scherrer:2007pu,Scherrer:2008be,Gupta:2014uea,Dalmazi:2006ws,Chiba:2012cb,Pantazis:2016nky,Roy:2013wqa,Roy:2014yta,Sahni:2006pa,Carvalho:2006fy}.

\begin{itemize}

\item \textbf{\textit{Phantom field model}}
\end{itemize}
The idea of the phantom field in the realm of dark energy was initially introduced by Caldwell \cite{Caldwell:1999ew}. What sets the phantom field apart from the quintessence field is its distinctive feature of possessing negative kinetic energy($X$). Therefore, the Lagrangian density for phantom field $\phi$ is, 
$\mathcal{L}_p=-X-V(\phi)=-\partial_\mu\phi\partial^\mu\phi/2-V(\phi)$ (with metric signature $+ - - -$). The corresponding action governing the behaviour of the phantom field can be expressed as follows,
\begin{equation}
\mathcal{S}=\int d^4x\sqrt{-g}[-X-V(\phi)].
\end{equation}
The phantom field energy density and pressure are respectively given as,
\begin{equation}
\rho_p=-\frac{\dot{\phi}^2}{2}+V(\phi),
\end{equation}
and
\begin{equation}
p_p=-\frac{\dot{\phi}^2}{2}-V(\phi).
\end{equation}
The equations of motion are,
\begin{equation}
\ddot{\phi}+3H\dot{\phi}=+V_{,\phi}.
\end{equation}
The equation of state parameter of dark energy for the phantom field is,
\begin{equation}
w_{de}=\frac{p_p}{\rho_p}=\frac{\dot{\phi}^2+2V(\phi)}{\dot{\phi}^2-2V(\phi)}.
\end{equation}

For $w_{de}$ to be less than $-1$, $V(\phi)>>\dot{\phi}^2$. A phantom field, propelled by its negative kinetic energy  will tend to run up potential energy.
The consequence is an extremely swift expansion of the universe, reaching infinite extent within a finite time. This phenomenon is termed the \textit{Big Rip}. This is characterized by the  infinite growth of both the volume and the expansion rate.

A potential for the phantom field featuring a maximum value(for example $V(\phi)=V_0\left[\cosh \left(\frac{\alpha\phi}{m_{pl}}\right)\right]^{-1}$, where $\alpha$ is a constant) has the capability to avert the occurrence of the Big Rip. For $\dot{\phi}=0$,  the field eventually reaches its maximum position, following a damped oscillatory phase, causing the equation of state parameter, $w_{de}$, to attain a value of $-1$. Consequently, this behaviour can reinstate the scenario of the cosmological constant. \textit{Quintom models} is referred to the scalar field models, wherein the behaviour of the equation of state (EoS) parameter mirrors that of the phantom field \cite{Feng:2004ad,Cai:2006dm,Cai:2007gs,Cai:2008gk}.

\begin{itemize}

\item \textbf{\textit{Chaplygin Gas model}}
\end{itemize}

The Chaplygin gas model, named after the Russian mathematician Sergei Alekseevich Chaplygin, is a theoretical model that was originally introduced by Chaplygin in 1904 in aerodynamics to describe the lift force on an object moving through a gas.

 This model was reintroduced in cosmology by Kamenshchik, Moschella and Pasquir  \cite{Kamenshchik:2001cp} in the context of cosmic acceleration. The Chaplygin gas model is characterized by an equation of state given by:
\begin{equation}
p=-\frac{A}{\rho},
\end{equation}
where $A$ is a positive constant and $p$ and $\rho$ are respectively pressure and energy density in a comoving  frame with $\rho>0$.  This equation  

The  Chaplygin gas model provides a smooth transition between different phases of cosmic evolution. Specifically,  it can smoothly interpolate between:
\begin{itemize}
\item A dust dominated phase where $\rho \sim \sqrt{B}a^{-3}$.
\item A de Sitter phase, where the pressure $p\sim -\rho$.
\item An intermediate regime with the equation of state for stiff matter the, where $p=\rho$. 
\end{itemize}
In this model, once a universe undergoing expansion enters a phase of acceleration, it is incapable of reverting to a state of deceleration.

 This model has been generalized by Bento, Bertolami and Sen \cite{Bento:2002ps}. The authors proposed that the change in the characteristics of the elusive energy density could be governed by alterations in the equation of state of the underlying fluid, rather than relying on adjustments to the potential. This approach offers a means to circumvent the intricate fine-tuning problem. This is achieved by introducing a more flexible equation of state that allows for a wider range of behaviours. The generalized equation of state is given as:
\begin{equation}
p=-\left(\frac{1}{\rho}\right)^\alpha,
\end{equation}
where $\alpha$ is the parameter with $0<\alpha<1$, that determines the behaviour of the gas. The energy density $\rho$ evolves with the scale factor $a$ of the universe as,
\begin{equation}
\rho=\left(A+\frac{B}{a^{3(1+\alpha)}}\right)^{\frac{1}{1+\alpha}},
\end{equation}
where $B$ is the integration constant. In this scenario, instead of stiff matter, the intermediate phase is soft matter with equation of state, $p=\alpha\rho$ ($\alpha\neq 1$).

This model has been explored in cosmology as a way to unify dark matter and dark energy within a single framework and has connections to brane theory and supersymmetry in theoretical physics \cite{Jackiw:2000mm}. While the Chaplygin gas models have been invalidated by data related to temperature fluctuations in the cosmic microwave background \cite{LAmendola_2003,Bento:2002yx}, there is a limited range of parameter values, specifically $0 \leq \alpha\leq 0.2$, within which the generalized Chaplygin gas models are still considered plausible \cite{LAmendola_2003}.

\begin{itemize}

\item \textbf{\textit{K-essence model}}
\end{itemize}
The K-essence model is a theoretical framework in cosmology  that introduces a scalar field with a non-standard kinetic term to explain the dynamics of dark energy and, in some cases, an early inflation as well. It offers an alternative to more traditional scalar field models, such as quintessence, where the potential energy of the scalar field plays a central role. In the K-essence model, the kinetic energy of the scalar field dominates, hence the name \textit{K-essence} (kinetic essence). At its inception, it was introduced in the context of inflation \cite{Armendariz-Picon:1999hyi,Garriga:1999vw}. Later, Chiba \textit{et al.} introduced it when considering late-time acceleration \cite{Chiba:1999ka}. The generalized k-essence for dark energy was proposed by  Armendariz-Picon \textit{et al.} \cite{Armendariz-Picon:2000nqq,Armendariz-Picon:2000ulo}. In this model, the Lagrangian density is written  in the form of pressure as $P(\phi,X)$, where $X=-\frac{1}{2}(\nabla\phi)^2$ is the kinetic energy for the scalar field $\phi$. Therefore, the action is given as, 
\begin{equation}
\mathcal{S}=\int d^4x\sqrt{-g}P(\phi,X).
\end{equation}
The energy density is given by, 
\begin{equation}
\rho=2X\frac{\partial P}{\partial X}-P.
\end{equation}

Therefore, EoS parameter can be written as, 
\begin{equation}
w=\frac{P}{2X\frac{\partial P}{\partial X}-P}.
\end{equation} 

For $2X\frac{\partial P}{\partial X}=0$, this model reduces to cosmological constant model with $w=-1$ \cite{NimaArkani-Hamed_2004}. More detailed knowledge about the K-essence model can be found in the references \cite{Barger:2000jg,Li:2002wd,Malquarti:2003nn,Malquarti:2003hn}.

\begin{itemize}

\item \textbf{\textit{Tachyon field model}}
\end{itemize}
 Tachyons are hypothetical particles that are often characterized by having imaginary mass. During the decay process of D-branes, a state emerges that behaves like a gas lacking pressure but possessing finite energy density. It resembles classical dust \cite{Sen:1999mh,Sen:2002in,Sen:2002nu,Garousi:2000tr,Bergshoeff:2000dq,Kluson:2000iy,Garousi:2002wq,Garousi_2003}. Tachyons exhibit an equation of state (EoS) parameter that smoothly varies within the range of $-1<w<0$. This has piqued the interest of cosmologists, prompting them to consider tachyons as a plausible contender of dark energy \cite{Gibbons:2002md}. For a comprehensive analysis of tachyonic dark energy models, with a focus on their ability to generate late-time cosmic acceleration, refer to \cite{Padmanabhan:2002cp,Bagla:2002yn,Abramo:2003cp,Aguirregabiria:2004xd,Zong-KuanGuo_2004,Copeland:2004hq}.
The tachyon's squared mass is intrinsically negative and stabilizes at the peaks of its associated scalar field potential. This state experiences infinitesimal perturbations, ultimately resulting in a tachyon condensation process characterized by a descent from the peaks, leading to the attainment of a real mass. The action for tachyon field ($\psi$) with potential $V(\psi)$ can be expressed as follows:
\begin{equation}
\mathcal{S}=-\int d^4xV(\psi)\sqrt{-\mbox{det}(g_{\mu\nu}+\partial_\mu\psi\partial_\nu\psi)}. 
\end{equation}
The wave equation for this field is, 
\begin{equation}
\frac{\ddot{\psi}}{1-\dot{\psi}^2}+3H\dot{\psi}+\frac{1}{V}\frac{dV}{d\psi}=0.
\end{equation}
The energy density and pressure are respectively given as, 
\begin{align}
\rho_\psi &= \frac{V(\psi)}{\sqrt{1-\dot{\psi}^2}},\\
p_\psi &= -V(\psi)\sqrt{1-\dot{\psi}^2}.
\end{align}
We can write the EoS parameter as, 
\begin{equation}
w=\frac{p_\psi}{\rho_\psi}=\dot{\psi}^2-1.
\end{equation}
Therefore, the criterion for the universe to experience accelerated expansion is when $\dot{\psi}^2<2/3$.

\begin{itemize}
\item \textbf{\textit{Holographic Dark Energy}}
\end{itemize}
The holographic dark energy (HDE) draws its inspiration from the holographic principle, a concept rooted in quantum gravity theory. This model aims to shed light on the mysterious nature of dark energy and its role in the accelerated expansion of the universe by establishing a connection between dark energy and the information content residing on the boundary. The holographic principle, originally proposed by 't Hooft \cite{tHooft:1993dmi} and Susskind \cite{Susskind:1994vu}, suggests that the physical properties and degrees of freedom within a given region of space can be entirely encoded on its boundary rather than within the volume itself. This concept is closely linked to the notion of entropy and has its roots in the work of Bekenstein on black hole entropy bounds \cite{Bekenstein:1973ur,Bekenstein:1993dz}. According to this bound, there exists a connection between the short-distance ultraviolet (UV) cut-off and the long-distance infrared (IR) cut-off due to the constraint that the total quantum zero-point energy of a system should not exceed the mass of black holes of the same size \cite{Cohen:1998zx}. This constraint can be expressed as:
\begin{equation}
L^3\rho_\Lambda \leq LM^2_p,
\end{equation}
where $M_p=(8\pi G)^{-1/2}$  is the reduced Planck mass, $\rho_\Lambda$ represents the quantum zero-point energy density determined by the UV cut-off, and $L$ is the length scale of the system size. The IR cut-off is the length for which this inequality saturates. The largest allowable value for $L$ is the one that makes this inequality reach its limit. Therefore,
\begin{equation}
\rho_\Lambda=3C^2M^2_pL^{-2}.
\end{equation}
In the context of dark energy, the holographic principle is applied by introducing the holographic energy density ($\rho_H$) with the following expression \cite{Li:2004rb},
\begin{equation}
\rho_H=3c^2M^2_pL^{-2}.
\end{equation}
Here system size is the size of the observable universe and $C^2$ is a dimensionless coupling parameter. The cosmic horizon serves as the IR (infrared) cut-off. Similar concepts and ideas were explored in references \cite{Horava:2000tb,Thomas:2002pq}. Various approaches can be found in the literature, each with its own choice of the IR cut-off length scale. Such examples include the particle horizon \cite{Fischler:1998st,Cataldo:2001bn}, the future event horizon
\cite{Li:2004rb,Guberina:2005mp,Wang:2005jx,Qing-GuoHuang_2004,Gong:2004cb,Nojiri:2005pu,Wang:2005ph} and the Hubble horizon \cite{Pavon:2005yx,Zimdahl:2007zz,LixinXu_2009} as IR cut-off.

\section{Outline of the thesis}

The primary focus of this thesis centers on delving into the thermodynamic characteristics of diverse cosmological models.  An assessment of the models' viability has been carried out by applying the Generalized Second Law (GSL), which asserts that the total entropy comprising both the horizon and the encompassed fluid must never decrease. Given the evolving nature of the universe, our approach has involved working with the apparent horizon. We have assumed a state of thermodynamic equilibrium between this apparent horizon and the fluid inside the horizon. In this equilibrium state, we have adopted the Hayward-Kodama temperature as the horizon temperature. Chapters \ref{Chapter2}-\ref{Chapter5} contain the main research work. Chapters \ref{Chapter2},\ref{Chapter3} and \ref{Chapter4} are focused on GSL test. Chapter \ref{Chapter5} is focused on the stability analysis of a cosmological model. \\

In the second chapter, we have discussed about apparent horizon, blackhole thermodynamics and their analogies in the dynamical apparent horizon in cosmology. A thorough description of how to calculate the rate of change of entropy has been prescribed. Also, we have touched upon the subject of the thermodynamic stability of the cosmological model.\\

In the third chapter, our study conducts a thorough comparative analysis between thawing and freezing models, particularly with regard to their adherence to the Generalized Second Law (GSL) of thermodynamics. In our comprehensive evaluation of the total entropy ($S_{\text{tot}}$), we incorporate both the entropy of the horizon and the entropy stemming from the matter enclosed within the horizon. To facilitate this investigation, we employ a simple ansatz proposed by Carvalho \textit{et al.} \cite{Carvalho:2006fy} to model the dynamic evolution of the energy density within the quintessence field, allowing us to pinpoint the parameter range ($\alpha$) associated with thawing and freezing behaviours. Our findings reveal a common inconsistency with the GSL in both type of models.  In the context of freezing models, this GSL breakdown is traced back to a remote past, corresponding to a redshift of approximately $z \sim 10^{4}$. During this distant epoch, a quintessence model along with cold dark matter fails to adequately account for the evolution of the universe,  as the dominant contribution comes from radiation distribution. This suggests that the GSL breakdown may not hold true under these circumstances. Conversely, for thawing models, this unusual violation of the GSL is anticipated to manifest in a finite future. The key implication here is that freezing models appear to enjoy better thermodynamic favourability when compared with their thawing counterparts. \\
 
In the fourth chapter, we explore GSL to models of the universe filled with radiation and dust, assuming the universe is flat, homogeneous and isotropic in the framework of Brans-Dicke theory in \textit{Einstein frame}. 
 When it comes to a universe dominated by radiation, the solutions in Brans-Dicke theory with a positive value for the BD parameter $\omega$, do not follow GSL. But when $\omega$ has a negative value within a certain range, it does match the thermodynamic requirements. And that is exciting because for cosmic acceleration, as per observation, one needs this negative $\omega$ value. Now, if we switch our focus to a universe dominated by dust (like galaxies and matter), the model does satisfy GSL when $\omega$ is a small negative number within a specific range.  This range of $\omega$ significantly overlaps with the range necessary for achieving accelerated expansion without the need for any exotic matter.
 

In the fifth chapter, we studied the thermodynamic viability of some dark energy models reconstructed through the cosmological jerk parameter. 
 As the deceleration parameter, $q$ evolves, we become interested in the next-order derivative called the \textit{jerk parameter}, represented as $j$. It tells us how $q$ changes over time.  In this study, we select some of these models from existing literature and evaluate them in terms of their thermodynamic viability. By reconstructing the jerk parameter, it is entirely possible to find models that satisfy the laws of thermodynamics. Among the four models tested for a non-interacting scenario, only one (model IV), which has an inverse relationship with ($1+z$), shows a decrease in entropy in the future (at $z<0$). This decrease is particularly significant near the present epoch ($z=0$). All the other models, including one that allows for interaction in the dark sector, pass the GSL. 
 Model I, lacking an explicit dependence on $z$, satisfies the GSL but undergoes a sudden entropy surge in the future ($z<0$). However, any analysis in terms of $z$ is not very sound for negative values of $z$. Models II and III, dependent on $(1+z)$ and $(1+z)^2$ respectively, as well as Model V with interaction in the matter sector, exhibit favorable behaviour across a broad parameter range within a $3\sigma$ confidence interval. Overall, from the past to the present epoch, all models demonstrate satisfactory behaviour.\\

  In the sixth chapter, we conducted a thermodynamic stability analysis on a model designed to mimic the $\Lambda$CDM model for the current state of the universe. In this analysis, we took into account the evolving horizon and considered the Hayward-Kodama temperature as the temperature of the horizon. In thermodynamics, the stability of an equilibrium system hinges on having a positive thermal capacity and compressibility. This principle also applies to the matter content within a cosmological system. However, in our current scenario, the specific heat capacity ($C_V$) turns out to be negative. This suggests that the model is likely to exhibit thermodynamic instability. The significant outcome of our investigation is that the matter content undergoes a phase transition as the universe transitions from a decelerated to an accelerated state of expansion. This phase transition is notably a second-order transition, evident by the discontinuity in $C_V$. The deceleration parameter, $q$ serves as the order parameter in this context. This feature goes missing if we consider Hawking temperature, which neglects the fact that the apparent horizon is evolving over time.
  
  The final chapter, chapter \ref{Chapter6}, concludes the research in this thesis and provides some future aspects.


%% file: Chapters/Chapterthermo.tex
\chapter{Cosmological Apparent Horizon and Thermodynamics} 
\label{Chapterthermo}
\section{Cosmological Apparent Horizon}
 We shall discuss the cosmological apparent horizon by drawing an analogy with black hole apparent horizon. Therefore, let us first take some notes on black hole apparent horizon and then delve into the apparent horizon in FRW cosmologies.
\subsection{Apparent Horizon} \label{Apparenthorizon}

In simple words, a horizon can be described as "a frontier between things observable and things unobservable" \cite{Rindler:1956yx}. The key feature of a black hole spacetime is the presence of an event horizon, a boundary that separates the black hole from the external observers and conceals internal events. This hypersurface consists of a congruence of  null geodesics, or null generators, which are crucial in comprehending the behavior of the  horizon. To grasp the overall behavior of the horizon, it becomes essential to study how these generators behave. In this thesis, this topic shall be briefly discussed. For more details, we refer to\cite{faraoni2015horizons}. 

\subsubsection{Congruence of Null Geodesics}

\begin{itemize}
\item \textbf{\textit{Affinely parametrized null geodesic congruence:}}
\end{itemize}
A null geodesic refers to a path followed by a massless particle (such as a photon, which has zero rest mass) in the spacetime described by relativistic theory of gravity. Therefore, on spacetime manifold, its tangent, denoted by $l^\mu$ is null,i.e., $l^\mu l_\mu=0$. The geodesic equation it satisfies is given by the equation, 

\begin{equation}
l^\mu;_{\nu} l^\nu=\lambda l^\mu,
\end{equation}
here $\lambda$ serves as a parameter along the curve. The geodesic equation signifies that the tangent is parallelly transported to itself as one follows the geodesic. The choice of the parameter $\lambda$ allows for simplification of the geodesic equation to a more convenient form without loss of generality as, 
\begin{equation}
l^\mu;_{\nu} l^\nu=0.
\end{equation}

In terms of Christoffel symbols, the above equation can be written as,
\begin{equation}\label{affine}
\frac{dx^\alpha}{d\lambda^2}+\Gamma^\alpha_{\mu\nu}\frac{dx^\mu}{d\lambda}\frac{dx^\nu}{d\lambda}=0.
\end{equation} 

The variable $\lambda$ is therefore an affine parameter of the affinely parametrized geodesic equation\eqref{affine}. \\
Consider an open region $\mathcal{M}$ in the spacetime manifold. A congruence of curves represents a family of curves in which every point within $\mathcal{M}$ is traversed by one and only one curve from the family. The tangents to these curves define a vector field on $\mathcal{M}$ and conversely, any continuous vector field within $\mathcal{M}$ generates a congruence of curves with the tangents of the field. When the field of tangents is smooth, we refer to the congruence as smooth. In particular, we can focus on a smooth congruence of null geodesics whose tangents are represented by $\lambda$  within the open region $\mathcal{M}$. To mark different geodesics within the congruence in $\mathcal{M}$, we consider another parameter $\zeta$. We now define the deviation vector with components $\eta^\mu\equiv \frac{\partial x^\mu}{\partial\zeta}$.

Though by construction, we see $l^\mu\eta_\mu=0$, but that does not mean $\eta^\mu$ is orthogonal to the curve, since $l^\mu$ is a null vector. However, we can limit to deviation vectors that are considered equivalent if they vary solely by a component along $l^\mu$. The tangent space, comprising all vectors that are orthogonal in this manner to $l^\mu$, forms a 2-dimensional vector space. We can also explore its dual space and the set of tensors constructed using these vectors. The geodesic deviation vector follows the geodesic deviation equation, given by,
\begin{equation}
(D_\lambda)^2\eta^\mu = -R^\mu_{\nu\alpha\beta}u^\nu\eta^\alpha u^\beta,
\end{equation}

where $R^\mu_{\nu\alpha\beta}$ is the Riemann tensor. $D_\lambda$ is covariant derivative operator, $D_\lambda\eta^\mu=\frac{d}{d\lambda}\eta^\mu+\Gamma^\mu_{\alpha\beta}\frac{dx^\alpha}{d\lambda}\eta^\beta.$
This mathematical equation expresses how the behaviour of geodesics, rather the deviations of nearby geodesics is impacted by the spacetime curvature.  \\
We can obtain the famous \textit{Raychaudhuri equation} for null geodesics by deriving the rate of change of the gradient $l^\mu;_{\mu}$ along the geodesic. Let us consider a tensor field \cite{Wald:1984rg,Poisson:2009pwt},
\begin{equation}
B_{\mu\nu}\equiv l_\mu;_{\nu}.
\end{equation}
$B_{\mu\nu}$ satisfies $l^\mu;_{\mu}\eta^\nu=B^\mu_\nu\eta^\mu$. $B^{\mu\nu}$ is orthogonal to $l^\mu$, therefore, $B^{\mu\nu}l_\mu=B^{\mu\nu}l_\nu=0$. So it has components only along the transverse of $l^\mu$. 

Now the spatial metric in 2-space orthogonal to $l^\mu$ can be defined as, 
\begin{equation}
h_{\mu\nu}\equiv g_{\mu\nu}+l_\mu n_\nu+l_\nu n_\mu,
\end{equation}
where $n^\alpha$ is another null vector and normalized as $l^\alpha n_\alpha=-1$. The selection of $n^\alpha$ is not unique. The only fixed quantity is the null congruence with tangent $l^\alpha$. However, geometric and physically relevant quantities remain independent of the choice of $n^\alpha$.  

The trace of the tensor $B_{\mu\nu}$,
\begin{equation}
\theta\equiv B^\mu_\mu=g^{\mu\nu}B_{\mu\nu}=l^\mu;_{\mu},
\end{equation}
is called \textit{expansion} of the affinely-parametrized congruence.
The expansion tensor is defined as, 
\begin{equation}
\theta_{\alpha\beta}\equiv \frac{\theta}{2}h_{\alpha\beta}.
\end{equation}
The transverse tensor $\tilde{B}_{\alpha\beta}$ can be written as sum of symmetric and antisymmetric parts,
\begin{equation}
\tilde{B}_{\alpha\beta}=\tilde{B}_{(\alpha\beta)}+\tilde{B}_{[\alpha\beta]}.
\end{equation}
It can be further decomposed into trace and traceless parts as,
\begin{equation}
\tilde{B}_{\alpha\beta}=\left(\frac{\theta}{2}h_{\alpha\beta}+\sigma_{\alpha\beta} \right)+\omega_{\alpha\beta}.
\end{equation}
In the above equation, $\sigma_{\alpha\beta}$ is known as the shear tensor, 
\begin{equation}
\sigma_{\alpha\beta}\equiv \tilde{B}_{(\alpha\beta)}-\frac{\theta}{2}h_{\alpha\beta},
\end{equation}
and $\omega_{\alpha\beta}$ is known as the vorticity tensor,
\begin{equation}
\omega_{\alpha\beta}\equiv \tilde{B}_{[\alpha\beta]}. 
\end{equation}
The shear and vorticity scalars are,
\begin{align}
\sigma^2=\sigma_{\alpha\beta}\sigma^{\alpha\beta}, \hspace{3em} \omega^2 = \omega_{\alpha\beta}\omega^{\alpha\beta}.
\end{align}
The Raychaudhuri equation that governs the expansion along affinely parametrized null geodesic,
\begin{equation}
\frac{d\theta}{d\lambda}=-\frac{\theta^2}{2}-\sigma^2-\omega^2-R_{\alpha\beta}l^\alpha l^\beta.
\end{equation}

The other choices of $n^{\alpha}$ do not affect this equation.
The term $\frac{d\theta}{d\lambda}<0$ implies that expansion decreases with the evolution of congruence, which means null rays will be focused, and  $\frac{d\theta}{d\lambda}>0$ implies that expansion increases with the evolution of congruence, which means null rays will be defocused. Therefore, this equation illustrates how the focusing or defocusing of null rays occurs due to the combined effects of expansion, shear, vorticity, and matter. 

\begin{itemize}
\item \textbf{\textit{Non-affinely parametrized null geodesic congruence:}}
\end{itemize}
The geodesic equation for non-affinely parametrized null geodesic congruence is, 
\begin{equation}
l^{\mu};_{\nu}l^{\nu} =\kappa l^{\mu}.
\end{equation}
The expansion scalar takes the form,
\begin{equation}
\theta=l^{\mu};_{\mu}-\kappa.
\end{equation}
The Raychaudhuri equation transforms into the modified form as,
\begin{equation}
\frac{d\theta}{d\lambda}=\kappa\theta-\frac{\theta^2}{2}-\sigma^2-\omega^2-R_{\alpha\beta}l^\alpha l^\beta.
\end{equation}

A compact and orientable surface possesses two distinct directions perpendicular to it, representing outgoing and ingoing null rays. When a spherical symmetry is present, it naturally prompts an examination of congruences formed by radial  outgoing and ingoing null geodesics with their respective tangent fields $l^\alpha$ and $n^\alpha$. 
For non-affinely parametrized congruence of null geodesics, the expansion of null rays $l^\alpha$ is given by the following equation,
\begin{align}\label{expansion}
\theta_l\equiv h^{\alpha\beta}l^\beta;_{\alpha}=\left[g^{\alpha\beta}+\frac{l^\alpha n^\beta+l^\beta n^\alpha}{\left(l^\mu n^\nu g_{\mu\nu}\right)}\right]l^\beta;_{\alpha}.
\end{align}

In the above mathematical expression, $h^{\alpha\beta}$ serves as a projection metric onto 2-D hypersurface.

\subsubsection{Definitions pertain to closed 2-surfaces}
In the subsequent definitions, $\theta_l$ and $\theta_n$ represent the outgoing and ingoing future directed null geodesic congruences respectively.
 \begin{itemize}
 \item \textit{Normal Surface}: Future-directed outgoing null ray is diverging but ingoing null ray is converging,i.e., $\theta_l>0$ and $\theta_n<0$.
 \item \textit{Trapped surface}: Both the future-directed  null rays are converging,i.e., $\theta_l<0$ and $\theta_n<0$.
 \item \textit{Marginally outer trapped surface(MOTS)}: $\theta_l=0$ and $\theta_n<0$.
 \item\textit{Anti-trapped surface}: Both the future-directed  null rays are diverging,i.e., $\theta_l>0$ and $\theta_n>0$.
 \item \textit{Untrapped surface}: $\theta_l\theta_n<0$.
 \item \textit{Marginally outer trapped tube}: 3-D surface foliated by MOTSs.

\end{itemize}

The closure of a surface, often a three-surface that is foliated by marginal surfaces, is what is known as a future apparent horizon. The future apparent horizon is characterized by specific conditions,
\begin{align}
\theta_l &= 0, \nonumber\\
\theta_n &<0.
\end{align}
Therefore, on the apparent horizon, the future-directed outgoing null geodesics stop transmitting outward. This differs from the event horizon in the sense that, it is a quasi-local horizon. It is to be noted that, the inner trapping horizon can be obtained by switching $l^\alpha$, $n^\alpha$ and reversing the inequality sign.

\subsubsection{Apparent Horizon in FRW Cosmologies}

The use of notation and terminology can become perplexing when transitioning between black hole horizons and cosmological horizons. It also confuses readers when shifting from observations made by external observers to those positioned within a horizon's domain. For an expanding FLRW space and observer inside the apparent horizon, the notations modify as,
\begin{itemize}
\item \textit{Normal Surface}: Outgoing null ray is diverging but ingoing null ray is converging,i.e., $\theta_l>0$ and $\theta_n<0$.
 \item \textit{Trapped surface}: Both the null rays are diverging,i.e., $\theta_l>0$ and $\theta_n>0$.
 \item \textit{Marginally trapped (past) surface}: $\theta_l>0$ and $\theta_n=0$.
\end{itemize}
The apparent horizon for expanding FRW cosmologies is characterized by the conditions, 

\begin{align} \label{defah}
\theta_l &> 0, \nonumber\\
\theta_n &=0.
\end{align}

In comoving co-ordinates, the tangent of outgoing and ingoing radial null geodesics are, 
\begin{align}\label{3.48}
l^\alpha &= \left(1,\frac{\sqrt{1-kr^2}}{a(t)},0,0\right), \nonumber \\ 
n^\alpha &= \left(1,-\frac{\sqrt{1-kr^2}}{a(t)},0,0\right),
\end{align}
respectively.
Therefore, the covariant derivatives of these null geodesics are, 
\begin{align}
\nabla_{\alpha}l^\alpha&=\frac{1}{\sqrt{-g}}\partial_\mu\left(\sqrt{-g}l^\mu\right)\nonumber\\ 
&=\frac{\left(1-kr^2\right)^{1/2}}{a^3r^2}\left(\frac{3a^2\dot{a}r^2}{\left(1-kr^2\right)^{1/2}}+2a^2r\right)\nonumber\\
&=3H+\frac{2}{ar}\sqrt{1-kr^2},\\
\nabla_{\alpha}n^\alpha&=\frac{1}{\sqrt{-g}}\partial_\mu\left(\sqrt{-g}n^\mu\right)\nonumber\\ &=\frac{\left(1-kr^2\right)^{1/2}}{a^3r^2}\left(\frac{3a^2\dot{a}r^2}{\left(1-kr^2\right)^{1/2}}-2a^2r\right)\nonumber\\
&=3H-\frac{2}{ar}\sqrt{1-kr^2}.
\end{align}
Using the equation \eqref{expansion}, we get the expansions of these null geodesic congruences as,

\begin{equation}\label{outgoingfrw}
\theta_l = \frac{2\left(\dot{a}r+\sqrt{1-kr^2}\right)}{ar}=2\left(H+\frac{1}{R}\sqrt{1-\frac{kR^2}{a^2}}\right),
\end{equation} 
 
 \begin{equation}\label{ingoingfrw}
\theta_n = \frac{2\left(\dot{a}r-\sqrt{1-kr^2}\right)}{ar}=2\left(H-\frac{1}{R}\sqrt{1-\frac{kR^2}{a^2}}\right).
\end{equation} 
 
 Now, from the definition \eqref{defah}, we get the cosmological apparent horizon is located at, 
 \begin{equation}\label{radah}
 r_{h} = \frac{1}{\sqrt{\dot{a}^2+k}}.
 \end{equation}
 In terms of proper radius $R\equiv ar$, the above equation\eqref{radah} can be written as,
 \begin{equation}\label{rApFRW}
 R_{h} = \frac{1}{\sqrt{H^2+k/a^2}}.
 \end{equation}
 
It is important to highlight that the definition of the apparent horizon relies exclusively on null geodesic congruences and their expansions, without any consideration of the global causal structure. At times, there may be a temptation to approximate the location of apparent horizons by speculating where the outward radial null rays come to a halt, essentially setting $l^r=0$. This simplified approach may yield accurate results on occasion, especially with spherically symmetric metrics in Painlevé-Gullstrand coordinates \cite{Nielsen:2005af}. But it should not be regarded as a substitute for the proper procedure, which entails identifying surfaces where $\theta_l=0$ and $\theta_n\neq 0$.  \vspace{0.5em}

 If we were to apply equation\eqref{3.48} and assume that $n^1$ is equal to zero, it would lead to an incorrect inference, indicating the absence of apparent or trapping horizons in FRW cosmologies with $k=0$ or $k=-1$. It would yield an inaccurate value for $R_{h}$ in the case of $k=1$ in FRW cosmologies. Clearly, this is not the case. With the exception of the scenario where $H=0$, apparent horizons consistently exist and can be determined using equation\eqref{rApFRW}. The event horizon may not exist in some cases. Therefore, the radial null geodesic congruences within FRW cosmologies provide a case that contradicts this approach. \vspace{0.5em} 
 
Unlike the event and particle horizons, the apparent horizon is typically not a null surface. Similar to horizons in flat space, the cosmological apparent horizon is observer-dependent. It operates as a spherical barrier encircling the observer and veiling information. 

\section{Thermodynamics}

This section commences with an exploration of the four laws governing the mechanics of black holes and their intricate relationship with the four fundamental laws of thermodynamics. Subsequently, we will delve into the extension of this connection to include the apparent horizon within the context of the FRW cosmological framework.
\subsection{Black hole thermodynamics}\label{BHT}

Black hole thermodynamics is a fascinating field that explores the connection between black holes, which are objects with extremely strong gravitational forces, and the laws of thermodynamics, which describe the behavior of energy and entropy in various physical systems. This connection was first suggested by Jacob Bekenstein in 1973 \cite{Bekenstein:1973ur} and later developed by Stephen Hawking and others.

In 1973, Jim Bardeen, Brandon Carter, and Stephen Hawking formulated a set of four laws governing the behaviour of black holes \cite{Bardeen:1973gs}. These laws of black-hole mechanics bear a striking resemblance to the four laws of thermodynamics. While this analogy was at first perceived to be purely formal and coincidental, it soon became clear that black holes do indeed behave as thermodynamic systems. The crucial step in this realization was Hawking's remarkable discovery of 1974 \cite{Hawking:1974rv} that quantum processes allow a black hole to emit a thermal flux of particles. It is thus possible for a black hole to be in thermal equilibrium with other thermodynamic systems. The laws of black-hole mechanics, therefore, are nothing but a description of the thermodynamics of black holes.

The four laws of black hole mechanics are \cite{Bardeen:1973gs,Wald:1984rg,Wald:1999vt,Poisson:2009pwt},
\begin{itemize}
\item \textit{Zeroth Law}: The surface gravity($\kappa$) of a stationary black hole is uniform over the entire event horizon (see \cite{Carter:1973rla,Bardeen:1973gs,Hawking:1973qla}).
\item \textit{First Law}: We consider a quasi-static process during which a stationary black hole of mass
$M$, angular momentum $J$, and surface area $A$ is taken to a new stationary black hole with parameters $M + \delta M$, $J + \delta J$, and $A + \delta A$. The first law of black-hole
mechanics states that the changes in mass, angular momentum, and surface area are related by,
\begin{equation}
\delta M =\frac{\kappa}{8\pi}\delta A+\Omega_H\delta J.
\end{equation}
\item \textit{Second Law}: It states that if the null energy condition is satisfied, then the surface area of a black hole can never decrease \cite{Hawking:1971vc}, i.e., $\delta A \geq 0$. 

\item \textit{Third Law}: It states that if the stress-energy tensor is
bounded and satisfies the weak energy condition, then the surface gravity of a black hole cannot be reduced to zero within a finite advanced time. A precise formulation of this law was given by Werner Israel in 1986.
\end{itemize}

These statements are quoted from reference \cite{Poisson:2009pwt}.

The intriguing connection between the four fundamental laws governing black hole mechanics and the well-established principles of thermodynamics has not gone unnoticed. In this relationship, the parameter $\kappa$ takes on the role of temperature, while $A$ mirrors the concept of entropy, and $M$ serves as an analogy to internal energy. The discovery of Hawking radiation evinced that black holes indeed possess a distinct temperature, which correlates directly with their surface gravity,
\begin{equation}
T=\frac{\hbar}{2\pi}\kappa.
\end{equation}

In a fascinating parallel, the zeroth law can be regarded as a specific manifestation of the analogous thermodynamic concept, signifying that a system achieves thermal equilibrium when it maintains a consistent temperature throughout. Similarly, when we view the first law through the lens of its thermodynamic counterpart, it logically implies that the entropy associated with a black hole should be defined as follows \cite{Hawking:1975vcx,Majumdar:1998xv},
\begin{equation}
S=\frac{1}{4\hbar}A.
\end{equation}

The second law, likewise, emerges as a specific instantiation of its thermodynamic counterpart, asserting the principle that the entropy of an isolated system never experiences a decrease. It is pertinent to acknowledge that Hawking radiation \cite{Hawking:1974rv} leads to a reduction in the black hole's surface area, a seeming contradiction to the area theorem due to the non-compliance of the radiation's stress-energy tensor with the null energy condition. However, it is important to emphasize that the process of black hole evaporation \cite{Susskind:2005js} remains in harmony with the generalized second law, which dictates that the overall entropy, encompassing the entropies of both radiation and black holes, remains conserved. We shall discuss the generalized second law of thermodynamics later in more detail.
\begin{table}

\begin{center}
\begin{tabular}{ | m{2.4cm} | m{5.4cm}| m{5.4cm} | } 
  \hline
  \textbf{Laws}& \textbf{Thermodynamics} & \textbf{Black holes} \\ 
  \hline
  Zeroth Law & When a system achieves thermal equilibrium, it maintains a uniform $T$ throughout its entirety & For a stationary black hole, $\kappa$ exhibits uniformity across the entire expanse of its event horizon \\ 
 \vspace{1em} First Law & \vspace{1em} $TdS= dU$ + work term &\vspace{1em}  $dM =\frac{\kappa}{8\pi}\delta A+\Omega_H\delta J$ \\ 
  \vspace{1em}Second Law & \vspace{1em} $\delta S \geq 0$ in any process &\vspace{1em} $\delta A\geq 0$ in any process \\
 \vspace{1em} Third Law & \vspace{1em} $T=0$ is an unattainable state through any physical process & \vspace{1em} Achieving $\kappa=0$ through physical means is an unattainable feat\\
  \hline
\end{tabular}
\end{center}
\caption{Laws of Black Holes and Thermodynamics}
\end{table}
The revelation that black holes adhere to thermodynamic principles unveils a profound interplay between seemingly distinct realms of science, namely, gravitation, quantum mechanics, and thermodynamics. Remarkably, this intricate relationship continues to be a subject of active investigation and comprehension in contemporary scientific exploration.

\subsection{Thermodynamics of Apparent Horizons in FLRW Space}\label{Apparenthorizonthermo}

At its inception, black hole thermodynamics primarily centered around stationary event horizons. However, its evolution has led to a more comprehensive exploration, now encompassing various horizon types like apparent, trapping, isolated, dynamical, and slowly evolving horizons. Early on, it was recognized that cosmological horizons, with their origins traced back to the static event horizon of de Sitter space \cite{Gibbons:1977mu}, possess thermodynamic properties. A substantial amount of research is dedicated to investigating the thermodynamic characteristics of de Sitter spacetime (for thermodynamics applied in de Sitter space see references \cite{Gibbons:1977mu,Cai:2001tv,Frolov:2002va,Gibbons:1976ue,Redmount:1988pg}). Additionally, endeavors have been undertaken to broaden these inquiries beyond the confines of de Sitter space \cite{Davies:1987ti,Davies:1986wx,GianlucaCalcagni_2005}. Researchers in this field have put forth the argument that this thermodynamic framework is also applicable to FLRW apparent horizons (such as in \cite{DongsuBak_2000,Akbar:2006kj}).

  A large number of investigations adapted the originally formulated thermodynamic equations, tailored for the de Sitter event horizon and the apparent horizon, to suit the dynamic nature of the non-static apparent horizon within FLRW space. It is imperative to emphasize that this apparent horizon does not necessarily align with the event horizon, which may not even have a presence in this specific context. The apparent horizon is frequently regarded as a causal boundary linked to gravitational temperature, entropy, and surface gravity within the evolving spacetime framework. These intriguing concepts have been thoroughly explored in a body of research \cite{DongsuBak_2000,Bousso:2004tv,Collins:1992eca,JoseTomasGalvezGhersi_2011,Hayward:1997jp,Hayward:1998ee,Nielsen:2008kd}.

In contrast, previous investigations \cite{Bousso:2004tv,JoseTomasGalvezGhersi_2011,Davies:1988dk,AndreiVFrolov_2003,Wang:2005pk} contend that the application of thermodynamics to the event horizon of FRW space encounters challenges in establishing a coherent framework except however in de Sitter space.

Moreover, there have been concerted attempts to calculate the Hawking radiation emanating from the apparent horizon within the framework of FLRW cosmology. The studies conducted by Jiang and Zhu \cite{Jiang:2009kzr,Zhu:2008hn} and the work by Medved \cite{Medved:2002zj} and Cai \cite{Cai:2008gw}, have delved into this area. Various techniques, including the Hamilton-Jacobi method \cite{MarcoAngheben_2005,Nielsen:2005af,Visser:2001kq}, along with the Parikh-Wilczek approach, initially crafted for the analysis of black hole horizons \cite{Parikh:1999mf}.

There exists an extensive corpus of literature dedicated to exploring the thermodynamic aspects of FRW cosmologies \cite{DiCriscienzo:2009kun,Akbar:2006kj,Rong-GenCai_2005,Verlinde:2000wg,Cai:2002ub,Pavon:1990qf,Akbar:2006er}. For a comprehensive assessment of the thermodynamic attributes associated with the FLRW apparent horizon, one may refer to Reference \cite{DiCriscienzo:2009kun}. In classical terms, surface gravity finds its definition grounded in the geometric characteristics of the metric tensor. However, it takes on a significant role in the realm of black hole thermodynamics, acting as the constant factor that relates changes in black hole mass (representing internal energy) to variations in the area of the event horizon (directly proportional to entropy)(see section \ref{BHT} in this chapter). The ongoing debate surrounding the appropriate definition of black hole mass in non-trivial backgrounds, as delineated in the review \cite{Szabados:2004xxa}, naturally extends to the definition of surface gravity. Moreover, surface gravity also emerges in semiclassical contexts, as it essentially corresponds, with minor numerical adjustments, to the Hawking temperature of a black hole. The conventional characterization of surface gravity pertains to a Killing horizon in the context of stationary spacetimes \cite{Wald:1984rg,Wald:1999vt}. In static and stationary scenarios, a timelike Killing vector field exists beyond the horizon and becomes null at the horizon itself. In these situations, the definitions of surface gravity align, and they are familiar concepts derived from the analysis of Kerr-Newman black holes in GTR. The surface gravity ($\kappa$) is defined in relation to the Killing vector ($\xi^a$) using the equation \cite{Wald:1984rg,Wald:1999vt}:
\begin{equation}\label{sgravstatic}
\xi^a\xi^b;_a = \kappa\xi^b.
\end{equation}

In dynamic situations, the presence of a timelike Killing vector field is absent. Hence the applicability of Killing horizons does not extend to general non-stationary contexts, particularly when dealing with quasi-local horizons rather than event or Killing horizons. In such situations, a suitable notion of surface gravity becomes essential. Multiple definitions of surface gravity can be found in the literature, and it is important to recognize that these definitions are not synonymous. Within the framework of spherical symmetry, the Kodama vector successfully replaces a Killing vector in an evolving system, ultimately leading to the generation of a conserved current and surface gravity.

The Kodama vector \cite{Kodama:1979vn} broadens the applicability of a Killing vector field to spacetimes that lack one, making it a viable substitute for a Killing vector in the thermodynamics associated with evolving horizons. It is essential to emphasize that the Kodama vector is only defined within the context of spherical symmetry (an approach to generalization in non-spherical symmetric spacetimes can be found in reference \cite{Tung:2007vq}). The spacetime metric can be represented as,
\begin{equation}
dS^2=h_{ab}dx^a dx^b+R^2 d\Omega^2_{(2)}.
\end{equation} 
Here $a,b=0,1$ and $R$ is the areal radius. Let us denote the volume of 2-metric $h_{ab}$ by $\epsilon_{ab}$ \cite{Wald:1984rg}. The definition of Kodama vector \cite{Kodama:1979vn} is,

\begin{equation}
K_a\equiv\epsilon_{ab} R;_b,
\end{equation}
with $K^\theta=K^\phi=0$. The Kodama vector lies in the (t,R) plane that is orthogonal to the 2-spheres of symmetry. The antisymmetric nature of $\epsilon^{ab}$ and the symmetric nature of $R;_a R;_b$ makes $K^a R;_a=\epsilon^{ab}R;_a R;_b$ vanish. In a static spacetime, the Kodama vector is parallel to the timelike Killing vector. The Kodama vector exhibits zero divergence \cite{Kodama:1979vn,Abreu:2010ru},
\begin{equation}
K^a;_a=0.
\end{equation}
Consequently, the Kodama energy current($J^a$), defined as,
\begin{equation}
J^a\equiv G^{ab}K_b,
\end{equation}
becomes conserved covariantly.
Therefore, 
\begin{equation}
J^a;_a=0.
\end{equation}
This conservation occurs even if there is no timelike killing vector. This remarkable attribute is at times humorously coined as the \textit{Kodama miracle} \cite{Kodama:1979vn,Abreu:2010ru}.

The Hayward proposition for surface gravity, tailored for spherical symmetry \cite{Hayward:1997jp}, employs the Kodama vector $K^a$, which is consistently applicable in scenarios involving spherical symmetry. This future-directed vector
satisfies,
\begin{equation}
(K_aT^{ab});_b=0.
\end{equation}

The Hayward concept of surface gravity ($\kappa_{\mbox{ko}}$) associated with a trapping horizon is expressed as follows:

\begin{equation}\label{HK surgrav}
\frac{1}{2}g^{ab}K^c( K_a;_c- K_c;_a)= \kappa_{\mbox{ko}} K^b.
\end{equation}
This definition stands out due to the uniqueness of the Kodama vector. An equivalent expression to equation \eqref{HK surgrav} is, 
\begin{equation}\label{HK surgrav def}
\kappa_{\mbox{ko}}=\frac{1}{2\sqrt{-h}}\partial_\alpha\left(\sqrt{-h}h^{\alpha\beta}\partial_\nu R \right),
\end{equation}

where $h$ denotes the determinant of the 2-metric $h_{ab}$. This surface gravity is popularly known as the Hayward-Kodama surface gravity.

In FRW cosmology, this Hayward-Kodama surface gravity, given in equation \eqref{HK surgrav def}, becomes,
\begin{equation}\label{HKgravfrw}
\kappa_{\mbox{ko}} = -\frac{1}{2H}\left(\dot{H}+2H^2+\frac{k}{a^2}\right).
\end{equation}
 For spatially flat FRW metric ($k=0$) this equation takes the form,
 \begin{equation}\label{HKgravflatfrw}
\kappa_{\mbox{ko}} = -\frac{1}{2H}(\dot{H}+2H^2).
\end{equation}
The Hayward-Kodama dynamic surface gravity (\eqref{HKgravfrw} and \eqref{HKgravflatfrw}) becomes null when the scale factor($a(t)$) exhibits the specific condition,
\begin{equation}
a(t)=\sqrt{\alpha t^2+\beta t+\gamma},
\end{equation}
where $\alpha, \beta, \gamma$ are constants. One such example would be the universe purely dominated by radiation.

Thus, based on this surface gravity, we can derive the Hayward-Kodama temperature for spatially flat FRW cosmologies in the following manner:
\begin{align}\label{hktempdef}
T &= \frac{\mid{\kappa_{\mbox{ko}}\mid}}{2\pi} \vspace{0.em} \nonumber\\
&= \frac{2H^2+\dot{H}}{4\pi H}.
\end{align}

\begin{itemize}
\item\textbf{HK temperature in de Sitter space}
\end{itemize}
In Schwarzschild-like coordinates, the spherically symmetric metric is,
\begin{equation}
dS^2=-A(t,R)dt^2+B(t,R)dR^2+R^2d\Omega^2_{(2)}.
\end{equation}
In this metric, the kodama vector is written as \cite{Kodama:1979vn,Racz:2005pm}, 
\begin{equation}\label{kod}
K^a=-\frac{1}{\sqrt{AB}}\left(\frac{\partial}{\partial t}\right)^a.
\end{equation}
In static Schwarzschild-like coordinates, de Sitter metric is written as,
\begin{equation}\label{desittermetric}
dS^2=-(1-H^2R^2)dT^2+\frac{dR^2}{1-H^2R^2}+R^2d\Omega^2_{(2)}.
\end{equation}
Therefore, from \eqref{kod} and \eqref{desittermetric}, the kodama vector can be written as, 
\begin{equation}
K^a= \left(\frac{\partial}{\partial t}\right)^a.
\end{equation} 
It is important to observe that the Kodama vector aligns with the timelike Killing vector  in de Sitter space. The surface gravity induced by this Killing-Kodama field is, $\kappa=H=\sqrt{\frac{\Lambda}{3}}$. As a consequence, the temperature corresponding to this scenario is, $T=\frac{H}{2\pi}$, which nothing but the famous \textit{Hawking temperature} \cite{Hawking:1974rv}.

\subsection{Generalized Second Law of Thermodynamics} \label{GSL}
 Bekenstein proposed generalization of second law of black hole thermodynamics \cite{Bekenstein:1972tm,Bekenstein:1973ur,Bekenstein:1974ax}. He formulated the GSL as, "The sum of the blackhole entropy and the common (ordinary) entropy
in the black-hole exterior never decreases". Bekenstein delved into the realm of black hole physics through the lens of information theory \cite{shannon1948mathematical,shannon1949mathematical,tribus1971energy,PhysRev.106.620,PhysRev.108.171}. He provided an argument rooted in information theory that lends support to the credibility of the generalized second law. An illustrative instance of the connection between an increase in information and a reduction in entropy can be observed in a common scenario. Consider an ideal gas confined within a container undergoing an isothermal compression process. As the compression takes place, the entropy of the gas notably diminishes, a widely acknowledged fact. However, simultaneously, our knowledge regarding the internal arrangement of the gas increases. Following the compression, the gas molecules become more tightly concentrated, leading to a heightened precision in our awareness of their positions compared to the state before compression. Therefore, though the entropy of the system decreases, the total entropy, i.e., the sum of system's entropy and the entropy of the surrounding does not decrease. \vspace{0.5em}

In case of black hole, consider a scenario in which an entity carrying a certain measure of conventional entropy descends into a black hole. During this process, the entropy within the observable universe diminishes. This situation appears to challenge the second law of thermodynamics, as an external observer cannot directly confirm whether the overall entropy of the entire universe remains unchanged during this event. Nonetheless, according to insights from the literature, we understand that the black hole's area compensates for the vanishing entity by undergoing an irreversible increase. As a result, it appears reasonable to hypothesize that the second law remains intact, albeit in a more generalized formulation: the aggregate entropy within the external vicinity of the black hole, along with the entropy possessed by the black hole itself, consistently maintains a non-decreasing trend. \vspace{0.5em}
    
    Researchers have expanded the scope of the Generalized Second Law (GSL) to realms beyond the domain of black hole physics. In earlier research, a second law of thermodynamics applicable to the de Sitter horizon was established by Gibbons and Hawking \cite{Gibbons:1977mu}, and this principle was revisited in \cite{Mottola:1985qt}. Davies \cite{Davies:1988dk} explored the event horizon within the context of FLRW space, particularly in the context of General Relativity with a perfect fluid acting as the source. In FRW spacetime, we consider apparent horizon to work with and hence the GSL shapes as: in any physical process, the combined entropy of matter and the horizon must remain constant or increase, i.e., 
    \begin{equation}
    \delta S_{\mbox{tot}}=\delta S_{\mbox{matter}}+\delta S_{\mbox{h}} \geq 0,
    \end{equation}
where $S_{\mbox{tot}}$ denotes the total entropy and $S_{\mbox{matter}}$, $S_{\mbox{h}}$ denotes entropy of matter bounded by the horizon and apparent horizon respectively. 

For spatially flat FLRW metric, the field equations are \eqref{F1}-\eqref{F2}, 

\begin{align}
 &H^2=\frac{\dot{a}^2}{a^2}=\frac{8\pi G}{3}\rho,\\
&2 \frac{\ddot{a}}{a} + \Big(\frac{\dot{a}}{a}\Big)^2 =-8\pi Gp.
\end{align}
The entropy of the apparent horizon is assumed to be proportional to area($A$), and given by,  
\begin{equation}
S_{\mbox{h}}=\frac{A}{4G}=2\pi A \hspace{3mm}\text{(in geometrized unit)}.
\end{equation}
The area of the apparent horizon is given by, 
\begin{align}
A &= 4\pi R^2_{h}\nonumber\\
  &= \frac{4\pi}{H^2},
\end{align}
where we have substituted $R_{h}$ from equation \eqref{rApFRW}, and put $k=0$.
Therefore, the rate of change of entropy of the apparent horizon is given by, 
\begin{equation}\label{shordot}
\dot{S}_{\mbox{h}}=-16\pi^2\frac{\dot{H}}{H^3}.
\end{equation}

For the fluid inside the horizon, first law of thermodynamics applied to a hydrostatic system  looks like,  
\begin{equation}\label{Gibbs1}
TdS_{\mbox{in}}=dU+pdV,
\end{equation}
where $S_{\mbox{in}}$, $U$ and $V$ denote the entropy, the internal energy and the volume of the fluid inside the horizon respectively. $V$ is bounded by the apparent horizon, 
\begin{align}\label{vol}
V &=\frac{4}{3}\pi R_{h}^3 \nonumber \\ 
&=\frac{4}{3}\pi \frac{1}{H^3}.
\end{align}

Rate of change of entropy of fluid inside the horizon is,
\begin{align}\label{ent}
\dot{S}_{\mbox{in}}&= \frac{1}{T}\left[(\rho+p)\dot{V}+\dot{\rho}V\right]\nonumber\\
&= \frac{1}{T}(\rho+p)(\dot{V}-3HV).
\end{align}
Now inserting $T$ from equation \eqref{hktempdef} and $V$ from equation \eqref{vol} in \eqref{ent}, one obtains the expression of $\dot{S}_{\mbox{in}}$ as,
\begin{align}\label{ent-rate1}
\dot{S}_{\mbox{in}}  = 16\pi^2\frac{\dot{H}}{H^3}\left(1+\frac{\dot{H}}{2H^2+\dot{H}} \right).
\end{align}

Therefore rate of change of the total entropy is,
\begin{align}\label{ent-rate22}
\dot{S}_{\mbox{tot}}&=\dot{S}_{\mbox{h}}+\dot{S}_{\mbox{in}} \nonumber\\
&= 16\pi^2\frac{\dot{H}^2}{H^3}\left(\frac{1}{2H^2+\dot{H}}\right).
\end{align}
 
Therefore we can find $\dot{S}$ for different cosmological models if we know the expression of scale factor $a(t)$.

\subsection{Thermodynamic Stability}\label{Stability}

  In a hydrodynamic system, a stable equilibrium can be achieved by minimizing the thermodynamic potentials i.e., the internal energy, enthalpy, Helmholtz free energy, and Gibbs free energy. On the other hand, the entropy principle, which says that in any process total entropy is a non-decreasing function, leads to an increase in entropy as the system approaches equilibrium, characterized by the fields becoming time-independent. Therefore, to achieve thermal equilibrium, total entropy is to be maximized. In other words, entropy must be a concave function if the system is to be in stable equilibrium \cite{callen1998thermodynamics,kubo1968thermodynamics, carter2000classical,Muller1985}. To achieve concavity, the hessian matrix defined as, 
  
  \begin{equation}
W=
\begin{bmatrix}
\frac{\partial^2{S_{\mbox{in}}}}{\partial U^2} & \frac{\partial^2{S_{\mbox{in}}}}{\partial U\partial V}\\ \vspace{0.005mm}\\
\frac{\partial^2{S_{\mbox{in}}}}{\partial V\partial U} & \frac{\partial^2{S_{\mbox{in}}}}{\partial V^2}
\end{bmatrix},
\end{equation}    
has to be semi-negative definite. Therefore, all the $k^{\text{th}}$ order principle minors of the matrix $W$ are $\leq$ 0 if $k$ is odd and $\geq$ 0 if $k$ is even. Hence, the thermodynamic stability requires that the conditions
\begin{align} 
(i)\frac{\partial^2{S_{\mbox{in}}}}{\partial U^2}\leq 0, \label{condin1} \\
(ii)\frac{\partial^2{S_{\mbox{in}}}}{\partial U^2}\frac{\partial^2{S_{\mbox{in}}}}{\partial V^2}- \left(\frac{\partial^2{S_{\mbox{in}}}}{\partial U\partial V}\right)^2\geq 0, \label{condin2}
\end{align}

are satisfied together. \\

Using this we explore the thermodynamic stability of a cosmological model that mimics a $\Lambda$CDM model. We also find that the transition from the decelerated to the accelerated expansion of the universe is a second-order thermodynamic phase transition for the matter content of the universe.

%% file: Chapters/Chapter2.tex
\chapter{Thermodynamics of Thawing and Freezing Quintessence Models} 
\label{Chapter2}
\chaptermark{Therm. of Thawing and Freezing Quintessence Models}
\section{Introduction:}
In chapter \ref{Chapter1}, we discussed different dark energy models. In this chapter, we shall consider the two broad classes of quintessence fields as dark energy models and try to understand their thermodynamic behaviour in the context of the Generalized Second Law of Thermodynamics (GSL).

The quintessence can be broadly categorized into two main groups known as \textit{thawing} and \textit{freezing} models. The behaviour of these models is distinguished by the way the effective equation of state parameter ($w$), evolves over time.
 The thawing model exhibits an effective EoS parameter ($w$) that initially behaves as almost constant and close to $-1$. However, as the universe evolves, $w$ undergoes a transformation, transitioning into an evolving state. 
 On the other hand, the freezing model behaves differently. In this model, from an evolving state at the initial stage, $w$  \textit{freeze} at a constant value $-1$ in late-time evolution. Among the freezing models, one particularly interesting group is known as the \textit{trackers}.
To get a concise and comprehensible overview of these models refer to \cite{Scherrer:2007pu}. 

The comparative studies of these models have been done concerning how well they align with observational data \cite{Thakur:2012rp} and from the viewpoint of stability \cite{Chakraborty:2018nrk}. The cluster number count comparison of these models has been studied \cite{Devi:2014rva}. However the outcome does not provide solid evidence to favor any of these models. In this chapter, we have done comparative studies of thawing and freezing models from the perspective of the generalized second law(GSL) of thermodynamics.

We employ a straightforward definition of freezing and thawing models and graph the rate of total entropy change during evolution. The expectation is that this rate remains positive, as per the GSL where total entropy never decreases. Surprisingly, the outcomes demonstrate that both freezing and thawing models encounter a violation of the GSL. We conducted tests using a pure quintessence, followed by a quintessence along with a cold dark matter (CDM), yielding similar results, highlighting the inherent thermodynamic non-compliance of the quintessence field. However, freezing models possess an advantage over thawing models, as the GSL breakdown occurs significantly further back in the past.

\section{Quintessence Models:}

\begin{itemize}
\item \textbf{\textit{Action and Stress-Energy Tensor:}}
We consider that the universe consists of a perfect fluid and the dark energy contribution comes from a scalar field $\Phi$ with potential $V(\Phi)$. Therefore, the action takes the form,
\begin{equation} \label{Qaction}
S^{field} = \int \mathrm{d}^4 x \sqrt{-g}\Big[ \frac{R}{16\pi G}+\mathcal{L}_m-\frac{g^{\mu\nu}}{2} \mathrm{\partial} _\mu \Phi \mathrm{\partial} _\nu \Phi-V(\Phi)\Big],
\end{equation} 

The Einstein field equation takes the form, 
\begin{equation}
G_{\mu\nu} = 8\pi G \left(T^{(m)} _{\mu\nu}+T^{(q)} _{\mu\nu}\right).
\end{equation}

Throughout this chapter, quantities with the superscripts $m$ and $q$ denote the quantities related to matter and the quintessence field respectively.

Stress-energy tensor of the matter part is given by the equation,
\begin{equation} \label{em-f}
 T^{(m)} _{\mu\nu}= (\rho+p)u_\mu u_\nu+pg_{\mu\nu}, 
\end{equation}
where $\rho, p$ are the density and pressure of the fluid respectively.
Stress-energy tensor of the quintessence field is given by the equation,
\begin{equation}\label{em-q}
T_{\mu\nu}^{(q)} = \Big[\mathrm{\partial} _\mu \Phi \mathrm{\partial} _\nu \Phi-\frac{1}{2}g_{\mu\nu}g^{\alpha\beta} \mathrm{\partial} _\alpha \Phi \mathrm{\partial} _\beta \Phi - 2g_{\mu\nu} V(\Phi)\Big].
\end{equation} 

\item \textbf{\textit{Metric and Field equations:}}

According to the \textit{cosmological principle}, the universe is spatially isotropic, homogeneous in the large-scale. Here we assume that the universe is spatially flat. Therefore, we consider the  spacetime metric is given by  spatially flat (k=0) Friedmann-Robertson-Walker metric(FRW metric),
\begin{equation} \label{metric}
\mathrm{d}s^2 = -\mathrm{d}t^2+{a(t)}^2 [\mathrm{d}r^2+r^2 \mathrm{d}\theta^2+r^2\sin^2\theta \mathrm{d}\phi^2].
\end{equation}
Here $\Phi$ is solely a function of cosmic time $t$.
Now the field equations can be explicitly written as, 
\begin{equation}\label{fe1}
 3 H^2 = 3 \Big(\frac{\dot{a}}{a}\Big)^2 = 8\pi G (\rho + \rho_\Phi),
\end{equation}

\begin{equation}\label{fe2}
 2 \frac{\ddot{a}}{a} + \Big(\frac{\dot{a}}{a}\Big)^2 = - 8\pi G (p + p_\Phi).
\end{equation}

In equations \eqref{fe1} and \eqref{fe2}, 
$\rho_\Phi, p_\Phi$ represent energy density and pressure due to the quintessence field. 

\begin{equation}\label{q-den2}
\rho_\Phi \equiv T^{0(q)}_0= \frac{1}{2}(\dot\Phi)^2+V(\Phi),
\end{equation} 
and
\begin{equation}\label{q-press2}
p_\Phi \equiv T^{1(q)}_1= T^{2(q)}_2=T^{3(q)}_3=\frac{1}{2}(\dot\Phi)^2-V(\Phi).
\end{equation}

 The conservation of the energy-momentum tensors of the cosmic fluid leads to, 

\begin{equation}\label{mat-cons}
 \dot{\rho} + 3H(\rho + p) = 0,
\end{equation}

and the conservation related to the quintessence field leads to the Klein-Gordon (KG) equation for $\Phi$ as follows,

\begin{equation}\label{kg-phi2}
\ddot{\Phi}+3H\dot{\Phi}+V^\prime(\Phi)=0.
\end{equation} 
  These equations, \eqref{mat-cons} and \eqref{kg-phi2}, are not independent from the field equations \eqref{fe1}and \eqref{fe2}. One of the equations \eqref{mat-cons} and \eqref{kg-phi2}, along with equations \eqref{fe1}and \eqref{fe2}, will yield the other because of Bianchi identity.
\end{itemize}

\section{Generalized Second Law of Thermodynamics:}
According to Generalized Second Law (GSL) of thermodynamics, the total entropy of the universe, i.e., sum  of the horizon entropy and entropy of the fluid inside the horizon does not decrease with time \cite{Bekenstein:1972tm,Bekenstein:1973ur,Bekenstein:1974ax}.
In mathematical expression, 
\begin{align}
\dot{S}_{tot}=\dot{S}_h+\dot{S}_{in}>0,
\end{align}
where total entropy is denoted by $S_{tot}$, entropy of the horizon and fluid inside the horizon is denoted respectively by $S_h$ and $S_{in}$. An overhead dot indicates derivative with respect to the cosmic time $t$.

When studying the kinematics of the universe, it is more sensible to focus on the entropy of the dynamic apparent horizon instead of the teleological event horizon. We have already discussed about the apparent horizon in a concise manner in chapter \ref{Chapterthermo}. The following equation represents the entropy of the apparent horizon,
\begin{equation}
S_h=\frac{A}{4G},
\end{equation}
where $A = 4\pi R_h^2$ is the area of the apparent horizon. In section \ref{Apparenthorizon}, we have computed the apparent horizon radius $R_h$. In case a spatially flat FRW spacetime (putting $k=0$), we get,
\begin{equation}\label{hor-rad}
R_h = \frac{1}{H}.
\end{equation} 
Hence, the rate at which the horizon entropy changes is,
\begin{align} \label{hor-entropy}
\dot{S}_h =-\frac{2\pi}{G}\Big(\frac{\dot{H}}{H^3}\Big).
\end{align}
The mathematical equation that expresses the 1st law of thermodynamics as applied to the matter content by the horizon is,
\begin{equation} \label{gibbs}
T_{in}\mathrm{d}S_{in} = \mathrm{d}E_{in}+p_{tot} \mathrm{d}V_h,
\end{equation}
where $V_h$ represents the volume and heres $V_h=\frac{4}{3}\pi R_h^3$.
From the above equation\eqref{gibbs}, we can write the rate at which the entropy of the fluid within the horizon changes as, 
\begin{equation}\label{s-matter-dot}
\dot{S}_{in}=\frac{1}{T_{in}} [(\rho_{tot} + p_{tot} )\dot{V}_h+\dot{\rho}_{tot} V_h],
\end{equation} 
where $\rho_{tot}$ and $p_{tot}$ denotes respectively the total energy density and pressure.
We assume that the fluid within the horizon is in a state of thermal equilibrium with the horizon. Therefore, the temperature ($T_{in}$) will be equivalent to the apparent horizon temperature ($T_h$).
Here, we have considered the \textbf{\textit{Hayward-Kodama temperature}} as the temperature of the dynamic apparent horizon \cite{faraoni2015horizons, Helou:2015yqa,Rani:2018kkx,DiCriscienzo:2007pcr, Rong-GenCai_2005}. The reason is discussed in chapter \ref{Chapterthermo}. The temperature of the dynamic apparent horizon is,
\begin{equation} \label{hor-temp}
T_{h} = \frac{2H^2+\dot{H}}{4\pi H}.
\end{equation}

Now putting the temperature \eqref{hor-temp} in the equation \eqref{s-matter-dot}, we obtain,
\begin{align}\label{s-matter-dot2}
\dot{S}_{in} = \frac{(\rho_\Phi+p_\Phi)}{T_{in}}4\pi R_h^2\big[\dot{R_h}-HR_h\big].
\end{align}
 This equation can further be simplified using the fields equations\eqref{fe1}and\eqref{fe2} and written as, 
\begin{equation}\label{s-matter-dot3}
\dot{S}_{in} =\frac{2\pi}{G}\Big(\frac{\dot{H}}{H^3}\Big)\Big(1+\frac{\dot{H}}{2H^2+\dot{H}}\Big).
\end{equation}

Summing up the equations \eqref{hor-entropy} and \eqref{s-matter-dot3}, we can express the rate of change of the total entropy as,
\begin{equation}\label{s-tot-dot}
\dot{S}_{tot} = \dot{S}_h + \dot{S}_{in} = \frac{2\pi}{G}\Big(\frac{\dot{H}^2}{H^3}\Big)\Big(\frac{1}{2H^2+\dot{H}}\Big).
\end{equation}

\section{A Pure Quintessence:}
At first, only pure quintessence model is taken into account. This implies that the fluid inside the horizon is considered to be the quintessence field only, devoid of baryonic and dark matter($p=\rho=0$). Therefore there are no contributions of energy density and pressure of these matters in the field equation \eqref{fe1}and\eqref{fe2}.

At this point, we have two equations to determine three unknowns: $a, \phi, V$, since the Klein-Gordon (KG) equation is not an independent equation. To complete the system of equations, we adopt the ansatz for $\rho$ proposed by Carvalho {\it et al} \cite{Carvalho:2006fy}. The ansatz says the divergence of the logarithm of energy density is power law dependent on scale factor, 
\begin{equation} \label{rhophi-ansatz}
\frac{1}{\rho_\Phi}\frac{\mathrm{\partial}\rho_\Phi}{\mathrm{\partial} a} = -\frac{\lambda}{a^{1-2\alpha}},
\end{equation}
where $\lambda$ and $\alpha$ are parameters. $\lambda$ is chosen to be positive, but the parameter $\alpha$ can take both positive or negative values. Integrating the above equation, we obtain,

 \begin{equation}\label{rhophi}
  \rho_\Phi(a)= {\rho_{\Phi,0}} \exp\left[-\frac{\lambda}{2\alpha}(a^{2\alpha}-1)\right],
 \end{equation}
where the present value of the scale factor is taken to be 1. Since $\lambda$ is positive, the energy density of the quintessence field decreases with the evolution. The quintessence energy density reduces to a power law, $\rho_\Phi(a)\propto a^{-\lambda}$ in the limit $\alpha\rightarrow 0 $. Using equation \eqref{kg-phi2}, the potential $V$ is obtained as (for more details see \cite{Carvalho:2006fy}),

\begin{equation}\label{potential}
V(\Phi) = \left[1-\frac{\lambda}{6}(1+\alpha\sqrt{\sigma}\Phi)^2\right]\rho_{\Phi,0}\exp\left[-\lambda\sqrt{\sigma}(\Phi+\frac{\alpha\sqrt{\sigma}\Phi^2}{2})\right].
\end{equation}

The equation of state parameter (EoS) for the scalar-field is defined by $w_\Phi \equiv \frac{p_\Phi}{\rho_\Phi}$. In this model, we get $w_\Phi$ in terms of scale factor as, 
\begin{equation}\label{14}
w_\Phi=-1+\frac{\lambda}{3}a^{2\alpha}.
\end{equation}
 The above EoS is not time-independent quantity (for $ \alpha\neq 0$). The dependency of time in $w_\Phi$ comes through the scale factor $a$. It is to be noticed that, it reduces to a constant, $w_\Phi = -1+\frac{\lambda}{3}$ in the limit $\alpha \rightarrow0$. \\

\begin{figure}
\centering     
\subfigure[]{\label{fig:1a}\boxed{\includegraphics[width=82mm]{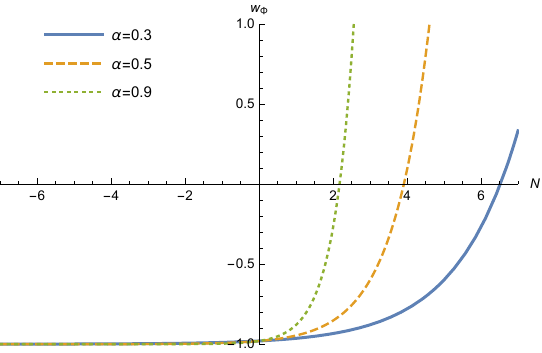}}}
\subfigure[]{\label{fig:1b}\boxed{\includegraphics[width=82mm]{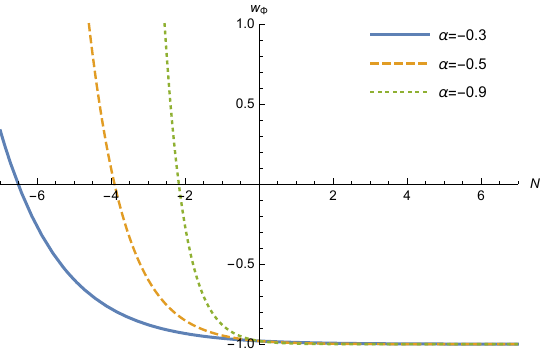}}}
\caption{$w_\Phi$ is plotted as a function of $N$. (a) For positive values of $\alpha$, EoS increases from $~-1$, i.e., it exhibits a thawing behaviour. (b) For negative values of $\alpha$, EoS $>-1$ decreases to a more negative value, i.e., it exhibits a freezing behaviour.}
\end{figure}

Now let us see the behaviour of evolution of $w_\Phi$ pictorially. We define a function $N$ as $N=\ln(\frac{a}{a_0})$, and plot $w_\Phi$ against $N$ (fig:\ref{fig:1a} \& \ref{fig:1b}). From the definition we see that, here $N=0$  implies the present epoch, positive and negative $N$ imply the future and past respectively. In the fig:\ref{fig:1a}, we plot $w_\Phi$ vs $N$ for positive  values of the parameter $\alpha$, which shows \textit{thawing} behaviour. And \textit{freezing} behaviour is shown in the fig:\ref{fig:1b}, where we plot $w_\Phi$ vs $N$ for positive  values of the parameter $\alpha$. 
Since $\lambda$ is positive, $w_\Phi\geqslant -1$ for all values of $a$ irrespective of the sign of $\alpha$ \cite{Carvalho:2006fy}. \\ 
 The equation (\ref{rhophi-ansatz}) serves as a comprehensive ansatz, encompassing both thawing and freezing behaviours through the parameter $\alpha$. For $\alpha > 0$, the scalar field exhibits a thawing behaviour, where the equation of state parameter $w_{\Phi}$ starts with a nearly flat value close to $-1$ in the past and gradually transitions to less negative values. Conversely, for $\alpha < 0$, the scalar field demonstrates a freezing behaviour, wherein $w_{\Phi}$ decreases towards more and more negative values and eventually settles into a plateau near $-1$ in the future.

It is essential to clarify that we have adopted the ansatz (\ref{rhophi-ansatz}) from the work by Carvalho {\it et al} \cite{Carvalho:2006fy}, and we have employed the parameter values $\alpha, \lambda$, and others from the reference \cite{Devi:2014rva}. 

Now, substituting $\rho_\Phi$ from equation \eqref{rhophi} in the Friedmann equation \eqref{fe1} with $\rho =0$ we obtain the solution for the scale factor as,
\begin{equation}\label{solution-1}
\frac{1}{2\alpha}\big[\Gamma(0,-\frac{\lambda}{2\alpha})-\Gamma(0,-\frac{\lambda}{4\alpha}a^{2\alpha}) \big] =\gamma (t-t_0),
\end{equation}

where $\Gamma(a,x)$ is the well-known upper incomplete gamma function and is defined as, $\Gamma(a,x)=\int^\infty_x z^{a-1}\exp(-z)dz$. And $\gamma=\sqrt{\frac{8\pi G }{3}\rho_{\Phi,0}}\exp\big(\frac{\lambda}{4\alpha}\big)$ is a constant term. While the solution may be intricate, it serves the purpose of assessing the thermodynamic feasibility of the model. \\
By employing the solution (\ref{solution-1}), it becomes possible to express $H$ and $\dot{H}$ in terms of $a$. Consequently, the equation (\ref{s-tot-dot}) takes the form,
\begin{equation}\label{s-tot-dot1}
\dot{S}_{tot} =\frac{\pi \lambda^2}{G}\Big(\frac{1}{\sqrt{\frac{8\pi G}{3}\rho_{\Phi,0}}\exp\big(-\frac{\lambda}{4\alpha}(a^{2\alpha}-1)\big)}\Big)\Big(\frac{a^{4\alpha}}{4-\lambda a^{2\alpha}}\Big).
\end{equation}
In order to assess the thermodynamic feasibility of the model, we proceed to plot $\dot{S}_{\text{tot}}$ as a function of the cosmic e-folding factor $N$.
\begin{figure}[h!]
\centering     
\subfigure[]{\label{fig:2a}\boxed{\includegraphics[width=82mm]{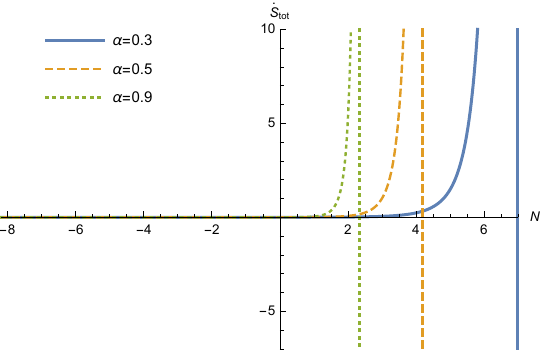}}}
\subfigure[]{\label{fig:2b}\boxed{\includegraphics[width=82mm]{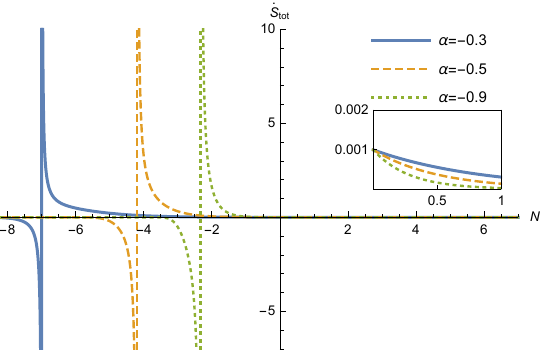}}}
\caption{ $\dot{S}_{tot}$ is plotted as a function of $N$. (a) in thawing scenario (for  $\alpha=0.3, 0.5$ and $0.9$) (b) in freezing scenario (for $\alpha=-0.3, -0.5$ and $-0.9$ )}
\end{figure}

As shown in Fig-\ref{fig:2a}, in the case of thawing quintessence ($\alpha > 0$), the total entropy experiences an initial increase up to a certain time, after which it deviates from obeying the Generalized Second Law (GSL). Notably, $\dot{S_{\text{tot}}}$ rises sharply to an infinitely large value and then abruptly drops to an infinitely large negative value. This pattern is similar for all permissible values of $\alpha$, with only the range of $N$ indicating the onset of this abnormal behaviour varies. On the other hand, freezing quintessence ($\alpha < 0$) demonstrates an opposing behaviour. It satisfies the GSL for the future, as the net entropy increases and eventually stabilizes at a constant value when $\dot{S_{\text{tot}}}$ approaches zero (as observed in Fig-\ref{fig:2b}). However, this convergence is not as rapid as depicted in the figure but is likely to occur asymptotically, as indicated by the zoomed-in version in the inset. Nevertheless, the model fails to comply with GSL in the past. A discontinuity in $\dot{S_{\text{tot}}}$ is evident, and it takes on negative values, signifying a decrease in $S_{\text{tot}}$ during that period.\\
The parameter $\lambda$, which governs the rate at which the dark energy density $\rho_\Phi$ declines (as shown in equation \ref{rhophi-ansatz}), possesses a small positive value as indicated in a prior study \cite{Devi:2014rva}. Throughout all the figures, we maintain a consistent value of $\lambda = 0.06$. We have also tested the model using significantly lower ($\lambda = 0.01$) and higher ($\lambda = 0.1$) values. However, since these variations do not significantly impact the qualitative characteristics, we have omitted them to avoid redundant information.

\section{Quintessence with cold dark matter:}

In this section, we consider a more realistic model, incorporating a pressureless fluid ($p=0$) comprising baryonic matter and cold dark matter in addition to the quintessence matter. Consequently, the field equations (\ref{fe1}, \ref{fe2}) will have the input $p=0$. 

Directly integrating equation (\ref{mat-cons}) leads to,
\begin{equation}\label{rho-mat}
 \rho =\frac{\rho_{0}}{a^3},
 \end{equation}
 where $\rho_{0}$ is the energy density of matter at the present epoch. Since the Klein-Gordon equation is not an independent equation, we are left with two equations (\ref{fe1}, \ref{fe2}), to solve for three variables: $a, \Phi, V$. To close the system of equations, we employ the same ansatz \eqref{rhophi-ansatz} as used in the previous section.

\begin{figure}[h!]
\centering     
\subfigure[]{\label{fig:3a}\boxed{\includegraphics[width=82mm]{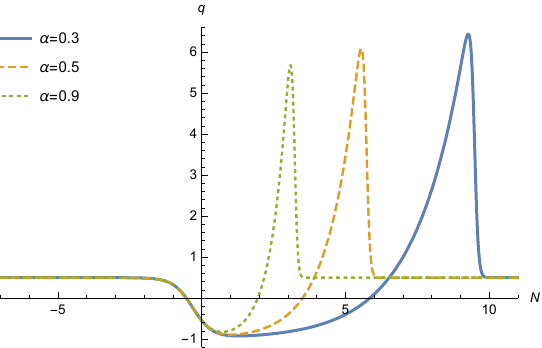}}}
\subfigure[]{\label{fig:3b}\boxed{\includegraphics[width=82mm]{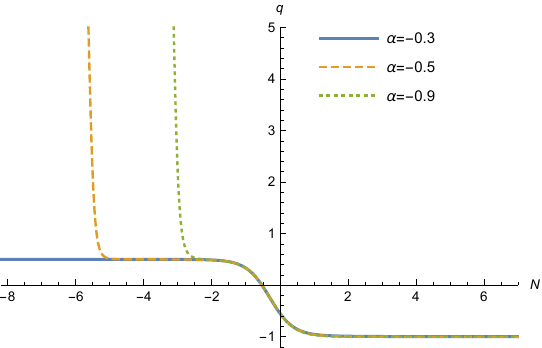}}}
\caption{$q$ is plotted as a function of $N$ . (a)  In Thawing scenario, transient acceleration is a possibility,(b) Freezing quintessence leads the universe into an eternal accelerating phase. }
\end{figure}

The expression for the deceleration parameter $q$, defined by $q\equiv-\frac{\dot{H}+H^2}{H^2}$ is given by,
\begin{align} \label{dec-par}
q(a)= \frac{3(-1+\frac{\lambda}{3}a^{2\alpha})+1+\frac{\frac{\rho_{m,0}}{a^3}}{{\rho_{\Phi,0}} \exp\big[-\frac{\lambda}{2\alpha}(a^{2\alpha}-1)\big]}}{2(1+\frac{\frac{\rho_{m,0}}{a^3}}{{\rho_{\Phi,0}} \exp\big[-\frac{\lambda}{2\alpha}(a^{2\alpha}-1)\big]})}.
\end{align}
Fig-\ref{fig:3a} reveals an intriguing possibility for a thawing model that the current accelerated expansion of the universe may only be a temporary phenomenon. According to the plot, the universe is expected to transit back to a decelerated phase in a finite future. This characteristic has been observed previously such as by Carvalho {\it et al} \cite{Carvalho:2006fy} and Devi {\it et al} \cite{Devi:2014rva}. However, in the freezing model, the universe eventually settles into a phase of accelerated expansion after passing through the decelerated phase.\\

It is important to note that these features have already been well-documented in the literature. However, the primary objective of this study is to assess the thermodynamic viability of the model.

 In this analysis, we assume that the fluid inside the horizon is in thermal equilibrium with the apparent horizon. To evaluate the rate of change of the total entropy, we use the Hayward-Kodama temperature, as given in equation (\ref{hor-temp}), following the same approach as in the previous section. Now, the total energy density is represented as $\rho_{\text{tot}} = \rho + \rho_\Phi$. The general form of the rate of change of the total entropy remains unchanged, as expressed in equation \eqref{s-tot-dot}. However, due to the modification of the Friedmann equations for the contribution of cold dark matter, the explicit form of ${\dot{S}}_{\text{tot}}$ can be obtained as,
 \begin{align}\label{s-tot-dot3}
\dot{S}_{tot} &=\frac{3\sqrt{3\pi}}{2\sqrt{2}G^{3/2}}\frac{{\big[\rho_{\Phi,0}\exp[-\frac{\lambda}{2\alpha}(a^{2\alpha}-1)](\frac{\lambda}{3}a^{2\alpha})+\frac{\rho_{m,0}}{a^3}\big]}^2}{{\big[\rho_{\Phi,0}\exp[-\frac{\lambda}{2\alpha}(a^{2\alpha}-1)]+\frac{\rho_{m,0}}{a^3}\big]}^{3/2}} \nonumber \\
&\times \frac{1}{\big[\rho_{\Phi,0}\exp[-\frac{\lambda}{2\alpha}(a^{2\alpha}-1)](4-\lambda a^{2\alpha})+\frac{\rho_{m,0}}{a^3}\big]}.
\end{align}

The qualitative characteristics remain similar to those observed in the case of pure quintessence. As demonstrated in Figure 4(a), the thawing models ($\alpha > 0$) will eventually violate the Generalized Second Law (GSL) in some finite future, since $\dot{S}_{\text{tot}}$ becomes negative at a certain value of $N$, indicating a decrease in entropy. On the other hand, the freezing models ($\alpha < 0$) adhere to GSL in the future. Despite gradually decreasing, $\dot{S}{\text{tot}}$ remains positive (as seen in Figure 4(b)), signifying that the entropy is continuously increasing and will eventually stabilize at a constant value in the future, as illustrated in Figure 4(b).

However, there is a concern in connection with the past evolution. The rate of entropy change becomes negative at a certain finite time in the past. Nonetheless, this issue can be mitigated by adjusting the parameter $\alpha$. For instance, by choosing $\alpha = -0.3$, the discrepancy is observed at $z \sim 10^{4}$ (see Figure 5(a)), which occurs before the onset of matter domination over radiation, rendering this system of equations inapplicable to describe the dynamics of the universe at that stage. Similarly, for $\alpha = -0.1$, this discrepancy is observed at $z \sim 10^{12}$, which is far beyond the regime of quintessence along with CDM (see Figure 5(b)).\\
Upon careful inspection of Eq. \eqref{s-tot-dot}, it becomes evident that the term $2H^2+\dot{H}$ determines the thermodynamic viability of the models. If $\dot{H} + 2H^2 < 0$ (i.e., $q \geq 1$), the model fails to meet the thermodynamic requirements.

Thus, it is evident that freezing models exhibit stronger thermodynamic viability, at least within the relevant period.\\
Similar to the previous section, in this analysis, we have maintained the same value of $\lambda = 0.06$ for the figures. The values of $\lambda = 0.01$ and $0.1$ have been excluded since they do not alter the qualitative aspects. The only noticeable difference is a slight shift in epochs, such as the onset of GSL violation for different values of $\lambda$.

\begin{figure}[h!]
\centering     
\subfigure[]{\label{fig:4a}\boxed{\includegraphics[width=82mm]{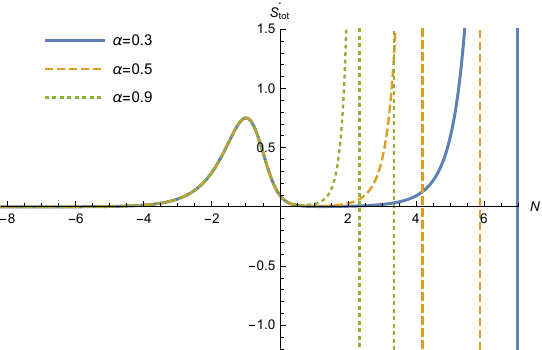}}}
\subfigure[]{\label{fig:4b}\boxed{\includegraphics[width=82mm]{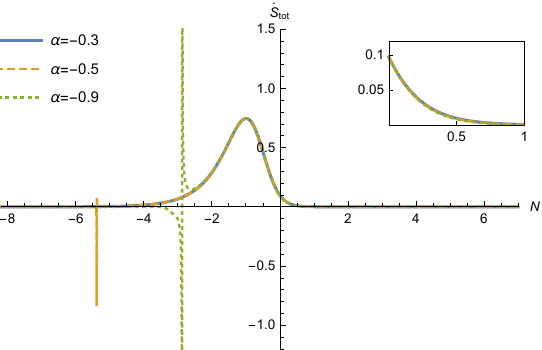}}}
\caption{$\dot{S}_{tot}$ is plotted as a function of $N$, (a) in thawing scenario (for  $\alpha=0.3 , 0.5$ and $0.9$) (b) in freezing scenario (for $\alpha=-0.3, -0.5$ and $-0.9$. ) }
\end{figure}
\begin{figure}[h!]
\centering     
\subfigure[]{\label{fig:5a}\boxed{\includegraphics[width=82 mm]{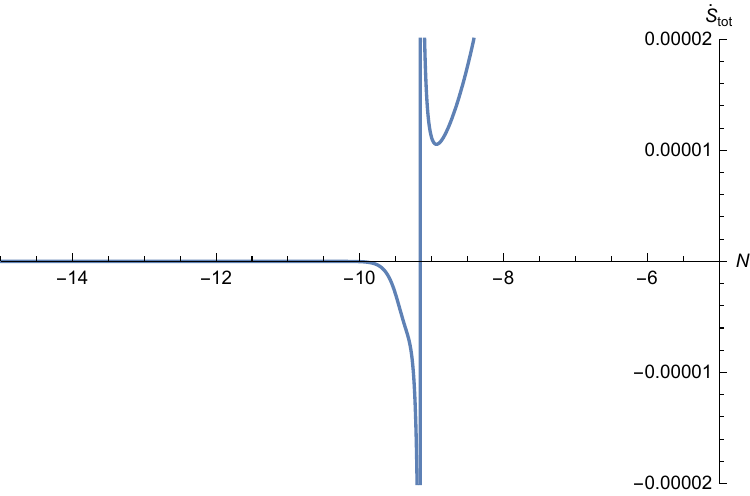}}}
\subfigure[]{\label{fig:5b}\boxed{\includegraphics[width=82 mm]{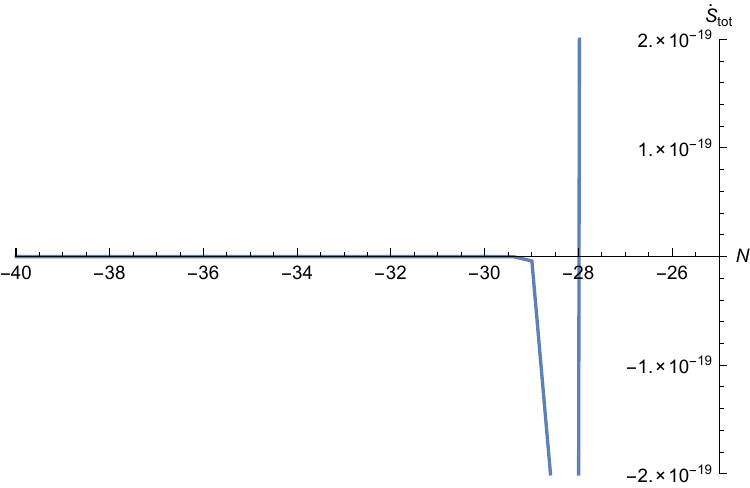}}}
\caption{$\dot{S}_{tot}$ is plotted as a function of N in Freezing scenario for (a) $\alpha=-0.3$ and (b) $\alpha=-0.1$. }
\end{figure}

\section{Summary and Discussion:}

In this study, we conduct a comparison between thawing and freezing models regarding their adherence to the Generalized Second Law (GSL) of thermodynamics. To assess the total entropy ($S_{\text{tot}}$), we combine the horizon entropy with the entropy of the matter enclosed within the horizon. We adopt a simple ansatz \cite{Carvalho:2006fy} to model the evolution of the energy density of the quintessence field. By doing so, we can readily identify the range of parameter values ($\alpha$) that correspond to the thawing and freezing behaviour of the field.\\

We find that both types of models exhibit an incompatibility with GSL. There are instances where the entropy ($S$) actually decreases and does so at a rapid pace. For the freezing models, this breakdown in GSL can occur in a distant past, corresponding to a redshift of $z \sim 10^{4}$. At such a distant past, a quintessence model in conjunction with cold dark matter does not adequately account for the evolution of the universe, and a dominant contribution from a radiation distribution would be necessary. Hence, this breakdown of GSL may not hold true in such a scenario.\\

On the other hand, for the thawing models, this pathological breakdown of GSL is predicted to happen in a finite future. 

Thus the major implication here is that the freezing models seem to be more favorable compared to the thawing models from the perspective of thermodynamic viability.

%% file: Chapters/Chapter3.tex
\chapter{Thermodynamics of Brans-Dicke Cosmology} 
\label{Chapter3}
\section{Introduction:}
In this chapter, we discuss the thermodynamic feasibility, especially the viability of GSL in Brans-Dicke Cosmology.
As we have discussed in chapter \ref{Chapter1}, the Brans-Dicke theory (BDT) of gravity \cite{Brans:1961sx} is widely recognized and frequently discussed due to its potential in addressing a lot of cosmological issues. Two notable examples include extended inflation \cite{Mathiazhagan:1984vi, La:1989za}, which solves the problem of graceful exit in the inflationary scenario, and the potential for driving the late-time acceleration of the universe even in the absence of dark energy \cite{Banerjee:2000mj}. However, BDT has its limitations. One limitation is that the characteristic coupling constant of the theory, denoted as $\omega$, needs to have a very high value to be consistent with local astronomical tests \cite{will_1993}. On the other hand, for successfully addressing cosmological problems, a small value of $\omega$ is required. Despite these challenges, the theory continues to attract attention due to its relevance in various real cosmological problems and its formal resemblance to other gravity theories with nonminimal coupling. To make the equations more manageable, we examine them in what is known as the Einstein frame, achieved through a suitable conformal transformation  \cite{Dicke:1961gz}. This results in a scenario where the evolution of matter distribution becomes intertwined with the BD scalar field. Consequently, the Bianchi identity enforces the conservation of both the BD field and the matter distribution as a whole. This characteristic aligns with the notion that there is no inherent reason to assume that dark matter and dark energy evolve independently, without any non-gravitational interaction between them \cite{Wang:2016lxa}. Some studies have already explored situations in which the BD scalar field interacts with matter \cite{Das:2005yg, Banerjee:2006rp}.\\

 The thermodynamic viability during the matter-dominated phase of the universe, encompassing the transition from decelerated to accelerated expansion is the primary motivation. However, for a comprehensive analysis, the radiation-dominated phase is also taken into consideration. A similar thermodynamic analysis of cosmological models was carried out by Bhattacharya and Debnath  \cite{Bhattacharya:2010iv}, albeit in a different context. They examined accelerated expansion in an extended version of BDT, introducing a potential and allowing the BD parameter $\omega$ to be a function of the scalar field. In contrast, the present study focuses on the original BDT, capable of driving accelerated expansion on its own. The results of this investigation are highly promising. Specifically, the models, especially those for a universe containing a pressureless fluid, align with exclusively negative values of the parameter $\omega$, which fall within the range where BDT can independently facilitate accelerated expansion \cite{Banerjee:2000mj}.

\section{Brans-Dicke Theory in Einstein frame}
\begin{itemize}
\item \textbf{\textit{Action:}}
The action in the BD theory in the Jordan frame is given by \cite{Brans:1961sx},
\begin{equation}\label{action1}
 \mathcal{S} = \frac{1}{16\pi G_0}\int\sqrt{-g}\left[\phi R-\omega\frac{\phi_{,\alpha}\phi^{,\alpha}}{\phi}+\mathcal{L}_M\right] d^4x.
\end{equation}
The action describes the dynamics of gravity and the Brans-Dicke scalar field, as well as their interaction with matter. 
In this equation, $\phi$ is the BD scalar field. The second term represents the kinetic term of the scalar field.
Now let us employ the following conformal transformation \cite{Dicke:1961gz},
\begin{equation}\label{conformal}
 \bar{g}_{\mu\nu} = \phi g_{\mu\nu}.
\end{equation}

Then the action takes the form, 
\begin{equation}\label{action2}
 \bar{\mathcal{S}} =\frac{1}{16\pi G_0} \int \sqrt{-\bar{g}}\left[\phi_0\left(\bar{R}-\frac{2\omega+3}{2}\psi_{,\alpha}\psi_{,\beta}\bar{g}^{\alpha\beta}\right)+\bar{\mathcal{L}_M}\right]d^4x ,
\end{equation} 
where $\psi = \ln (\frac{\phi}{\phi_0})$ and $\phi_0$ is a constant. The overhead bar will be omitted henceforth, as the rest of the work is in the revised frame.

In this modified version, commonly referred to as the \textit{Einstein frame}, the field equations take on a notably simplified appearance as follows,
\begin{align}
G_{\alpha\beta} = T_{\alpha\beta}+\frac{2\omega+3}{2}(\psi_{,\alpha}\psi_{,\beta}-\frac{1}{2}g_{\alpha\beta}\psi^{,\mu}\psi_{,\mu}).
\end{align}
This is expressed using units where the quantity $8\pi G_0$ is set to unity. This conformal transformation reconfigures the mathematical representation of the theory, leading to a formulation in which the complexity of the field equations is significantly reduced. As a result, the Einstein frame becomes a preferred framework for analysis due to its enhanced mathematical tractability.

The equation governing the evolution of the scalar field $\psi$, derived by varying the action (\ref{action2}) with respect to $\psi$, results in the expression,
\begin{align}
\Box \psi =\frac{T}{2\omega+3},
\end{align}
where $T$ represents the trace of the energy-momentum tensor for the matter sector. Let us now consider that the universe is filled with perfect fluid. Therefore, the energy-momentum tensor is given by the equation \eqref{stresstensor}. For the sake of clarity, lets rewrite it down here,
\begin{align}\label{B3}
T_{\alpha\beta}= (\rho+p)u_\alpha u_\beta +p g_{\alpha\beta},
\end{align}
where $\rho, p$ are the energy density and pressure of the fluid respectively and $u_{\alpha}$ is the unit timelike vector,  $u^\mu u_\mu =-1$. In a comoving coordinate system, $v^\mu =\delta^\mu_0$.

It is important to note that $\rho$ and $p$ are presented in the revised units. 

\item \textbf{\textit{The Metric and The Field Equations:}}
  We consider a a spatially flat, homogeneous and isotropic FRW cosmology. The metric is given by,
\begin{equation}\label{metric2}
\mathrm{d}s^2 = -\mathrm{d}t^2+{a(t)}^2 [\mathrm{d}x^2+\mathrm{d}y^2+\mathrm{d}z^2].
\end{equation}

Here $a$ is the scale factor that characterizes the change in the size of the universe over time.
The form that the Einstein-Brans-Dicke field equations assume in the version obtained through conformal transformation is as follows, as outlined by Dicke \cite{Dicke:1961gz},
\begin{align}\label{B4}
 3\left(\frac{\dot{a}}{a}\right)^2 = \rho+\rho_\psi ,
\end{align}
\begin{align}\label{B5}
 2\left(\frac{\ddot{a}}{a}\right)+\left(\frac{\dot{a}}{a}\right)^2 = -p-p_\psi.
\end{align}

In these equations, the energy density and pressure of the scalar field is represented by $\rho_\psi$ and $p_\psi$. The connection between the energy density and pressure of the scalar field is expressed as,
$$\rho_\psi = p_\psi= \frac{2\omega+3}{4}\dot{\psi}^2.$$

The wave equation that describes the behavior and evolution of the scalar field $\psi$ is given by, 
 \begin{align}\label{B6}
 \ddot{\psi}+3 \frac{\dot{a}}{a}\dot{\psi}=\frac{-T}{2\omega+3}.
 \end{align}
 In this version, the equations take on a formal similarity to those describing a scenario with two types of matter components: one corresponds to a fluid, while the other represents a massless scalar field denoted as $\psi$. But the fact is that those do not evolve independently. Hence, we do not have separate conservation equations for these two matters. When considering both the fluid and scalar field together, the continuity equation assumes a form,
 \begin{align}\label{B7}
 \dot{\rho}_t= -3\frac{\dot{a}}{a}(\rho_t+p_t).
\end{align}
The subscript $t$ indicates the total quantity, which is composed of the combined contributions from the fluid component and the scalar field component. This equation is not self-standing; instead, it derives from the Bianchi identities. 
\end{itemize}

\section{Thermodynamic Quantities:}
In this chapter also, we check the viability of the model against the Generalised Second Law of Thermodynamics (GSL), according to which the total entropy of the universe can not decrease with time  \cite{Bekenstein:1972tm,Bekenstein:1973ur,Bekenstein:1974ax}. \\
We assume that the fluid achieves thermal equilibrium with the horizon.  In this scenario, the temperature of the fluid inside the horizon ($T_{in}$) is the same as the temperature of the dynamical apparent horizon ($T_h$). In the current section, we have adopted   Hayward-Kodama temperature \cite{Hayward:1997jp, Hayward:2008jq,DiCriscienzo:2009kun, faraoni2015horizons} as the temperature of the apparent horizon, which is expressed as,
\begin{equation} \label{T7}
T_{h} = \frac{2H^2+\dot{H}}{4\pi H}.
\end{equation}
In the unit of $8\pi G_0=1$, the entropy of the apparent horizon becomes \cite{DongsuBak_2000},
\begin{equation}\label{T2}
S_h=2\pi A.
\end{equation}

In flat FRW spacetime, the area $A$ is related to Hubble parameter  \cite{DongsuBak_2000} as,
\begin{equation}\label{T3}
A = \frac{4\pi}{H^2}.
\end{equation}

Therefore, rate of change in horizon entropy is 
\begin{align} \label{T4}
\dot{S}_h =-16\pi^2\frac{\dot{H}}{H^3}.
\end{align}

The rate of change of entropy of the fluid confined within the horizon can be expressed as,
\begin{equation}\label{T6}
\dot{S}_{in}=\frac{1}{T_{in}} [(\rho_t+p_t)\dot{V}_h+\dot{\rho}_t V_h],
\end{equation}
where in flat FRW spacetime the volume confined is, $V_h=\frac{4}{3}\pi \frac{1}{H^3}$.

By utilizing the equation labeled as Eq. \eqref{T7} along with the field equations provided in Eq. \eqref{T6}, we derive the expression for the rate of entropy change within the horizon as represented by the equation,
\begin{equation}\label{T9}
\dot{S}_{in} =16\pi^2 \frac{\dot{H}}{H^3}\left(1+\frac{\dot{H}}{2H^2+\dot{H}}\right).
\end{equation}
Combining the equations referenced as Eq. \eqref{T4} and Eq. \eqref{T9}, we arrive at the expression for the rate of total entropy change, which can be represented as,
\begin{equation}\label{T10}
\dot{S}_{t} =16\pi^2 \frac{\dot{H}^2}{H^3}\left(\frac{1}{2H^2+\dot{H}}\right).
\end{equation}
We have discussed this in detail in Chapter \ref{Chapterthermo}. Here we are rewriting these equations in the unit $8\pi G_0=1$, because throughout this chapter we have worked out everything in this unit.\vspace{0.5em}

In the following step, we will examine two distinct epochs in cosmic evolution, namely the radiation era and the dust era. We will proceed to analyze how the Generalized Second Law (GSL) of thermodynamics is upheld during these specific periods of the evolution of the universe.\\
It is worth noting that research by Mimoso and Diego \cite{Mimoso:2016jwg} highlights the difficulty of achieving thermal equilibrium between radiation and the cosmic horizon. This challenge stems from Wien's law, which consistently predicts a wavelength larger than the horizon radius at all times. However, there is a potential for nonrelativistic particles to reach equilibrium based on their masses.

\section{Radiation Era:}

In the scenario of radiation-dominated era, the equation of state of the fluid is characterized by the expression $p=\frac{1}{3}\rho$. As a result of this equation of state, the trace of the stress-energy tensor becomes null. As a consequence, the wave equation labeled as \eqref{B6} can be readily integrated, leading to the derivation of the subsequent relationship:
\begin{align}\label{r1}
\dot{\psi}= \frac{\alpha}{a^3},
\end{align}
where $\alpha$ is an integration constant.

Utilizing the field equations \eqref{B4} and \eqref{B5}, and incorporating the equation of state along with equation \eqref{r1}, it becomes possible to arrive at,

\begin{align}\label{r2}
\ddot{a}+ \frac{\dot{a}^2}{a}=-\frac{\beta}{a^5}, 
\end{align}
where $\beta$ is a constant, specifically defined as $\frac{2\omega+3}{12}\alpha^2$.

Upon integrating the aforementioned equation, we get,
\begin{align}\label{r3}
 \dot{a}^2= \frac{\sigma a^2+\beta}{a^4},
 \end{align}
 where $\sigma$ is the integration constant.
 
 The characteristics of the solutions to this equation exhibit variation contingent upon the positive or negative nature of both $\sigma$ and $\beta$. It is important to observe, as indicated in equation \eqref{r3}, that both $\sigma$ and $\beta$ cannot take on negative values concurrently.  
 
 \begin{itemize}
 \item \textbf{Case-I : when both $\sigma$ and $\beta$ positive}
 \end{itemize}
In this particular scenario, the solution to the wave equation \eqref{B6} can be expressed as follows,
\begin{align}\label{r4}
\psi+\psi_0 = \frac{\alpha}{2\sqrt{\beta}}\ln\left| \frac{\sqrt{a^2+\beta/\sigma}-\sqrt{\beta/\sigma}}{\sqrt{a^2+\beta/\sigma}+\sqrt{\beta/\sigma}} \right|.
\end{align}

By performing the integration of equation \eqref{r3}, we derive the connection that establishes the relationship between the scale factor and time as,
\begin{align}\label{r5}
t+t_0= \frac{1}{2\sqrt{\sigma}}a\sqrt{a^2+\beta/\sigma}-\frac{1}{2}\frac{\beta}{\sigma^{3/2}}\ln \left| a+\sqrt{a^2+\beta/\sigma}\right|.
\end{align}
 This equation is involved and it is difficult to write $a$ in terms of $t$ explicitly. A subscript zero, as usual, indicates the present value of the quantity.\\
These solutions for the scale factor have previously been established and documented in existing literature \cite{banerjee1985anisotropic, banerjee1985isotropic}. Our motive here is not to find the solution but to see how that goes with GSL. Upon substituting the expression of the Hubble parameter ($H$) and the second derivative of the scale factor ($\ddot{a}$) into the definition of the deceleration parameter, the resulting outcome is,
 \begin{align}\label{r6}
q = 1+\frac{\beta}{\beta+\sigma a^2} >0.
\end{align}
Consequently, this particular model gives rise to a universe that consistently experiences deceleration over time. This implies that the expansion of the universe slows down progressively as time advances.
By utilizing equations \eqref{T10},\eqref{r2}and\eqref{r3}, we derive the expression for the rate of total entropy change as follows,
\begin{align}\label{r8}
\dot{S}_{t}=-64\pi ^2 \frac{\sqrt{\sigma}}{\beta}\frac{a^3\left(a^2+3\beta/2\sigma\right)^2}{\left(a^2+\beta/\sigma\right)^{3/2}}.
\end{align}

Given that both $\beta$ and $\sigma$ possess positive values, it becomes evident that the rate of change of total entropy ($\dot{S}_{t}$) is negative. This scenario unequivocally indicates that the test of the Generalized Second Law (GSL) is not satisfied in this case. In essence, the observed decrease in total entropy contradicts the requirements of the GSL, highlighting a failure of the test under these conditions.

\begin{itemize}
\item \textbf{Case-II : $\sigma$ is positive but $\beta$ is negative}
\end{itemize}
When $\omega$ is less than -3/2, the value of $\beta$ becomes negative. We express this negative value of $\beta$ as $-\gamma^2$, where $\gamma$ is a real number. Consequently, equation \eqref{r3} transforms into the following form,
\begin{align}\label{r9}
\dot{a}^2=\frac{\sigma a^2-\gamma^2}{a^4}.
\end{align}
The equation presented above indicates the presence of a rebound effect at the value of the scale factor $a = \gamma/\sqrt{\sigma}$. Moreover, at this bounce point, it is necessary for the proper volume ($a^3$) to reach a minimum since the second derivative of the scale factor ($\ddot{a}$) is positive, given by $\ddot{a} = - \frac{\beta}{a^5}$. It is worth noting that the total density and pressure also remain finite at this juncture. Consequently, this rebound mechanism seems to circumvent the singularity commonly associated with the Big Bang model.

However, there is an important consideration to make. While the bounce occurs at $a = \gamma/\sqrt{\sigma}$, the equation \eqref{r9} imposes a constraint, disallowing smaller values of $a$. This limitation introduces a discontinuity in the model at this rebound point.\\
Under these circumstances, the solution to the wave equation \eqref{B6} is as follows,
\begin{align}\label{r10}
\psi+\psi_0 = \frac{\alpha}{\gamma}\arctan\left(\sqrt{\frac{\sigma a^2}{\gamma^2}-1}\right).
\end{align}
In this scenario, the solution for the scale factor can be expressed as,
\begin{align}\label{r11}
t+t_0 = \frac{1}{2\sqrt{\sigma}}a\sqrt{a^2-\gamma^2/\sigma}+\frac{1}{2}\frac{\gamma^2}{\sigma^{3/2}}\ln \left|a+\sqrt{a^2-\gamma^2/\sigma}\right|. 
\end{align}
Hence, the deceleration parameter can be calculated to yield,
\begin{align}\label{r12}
q=-\frac{2\gamma^2/\sigma-a^2}{a^2-\gamma^2/\sigma}.
\end{align}

Consequently, an accelerated phase emerges within the range of $\gamma^2/\sigma < a^2 < 2\gamma^2/\sigma$. Once the square of the scale factor surpasses $2\gamma^2/\sigma$, the universe transitions into a phase of decelerated expansion. Importantly, it is crucial to observe that the validity of the model breaks down when the square of the scale factor is smaller than $\gamma^2/\sigma$. This particular range of values is where the model does not hold true.

By utilizing equations \eqref{T10} and \eqref{r9}, we derive the rate of change of total entropy as,
\begin{align}\label{r13}
\dot{S}_{t}=64\pi ^2 \frac{\sqrt{\sigma}}{\gamma^2}\frac{a^3\left(a^2-3\gamma^2/2\sigma\right)^2}{\left(a^2-\gamma^2/\sigma\right)^{3/2}}.
\end{align}

Given that the scale factor ($a$) is greater than or equal to $\gamma/\sqrt{\sigma}$, it follows that the rate of total entropy change remains positive. Consequently, the Generalized Second Law (GSL) is upheld in both the phases of acceleration and deceleration. This implies that the increase in total entropy aligns with the criteria of the GSL throughout these distinct phases of the evolution of the universe.

\begin{itemize}
\item \textbf{Case-III : $\sigma$ is negative but $\beta$ is positive}
\end{itemize}


Let us look at the case where $\sigma$ is negative and is given by $\sigma=-\lambda^2$. As a result, equation \eqref{r3} undergoes a modification, transforming into the expression,
\begin{equation}\label{r14}
\dot{a}^2= \frac{\beta-\lambda^2a^2}{a^4}.
\end{equation}
The equation that dictates the variation in the scalar field is as follows,
\begin{align}\label{r15}
\psi+\psi_0 = \frac{\alpha}{2\sqrt{\beta}}\ln \left|\frac{\sqrt{\beta/\lambda^2-a^2}-\sqrt{\beta/\lambda^2}}{\sqrt{\beta/\lambda^2-a^2}+\sqrt{\beta/\lambda^2}}\right|.
\end{align}
This equation is derived by integrating the wave equation utilizing equation \eqref{r1}.

The solution for the scale factor can be obtained by a straightforward integration of equation \eqref{r14} as,
\begin{equation}\label{r16}
t+t_0 =- \frac{1}{2\lambda}a\sqrt{\beta/\lambda^2-a^2}+\frac{\beta}{2\lambda^3}\arcsin\left(\frac{a}{\sqrt{\beta/\lambda^2}}\right).
\end{equation}

According to equation \eqref{r14}, it is evident that $a$ is not greater than $\sqrt{\beta}/\lambda$. When $a$ reaches $a=\sqrt{\beta}/\lambda$, the rate of change $\dot{a}$ becomes zero, and the second derivative $\ddot{a}$ equals $-\lambda^5/\beta^{5/2}$. However, the validity of the model does not extend beyond $\sqrt{\beta}/\lambda$, rendering this point not a true maximum for the scale factor. In reality, the model lacks definition beyond this threshold, and its validity ceases. Due to the absence of a lower bound limit on $\ddot{a}$, it will inevitably approach a singularity when traced back in time. The deceleration parameter is obtained as,
\begin{equation}\label{r17}
q  = \frac{2\beta-\lambda^2 a^2}{\beta-\lambda^2 a^2}.
\end{equation}
As $a$ is not greater than $\sqrt{\beta}/\lambda$, it follows that q  takes on a positive value. Consequently, this model results in a deceleration phase.

Through the utilization of equations  \eqref{T10} and \eqref{r14}, the rate of change of overall entropy is obtained as,
\begin{equation}\label{r18}
\dot{S}_{t}= -\frac{64\pi^2 \lambda}{\beta}\frac{a^3\left(3\beta/2\lambda^2-a^2\right)^2}{\left(\beta/\lambda^2-a^2\right)^{3/2}}.
\end{equation}
The rate at which the total entropy changes experiences a decrease, as indicated by the fact that $a$ is not greater than $\sqrt{\beta}/\lambda$. Consequently, this particular model does not align with the Generalized Second Law (GSL) of thermodynamics, which asserts that the total entropy of a closed system either remains constant or increases over time.

\section{Dust Era:}

When the fluid is a pressureless dust, the field equations are expressed as,
\begin{align}\label{D1}
 3\left(\frac{\dot{a}}{a}\right)^2 = \rho+\rho_\psi ,
\end{align}
\begin{align}\label{D2}
2\left(\frac{\ddot{a}}{a}\right)+\left(\frac{\dot{a}}{a}\right)^2 = -p_\psi. 
\end{align}

Substituting $p=0$ into the wave equation \eqref{B6} yields the result,

\begin{equation}\label{D3}
 \ddot{\psi}+3 \frac{\dot{a}}{a}\dot{\psi}=\frac{\rho}{2\omega+3}.
\end{equation}

After taking the derivative of the field Equation \eqref{D1} and subsequently employing the field Equations \eqref{D1} through \eqref{D2}, along with the wave equation \eqref{D3}, we arrive at a revised form of the continuity equation, which can be expressed as,
\begin{equation}\label{D4}
\dot{\rho}+3\frac{\dot{a}}{a} \rho = -\frac{\rho\dot{\psi}}{2}.
\end{equation}
The presence of a non-zero term on the right-hand side of the aforementioned equation indicates that the energy densities of two distinct fluids, namely, dust and scalar field, do not independently maintain conservation. However, they do adhere to a conservation equation, expressed as,
\begin{align}\label{D5}
 \dot{\rho}_t= -3\frac{\dot{a}}{a}(\rho_t+p_t).
 \end{align}
This signifies that their combined energy conservation is satisfied. Obtaining this overall conservation equation from the field equations is a straightforward process.
Upon integrating equation \eqref{D4}, the result is as follows,
\begin{equation}\label{D6}
 \rho= \rho_0 \frac{\exp[-\frac{\psi}{2}]}{a^3}.
 \end{equation}
 
 By employing the wave equation to eliminate $\rho$, the integration of a composite set of field equations leads to,
 \begin{equation}\label{D7}
a^2\dot{a} = \frac{2\omega+3}{2}a^3\dot{\psi}+\xi. 
\end{equation}
In the above equation, we have introduced  $\xi$ as an integration constant. By plugging in the expression for $\dot{\psi}$ from the preceding equation into Equation \eqref{D2}, we derive,
\begin{equation}\label{D8}
2\ddot{a}=-\chi\frac{\dot{a}^2}{a}+\frac{2\xi}{2\omega+3}\frac{\dot{a}}{a^3}-\frac{\xi^2}{2\omega+3}\frac{1}{a^5}.
\end{equation}
Here, $\chi=\frac{2\omega+4}{2\omega+3}$ stands as a constant linked to the Brans-Dicke parameter $\omega$.
To continue our analysis, we can assume, without losing generality, that the constant of integration $\xi$ is set to zero.
Consequently, Equation \eqref{D8} transforms into,
\begin{equation}\label{D9}
2\ddot{a}=-\chi\frac{\dot{a}^2}{a}.
\end{equation}

Now, the equation is amenable to integration, yielding,
\begin{equation}\label{D10}
\dot{a}= \mu a^{-\chi/2}.
\end{equation}
Here, $\mu$ represents a constant arising from the process of integration.

Next, by utilizing Equation \eqref{D7}, we acquire,
\begin{equation}\label{D11}
\dot{\psi}= \frac{2\mu}{2\omega+3}a^{-(\frac{\chi}{2}+1)}.
\end{equation}
Integrating the aforementioned pair of equations leads to the solution for the system, expressed as,
\begin{align}\label{D12}
t+t_0 &=\frac{1}{\mu(\frac{\chi}{2}+1)}a^{\frac{\chi}{2}+1}, \\
\psi+\psi_0 &= \ln\left(a^{\frac{2}{2\omega+3}}\right).
\end{align}
Here, $t_0$ and $\psi_0$ represent integration constants.
Now, by employing the definition of the deceleration parameter, equation \eqref{D9}, and the definition of $\chi$, we deduce the expression for the deceleration parameter as follows,
\begin{equation}\label{D13}
q = \frac{\omega+2}{2\omega+3}.
\end{equation}
By observing equation \eqref{D12}, it becomes evident that for the purpose of modeling an expanding universe, both $\mu$ and $\frac{\chi}{2}+1$ need to take on positive values. The requirement for $\frac{\chi}{2}+1$ to be positive can be satisfied through two distinct approaches, and we will consider each of these methods separately.

\begin{itemize}
\item \textbf{1st way to achieve $\frac{\chi}{2}+1$ to be positive and explore GSL:}
\end{itemize}

The expression $\frac{\chi}{2}+1 = \frac{3\omega+5}{2\omega+3}$ must be greater than zero for our analysis. In the initial scenario, we examine the conditions under which both $3\omega+5$ and $2\omega+3$ are positive, leading to $\omega > -\frac{3}{2}$. As a result, we can conclude that the expansion of the universe is accompanied by a decelerating motion in this case. \\

Subsequently, our task involves examining the Generalized Second Law (GSL) within this context. By utilizing equation \eqref{T10}, we are able to determine the rate at which the total entropy changes. This equation provides insights into how the total entropy of a system evolves over time,

\begin{equation}\label{D14}
\dot{S}_{t} = 16\pi^2\left(\frac{\chi}{2}+1\right)^3\left(\frac{2}{2-\chi}\right)t.
\end{equation}
In order for the quantity $\dot{S}_{t}$ to exhibit a positive value, it is necessary that the value of $\omega$ exceeds -1. This condition on the Brans-Dicke parameter ensures that the total entropy of the system is increasing over time, contributing to a consistent interpretation of the physical processes at play.

\begin{itemize}
\item \textbf{2nd way to achieve $\frac{\chi}{2}+1$ to be positive and explore GSL:}
\end{itemize}


Another approach to satisfy the condition $\frac{\chi}{2}+1>0$ is when both $3\omega+5$ and $2\omega+3$ are negative, corresponding to $\omega<-5/3$. In this scenario, cosmic acceleration can be achieved within the range of $-2<\omega<-5/3$. However, it is important to note that for $\omega<-2$, the expansion of the universe remains decelerated, as implied by equation (\ref{D13}). The limitations on the potential values of $\omega$ align consistently with those deduced by Banerjee and Pavón \cite{Banerjee:2000mj}. \\

 From equation \eqref{D14}, it becomes apparent that in order for the rate of change $\dot{S}_{t}$ to exhibit a positive value, the parameter $\chi$ needs to be smaller than 2. Given that $\omega$ is constrained to be less than $-5/3$ in this scenario, it is assured that $\chi$ is indeed less than 2. As a result of this, we can confidently conclude that $\dot{S}_{t}>0$, indicating an increase in the total entropy. Consequently, this particular model adheres to the requirements of the Generalized Second Law (GSL). This alignment signifies that the model upholds a fundamental principle of thermodynamics and consistent physical behavior.

\section{Summary and Discussion:}
This research delves into the thermodynamic characteristics of cosmological models dominated by radiation and dust within the framework of Brans-Dicke theory. Specifically, we consider a spatially flat, homogeneous, and isotropic universe in what is referred to as the \textit{Einstein frame}, which is the conformally transformed version of Brans-Dicke theory in the so-called \textit{Jordan frame}. Our focus centers on investigating the validity of the generalized second law of thermodynamics, posited by Bekenstein in the early 1970s \cite{Bekenstein:1972tm,Bekenstein:1973ur,Bekenstein:1974ax}, which asserts that the combined entropy of the universe, the sum of matter entropy and horizon entropy, cannot decrease with time.\\

Our findings indicate that for a radiation-dominated universe, the solutions derived from Brans-Dicke theory with a positively definite parameter $\omega$ fail to uphold the principles of the generalized second law. However, intriguingly, when certain ranges of negative $\omega$ values are considered, the model indeed aligns with the requirements of thermodynamics. \\\

For a universe dominated by dust, the model does satisfy the generalized second law for specific small negative values of $\omega$. Notably, this range of $-2< \omega < -\frac{5}{3}$ substantially overlaps with the parameter values required for an accelerated expansion without the need for exotic forms of matter \cite{Banerjee:2000mj}.\\

Perhaps the most intriguing revelation from this investigation is that Brans-Dicke theory finds thermodynamic support precisely within the parameter range of $\omega$ that is associated with driving the alleged accelerated expansion of the universe, as opposed to general cases of decelerated expansion.

%% file: Chapters/Chapter4.tex
\chapter{Thermodynamic Analysis of Cosmological models reconstructed from jerk parameter} 
\label{Chapter4}
\chaptermark{Thermodyn. of Cosmological models reconstructed from $j$}
\section{Introduction:}
The progress in observational cosmology yielded a remarkable outcome, indicating that the current universe is expanding at an accelerated rate \cite{SupernovaSearchTeam:1998fmf, SupernovaCosmologyProject:1998vns, SupernovaSearchTeam:1998bnz, SupernovaCosmologyProject:2003dcn, SupernovaSearchTeam:2003cyd, Barris:2003dq, Hicken:2009dk, SupernovaCosmologyProject:2011ycw}. However, the force responsible for countering the attractive nature of ordinary matter and fueling this acceleration remains an enigma. The absence of any theoretical inclination towards any of the proposed agents, referred to as dark energy, has prompted a different perspective on the issue in recent times. Instead of constructing models based on the matter sector and deriving conclusions about the evolution of the universe, this alternative approach formulates the evolution history based on observations and endeavors to deduce the plausible matter sector from that vantage point. This approach, initially proposed by Ellis and Madsen \cite{Ellis:1990wsa} long ago, has gained attention as none of the suggested candidates decisively emerges as the dark energy.\\

Numerous endeavors have been made towards reconstructing cosmological models from physical quantities, such as the dark energy equation of state parameter \cite{Saini:1999ba, Sahlen:2005zw, Sahlen:2006dn, Scherrer:2007pu, Scherrer:2008be, Holsclaw:2010nb, Holsclaw:2010sk, Holsclaw:2011wi, RobertGCrittenden_2012, RemyaNair_2014}, the quintessence potential \cite{Starobinsky:1998fr, Huterer:1998qv, Huterer:2000mj}, and others. Additionally, attempts have been made to construct models based on kinematical quantities like the deceleration parameter $q=-\frac{a\ddot{a}}{{\dot{a}}^2}$ \cite{Gong:2006gs, WangYuTing2010,Mamon:2016dlv,Naik:2023yhl,Xu:2007gvk}, and the jerk parameter $j = -\frac{1}{aH^3}\frac{\mathrm{d}^3a}{\mathrm{d}t^3}$ \cite{Luongo:2013rba, Rapetti:2006fv, Zhai:2013fxa, Mukherjee:2016trt, Mukherjee:2016shl}, where $a$ denotes the scale factor and $H=\frac{\dot{a}}{a}$ represents the Hubble parameter. In many of these investigations concerning kinematical quantities, $q$ or $j$ is expressed parametrically, and the parameters are estimated using observational data. Among these cosmographic quantities, the Hubble parameter $H$ has been an observable and is known to evolve over time. With the evolution of the deceleration parameter $q$ itself being observable, attention has turned to the third-order derivative, namely the jerk parameter $j$.\\

Beyond depicting the evolutionary trajectory, a cosmological model must account for structure formation in terms of matter perturbation growth and must adhere to some fitness test, such as thermodynamic viability. The motivation behind this study is to assess the thermodynamic viability of some specific models reconstructed from the jerk parameter. We focus on two sets of reconstructed jerk models existing in the literature: one without any interaction in the matter sector \cite{Mukherjee:2016trt} and the other allowing for the possibility of interaction \cite{Mukherjee:2016shl}. The rationale for selecting these models lies in the fact that they deviate from the usual anstaz considered in parametrically reconstructing $j$. Typically, the ansatz is designed such that the jerk parameter $j$, as a function of the redshift parameter $z$, equals $-1$ at $z=0$, which aligns the model with the present-day $\Lambda$CDM model. However, the models in \cite{Mukherjee:2016trt} and \cite{Mukherjee:2016shl} permit any value of $j(z=0)$ initially, to be determined by actual observational datasets. Hence, these models are more versatile in that sense.\\

Given the reconstructed $j$, it is possible to find the first derivative of the Hubble parameter and then the Hubble parameter. This, in turn, allows estimation of the rate of change of total entropy, enabling an assessment of whether entropy  increases and thereby verifying adherence to the Generalized Second Law (GSL), which asserts that the total entropy of the universe cannot decrease. The total entropy is a combination of the entropy of the fluid enclosed by the horizon and that of the horizon itself \cite{Bekenstein:1972tm, Bekenstein:1973ur, Bekenstein:1974ax}.\\

A critical assumption in these calculations is that cosmic matter is in thermal equilibrium with the horizon, with the temperature of the latter determined by the Hayward-Kodama temperature \cite{Hayward:1997jp, Hayward:2008jq, faraoni2015horizons}, which we have taken in all the previous chapters. This thermal equilibrium becomes questionable if the matter sector involves a radiation component, as recently demonstrated by Mimoso and Pavón \cite{Mimoso:2016jwg}. Importantly, the reconstructed models examined in this study lack a radiation component; they consist of either cold dark matter or a similar matter component where thermal equilibrium holds, alongside a dark energy component driving the acceleration, for which equilibrium is approximately valid \cite{Mimoso:2016jwg}. 
\section{Kinematics:}
We examine a cosmological scenario characterized by a spatially flat, homogeneous, and isotropic FRW model. The metric is given by the  equation\eqref{metric}. Although the pertinent kinematic quantities are introduced and discussed in detail in Chapter \ref{Chapter1}, let us reiterate the definitions here for the sake of clarity.
\begin{itemize}
\item \textbf{Hubble parameter:}
\begin{equation}
H \equiv \frac{\dot{a}}{a}. \nonumber
\end{equation}
\item \textbf{Deceleration parameter:}
\begin{equation}
q\equiv -\frac{\ddot{a} a}{{\dot{a}}^2}. \nonumber
\end{equation}
\item \textbf{Jerk parameter:}
\begin{equation}
j \equiv - \frac{1}{aH^3} \frac{d^3 a}{dt^3}. \nonumber
\end{equation}
\end{itemize}

As usual the dot denotes differentiation in relation to cosmic time $t$, which is the argument of all functions. We will now express these quantities in terms of a dimensionless parameter, specifically the redshift parameter denoted as $z$, defined by $1+z \equiv \frac{a_0}{a}$. The subscript $0$ designates the current value of the quantity. The expressions for the deceleration parameter and the jerk parameter as functions of $z$ are as follows,
\begin{align}
  q(z) &= -1+\frac{1}{2}(1+z)\frac{[H^2(z)]^\prime}{H^2(z)} ,\\
  j(z) &= -1+(1+z)\frac{[H^2(z)]^\prime}{H^2(z)}-\frac{1}{2}(1+z)^2\frac{[H^2(z)]^{\prime\prime}}{H^2(z)} \label{jerk}.
  \end{align}
  In the entirety of this article, the symbol prime is employed to signify differentiation with respect to the redshift parameter $z$.

  \section{Fundamental Thermodynamic Principle}

As previously stated in the introduction, we will assess the thermodynamic viability of the models by subjecting them to a test based on the Generalized Second Law of Thermodynamics (GSL) \cite{Bekenstein:1972tm, Bekenstein:1973ur, Bekenstein:1974ax}. This principle asserts that the overall entropy of the universe remains constant or increases over time. The collective entropy comprises the entropy associated with the horizon and the entropy of the fluid enclosed by the horizon. In our analysis, we make the assumption that the fluid components are in a state of thermal equilibrium with the horizon.

\begin{itemize}
\item \textbf{Revisiting Hayward-Kodama Temperature:}
\end{itemize}

Complete analytic description of Hyaward-Kodama temperature has been provided in chapter \ref{Chapterthermo}, section \ref{Apparenthorizonthermo}. It seems requied to revisit the fact in brief, why we consider Hayward-Kodama temperature instead of Hawking temperature as the equilibrium temperature.

Within a spatially flat FRW cosmological framework, the surface gravity pertaining to the apparent horizon is given by,
\begin{equation}
\kappa_{ko} = -\frac{1}{2H}(\dot{H}+2H^2),
\end{equation}
where $\kappa_{ko} $ signifies the surface gravity.
Thus, based on the preceding analysis, we can determine the Hayward-Kodama temperature as follows,
\begin{align}\label{hktemp}
T &= \frac{\mid{\kappa_{ko}\mid}}{2\pi} \nonumber\\
 &= \frac{2H^2+\dot{H}}{4\pi H}.
\end{align}

Since our analysis will focus on a spacetime that changes with time, it's important to carefully consider which temperature assumption to use. In this context, we opt for the Hayward-Kodama temperature instead of the Hawking temperature. Moreover, we shall direct our attention to the concept of an apparent horizon, as distinct from an event horizon, for our subsequent discourse. This temperature is particularly relevant when analyzing spacetimes with evolving geometry and kinematics, as it takes into account the dynamic aspects of the system.\\
In thermal equilibrium, the horizon temperature and the temperature of the fluid within the enclosed region are equal. Therefore, temperature of the enclosed fluid is also the Hayward-Kodama temperature.\\
To delve deeper into the theoretical underpinnings of these concepts and their implications, it is beneficial to refer to the existing literature. Relevant works, such as the monograph by Faraoni \cite{faraoni2015horizons}, offer comprehensive discussions on spacetime thermodynamics, temperature concepts, and horizon properties. This reference provides a solid foundation for understanding the motivations and applications of employing the Hayward-Kodama temperature in our analysis and the significance of utilizing the apparent horizon in the context of thermal equilibrium scenarios.\\
In the preceding chapter (Chapter \ref{Chapterthermo}), we computed the rate of change in the total entropy. In the current chapter, we are merely restating the expression for subsequent calculations. The rate of change of the total entropy is,
\begin{align}
\dot{S}_{tot} = 16\pi^2\Big(\frac{\dot{H}^2}{H^3}\Big)\Big(\frac{1}{2H^2+\dot{H}}\Big). \nonumber 
\end{align}

To facilitate a dimensionless representation, we express the Hubble parameter in the following manner:

We introduce the dimensionless Hubble parameter $h(z)$, which is defined as the ratio of the Hubble parameter at a specific redshift $z$ to the current value of the Hubble parameter $H_0$. Mathematically, this can be written as:
\begin{align*}
h(z) =\frac{H(z)}{H_0}.
\end{align*}

This dimensionless form of the Hubble parameter offers a normalized perspective on the rate of expansion of the universe at different cosmic epochs. By comparing the Hubble parameter at a particular redshift to its present-day value, we gain insight into the relative pace of cosmic expansion across different eras. This representation aids in simplifying calculations and analyses by removing the explicit dependence on the absolute scale of the Hubble parameter, focusing instead on the relative changes with respect to the current state of the universe.

Therefore, now we write the rate of change of entropy in terms of h(z) as,

\begin{align}\label{s-tot-dot2}
\dot{S}_{tot}= \frac{4\pi^2}{H_0}\frac{(1+z)^2\left[(h^2)^\prime\right]^2}{h^3\left[2h^2-\frac{1}{2}(1+z)(h^2)^\prime\right]}.
\end{align}
Consequently, armed with an understanding of the progression of $h$ and its rate of change with respect to $z$, it becomes feasible to assess the adherence to the Generalized Second Law (GSL) for a specific model.

\section{Cosmological Models Devoid of an Interactions in the Matter Sector }
 The first subsection delves into an in-depth discussion regarding the specific model that has been selected for examination. Subsequently, the second subsection is dedicated to an exploration of how the Generalized Second Law (GSL) is assessed within the context of these models.
\subsection{The model}
We  consider existing dark energy models sourced from existing literature, where the reconstruction is accomplished using the jerk parameter $j$. Specifically, these models are adapted from the work presented in \cite{Mukherjee:2016trt}, which we shall brielfly summarize here. What distinguishes these models is that, unlike certain other methods, they don't require a pre-established condition, such as $j(z=0)=-1$, that accords with the $\Lambda$CDM model. These models do not rely on any specific theory of gravity as a foundational assumption. We select all four jerk parameter parametrizations outlined in \cite{Mukherjee:2016trt}, as our starting point. These are as follows,
\begin{align}
\text{I}. \hspace{2mm} j(z)&=-1+j_1 \frac{1}{h^2(z)},  \\
\text{II}. \hspace{2mm} j(z)&=-1+j_1 \frac{(1+z)}{h^2(z)}, \\
\text{III}. \hspace{2mm}j(z)&=-1+j_1 \frac{(1+z)^2}{h^2(z)} ,\\
\text{IV}. \hspace{2mm} j(z)&=-1+j_1 \frac{1}{(1+z)h^2(z)}. \label{ansatz1}
  \end{align}
  
  Here, $j_1$ is a constant value to be determined by observational data. The selection of these parametrizations initially appears arbitrary. However, there are specific reasons underlying these choices.
  In Model I, the dependence on $z$ is not explicitly present; the variation of $j$ is exclusively managed by the function $h(z)$. Models II and III, on the other hand, exemplify instances where $j$ exhibits a direct proportionality to uncomplicated positive powers of $(1+z)$. In the fourth model, however, the relationship takes a form where $j$ is inversely proportional to $(1+z)$. The rationale underlying these chosen parameterizations becomes evident when the respective parameters are derived from datasets, a process undertaken in \cite{Mukherjee:2016trt}. The fundamental notion driving this selection is to opt for straightforward functions of $z$. This approach aligns with the principles often used for parameterizations of quantities with more tangible physical significance, such as the equation of state parameter.
  These equations can be integrated to yield,
  \begin{align} 
  \text{I}. \hspace{2mm} h^2(z)&=c_1(1+z)^3+c_2+\frac{2}{3}j_1\ln(1+z),\label{h1}\\
  \text{II}. \hspace{2mm} h^2(z)&=c_1(1+z)^3+c_2+j_1(1+z),\label{h2}\\
  \text{III}. \hspace{2mm} h^2(z)&=c_1(1+z)^3+c_2+j_1(1+z)^2,\label{h3}\\
  \text{IV}. \hspace{2mm} h^2(z)&=c_1(1+z)^3+c_2-j_1\frac{1}{2(1+z)}, \label{hsquare}
  \end{align}
   correspondingly. Here, $c_1$ and $c_2$ denote integration constants. Nevertheless, it is important to note that these constants are subject to limitations imposed by $h_0 (z) = 1$, a condition derived from the definition of $h$.
  Consequently, $c_2$ can be substituted using the subsequent relationships,
  \begin{align}
  \text{I}. \hspace{2mm} c_2 &= 1-c_1,\\
   \text{II}. \hspace{2mm}c_2 &=1-j_1-c_1 ,\\
    \text{III}. \hspace{2mm}c_2 &=1-j_1-c_1,\\
     \text{IV}. \hspace{2mm} c_2 &=1+\frac{j_1}{2}-c_1.
 \end{align}
 The values of the two distinct parameters, namely $c_1$ and $j_1$, are determined through the utilization of observational data. This calibration is conducted using combinations of datasets originating from sources such as SNe (Supernovae), OHD (Observational Hubble Data), BAO (Baryon Acoustic Oscillations), and CMBShift (Cosmic Microwave Background Shift), as meticulously detailed in \cite{Mukherjee:2016trt}. These calibrated values are presented in Table 1. It is noteworthy that $c_1$ serves as $\Omega_{m_0}$, which signifies the matter density parameter at the current epoch. The determination of both $c_1$ and $j_1$ through empirical data plays a pivotal role in establishing the foundation for the subsequent analyses and interpretations of the chosen dark energy models.\\

The equations (\ref{h1}-\ref{hsquare}) readily reveal that upon establishing the correspondence between $c_1$ and the current matter density parameter ${\Omega}_{m0}$, the initial term in all four models inherently represents the independent evolution of pressureless cold matter. This distinctive feature substantiates the characterization of this particular group as non-interacting models, as highlighted in \cite{Mukherjee:2016trt}. It is worth emphasizing that both Model I and Model IV encounter challenges concerning their future evolution. This predicament arises due to the presence of singularities at $z=-1$, a fact discerned from the expressions in equations (\ref{h1}) and (\ref{hsquare}).\\
 An intriguing observation stems from equation (\ref{h3})is that the behavior of the so-called dark energy in Model III appears reminiscent of the spatial curvature term found in the Friedmann equations. However, it is crucial to recognize that the reconstruction detailed in \cite{Mukherjee:2016trt} exclusively takes into account a spatially flat metric as depicted in equation \eqref{metric}. Consequently, this similarity is merely a coincidental resemblance rather than a substantial connection.\\

\begin{table}[h!]
 \centering
\subfigure{\begin{tabular}[50pt]{|M{3cm}|M{3cm}|M{3cm}|N}
 \hline \textbf{Model}& {$\mathbf{c_1}$} & {$\mathbf{j_1}$}\\
 \hline 
 I. &   $0.309\pm0.012$ &  $-0.080\pm 0.222$ \\
 II. &   $0.310\pm0.012$ & $-0.051\pm0.093$\\
 III. &  $0.313\pm0.015$ & $-0.027\pm0.045$\\
 IV. &   $0.308\pm0.011$ & $-0.106\pm 0.237$\\
 \hline 
\end{tabular}}
\\(a)
 \hfill\\ \vspace{2em}
 
 \centering
\subfigure{\begin{tabular}[50pt]{|M{3cm}|M{3cm}|M{3cm}|N}
  \hline \textbf{Model}& \textbf{$c_1$} & \textbf{$j_1$}\\
 \hline 
 I. & $0.298\pm0.010$ & $0.078\pm 0.140$ \\
 II. & $0.299\pm0.008$ & $0.045\pm0.050$\\
 III. & $0.300\pm0.008$ & $0.017\pm0.015$\\
 IV. & $0.298\pm0.008$ & $0.112\pm 0.176$\\
 \hline 
 \end{tabular}}
 \\(b)
 \caption{Outcomes of the statistical analysis of observational data provided for two distinct datasets: (a) SNe+OHD+BAO data and (b) SNe+OHD+BAO+CMBShift data. These tabulated results have been extracted from \cite{Mukherjee:2016trt}.}\label{jtable}
\end{table}

While not within the scope of this current study, it is worth noting that the outcomes of the best fits along with their associated $1\sigma$ error margins, as presented in Table \ref{jtable}, were attained through a minimization process using ${\chi}^2$ in \cite{Mukherjee:2016trt}.

\subsection{Thermodynamic Analysis}\label{jsec4}

By substituting the solutions for $h(z)$ as provided by equations \eqref{h1}-\eqref{hsquare} into equation \eqref{s-tot-dot2}, we are able to derive the rate of change of total entropy for various models as,
\begin{align}
\text{I}. \hspace{2mm} \dot{S}_{tot} &=\frac{4\pi^2}{H_0} \frac{(1+z)^2\left[3c_1(1+z)^2+\frac{2}{3}j_1\frac{1}{(1+z)}\right]^2}{\left[c_1(1+z)^3+c_2+\frac{2}{3}j_1\ln(1+z)\right]^{3/2}}\nonumber\\&\hspace{1em}\times\frac{1}{2\left[c_1(1+z)^3+c_2+\frac{2}{3}j_1\ln(1+z) \right]-\frac{(1+z)}{2}\left[3c_1(1+z)^2+\frac{2}{3}j_1\frac{1}{(1+z)} \right]}, \label{ent1}\\
\text{II}. \hspace{2mm} \dot{S}_{tot} &=\frac{4\pi^2}{H_0}(1+z)^2 \frac{\left[3c_1(1+z)^2+j_1 \right]^2}{\left[c_1(1+z)^3+c_2+j_1(1+z)\right]^{3/2}}\nonumber\\&\hspace{1em}\times\frac{1}{\left[2\left\{c_1(1+z)^3+c_2+j_1(1+z) \right\}-\frac{(1+z)}{2}\left\{3c_1(1+z)^2+j_1\right\} \right]}, \label{ent2}\\
\text{III}. \hspace{2mm} \dot{S}_{tot} &=\frac{4\pi^2}{H_0}(1+z)^2\frac{\left[3c_1(1+z)^2+2j_1(1+z) \right]^2}{\left[c_1(1+z)^3+c_2+j_1(1+z)^2 \right]^{3/2}}\nonumber\\&\hspace{1em}\times\frac{1}{\left[2\left\{c_1(1+z)^3+c_2+j_1(1+z)^2 \right\}-\frac{(1+z)}{2}\left\{3c_1(1+z)^2+2j_1(1+z) \right\} \right]}, \label{ent3}\\
\text{IV}. \hspace{2mm} \dot{S}_{tot} &=\frac{4\pi^2}{H_0}(1+z)^2 \frac{\left[ 3c_1(1+z)^2+\frac{j_1}{2}\frac{1}{(1+z)^2}\right]^2}{\left[c_1(1+z)^3+c_2-\frac{j_1}{2}\frac{1}{(1+z)} \right]^{3/2}}\nonumber\\&\hspace{1em}\times\frac{1}{\left[2\left\{c_1(1+z)^3+c_2-\frac{j_1}{2}\frac{1}{(1+z)} \right\}-\frac{(1+z)}{2}\left\{3c_1(1+z)^2+\frac{j_1}{2}\frac{1}{(1+z)^2} \right\} \right]}. \label{ent4} 
\end{align}

Utilizing the values of $c_1$ and $j_1$ extracted from Table I, we generate plots that illustrate the relationship between $\dot{S}_{\text{tot}}$ and the redshift $z$. For these plots, we adopt a value of $H_0$ at 70 km s$^{-1}$ Mpc$^{-1}$. It is important to note that the plots are presented on an arbitrary scale, with the primary focus directed towards conveying the qualitative characteristics of the entropy alteration rates. In the depicted plots (with the exception of Model-IV for SNe+OHD+BAO data), a consistent methodology is employed. The thick solid line represents the $\dot{S}{\text{tot}}$ values corresponding to the best-fit $c_1$ and $j_1$ parameters. This selection aligns with the contour plots outlined in \cite{Mukherjee:2016trt}, capturing a $3\sigma$ confidence level.\\

Furthermore, within these plots, we maintain the best-fit $c_1$ value while systematically varying the $j_1$ parameter. This involves selecting the farthest permissible values of $j_1$ within the contours established at the $3\sigma$ confidence level, as detailed in \cite{Mukherjee:2016trt}. The same strategy is then applied when keeping $c_1$ fixed at its best-fit value $\pm 3\sigma$, while exploring the furthest allowable values for $j_1$.\\

In summary, these plots elucidate the variations in $\dot{S}_{\text{tot}}$ against the redshift $z$ using specific parameter values. The methodology employed ensures a comprehensive exploration of the parameter space, offering insights into the qualitative trends of entropy change rates for distinct models.

For Model-IV with the SNe+OHD+BAO data, an examination of the contour plot in \cite{Mukherjee:2016trt} unveils a notable observation: it is unfeasible to encompass all the data points within the $3\sigma$ confidence level. In response, we adopt a distinct approach to selecting the $(c_1, j_1)$ values. Within this context, the more specific procedure is outlined below.
\begin{enumerate}

\item The boldest line depicted in the plot corresponds to the best-fit $(c_1, j_1)$ value.
\item With $c_1$ set at its best-fit value, we identify the farthest permissible $j_1$ values as indicated by the contour plot featuring the $2\sigma$ confidence level.
\item Maintaining $c_1$ at its best-fit value but now at $-3\sigma$, we choose the outermost allowed $j_1$ values, adhering to the $3\sigma$ confidence level.
\item Subsequently, we set $c_1$ at its best-fit value $+3\sigma$ and identify two $j_1$ values, one corresponding to the $3\sigma$ confidence level and the other to the $2\sigma$ confidence level.
\end{enumerate}
By adopting this methodology, we assemble a total of seven $(c_1, j_1)$ value pairs, encompassing both the best-fit values and the permissible range according to the confidence levels. This approach offers a comprehensive perspective on the rate of entropy change within the region defined by the $(c_1, j_1)$ contour plots.

\begin{figure}
\centering   
\subfigure[]{\label{jfig1a}\boxed{\includegraphics[width=82mm]{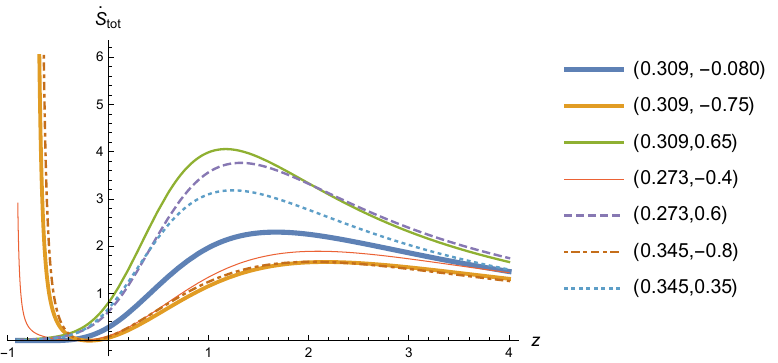}}}
\subfigure[]{\label{jfig1b}\boxed{\includegraphics[width=82mm]{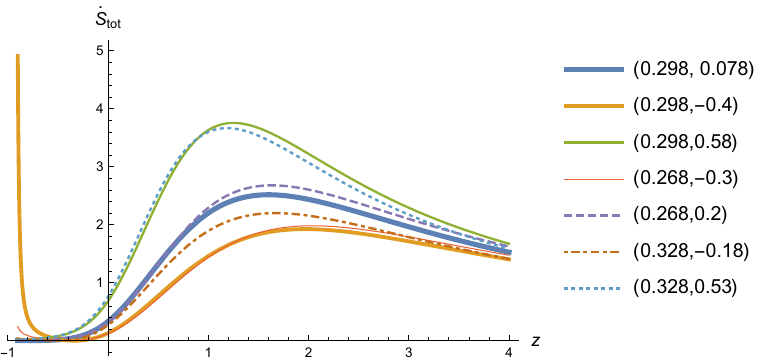}}}
 \caption{$\dot{S}_{tot}$ vs $z$ for Model-I. (a) SNe+OHD+BAO data, (b) SNe+OHD+BAO+CMBShift data. The parameter values are mentioned in the order $(c_1,j_1)$.}\label{jfig1}
\end{figure}

\begin{figure} [h]
 \centering 
 \subfigure[]{\label{jfig2a}\boxed{\includegraphics[width=82mm]{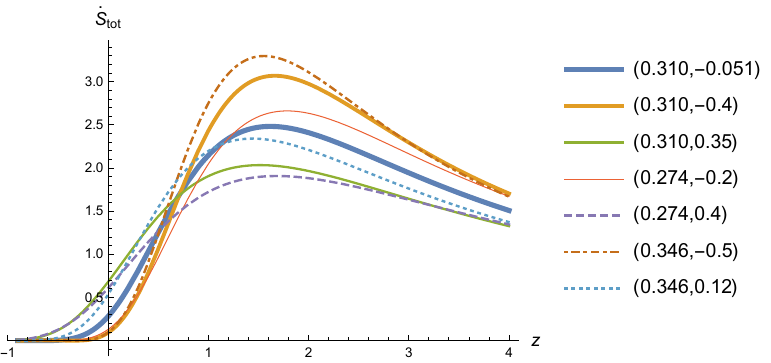}}}
 \subfigure[]{\label{jfig2b} \boxed{\includegraphics[width=82mm]{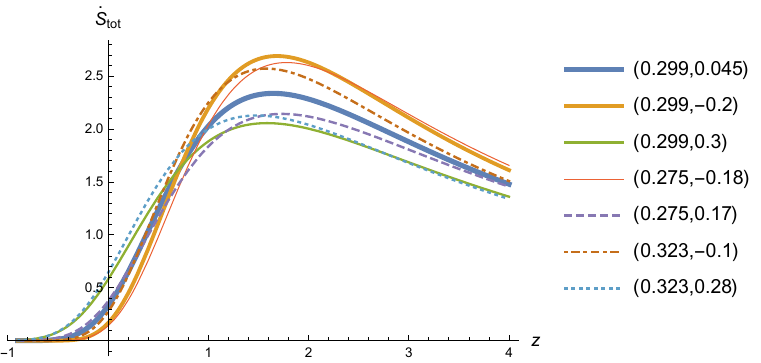}}}
 \caption{$\dot{S}_{tot}$ vs $z$ for Model-II. (a) SNe+OHD+BAO data, (b) SNe+OHD+BAO+CMBShift data. The parameter values are mentioned in the order $(c_1,j_1)$.}\label{jfig2}
 \end{figure}

 \begin{figure} [h]
 \centering 
\subfigure[]{\label{jfig3a} \boxed{\includegraphics[width=82mm]{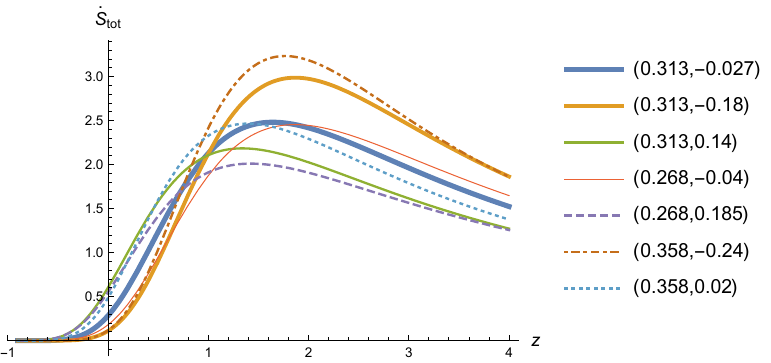}}}
 \subfigure[]{\label{jfig3b}\boxed{\includegraphics[width=82mm]{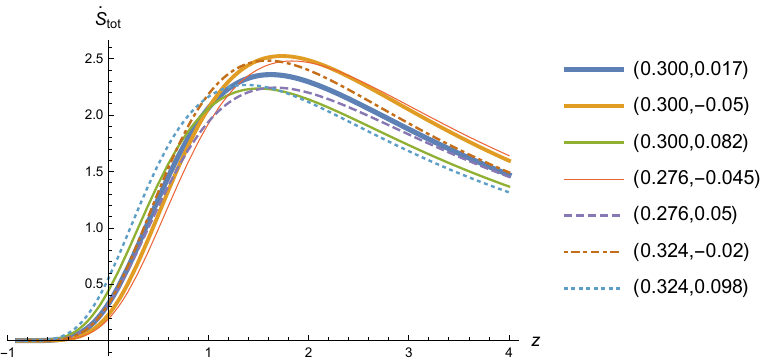}}}
 \caption{$\dot{S}_{tot}$ vs $z$ for Model-III. (a) SNe+OHD+BAO data, (b) SNe+OHD+BAO+CMBShift data. The parameter values are mentioned in the order $(c_1,j_1)$.}\label{jfig3}
  \end{figure}

 \begin{figure} [h!]
\centering 
\subfigure[]{\label{jfig4a} \boxed{\includegraphics[width=82mm]{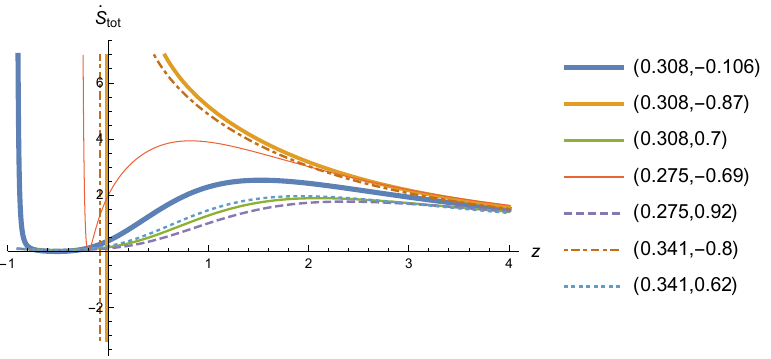}}}
 \subfigure[]{\label{jfig4b}\boxed{\includegraphics[width=82mm]{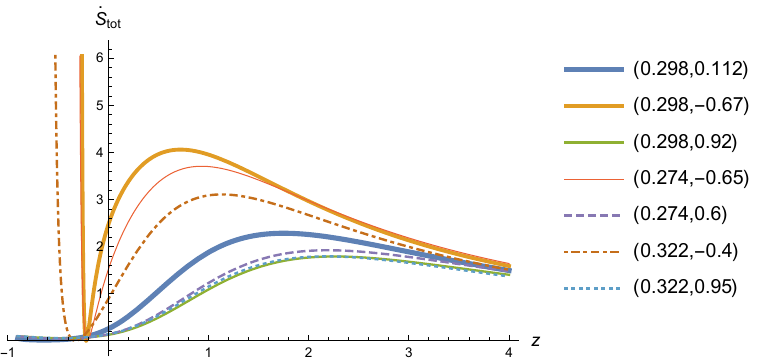}}}
 \caption{$\dot{S}_{tot}$ vs $z$ for Model-IV. (a) SNe+OHD+BAO data, (b) SNe+OHD+BAO+CMBShift data. The parameter values are mentioned in the order $(c_1,j_1)$.}\label{jfig4}
 \end{figure}

Figures \eqref{jfig1}, \eqref{jfig2}, and \eqref{jfig3} distinctly illustrate that the rate of entropy change remains consistently positive. At most, it can approach zero at some point in the future ($z<0$). In the case of Model I, there is a notable sharp surge in entropy in the future, observed for both data set combinations. Despite this surge, $\dot{S}_{\text{tot}}$ maintains a non-negative trajectory (Figure \eqref{jfig1}).\\
However, Model IV does not meet this test of compatibility with thermodynamic principles. For certain parameter selections within the permissible range, $\dot{S}_{\text{tot}}$ exhibits negative values in a future that is very close to $z=0$, particularly when the CMBShift data is not incorporated (Figure \eqref{jfig4}). This indicates a violation of the expected thermodynamic behavior and suggests a lack of compatibility with established principles.\\

Both Model I and Model IV exhibit singular behavior in the rate of entropy change as the redshift approaches $z\rightarrow -1$. This outcome arises from the chosen framework outlined in \cite{Mukherjee:2016trt}, evident from the expressions derived for $h^2$ as presented in equations \eqref{h1} and \eqref{hsquare}. These equations incorporate terms like $\ln (1+z)$ and $\frac{1}{1+z}$, leading to the observed singularities. However, it is important to note that $z=-1$ corresponds to $a\rightarrow \infty$, thereby signifying that these singularities do not manifest in any finite future. Consequently, there is no need for concern in this regard.

The sudden increase in $\dot{S}{\text{tot}}$ observed within the domain $0> z > -1$ for Model I is not a consequence of the chosen framework itself. Instead, this phenomenon emerges due to the specific model parameters that have been determined based on the available datasets. 

\section{Model Incorporating Interaction in the Matter Sector}
 The first subsection provides a detailed description of the chosen model which allows an interaction in the matter sector which is given in \cite{Mukherjee:2016shl}. Following that, the second subsection focuses on the evaluation of the Generalized Second Law (GSL) within the framework of these models.
\subsection{The model}

The initial terms in equations (\ref{h1}-\ref{hsquare}) experience changes in proportion to $(1+z)^3$. This characteristic strongly indicates the presence of cold dark matter (CDM) in all the models, which exhibits no interaction with other components in the matter sector. However, there exists a specific model that deviates from this non-interacting CDM constraint, as detailed in the reference \cite{Mukherjee:2016shl}. This model introduces a relaxation to the non-interacting CDM scenario by assuming that $j$ is a slowly varying parameter. Consequently, the differential equation \eqref{jerk} can be integrated under the assumption that $j$ remains constant. This integration results in the expression,
\begin{equation}\label{hcqg}
 h^2(z) = A (1+z)^{\frac{3+\sqrt{9-8(1+j)}}{2}}+(1-A) (1+z)^{\frac{3-\sqrt{9-8(1+j)}}{2}}.
 \end{equation}
 In this context, $A$ represents a dimensionless constant coefficient (potentially corresponding to $\Omega_{m_0}$), while the jerk parameter $j$ takes on the role of a model parameter. Through a statistical analysis of various combinations of SNe+OHD+BAO data sets, the subsequent outcomes for the parameters $A$ and $j$ have been derived (for further insights, refer to \cite{Mukherjee:2016shl})as,

\begin{align}
A &= 0.286\pm 0.015, \nonumber \\
j &= -1.027 \pm 0.037.
\end{align}

The best fit parameter values are acquired through a conventional minimization of ${\chi}^2$, and these values are presented at a $1\sigma$ confidence interval. This specific reconstruction process is elaborated upon in \cite{Mukherjee:2016shl}. We will refer to this as \textit{model V}. Importantly, it should be acknowledged that when $j$ equals $-1$, the model effectively simplifies to the well-known $\Lambda$CDM model.
    The initial term bears a striking resemblance to the evolution pattern of pressureless matter, albeit with subtle distinctions. These deviations from the conventional CDM behavior can be interpreted as resulting from an interaction with another constituent within the matter sector. This underlying concept forms the basis for characterizing this model as an \textit{interacting model} in \cite{Mukherjee:2016shl}.

\subsection{Thermodynamic Analysis}
\begin{figure}[h!]\centering
\boxed{ \includegraphics[width=82mm]{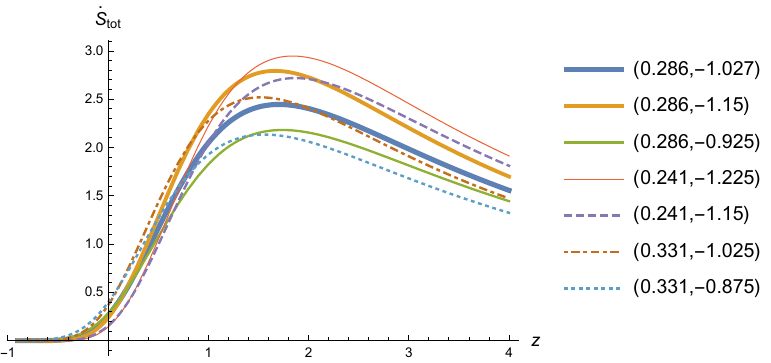}}
\caption{$\dot{S}_{tot}$ vs $z$ for Model-V. for SNe+OHD+BAO data. The parameter values are mentioned in the order $(A,j)$.}\label{jerkfig5}
 \end{figure}

Utilizing the derived solution for $h(z)$ as presented in equation \eqref{hcqg}, we can calculate the rate of change of total entropy for model V by applying equation \eqref{s-tot-dot2}, yielding:
{\scriptsize\begin{align}
\begin{split}
\text{V}. \hspace{2mm} \dot{S}_{tot} &= \frac{8\pi^2}{H_0}(1+z)^2 
\times\frac{\left[A\left(\frac{3+\sqrt{9-8(1+j)}}{2}\right)\left(1+z\right)^\frac{1+\sqrt{9-8(1+j)}}{2}
+\left(1-A\right)\left(\frac{3-\sqrt{9-8(1+j)}}{2}\right)\left(1+z\right)^\frac{1-\sqrt{9-8(1+j)}}{2}\right]^2}{\left[ A\left(1+z\right)^\frac{3+\sqrt{9-8(1+j)}}{2}+\left(1-A\right)\left(1+z\right)^\frac{3-\sqrt{9-8(1+j)}}{2}\right]^{3/2}}\\ &\hspace{2.5mm} \times \Bigg[4\left\{A\left(1+z\right)^\frac{3+\sqrt{9-8(1+j)}}{2}+\left(1-A\right)\left(1+z\right)^\frac{3-\sqrt{9-8(1+j)}}{2}\right\}-\left(1+z\right)\Bigg\{A\left(\frac{3+\sqrt{9-8(1+j)}}{2}\right)\left(1+z\right)^\frac{1+\sqrt{9-8(1+j)}}{2}
\nonumber\\
&\hspace{2.5mm}+\left(1-A\right)\left(\frac{3-\sqrt{9-8(1+j)}}{2}\right)\left(1+z\right)^\frac{1-\sqrt{9-8(1+j)}}{2}\Bigg\}\Bigg]^{-1}.
\end{split}
 \end{align}}
 Using the aforementioned optimal parameter values for $A$ and $j$, which were determined in \cite{Mukherjee:2016shl}, we proceed to generate a plot of the rate of entropy change against redshift ($z$), complete with error bars, as illustrated in Figure \eqref{jerkfig5}. Employing a comparable approach as outlined in Section \ref{jsec4} for models I, II, and III, we select the seven sets of $(A, j)$ values. This strategy allows for a comprehensive analysis of $\dot{S}_{\text{tot}}$, revealing that it remains consistently positive across the entire range. Furthermore, the Generalized Second Law (GSL) is upheld with considerable fidelity.

\section{Summary and Discussion}

Given the absence of a universally accepted candidate for dark energy or a specific modified theory of gravity that can fully explain the universe's supposed accelerated expansion, a new approach has surfaced. This approach involves reconstructing models to match observed characteristics of the universe based on the kinematical aspects.

However, it is essential for these reconstructed models to withstand certain critical evaluations. Among these tests, the examination of thermodynamic consistency is of utmost significance. Fortunately, it is noteworthy that the assessment of the Generalized Second Law (GSL) of thermodynamics can be effectively conducted by utilizing kinematical quantities such as the Hubble parameter and its corresponding derivative, as outlined in equation \eqref{s-tot-dot2}. This thermodynamic viability analysis ensures that the reconstructed models align with the principles of thermodynamics.


Our investigation reveals that it is indeed feasible to identify cosmological models that uphold thermodynamic consistency through the reconstruction of the jerk parameter. Notably, within the subset of four models that pertain to the non-interacting scenario, only one—specifically, model IV — exhibits a decline in entropy within the future epoch ($z<0$) due to its inverse dependence on $(1+z)$. This phenomenon occurs in close proximity to the present era ($z=0$), as depicted in figure \eqref{jfig4a}. On the other hand, all the remaining models, including the one that entertains the prospect of interaction within the dark sector (Model V), satisfactorily adhere to the Generalized Second Law (GSL) of thermodynamics. Model I, while it does conform to the GSL, displays an abrupt surge in entropy within the future ($z<0$). Models II and III exhibit well-behaved attributes in all respects, as does Model V, which permits an interaction within the dark sector. The graphical representations encompass a broad parameter space, encompassing the scope of variation within the $3\sigma$ range.

Nonetheless, it is important to highlight that the assessment of entropy evolution is predominantly centered around the redshift $z=0$, and this evaluation is particularly reliable for values of $z$ greater than zero. The anomalies and irregularities we identify predominantly manifest for values of $z$ that are less than zero. Until the current epoch, all the models exhibit commendable behavior without any significant issues.

It is worth highlighting that this study operates under the implicit assumption that the temperature of the horizon coincides with that of the fluid. However, this assumption might not hold true in cases involving a radiation distribution. Nevertheless, the scope of this investigation is limited to scenarios involving non-relativistic matter, with models I to IV encompassing a pressureless fluid and model V involving a similar component. In the case of the former, the equality of temperatures is indeed accurate, and even for the latter, it remains reasonable at the very least, as discussed in reference \cite{Mimoso:2016jwg}.

%% file: Chapters/Chapter5.tex
\chapter{A Possible Thermodynamic Phase Transition: Signature flip of the Deceleration Parameter} 
\label{Chapter5}
\chaptermark{Signature flip in $q$: Thermodynamic Phase Transition?}
\section{Introduction:}
In this chapter, we delve into a comprehensive examination of thermodynamics for an accelerating universe. The scale factor is modeled using a hyperbolic function ($a\sim {\sinh}^{\frac{2}{3}}(t/t_0)$). This chosen mathematical representation closely emulates the characteristic behaviour of a $\Lambda$CDM model. A notable distinction from the preceding chapters lies in the fact that our current study encompasses not only the application of the Generalized Second Law (GSL) test but also extends to an exploration of the thermodynamic stability of the system. This entails a deeper understanding of how the system evolves and behaves from a thermodynamic perspective. 

Our calculation has led to an exciting result. Through careful examination, it has emerged that despite the entropy ($S$) retaining its continuity, a discontinuity becomes evident in the thermal capacity at constant volume ($C_V$) at a specific value of redshift ($z$). This notable occurrence coincides with the point at which the cosmological evolution undergoes a shift from a decelerated state to an accelerated state of expansion. This discontinuity indicates a second-order phase transition. 

The deceleration parameter ($q$) plays a crucial role similar to an order parameter. This parameter encapsulates the essence of the transition between the two different expansion states. Moreover, the discontinuity seen in $C_V$ is characterized by an order of unity.

Essentially, our analysis highlights a striking connection between the thermodynamic features of the system and the evolution of the universe. The significant change in thermal capacity at a particular redshift indicates a transformative phase transition, supported by its alignment with an order parameter and the inherent order of the discontinuity. This finding deepens our understanding of the complex interplay between cosmological dynamics and thermodynamic behavior, bringing together different ideas into a cohesive framework that enhances our comprehension of the evolution of the universe.

As in the preceding chapters, we continue to center our analysis on the apparent horizon. There is a notable difference between the context of a stationary black hole and the evolving nature of the apparent horizon. This distinctive feature serves as the motivation for replacing the Hawking temperature with the Hayward-Kodama temperature \cite{Hayward:1997jp,Hayward:2008jq} to designate the temperature of the horizon. 

It is worth highlighting that recent investigations into evolving black holes within a de-Sitter spacetime have indicated the presence of a second-order phase transition \cite{Zhang:2022aqg,Zhao:2017vlj,Ali:2019rjn,Zhang:2020odg,Ma:2018hni}. This observation adds further credence to the potential significance of the connection we have identified between the signature flip in the deceleration parameter ($q$) and the occurrence of a second-order phase transition.

The stability of the model's thermodynamics can be determined by examining the characteristics of the second-order derivatives pertaining to the system's internal entropy. For examples that demonstrate this approach, readers may refer to the references provided in \cite{Ferreira:2015iaa} and \cite{mukherjee2021nonparametric}. The primary impetus behind the current study is to initially explore the thermodynamic stability of a cosmological model designed to replicate the late-stage evolution characteristics of a $\Lambda$CDM model. The methodology employed involves investigating the concavity of the entropy function associated with the matter content present in the universe.

To undertake this investigation, we delve into the intricacies of the behaviour of entropy and its relationship with the matter content. The key technique revolves around scrutinizing the concavity properties of the entropy function. This analysis hinges on the properties of the Hessian matrix, which encompasses the second-order derivatives of the entropy. For a more comprehensive understanding, we refer to Chapter \eqref{Chapterthermo}, Section \ref{Stability}. To provide a concrete example of this approach within the realm of cosmology, reference can be made to the work conducted by Bhandari, Haldar, and Chakraborty  \cite{bhandari2017interacting}. This framework allows us to establish a connection between the thermodynamic stability of the cosmological model and the underlying properties of the entropy function, thereby offering insights into the system's overall behaviour and evolution.

\section{A Model Mimicking the Characteristics of $\Lambda$CDM}

We consider a universe that is spatially flat, homogeneous, and isotropic, as described by the FRW metric presented in equation\eqref{metric}. Let us reiterate the metric formulation here,
\begin{equation}
ds^2=-dt^2+a^2(t)[dr^2+r^2 d\Omega^2],
\end{equation} 
$a(t)$ denotes the scale factor here.
The Einstein field equations associated with this context can be expressed as follows:
\begin{align} \label{fe}
3H^2 &= \rho , \\
2\dot{H}&= - (\rho+p).
\end{align}
In these equations, $\rho$ and $p$ symbolize the total energy density and pressure attributed to the matter constituents. The parameter $H=\frac{\dot a}{a}$ corresponds to the Hubble parameter, while a dot positioned above a variable signifies its derivative with respect to cosmic time $t$. We adopt units where $c=1$ and $8\pi G=1$. In this context, the radius of the apparent horizon, denoted as $R_{ah}$, is defined by the equation $g^{\mu\nu} R_{{ah}_{,\mu}} R_{{ah}{,\nu}}=0$. For a universe with spatial flatness ($k=0$), the radius of the apparent horizon is given by ${R_{ah}}=\frac{1}{H}$, as detailed in the work by Faraoni \cite{faraoni2015horizons}.

   We adopt a straightforward assumption for the scale factor, given by the expression:
\begin{equation}
\label{ansatz-a}
\frac{a}{a_0}\sim\frac{\sinh^{2/3}(t/t_0)}{\sinh^{2/3}(1)}.
\end{equation}
This choice results in an expansion that is accelerated during later periods while a decelerated expansion in the earlier era is dominated by matter. Here, it is important to note that we consider $a=a_0$ at the time $t=t_0$, and we have set $t_0=1$. Remarkably, this particular ansatz \eqref{ansatz-a} effectively captures the behaviour akin to the $\Lambda$CDM model during the later stages, which is currently favored as the model for our present universe \cite{Padmanabhan:2002ji}.

The equation \eqref{ansatz-a} can be utilized to express the cosmic time $t/t_0$ in terms of the redshift ($z$), wherein $z$ is defined as $1+z = \frac{a_0}{a}$. This relationship can be written as $t/t_0 = \arcsinh\left((\frac{1}{1+z})^{3/2}\sinh(1)\right)$, effectively establishing a connection between time and redshift, where $a_0$ signifies the present value of the scale factor.

The Hubble parameter can be expressed in terms of the redshift $z$ as follows,
\begin{align}\label{H}
H &=\frac{2}{3}\coth (t/t_0)\nonumber\\
&=\frac{2\csch(1)}{3}\frac{\sqrt{1+\frac{\sinh^2(1)}{(1+z)^3}}}{\left(\frac{1}{1+z}\right)^{3/2}}.
\end{align}

The deceleration parameter, which is defined as $q=-\left[1+\frac{\dot{H}}{H^2}\right]$, takes the form:
\begin{equation}\label{q}
q(z)= -1+\frac{3}{2\left(1+\frac{\sinh^2(1)}{(1+z)^3}\right)},
\end{equation}
when expressed in terms of the redshift $z$.

\begin{figure}
\centering
\boxed{\includegraphics[width=82mm]{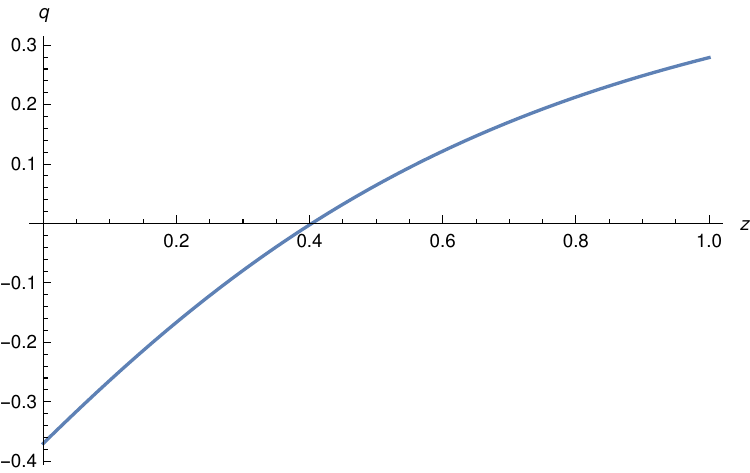}}
\caption{$q$ plotted as a function of $z$}
\end{figure}

In Figure (1), we observe the variation of the deceleration parameter ($q$) with respect to the redshift ($z$). Notably, at a specific redshift value of approximately $z=-1+2^{1/3}\sinh^{2/3}(1)\simeq 0.403$ the evolution of the universe transitions from a decelerating phase to an accelerating one. 

\section{Thermodynamic approach}
\begin{itemize}

\item \textbf{GSL:}
To perform the Generalized Second Law (GSL) test, it is essential to determine the rate at which the total energy changes within the model under consideration. We already provided a  calculation method for the evaluation of the rate of change of total entropy, specifically in terms of the Hubble parameter and its derivative, within the framework of FRW cosmology in chapter \ref{Chapterthermo}.

We have assumed that the fluid within the horizon is in a state of thermodynamic equilibrium with the horizon itself. The conclusions drawn from the research conducted by Mimoso and Pavón \cite{Mimoso:2016jwg} provide insights into the topic at hand. Their study establishes that achieving thermal equilibrium between radiation and the cosmic horizon remains an elusive endeavor. This challenge arises due to Wien's law, which consistently yields a wavelength exceeding the horizon radius across all temporal phases. However, it is possible for nonrelativistic particles to reach equilibrium, depending on their individual masses. In this regard, the notion of thermal equilibrium between dark energy and the horizon, advocated by various scholars, such as in \cite{Gong:2006ma,Izquierdo:2005ku,Gong:2006sn,Zhang:2007iw,Pereira:2008af,Wang:2005pk,MohseniSadjadi:2010nu}, finds justifiable ground.

In the present investigation, it is pertinent to note that we have excluded any radiation component from consideration. Consequently, the assumption of establishing thermodynamic equilibrium between the horizon and the fluid content remains valid and applicable in our study. This rationalizes our approach and lends support to the foundational assumptions guiding our analysis. 

We consider the temperature at equilibrium to be determined by the Hayward-Kodama temperature formula, as established in the references by  \cite{Hayward:1997jp,Hayward:2008jq}:
\begin{equation}\label{temp}
T= \frac{2H^2+\dot{H}}{4\pi H}.
\end{equation}

It is worth noting that the temperature becomes zero when the scale factor takes the specific form of $a(t)=\sqrt{\alpha t^2+\beta t+\gamma}$. Consequently, during a phase dominated solely by radiation (for which $a(t)\propto t^{1/2}$), equation (\ref{temp}) results in a temperature of zero. However, there is no cause for concern in the context of our present study, as we do not engage with radiation-related considerations in any manner.

 In Chapter \ref{Chapterthermo}, we derived the expression for the rate of change of the total entropy, which can be written as follows,
\begin{align}\label{ent-rate2}
\dot{S}&=\dot{S}_h+\dot{S}_{\mbox{in}} \nonumber\\
&= 16\pi^2\frac{\dot{H}^2}{H^3}\left(\frac{1}{2H^2+\dot{H}}\right).
\end{align}

\item \textbf{Thermodynamic Stability Conditions:}
To ensure thermodynamic stability, it is essential to maximize the entropy of the fluid contained within the horizon. This requirement can be translated into a condition involving the Hessian matrix of entropy, as explained in references \cite{callen1998thermodynamics,kubo1968thermodynamics,carter2000classical,Muller1985}. Specifically, all the principle minors of order $k$ in the matrix follow the pattern where they are $\leq 0$ when $k$ is odd and $\geq 0$ when $k$ is even.

The Hessian matrix, denoted as $W$, pertains to the entropy $S_{\text{in}}$ can be written as follows,
\begin{equation}
W=
\begin{bmatrix}
{S_{{\mbox{in}}_{\scriptsize{UU}}}} & {S_{{\mbox{in}}_{\scriptsize{UV}}}}\\
{S_{{\mbox{in}}_{\scriptsize{VU}}}} & {S_{{\mbox{in}}_{\scriptsize{VV}}}}
\end{bmatrix}.
\end{equation}
In the above matrix, a subscript indicates a partial derivative with respect to the specific variable ($U,V$).
Hence, if we express the Hessian matrix explicitly using the partial derivative notation, we obtain,
\begin{equation}
W=
\begin{bmatrix}
\frac{\partial^2{S_{\mbox{in}}}}{\partial U^2} & \frac{\partial^2{S_{\mbox{in}}}}{\partial U\partial V}\\ \vspace{0.005mm}\\
\frac{\partial^2{S_{\mbox{in}}}}{\partial V\partial U} & \frac{\partial^2{S_{\mbox{in}}}}{\partial V^2}
\end{bmatrix}.
\end{equation}

Therefore, in order to ensure thermodynamic stability, it becomes imperative the simultaneous satisfaction of the following conditions,
\begin{align} 
(i)&\hspace{3mm}{S_{{\mbox{in}}_{\scriptsize{UU}}}}\leq 0, \label{condi1}\\
(ii)&\hspace{3mm}{S_{{\mbox{in}}_{\scriptsize{UU}}}}{S_{{\mbox{in}}_{\scriptsize{VV}}}}-{S_{{\mbox{in}}_{\scriptsize{UV}}}^2} \geq 0. \label{condi2}
\end{align}

In terms of thermodynamic parameters, we get the conditions as,

\begin{equation}\label{fc}
 {S_{{\mbox{in}}_{\scriptsize{UU}}}}= -\frac{1}{T^2C_V},
\end{equation}

and

\begin{equation}\label{sc}
{S_{{\mbox{in}}_{\scriptsize{UU}}}}{S_{{\mbox{in}}_{\scriptsize{VV}}}}-{S_{{\mbox{in}}_{\scriptsize{UV}}}^2}=\frac{1}{C_VT^3V\beta_T} = \alpha.
\end{equation}

For the sake of conciseness, the second expression is denoted as $\alpha$. In these equations, $T$ represents temperature, $C_V$ stands for heat capacity at constant volume, and $\beta_T$ indicates isothermal compressibility.\\

Heat capacity at constant volume ($C_V$) is a thermodynamic property that quantifies the amount of heat energy required to produce a unit change in its temperature while keeping its volume constant. It is defined as, 
\begin{equation}\label{cv}
C_V= T\left(\frac{\partial S_{\mbox{in}}}{\partial T}\right)_V.
\end{equation}

Heat capacity at constant pressure ($C_P$) is a thermodynamic property that measures the amount of heat energy required to produce a unit change in its temperature while allowing it to expand or contract under constant pressure conditions. It is defined as, 
\begin{equation}\label{cp}
C_P= T\left(\frac{\partial S_{\mbox{in}}}{\partial T}\right)_P.
\end{equation}

Isothermal compressibility ($\beta_T$) is a thermodynamic property that quantifies a substance's responsiveness to changes in pressure while its temperature is held constant. It measures the fractional change in volume of a substance in response to a unit change in pressure, keeping the temperature constant. Isothermal compressibility is defined as,

\begin{equation}\label{bet}
\beta_T =-\frac{1}{V}\left(\frac{\partial V }{\partial P}\right)_T.
\end{equation}

\end{itemize}
\section{Thermodynamic analysis of the model}

We will examine whether the model adheres to the GSL and assess its thermodynamic stability. We will explore the characteristics exhibited by various thermodynamic parameters. We substitute the expression of $H$ (as given in equation \eqref{H}) into equation \eqref{ent-rate2}, resulting in the following expression,
\begin{equation}
\dot{S} = 108\pi^2\frac{\csch^4(t)}{\coth^3(t)}\frac{1}{\left(4\coth^2(t)-3\csch^2(t)\right)}.
\end{equation}

Based on the aforementioned mathematical expression, it is evident that entropy increases with time. Consequently, the GSL remains valid for the model.

When considering the fluid confined within the horizon, the expression of Gibbs' law takes the form:

\begin{equation}\label{Gibbs}
TdS_{\text{in}}=dU+pdV.
\end{equation}

 This equation serves as a cornerstone in comprehending the thermodynamic processes occurring within the confines of the event horizon.
By utilizing the equation \eqref{Gibbs}, one can compute the heat capacities and isothermal compressibility for the material enclosed within the event horizon.

\begin{align}\label{cv-expr}
 C_V &= V\left(\frac{\partial \rho}{\partial T}\right)_V \nonumber \\
 &= 32\pi^2\frac{\dot{H}}{2H^2\dot{H}+H\ddot{H}-\dot{H}^2}\nonumber\\
&= \frac{144\pi^2}{-2+(1+z)^3\csch^2(1)},
 \end{align}
 
 \begin{align}
 C_P &=  V\left(\frac{\partial \rho}{\partial T}\right)_P+(\rho+P)\left(\frac{\partial V}{\partial T}\right)_P \nonumber \\
 &= 32\pi^2\frac{H^2\dot{H}+\dot{H^2}}{H^2(2H^2\dot{H}+H\ddot{H}-\dot{H}^2)} \nonumber\\
 &= -72\pi^2 \frac{\sinh^2(1)}{(1+z)^3+\sinh^2(1)},
 \end{align}
 
\begin{align}\label{beta-expr}
 \beta_T &= \frac{3\dot{H}(2H^2+\dot{H})}{2(H^2+\dot{H})(2H^2\dot{H}+H\ddot{H}-\dot{H}^2)} \nonumber\\
 &=-\frac{27}{4}\frac{4+(1+z)^3\csch^2(1)}{\left(-2+(1+z)^3\csch^2(1)\right)}.
 \end{align}

 \begin{figure}
 \centering
\boxed{\includegraphics[scale=0.6]{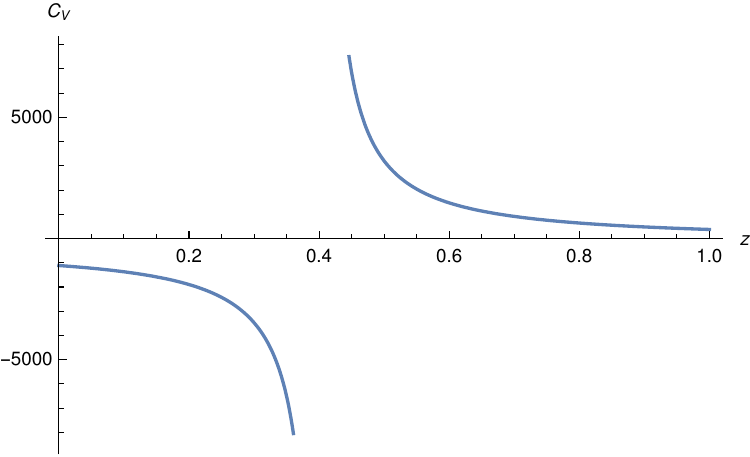}}
\caption{$C_V$ plotted as a function of $z$}\label{cvfig}
\end{figure}.

\begin{figure}
\centering
\boxed{\includegraphics[scale=0.6]{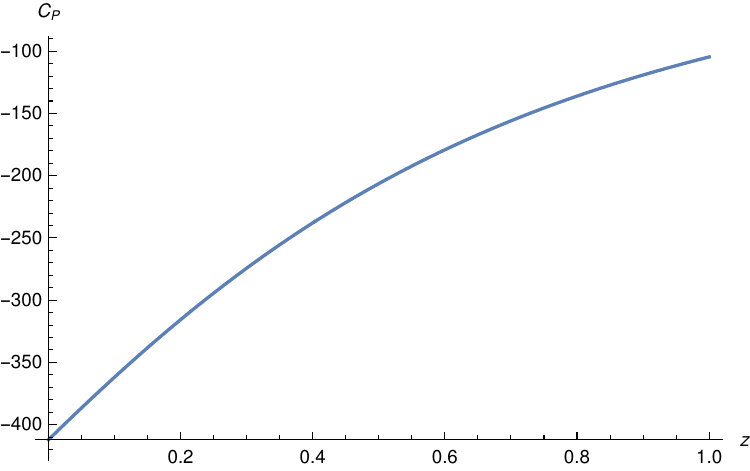}}
\centering\caption{$C_P$ plotted as a function of $z$}\label{cpfig}
\end{figure}

\begin{figure}
\centering
 \boxed{\includegraphics[scale=0.6]{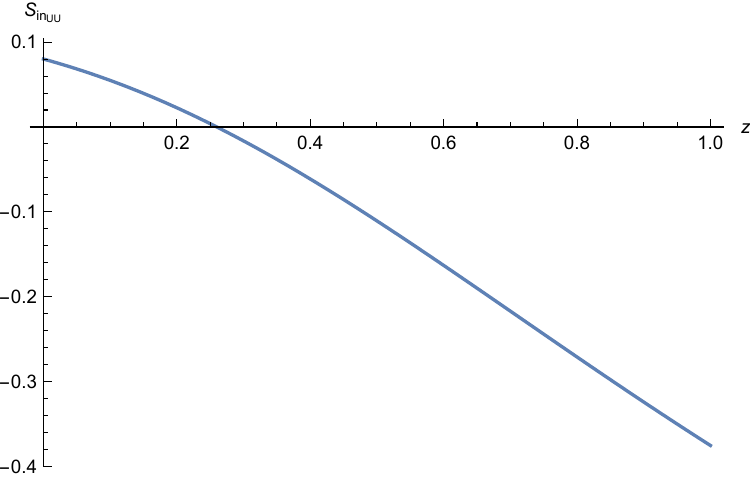}}
\caption{${S_{\mbox{in}}}_{UU}$  plotted as a function of $z$}\label{cond1}
 \end{figure}

\begin{figure}
 \centering
 \boxed{\includegraphics[scale=0.6]{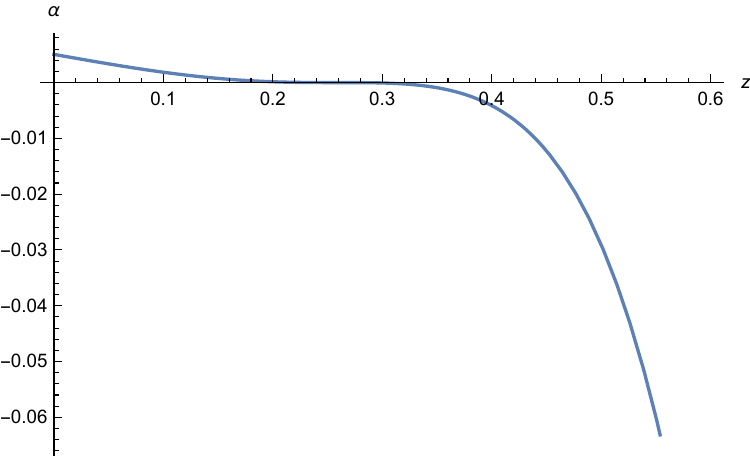}}
 \caption{ $\alpha$ plotted as a function of $z$}\label{cond2}
 \end{figure}

Figures (\ref{cvfig}) and (\ref{cpfig}) illustrate the behaviour of $C_V$ and $C_P$, respectively, with respect to the redshift $z$ within the range of low values ($0\leq z \leq 1$). Utilizing the expressions for $C_V$ and $\beta_T$ derived from equations \eqref{cv-expr} and \eqref{beta-expr} in equations \eqref{fc} and \eqref{sc}, it is possible to generate plots for $S_{\text{in}UU}$ and $\alpha$, as depicted in figures \eqref{cond1} and \eqref{cond2}, respectively. It is evident that the two conditions \eqref{condi1} and \eqref{condi2} are not simultaneously met within the specified low redshift range ($0\leq z \leq 1$). Consequently, the model does not exhibit thermodynamic stability within this particular redshift range.\\

The significant observation gleaned from figure \eqref{cvfig} is that heat capacity at constant volume ($C_V$) displays a noteworthy behaviour – specifically, it exhibits a divergence or discontinuity at a particular redshift value of $z=-1+2^{1/3}\sinh^{2/3}(1)\simeq 0.403$. This redshift value corresponds to the critical point where the expansion of the universe transitions from a decelerated phase to an accelerated one. Remarkably, this transition holds the characteristics of a thermodynamic phase transition.\\

The transition from decelerated to accelerated expansion, it turns out, aligns with a distinct thermodynamic phase transition. Notably, the entropy ($S_{mbox{in}}$) does not exhibit any corresponding discontinuity within the specified redshift range; rather, the discontinuity manifests in the behaviour of $C_V$. Consequently, this phase transition is unequivocally identified as a second-order phase transition.\\

It is noteworthy to highlight that $C_V$ assumes a negative value for the present universe, specifically for $z>-1+2^{1/3}\sinh^{2/3}(1)\simeq 0.403$. Nevertheless, the presence of a negative heat capacity in gravitational systems is not a surprising phenomenon (for a comprehensive overview, we refer to \cite{Padmanabhan:1989gm}). A study conducted by Luongo and Quevedo \cite{Luongo:2012dv} yielded a significant finding that in the context of a currently accelerating universe, a negative value for $C_V$ becomes a requisite.\\

By substituting equation \eqref{q} into equation \eqref{cv-expr}, it becomes possible to express $C_V$ in terms of the deceleration parameter ($q$) as follows,
\begin{equation}\label{cvq}
 C_V=24\pi^2\frac{1-2q}{q}.
\end{equation}

Therefore, it becomes evident that the origin of the discontinuity in $C_V$ arises from the presence of the deceleration parameter ($q$) in the denominator with a positive exponent of $1$. As a consequence, the deceleration parameter ($q$) serves as the order parameter, and the observed discontinuity is characterized by an order of unity.

\section{Summary and Discussion}

The focus of the thermodynamic stability analysis was directed toward a model designed to emulate the behaviour of the $\Lambda$CDM model, which faithfully represents the present universe. Given the dynamic nature of the evolving horizon, the Hayward-Kodama temperature was chosen as the appropriate measure of the temperature of the horizon. The analysis revealed that the thermal capacity exhibited a negative value, implying that the cosmic matter contained within the horizon lacks thermodynamic stability.

The significance of this study lies in the profound result indicating that the matter content experiences a phase transition precisely at the point where the universe undergoes a transition from decelerated expansion to an accelerated one. Remarkably, this phase transition follows the characteristics of a second-order transition, as evidenced by the discontinuity in the heat capacity ($C_V$). The deceleration parameter $q$ serves as the order parameter.

 Interestingly, the earlier studies failed to detect the second-order phase transition at the onset of the accelerated expansion of the universe could potentially be attributed to the utilization of the Hawking temperature of the horizon. This approach overlooked the fact that the apparent horizon is in a state of evolution, which the Hayward-Kodama temperature accurately addresses.\\
 
 Pavón and Wang \cite{Pavon:2007gt} showed that the dark matter and dark energy could potentially evolve independently. Therefore, they may not be in thermal equilibrium with each other. However, it is important to note that our approach considers a composite fluid in which distinct sectors are not explicitly differentiated, only the evolution history holds significance.

%% file: Chapters/Chapter6.tex
\chapter{Conclusions and Outlook} 
\label{Chapter6}
This dissertation centers on the exploration of thermodynamic analyses in the context of cosmology. This thermodynamic exploration involves two main aspects: assessment through the Generalized Second Law (GSL)(Chapters \ref{Chapter2},\ref{Chapter3},\ref{Chapter4}) and the analysis of thermodynamic stability (Chapter\ref{Chapter5}). For the purpose of our research, we have made the assumption that the fluid contained within the horizon is in a state of thermodynamic equilibrium with the horizon itself. Thermal equilibrium between radiation and the cosmic horizon is impossible due to Wien's law, which consistently yields a wavelength larger than the horizon radius over all time periods \cite{Mimoso:2016jwg}. Nevertheless, nonrelativistic particles can achieve equilibrium at a certain expansion point based on particle mass. Given this, the notion of thermal equilibrium between dark energy and the horizon, proposed by various researchers, has a valid basis. Our study excludes radiation (except for a part of chapter \ref{Chapter3}), affirming the validity of assuming thermodynamic equilibrium between the horizon and fluid content.\vspace{0.5em} 

As the universe is evolving, we have considered the dynamic apparent horizon, instead of the event horizon to work with. The temperature of the apparent horizon is considered to be the Hayward-Kodama temperature. Hayward proposed the definition of this temperature linked through an alternative definition of surface gravity that applies to dynamic, spherically symmetric spacetimes, which relies on the Kodama vector. This temperature, $T= \frac{2H^2+\dot{H}}{4\pi H}$ serves as the equilibrium temperature \cite{Hayward:1997jp,Hayward:2008jq}. \vspace{0.5em}

Initially, the generalised second law of thermodynamics was proposed by Bekenstein in the early 1970s \cite{Bekenstein:1972tm,Bekenstein:1973ur,Bekenstein:1974ax}.
This principle asserts that the total entropy of the universe,  the sum of matter entropy and horizon entropy, must never diminish as time progresses. Therefore, we have subjected various cosmological models to scrutiny using the GSL criterion, identifying which models successfully meet the test requirements or determining the constraints needed for passing the test. \vspace{0.5em}

Our next course of action involved conducting a thermodynamic stability analysis on a model. This involved utilizing the property of the concavity of the entropy, wherein we examined the Hessian matrix of entropy to ascertain whether it exhibited semi-negative definiteness. \vspace{0.5em}

In chapter \ref{Chapter2}, we have undertaken a comparative analysis between thawing and freezing models, focusing on their adherence to the fundamental principles of thermodynamics, specifically the GSL. Our work involves the assessment of the total entropy ($S_{\text{tot}}$), accomplished by adding up the entropy of the cosmic horizon with that of the enclosed matter within said horizon. To facilitate this study, we employ a straightforward ansatz, $\frac{1}{\rho_\Phi}\frac{\mathrm{\partial}\rho_\Phi}{\mathrm{\partial} a} = -\frac{\lambda}{a^{1-2\alpha}}$, proposed by Carvalho {\it et al} \cite{Carvalho:2006fy}, to formulate the evolution of the energy density of the quintessence field. The ansatz proposes that the power-law dependence of the scalar field is reflected in the divergence of the logarithm of energy density.  This approach enables us to effectively delineate the parameter space ($\alpha$) that corresponds to the distinct behaviors of thawing and freezing of the field. The findings reveal an intriguing incongruity between both model categories and the GSL. Notably, there are instances in which the entropy ($S$) experiences a decrease, and this descent occurs at an accelerated rate. For the freezing models, this contravention of the GSL is observed in a remote cosmic past, specifically during an epoch characterized by a redshift of $z \sim 10^{4}$. During this epoch, a cosmological model that combines quintessence and cold dark matter fails to satisfactorily elucidate the cosmic evolution, necessitating a dominant contribution from a radiation distribution. Consequently, the applicability of the GSL seems questionable in such a scenario. \vspace{0.5em}

Conversely, in the context of thawing models, our analysis predicts an anomalous breakdown of the GSL within a finite  future. This observation underscores a pivotal implication: freezing models exhibit a thermodynamically more tenable model when compared to their thawing counterparts. \vspace{0.5em}

In chapter \ref{Chapter3}, our research thoroughly investigates the thermodynamic properties of cosmological models governed by radiation and dust dominance. These models are analyzed in the framework of the Brans-Dicke theory. In particular, we delve into the properties of a spatially flat, homogenous, and isotropic universe within  the \textit{Einstein frame}. This frame represents the conformally transformed version of the Brans-Dicke theory. In a universe dominated by radiation, the solutions obtained from Brans-Dicke theory, with a positive parameter $\omega$, fail to satisfy the principles of the generalized second law. However, remarkably, when specific ranges of negative $\omega$ values are considered, the model harmonizes effectively with the requisites of thermodynamics. This finding is very promising because negative values of the parameter $\omega$ have been strongly associated with the phenomenon of accelerated cosmic expansion.

Shifting the focus to a universe governed predominantly by dust, the model successfully upholds the principles of the generalized second law, but notably, only for certain small negative values of $\omega$. Particularly noteworthy is the fact that this range, characterized by $-2< \omega < -\frac{5}{3}$, extensively overlaps with the parameter values necessary to elucidate an accelerated expansion of the universe, all without necessitating the introduction of exotic forms of matter \cite{Banerjee:2000mj}.\vspace{0.5em}

Chapter \ref{Chapter4} is about the thermodynamic assessment of the models reconstructed from the jerk parameter, proposed by Mukherjee \textit{et al.} \cite{Mukherjee:2016trt,Mukherjee:2016shl}. Encouragingly, the assessment of the Generalized Second Law (GSL) of thermodynamics can be effectively conducted using kinematical quantities like the Hubble parameter and its derivative, as described in equation \eqref{s-tot-dot}. This analysis of thermodynamic viability ensures that the reconstructed models adhere to thermodynamic principles, strengthening their credibility. \vspace{0.5em}

Our investigation reveals the feasibility of identifying cosmological models that maintain thermodynamic consistency through the reconstruction of the jerk parameter. Notably, among the subset of four non-interacting models, model IV—exhibits a future decline in entropy ($z<0$) due to its inverse dependence on $(1+z)$. This behavior occurs in proximity to the present era ($z=0$). However, all other models, including the one involving interaction within the dark sector (Model V), adhere to the Generalized Second Law (GSL) of thermodynamics. Model I, while in line with the GSL, demonstrates a significant and sudden entropy increase in the future ($z<0$). Models II and III exhibit consistent behavior, as does Model V, which allows interaction within the dark sector. \vspace{0.5em}

Chapter \ref{Chapter5} deviates slightly from the preceding chapters in that it encompasses a dual focus. Here, our exploration goes beyond solely examining the feasibility of the Generalized Second Law (GSL). Instead, we make an attempt to analyze the thermodynamic stability of an accelerating cosmological model. \vspace{0.5em}

The thermodynamic stability analysis focused on a model crafted to mimic the behavior of the $\Lambda$CDM model, favoured model for the portrayal of the current state of the universe. In light of the dynamic nature of the evolving horizon, the Hayward-Kodama temperature emerged as the suitable parameter to gauge the temperature of the horizon. The findings of the analysis brought forth a significant revelation, the thermal capacity exhibited a negative value, indicating an absence of thermodynamic stability within the cosmic matter confined by the horizon. \vspace{0.5em} 

The interesting outcome lies in the fact that the thermodynamic phase transition of matter content is intimately connected with the nature of the cosmic evolution.  The results suggest a thermodynamic phase transition that aligns perfectly with the shift from a decelerated cosmic expansion to an accelerated one. Notably, this phase transition demonstrates the characteristics of a second-order transition, as evident from the discontinuity observed in the heat capacity ($C_V$). The deceleration parameter ($q$) serves as the order parameter in this intricate connection. \vspace{0.5em}

The aim of this investigation was merely to align the observed value of $z$ with $q=0$, but rather to delve into the qualitative nature of the thermodynamic aspect accompanying the pivotal shift in $q$. Apparently, the prior investigations failed to detect the second-order phase transition at the onset of the accelerated expansion of the universe because the dynamical nature of the horizon had been ignored and Hawking temperature was relied upon. The use of Hayward-Kodama temperature brings out the remarkable feature.

\section*{Future Prospects}

In this thesis, we have explored the thermodynamics of some cosmological models, especially those that give rise to a late-time cosmic acceleration. We have examined the viability of these models via GSL. We have considered the thermodynamic equilibrium between the apparent horizon and the matter bounded by the horizon. Haward-Kodama (HK) temperature has been considered as the horizon temperature. It will be interesting to proceed with the thermodynamic analysis by considering another horizon temperature such as Cai-Kim (CK) temperature  \cite{Cai:2005ra,Cai:2008gw} instead of the Hayward-Kodama temperature. One can also do a comparative analysis between the outcomes of using the HK temperature and CK temperature.  One may take the thermodynamic analysis to the next step and explore GSL in non-equilibrium thermodynamics. As shown in the article \cite{Pavon:2007gt}, dark matter and dark energy may evolve independently and therefore they may not be in thermal equilibrium; hence it will be worthwhile to consider these two fluids separately and then proceed with the thermodynamic analysis.  \vspace{0.5em}

Thermodynamic stability analysis can also be conducted for other models. It will be interesting to see what thermodynamic features these models reveal. We have tried to see if the transition from decelerated to accelerated cosmic expansion is connected to the phase transition without considering any model. But we have not progressed much in this project.

%% file: combibib.bib
@Inbook{Tsujikawa2010,
author="Tsujikawa, S.",
editor="Wolschin, Georg",
title="Modified Gravity Models of Dark Energy",
bookTitle="Lectures on Cosmology: Accelerated Expansion of the Universe",
year="2010",
publisher="Springer Berlin Heidelberg",
address="Berlin, Heidelberg",
pages="99--145",
}

@article{Sotiriou:2008rp,
    author = "Sotiriou, Thomas P. and Faraoni, Valerio",
    title = "{f(R) Theories Of Gravity}",
    eprint = "0805.1726",
    archivePrefix = "arXiv",
    primaryClass = "gr-qc",
    doi = "10.1103/RevModPhys.82.451",
    journal = "Rev. Mod. Phys.",
    volume = "82",
    pages = "451--497",
    year = "2010"
}

@article{DeFelice:2010aj,
    author = "De Felice, Antonio and Tsujikawa, Shinji",
    title = "{f(R) theories}",
    eprint = "1002.4928",
    archivePrefix = "arXiv",
    primaryClass = "gr-qc",
    doi = "10.12942/lrr-2010-3",
    journal = "Living Rev. Rel.",
    volume = "13",
    pages = "3",
    year = "2010"
}

@article{Maartens:2010ar,
    author = "Maartens, Roy and Koyama, Kazuya",
    title = "{Brane-World Gravity}",
    eprint = "1004.3962",
    archivePrefix = "arXiv",
    primaryClass = "hep-th",
    doi = "10.12942/lrr-2010-5",
    journal = "Living Rev. Rel.",
    volume = "13",
    pages = "5",
    year = "2010"
}

@article{Clifton:2011jh,
    author = "Clifton, Timothy and Ferreira, Pedro G. and Padilla, Antonio and Skordis, Constantinos",
    title = "{Modified Gravity and Cosmology}",
    eprint = "1106.2476",
    archivePrefix = "arXiv",
    primaryClass = "astro-ph.CO",
    doi = "10.1016/j.physrep.2012.01.001",
    journal = "Phys. Rept.",
    volume = "513",
    pages = "1--189",
    year = "2012"
}

@article{Joyce:2014kja,
    author = "Joyce, Austin and Jain, Bhuvnesh and Khoury, Justin and Trodden, Mark",
    title = "{Beyond the Cosmological Standard Model}",
    eprint = "1407.0059",
    archivePrefix = "arXiv",
    primaryClass = "astro-ph.CO",
    doi = "10.1016/j.physrep.2014.12.002",
    journal = "Phys. Rept.",
    volume = "568",
    pages = "1--98",
    year = "2015"
}

@article{Koyama_2016,
doi = {10.1088/0034-4885/79/4/046902},
url = {https://dx.doi.org/10.1088/0034-4885/79/4/046902},
year = {2016},
month = {mar},
publisher = {IOP Publishing},
volume = {79},
number = {4},
pages = {046902},
author = {Kazuya Koyama},
title = {Cosmological tests of modified gravity},
journal = {Reports on Progress in Physics},
}

@article{Capozziello:2002rd,
    author = "Capozziello, Salvatore",
    title = "{Curvature quintessence}",
    eprint = "gr-qc/0201033",
    archivePrefix = "arXiv",
    doi = "10.1142/S0218271802002025",
    journal = "Int. J. Mod. Phys. D",
    volume = "11",
    pages = "483--492",
    year = "2002"
}

@article{Capozziello:2003gx,
    author = "Capozziello, S. and Cardone, V. F. and Carloni, S. and Troisi, A.",
    title = "{Curvature quintessence matched with observational data}",
    eprint = "astro-ph/0307018",
    archivePrefix = "arXiv",
    doi = "10.1142/S0218271803004407",
    journal = "Int. J. Mod. Phys. D",
    volume = "12",
    pages = "1969--1982",
    year = "2003"
}

@article{Carroll:2003wy,
    author = "Carroll, Sean M. and Duvvuri, Vikram and Trodden, Mark and Turner, Michael S.",
    title = "{Is cosmic speed - up due to new gravitational physics?}",
    eprint = "astro-ph/0306438",
    archivePrefix = "arXiv",
    reportNumber = "FERMILAB-PUB-03-263-A, SU-GP-03-6-2",
    doi = "10.1103/PhysRevD.70.043528",
    journal = "Phys. Rev. D",
    volume = "70",
    pages = "043528",
    year = "2004"
}

@article{Carroll:2004de,
    author = "Carroll, Sean M. and De Felice, Antonio and Duvvuri, Vikram and Easson, Damien A. and Trodden, Mark and Turner, Michael S.",
    title = "{The Cosmology of generalized modified gravity models}",
    eprint = "astro-ph/0410031",
    archivePrefix = "arXiv",
    reportNumber = "FERMILAB-PUB-04-335-A",
    doi = "10.1103/PhysRevD.71.063513",
    journal = "Phys. Rev. D",
    volume = "71",
    pages = "063513",
    year = "2005"
}

@article{Nojiri:2003ft,
    author = "Nojiri, Shin'ichi and Odintsov, Sergei D.",
    title = "{Modified gravity with negative and positive powers of the curvature: Unification of the inflation and of the cosmic acceleration}",
    eprint = "hep-th/0307288",
    archivePrefix = "arXiv",
    doi = "10.1103/PhysRevD.68.123512",
    journal = "Phys. Rev. D",
    volume = "68",
    pages = "123512",
    year = "2003"
}

@article{Nojiri:2006gh,
    author = "Nojiri, Shin'ichi and Odintsov, Sergei D.",
    title = "{Modified f(R) gravity consistent with realistic cosmology: From matter dominated epoch to dark energy universe}",
    eprint = "hep-th/0608008",
    archivePrefix = "arXiv",
    doi = "10.1103/PhysRevD.74.086005",
    journal = "Phys. Rev. D",
    volume = "74",
    pages = "086005",
    year = "2006"
}

@article{Nojiri:2007jr,
    author = "Nojiri, Shin'ichi and Odintsov, Sergei D.",
    title = "{Newton law corrections and instabilities in f(R) gravity with the effective cosmological constant epoch}",
    eprint = "0706.1378",
    archivePrefix = "arXiv",
    primaryClass = "hep-th",
    doi = "10.1016/j.physletb.2007.07.039",
    journal = "Phys. Lett. B",
    volume = "652",
    pages = "343--348",
    year = "2007"
}

@article{Nojiri:2007as,
    author = "Nojiri, Shin'ichi and Odintsov, Sergei D.",
    title = "{Unifying inflation with LambdaCDM epoch in modified f(R) gravity consistent with Solar System tests}",
    eprint = "0707.1941",
    archivePrefix = "arXiv",
    primaryClass = "hep-th",
    doi = "10.1016/j.physletb.2007.10.027",
    journal = "Phys. Lett. B",
    volume = "657",
    pages = "238--245",
    year = "2007"
}

@article{Vollick:2003aw,
    author = "Vollick, Dan N.",
    title = "{1/R Curvature corrections as the source of the cosmological acceleration}",
    eprint = "astro-ph/0306630",
    archivePrefix = "arXiv",
    doi = "10.1103/PhysRevD.68.063510",
    journal = "Phys. Rev. D",
    volume = "68",
    pages = "063510",
    year = "2003"
}

@article{Mena:2005ta,
    author = "Mena, Olga and Santiago, Jose and Weller, Jochen",
    title = "{Constraining inverse curvature gravity with supernovae}",
    eprint = "astro-ph/0510453",
    archivePrefix = "arXiv",
    reportNumber = "FERMILAB-PUB-05-466-T",
    doi = "10.1103/PhysRevLett.96.041103",
    journal = "Phys. Rev. Lett.",
    volume = "96",
    pages = "041103",
    year = "2006"
}

@article{Banerjee:1996iy,
    author = "Banerjee, N. and Sen, S.",
    title = "{Does Brans-Dicke theory always yield general relativity in the infinite omega limit?}",
    reportNumber = "PRINT-97-099 (JADAVPUR)",
    doi = "10.1103/PhysRevD.56.1334",
    journal = "Phys. Rev. D",
    volume = "56",
    pages = "1334--1337",
    year = "1997"
}

@article{Amendola:1999qq,
    author = "Amendola, Luca",
    title = "{Scaling solutions in general nonminimal coupling theories}",
    eprint = "astro-ph/9904120",
    archivePrefix = "arXiv",
    doi = "10.1103/PhysRevD.60.043501",
    journal = "Phys. Rev. D",
    volume = "60",
    pages = "043501",
    year = "1999"
}

@article{Uzan:1999ch,
    author = "Uzan, Jean-Philippe",
    title = "{Cosmological scaling solutions of nonminimally coupled scalar fields}",
    eprint = "gr-qc/9903004",
    archivePrefix = "arXiv",
    reportNumber = "UGVA-DPT-1998-12-1022",
    doi = "10.1103/PhysRevD.59.123510",
    journal = "Phys. Rev. D",
    volume = "59",
    pages = "123510",
    year = "1999"
}

@article{Chiba:1999wt,
    author = "Chiba, Takeshi",
    title = "{Quintessence, the gravitational constant, and gravity}",
    eprint = "gr-qc/9903094",
    archivePrefix = "arXiv",
    reportNumber = "UTAP-321",
    doi = "10.1103/PhysRevD.60.083508",
    journal = "Phys. Rev. D",
    volume = "60",
    pages = "083508",
    year = "1999"
}

@article{Amendola:1999er,
    author = "Amendola, Luca",
    title = "{Coupled quintessence}",
    eprint = "astro-ph/9908023",
    archivePrefix = "arXiv",
    doi = "10.1103/PhysRevD.62.043511",
    journal = "Phys. Rev. D",
    volume = "62",
    pages = "043511",
    year = "2000"
}

@article{Bartolo:1999sq,
    author = "Bartolo, Nicola and Pietroni, Massimo",
    title = "{Scalar tensor gravity and quintessence}",
    eprint = "hep-ph/9908521",
    archivePrefix = "arXiv",
    reportNumber = "DFPD-99-TH-40",
    doi = "10.1103/PhysRevD.61.023518",
    journal = "Phys. Rev. D",
    volume = "61",
    pages = "023518",
    year = "2000"
}

@article{Perrotta:1999am,
    author = "Perrotta, Francesca and Baccigalupi, Carlo and Matarrese, Sabino",
    title = "{Extended quintessence}",
    eprint = "astro-ph/9906066",
    archivePrefix = "arXiv",
    doi = "10.1103/PhysRevD.61.023507",
    journal = "Phys. Rev. D",
    volume = "61",
    pages = "023507",
    year = "1999"
}

@article{Riazuelo:2001mg,
    author = "Riazuelo, Alain and Uzan, Jean-Philippe",
    title = "{Cosmological observations in scalar - tensor quintessence}",
    eprint = "astro-ph/0107386",
    archivePrefix = "arXiv",
    reportNumber = "SACLAY-SPHT-T02-011",
    doi = "10.1103/PhysRevD.66.023525",
    journal = "Phys. Rev. D",
    volume = "66",
    pages = "023525",
    year = "2002"
}

@article{Banerjee:2000mj,
    author = "Banerjee, Narayan and Pavon, Diego",
    title = "{Cosmic acceleration without quintessence}",
    eprint = "gr-qc/0012048",
    archivePrefix = "arXiv",
    doi = "10.1103/PhysRevD.63.043504",
    journal = "Phys. Rev. D",
    volume = "63",
    pages = "043504",
    year = "2001"
}

@article{Banerjee:2000gt,
    author = "Banerjee, Narayan and Pavon, Diego",
    title = "{A Quintessence scalar field in Brans-Dicke theory}",
    eprint = "gr-qc/0012098",
    archivePrefix = "arXiv",
    doi = "10.1088/0264-9381/18/4/302",
    journal = "Class. Quant. Grav.",
    volume = "18",
    pages = "593",
    year = "2001"
}

@article{Sen:2000zk,
    author = "Sen, S. and Sen, A. A.",
    title = "{Late time acceleration in Brans-Dicke cosmology}",
    eprint = "gr-qc/0010092",
    archivePrefix = "arXiv",
    reportNumber = "MRI-P-001002",
    doi = "10.1103/PhysRevD.63.124006",
    journal = "Phys. Rev. D",
    volume = "63",
    pages = "124006",
    year = "2001"
}

@article{Mota:2003tm,
    author = "Mota, David F. and Barrow, John D.",
    title = "{Local and global variations of the fine structure constant}",
    eprint = "astro-ph/0309273",
    archivePrefix = "arXiv",
    doi = "10.1111/j.1365-2966.2004.07505.x",
    journal = "Mon. Not. Roy. Astron. Soc.",
    volume = "349",
    pages = "291",
    year = "2004"
}

@article{Mota:2003tc,
    author = "Mota, David F. and Barrow, John D.",
    title = "{Varying alpha in a more realistic Universe}",
    eprint = "astro-ph/0306047",
    archivePrefix = "arXiv",
    doi = "10.1016/j.physletb.2003.12.016",
    journal = "Phys. Lett. B",
    volume = "581",
    pages = "141--146",
    year = "2004"
}

@article{Mukherjee:2019see,
    author = "Mukherjee, Purba and Chakrabarti, Soumya",
    title = "{Exact solutions and accelerating universe in modified Brans\textendash{}Dicke theories}",
    eprint = "1908.01564",
    archivePrefix = "arXiv",
    primaryClass = "gr-qc",
    reportNumber = "EPJC-19-06-153.R1",
    doi = "10.1140/epjc/s10052-019-7201-x",
    journal = "Eur. Phys. J. C",
    volume = "79",
    number = "8",
    pages = "681",
    year = "2019"
}

@article{Bergmann:1968ve,
    author = "Bergmann, Peter G.",
    title = "{Comments on the scalar tensor theory}",
    doi = "10.1007/BF00668828",
    journal = "Int. J. Theor. Phys.",
    volume = "1",
    pages = "25--36",
    year = "1968"
}

@article{Dvali:2000hr,
    author = "Dvali, G. R. and Gabadadze, Gregory and Porrati, Massimo",
    title = "{4-D gravity on a brane in 5-D Minkowski space}",
    eprint = "hep-th/0005016",
    archivePrefix = "arXiv",
    reportNumber = "NYU-TH-00-04-01",
    doi = "10.1016/S0370-2693(00)00669-9",
    journal = "Phys. Lett. B",
    volume = "485",
    pages = "208--214",
    year = "2000"
}

@article{Deffayet:2001pu,
    author = "Deffayet, Cedric and Dvali, G. R. and Gabadadze, Gregory",
    title = "{Accelerated universe from gravity leaking to extra dimensions}",
    eprint = "astro-ph/0105068",
    archivePrefix = "arXiv",
    reportNumber = "NYU-TH-01-04-03, TPI-MINN-01-17, UMN-TH-2003",
    doi = "10.1103/PhysRevD.65.044023",
    journal = "Phys. Rev. D",
    volume = "65",
    pages = "044023",
    year = "2002"
}

@article{Cai_2016,
doi = {10.1088/0034-4885/79/10/106901},
url = {https://dx.doi.org/10.1088/0034-4885/79/10/106901},
year = {2016},
month = {sep},
publisher = {IOP Publishing},
volume = {79},
number = {10},
pages = {106901},
author = {Yi-Fu Cai and Salvatore Capozziello and Mariafelicia De Laurentis and Emmanuel N Saridakis},
title = {f(T) teleparallel gravity and cosmology},
journal = {Reports on Progress in Physics},
}

@article{Nicolis:2008in,
    author = "Nicolis, Alberto and Rattazzi, Riccardo and Trincherini, Enrico",
    title = "{The Galileon as a local modification of gravity}",
    eprint = "0811.2197",
    archivePrefix = "arXiv",
    primaryClass = "hep-th",
    doi = "10.1103/PhysRevD.79.064036",
    journal = "Phys. Rev. D",
    volume = "79",
    pages = "064036",
    year = "2009"
}

@article{Nojiri:2005vv,
    author = "Nojiri, Shin'ichi and Odintsov, Sergei D. and Sasaki, Misao",
    title = "{Gauss-Bonnet dark energy}",
    eprint = "hep-th/0504052",
    archivePrefix = "arXiv",
    reportNumber = "YITP-05-14",
    doi = "10.1103/PhysRevD.71.123509",
    journal = "Phys. Rev. D",
    volume = "71",
    pages = "123509",
    year = "2005"
}

@article{Nojiri:2005jg,
    author = "Nojiri, Shin'ichi and Odintsov, Sergei D.",
    title = "{Modified Gauss-Bonnet theory as gravitational alternative for dark energy}",
    eprint = "hep-th/0508049",
    archivePrefix = "arXiv",
    doi = "10.1016/j.physletb.2005.10.010",
    journal = "Phys. Lett. B",
    volume = "631",
    pages = "1--6",
    year = "2005"
}

@article{Nojiri:2006je,
    author = "Nojiri, Shin'ichi and Odintsov, Sergei D. and Sami, M.",
    title = "{Dark energy cosmology from higher-order, string-inspired gravity and its reconstruction}",
    eprint = "hep-th/0605039",
    archivePrefix = "arXiv",
    doi = "10.1103/PhysRevD.74.046004",
    journal = "Phys. Rev. D",
    volume = "74",
    pages = "046004",
    year = "2006"
}

@article{Bamba:2013aca,
    author = "Bamba, K. and Hossain, Md. Wali and Myrzakulov, R. and Nojiri, S. and Sami, M.",
    title = "{Cosmological investigations of (extended) nonlinear massive gravity schemes with nonminimal coupling}",
    eprint = "1309.6413",
    archivePrefix = "arXiv",
    primaryClass = "hep-th",
    doi = "10.1103/PhysRevD.89.083518",
    journal = "Phys. Rev. D",
    volume = "89",
    number = "8",
    pages = "083518",
    year = "2014"
}

@article{Hossain:2014xha,
    author = "Hossain, Md. Wali and Myrzakulov, R. and Sami, M. and Saridakis, Emmanuel N.",
    title = "{Variable gravity: A suitable framework for quintessential inflation}",
    eprint = "1402.6661",
    archivePrefix = "arXiv",
    primaryClass = "gr-qc",
    doi = "10.1103/PhysRevD.90.023512",
    journal = "Phys. Rev. D",
    volume = "90",
    number = "2",
    pages = "023512",
    year = "2014"
}

@article{Starobinsky:1980te,
    author = "Starobinsky, Alexei A.",
    editor = "Khalatnikov, I. M. and Mineev, V. P.",
    title = "{A New Type of Isotropic Cosmological Models Without Singularity}",
    doi = "10.1016/0370-2693(80)90670-X",
    journal = "Phys. Lett. B",
    volume = "91",
    pages = "99--102",
    year = "1980"
}

@article{Kerner:1982yg,
    author = "Kerner, R",
    title = "{Cosmology without singularity and nonlinear gravitational Lagrangians}",
    doi = "10.1007/BF00756329",
    journal = "Gen. Rel. Grav.",
    volume = "14",
    pages = "453--469",
    year = "1982"
}

@article{Duruisseau:1986ga,
    author = "Duruisseau, J. P. and Kerner, R.",
    title = "{The Effective Gravitational Lagrangian and the Energy Momentum Tensor in the Inflationary Universe}",
    doi = "10.1088/0264-9381/3/5/012",
    journal = "Class. Quant. Grav.",
    volume = "3",
    pages = "817--824",
    year = "1986"
}

@article{Nojiri:2003ni,
    author = "Nojiri, Shinichi and Odintsov, Sergei D.",
    title = "{Modified gravity with in R terms and cosmic acceleration}",
    eprint = "hep-th/0308176",
    archivePrefix = "arXiv",
    doi = "10.1023/B:GERG.0000035950.40718.48",
    journal = "Gen. Rel. Grav.",
    volume = "36",
    pages = "1765--1780",
    year = "2004"
}

@article{Das:2005bn,
    author = "Das, Sudipta and Banerjee, Narayan and Dadhich, Naresh",
    title = "{Curvature driven acceleration : a utopia or a reality ?}",
    eprint = "astro-ph/0505096",
    archivePrefix = "arXiv",
    doi = "10.1088/0264-9381/23/12/012",
    journal = "Class. Quant. Grav.",
    volume = "23",
    pages = "4159--4166",
    year = "2006"
}

@article{Bean:2006up,
    author = "Bean, Rachel and Bernat, David and Pogosian, Levon and Silvestri, Alessandra and Trodden, Mark",
    title = "{Dynamics of Linear Perturbations in f(R) Gravity}",
    eprint = "astro-ph/0611321",
    archivePrefix = "arXiv",
    doi = "10.1103/PhysRevD.75.064020",
    journal = "Phys. Rev. D",
    volume = "75",
    pages = "064020",
    year = "2007"
}

@article{Pogosian:2007sw,
    author = "Pogosian, Levon and Silvestri, Alessandra",
    title = "{The pattern of growth in viable f(R) cosmologies}",
    eprint = "0709.0296",
    archivePrefix = "arXiv",
    primaryClass = "astro-ph",
    doi = "10.1103/PhysRevD.77.023503",
    journal = "Phys. Rev. D",
    volume = "77",
    pages = "023503",
    year = "2008",
    note = "[Erratum: Phys.Rev.D 81, 049901 (2010)]"
}

@article{Evans:2007ch,
    author = "Evans, Jonathan D. and Hall, Lisa M. H. and Caillol, Philippe",
    title = "{Standard Cosmological Evolution in a Wide Range of f(R) Models}",
    eprint = "0711.3695",
    archivePrefix = "arXiv",
    primaryClass = "astro-ph",
    doi = "10.1103/PhysRevD.77.083514",
    journal = "Phys. Rev. D",
    volume = "77",
    pages = "083514",
    year = "2008"
}

@article{Mukherjee:2014fna,
    author = "Mukherjee, Ankan and Banerjee, Narayan",
    title = "{Acceleration of the Universe in $f(R)$ Gravity Models}",
    eprint = "1405.6788",
    archivePrefix = "arXiv",
    primaryClass = "gr-qc",
    doi = "10.1007/s10509-014-1949-0",
    journal = "Astrophys. Space Sci.",
    volume = "352",
    pages = "893--898",
    year = "2014"
}

@article{Amendola:2006kh,
    author = "Amendola, Luca and Polarski, David and Tsujikawa, Shinji",
    title = "{Are f(R) dark energy models cosmologically viable ?}",
    eprint = "astro-ph/0603703",
    archivePrefix = "arXiv",
    doi = "10.1103/PhysRevLett.98.131302",
    journal = "Phys. Rev. Lett.",
    volume = "98",
    pages = "131302",
    year = "2007"
}

@article{Amendola:2006we,
    author = "Amendola, Luca and Gannouji, Radouane and Polarski, David and Tsujikawa, Shinji",
    title = "{Conditions for the cosmological viability of f(R) dark energy models}",
    eprint = "gr-qc/0612180",
    archivePrefix = "arXiv",
    doi = "10.1103/PhysRevD.75.083504",
    journal = "Phys. Rev. D",
    volume = "75",
    pages = "083504",
    year = "2007"
}

@article{DeFelice:2007ez,
    author = "De Felice, Antonio and Mukherjee, Pia and Wang, Yun",
    title = "{Observational Bounds on Modified Gravity Models}",
    eprint = "0706.1197",
    archivePrefix = "arXiv",
    primaryClass = "astro-ph",
    doi = "10.1103/PhysRevD.77.024017",
    journal = "Phys. Rev. D",
    volume = "77",
    pages = "024017",
    year = "2008"
}

@article{Capozziello:2009ka,
    author = "Capozziello, S. and Salzano, V.",
    title = "{Cosmography and large scale structure by f(R) gravity: new results}",
    eprint = "0902.0088",
    archivePrefix = "arXiv",
    primaryClass = "astro-ph.CO",
    doi = "10.1155/2009/217420",
    journal = "Adv. Astron.",
    volume = "2009",
    pages = "217420",
    year = "2009"
}

@article{Babichev:2012qs,
    author = "Babichev, Eugeny",
    title = "{Plane waves in the generalized Galileon theory}",
    eprint = "1207.4764",
    archivePrefix = "arXiv",
    primaryClass = "gr-qc",
    doi = "10.1103/PhysRevD.86.084037",
    journal = "Phys. Rev. D",
    volume = "86",
    pages = "084037",
    year = "2012"
}

@article{Gao:2011qe,
    author = "Gao, Xian and Steer, Daniele A.",
    title = "{Inflation and primordial non-Gaussianities of 'generalized Galileons'}",
    eprint = "1107.2642",
    archivePrefix = "arXiv",
    primaryClass = "astro-ph.CO",
    doi = "10.1088/1475-7516/2011/12/019",
    journal = "JCAP",
    volume = "12",
    pages = "019",
    year = "2011"
}

@article{Horndeski:1974wa,
    author = "Horndeski, Gregory Walter",
    title = "{Second-order scalar-tensor field equations in a four-dimensional space}",
    doi = "10.1007/BF01807638",
    journal = "Int. J. Theor. Phys.",
    volume = "10",
    pages = "363--384",
    year = "1974"
}

@article{Barrow:2012ay,
    author = "Barrow, John D. and Thorsrud, Mikjel and Yamamoto, Kei",
    title = "{Cosmologies in Horndeski's second-order vector-tensor theory}",
    eprint = "1211.5403",
    archivePrefix = "arXiv",
    primaryClass = "gr-qc",
    doi = "10.1007/JHEP02(2013)146",
    journal = "JHEP",
    volume = "02",
    pages = "146",
    year = "2013"
}

@article{Koyama:2013paa,
    author = "Koyama, Kazuya and Niz, Gustavo and Tasinato, Gianmassimo",
    title = "{Effective theory for the Vainshtein mechanism from the Horndeski action}",
    eprint = "1305.0279",
    archivePrefix = "arXiv",
    primaryClass = "hep-th",
    doi = "10.1103/PhysRevD.88.021502",
    journal = "Phys. Rev. D",
    volume = "88",
    pages = "021502",
    year = "2013"
}

@article{Park:1997zf,
    author = "Park, D. H.",
    title = "{General BPS solution in Einstein-Maxwell-dilaton theory}",
    journal = "J. Korean Phys. Soc.",
    volume = "31",
    pages = "894--897",
    year = "1997"
}

@article{Park:2000an,
    author = "Park, D. H.",
    title = "{Geometric construction of static dyons in scalar-vector-tensor theory}",
    journal = "J. Korean Phys. Soc.",
    volume = "37",
    pages = "177--182",
    year = "2000"
}

@article{Cho:1987yc,
    author = "Cho, Y. M.",
    title = "{Effective Couplings in {Kaluza-Klein} Unification}",
    reportNumber = "ITP-SB-87-31",
    doi = "10.1016/0370-2693(87)90933-6",
    journal = "Phys. Lett. B",
    volume = "199",
    pages = "358--362",
    year = "1987"
}

@article{Fujii:1971vv,
    author = "Fujii, Yasunori",
    title = "{Dilaton and possible non-newtonian gravity}",
    reportNumber = "UT-KOMABA-71-15",
    journal = "Nature",
    volume = "234",
    pages = "5--7",
    year = "1971"
}

@article{Sugimoto:1972zx,
    author = "Sugimoto, D.",
    title = "{Astrophysical test for dilaton theory of non-newtonian gravity}",
    doi = "10.1143/PTP.48.699",
    journal = "Prog. Theor. Phys.",
    volume = "48",
    pages = "699--700",
    year = "1972"
}

@article{Capozziello:2009nq,
    author = "Capozziello, Salvatore and De Laurentis, Mariafelicia and Faraoni, Valerio",
    title = "{A Bird's eye view of f(R)-gravity}",
    eprint = "0909.4672",
    archivePrefix = "arXiv",
    primaryClass = "gr-qc",
    doi = "10.2174/1874381101003020049",
    journal = "Open Astron. J.",
    volume = "3",
    pages = "49",
    year = "2010"
}

@book{Will:2018bme,
    author = "Will, Clifford M.",
    title = "{Theory and Experiment in Gravitational Physics}",
    isbn = "978-1-108-67982-4, 978-1-107-11744-0",
    publisher = "Cambridge University Press",
    month = "9",
    year = "2018"
}

@article{Reasenberg:1979ey,
    author = "Reasenberg, R. D. and Shapiro, I. I. and MacNeil, P. E. and Goldstein, R. B. and Breidenthal, J. C. and Brenkle, J. P. and Cain, D. L. and Kaufman, T. M. and Komarek, T. A. and Zygielbaum, A. I.",
    title = "{Viking relativity experiment: Verification of signal retardation by solar gravity}",
    doi = "10.1086/183144",
    journal = "Astrophys. J. Lett.",
    volume = "234",
    pages = "L219--L221",
    year = "1979"
}

@article{Bertotti:2003rm,
    author = "Bertotti, B. and Iess, L. and Tortora, P.",
    title = "{A test of general relativity using radio links with the Cassini spacecraft}",
    doi = "10.1038/nature01997",
    journal = "Nature",
    volume = "425",
    pages = "374--376",
    year = "2003"
}

@book{Weinberg:1972kfs,
    author = "Weinberg, Steven",
    title = "{Gravitation and Cosmology}: {Principles and Applications of the General Theory of Relativity}",
    isbn = "978-0-471-92567-5, 978-0-471-92567-5",
    publisher = "John Wiley and Sons",
    address = "New York",
    year = "1972"
}

@article{Faraoni:1999yp,
    author = "Faraoni, Valerio",
    title = "{Illusions of general relativity in Brans-Dicke gravity}",
    eprint = "gr-qc/9902083",
    archivePrefix = "arXiv",
    doi = "10.1103/PhysRevD.59.084021",
    journal = "Phys. Rev. D",
    volume = "59",
    pages = "084021",
    year = "1999"
}

@article{Brans:1962zz,
    author = "Brans, C. H.",
    title = "{Mach's Principle and a Relativistic Theory of Gravitation. II}",
    doi = "10.1103/PhysRev.125.2194",
    journal = "Phys. Rev.",
    volume = "125",
    pages = "2194--2201",
    year = "1962"
}

@article{Bhadra:2001my,
    author = "Bhadra, A. and Nandi, K. K.",
    title = "{Brans type II-IV solutions in the Einstein frame and physical interpretation of constants in the solutions}",
    doi = "10.1142/S0217732301005539",
    journal = "Mod. Phys. Lett. A",
    volume = "16",
    pages = "2079--2087",
    year = "2001"
}

@article{Bhadra:2005mc,
    author = "Bhadra, Arunava and Sarkar, Kabita",
    title = "{On static spherically symmetric solutions of the vacuum Brans-Dicke theory}",
    eprint = "gr-qc/0505141",
    archivePrefix = "arXiv",
    doi = "10.1007/s10714-005-0181-1",
    journal = "Gen. Rel. Grav.",
    volume = "37",
    pages = "2189--2199",
    year = "2005"
}

@article{Campanelli:1993sm,
    author = "Campanelli, Manuela and Lousto, C. O.",
    title = "{Are black holes in Brans-Dicke theory precisely the same as a general relativity?}",
    eprint = "gr-qc/9301013",
    archivePrefix = "arXiv",
    reportNumber = "PRINT-93-0104 (KONSTANZ)",
    doi = "10.1142/S0218271893000325",
    journal = "Int. J. Mod. Phys. D",
    volume = "2",
    pages = "451--462",
    year = "1993"
}

@article{Ruffini:1971bza,
    author = "Ruffini, Remo and Wheeler, John A.",
    title = "{Introducing the black hole}",
    doi = "10.1063/1.3022513",
    journal = "Phys. Today",
    volume = "24",
    number = "1",
    pages = "30",
    year = "1971"
}

@article{Faraoni:2016ozb,
    author = "Faraoni, Valerio and Hammad, Fay\c{c}al and Belknap-Keet, Shawn D.",
    title = "{Revisiting the Brans solutions of scalar-tensor gravity}",
    eprint = "1609.02783",
    archivePrefix = "arXiv",
    primaryClass = "gr-qc",
    doi = "10.1103/PhysRevD.94.104019",
    journal = "Phys. Rev. D",
    volume = "94",
    number = "10",
    pages = "104019",
    year = "2016"
}

@article{Bhadra:2001fx,
    author = "Bhadra, A. and Nandi, K. K.",
    title = "{On the equivalence of the Buchdahl and the Janis-Newman-Winnicour solutions}",
    doi = "10.1142/S0217751X01005328",
    journal = "Int. J. Mod. Phys. A",
    volume = "16",
    pages = "4543--4545",
    year = "2001"
}

@article{Agnese:1995kd,
    author = "Agnese, A. G. and La Camera, M.",
    title = "{Wormholes in the Brans-Dicke theory of gravitation}",
    doi = "10.1103/PhysRevD.51.2011",
    journal = "Phys. Rev. D",
    volume = "51",
    pages = "2011--2013",
    year = "1995"
}

@article{Nandi:1997mx,
    author = "Nandi, Kamal K. and Islam, Anwarul and Evans, James",
    title = "{Brans wormholes}",
    eprint = "0906.0436",
    archivePrefix = "arXiv",
    primaryClass = "gr-qc",
    doi = "10.1103/PhysRevD.55.2497",
    journal = "Phys. Rev. D",
    volume = "55",
    pages = "2497--2500",
    year = "1997"
}

@article{Anchordoqui:1997yb,
    author = "Anchordoqui, Luis A. and Grunfeld, A. G. and Torres, Diego F.",
    title = "{Vacuum static Brans-Dicke wormhole}",
    eprint = "gr-qc/9707025",
    archivePrefix = "arXiv",
    journal = "Grav. Cosmol.",
    volume = "4",
    pages = "287--290",
    year = "1998"
}

@article{PhysRevD.65.084022,
  title = {New Brans-Dicke wormholes},
  author = {He, Feng and Kim, Sung-Won},
  journal = {Phys. Rev. D},
  volume = {65},
  issue = {8},
  pages = {084022},
  numpages = {4},
  year = {2002},
  month = {Apr},
  publisher = {American Physical Society},
  doi = {10.1103/PhysRevD.65.084022},
  url = {https://link.aps.org/doi/10.1103/PhysRevD.65.084022}
}

@article{Bhadra:2004hu,
    author = "Bhadra, Arunava and Simaciu, Ion and Nandi, Kamal Kanti and Zhang, Yuan-Zhong",
    title = "{Comment on `New Brans-Dicke wormholes'}",
    eprint = "gr-qc/0406014",
    archivePrefix = "arXiv",
    doi = "10.1103/PhysRevD.71.128501",
    journal = "Phys. Rev. D",
    volume = "71",
    pages = "128501",
    year = "2005"
}

@article{Nandi:2004ha,
    author = "Nandi, K. K. and Zhang, Yuan-Zhong",
    title = "{On traversable Lorentzian wormholes in the vacuum low energy effective string theory in Einstein and Jordan frames}",
    eprint = "gr-qc/0405051",
    archivePrefix = "arXiv",
    doi = "10.1103/PhysRevD.70.044040",
    journal = "Phys. Rev. D",
    volume = "70",
    pages = "044040",
    year = "2004"
}

@article{Bronnikov:1998hm,
    author = "Bronnikov, K. A. and Clement, G. and Constantinidis, C. P. and Fabris, J. C.",
    title = "{Cold scalar tensor black holes: Causal structure, geodesics, stability}",
    eprint = "gr-qc/9804064",
    archivePrefix = "arXiv",
    journal = "Grav. Cosmol.",
    volume = "4",
    pages = "128--138",
    year = "1998"
}

@article{Nandi:2000gt,
    author = "Nandi, K. K. and Bhadra, A. and Alsing, P. M. and Nayak, T. B.",
    title = "{Tidal forces in cold black hole space-times}",
    eprint = "gr-qc/0008025",
    archivePrefix = "arXiv",
    doi = "10.1142/S0218271801001050",
    journal = "Int. J. Mod. Phys. D",
    volume = "10",
    pages = "529--538",
    year = "2001"
}

@article{OHanlon:1972ysn,
    author = "O' Hanlon, J. and Tupper, B. O. J.",
    title = "{Vacuum-field solutions in the Brans-Dicke theory}",
    doi = "10.1007/BF02743602",
    journal = "Nuovo Cim. B",
    volume = "7",
    pages = "305--312",
    year = "1972"
}

@article{10.1143/PTP.40.49,
    author = {Nariai, Hidekazu},
    title = "{On the Green's Function in an Expanding Universe and Its Role in the Problem of Mach's Principle}",
    journal = {Progress of Theoretical Physics},
    volume = {40},
    number = {1},
    pages = {49-59},
    year = {1968},
    month = {07},
}

@article{gurevich1973problem,
  title={On the problem of the initial state in the isotropic scalar-tensor cosmology of Brans-Dicke},
  author={Gurevich, LE and Finkelstein, AM and Ruban, VA},
  journal={Astrophysics and Space Science},
  volume={22},
  number={2},
  pages={231--242},
  year={1973},
  publisher={Springer}
}

@article{lorenz1984exact,
  title={Exact perfect fluid solutions in the Brans-Dicke-theory},
  author={Lorenz-Petzold, Dieter},
  journal={Astrophysics and space science},
  volume={98},
  pages={249--254},
  year={1984}
}

@article{Morganstern:1971dz,
    author = "Morganstern, R. E.",
    title = "{Exact Solutions to Radiation-Filled Brans-Dicke Cosmologies}",
    doi = "10.1103/PhysRevD.4.282",
    journal = "Phys. Rev. D",
    volume = "4",
    pages = "282--286",
    year = "1971"
}

@article{La:1989za,
    author = "La, Daile and Steinhardt, Paul J.",
    title = "{Extended Inflationary Cosmology}",
    reportNumber = "PRINT-89-0009 (PENN)",
    doi = "10.1103/PhysRevLett.62.376",
    journal = "Phys. Rev. Lett.",
    volume = "62",
    pages = "376",
    year = "1989",
    note = "[Erratum: Phys.Rev.Lett. 62, 1066 (1989)]"
}

@article{Romero:1992xx,
    author = "Romero, C. and Barros, A.",
    title = "{Brans-Dicke cosmology and the cosmological constant: the spectrum of vacuum solutions}",
    doi = "10.1007/BF00684484",
    journal = "Astrophys. Space Sci.",
    volume = "192",
    pages = "263--274",
    year = "1992"
}

@article{Liddle:1993fq,
    author = "Liddle, Andrew R. and Lyth, David H.",
    title = "{The Cold dark matter density perturbation}",
    eprint = "astro-ph/9303019",
    archivePrefix = "arXiv",
    reportNumber = "SUSSEX-AST-92-8-1-REV, LANC-TH-8-92-REV",
    doi = "10.1016/0370-1573(93)90114-S",
    journal = "Phys. Rept.",
    volume = "231",
    pages = "1--105",
    year = "1993"
}

@article{PhysRevD.47.5329,
  title = {Scalar-tensor cosmologies},
  author = {Barrow, John D.},
  journal = {Phys. Rev. D},
  volume = {47},
  issue = {12},
  pages = {5329--5335},
  numpages = {0},
  year = {1993},
  month = {Jun},
  publisher = {American Physical Society},
  doi = {10.1103/PhysRevD.47.5329},
  url = {https://link.aps.org/doi/10.1103/PhysRevD.47.5329}
}

@article{Dehnen:1971zz,
    author = "Dehnen, H and Obregen, O.",
    title = "{Exact Cosmological Solutions in Brans and Dicke s Scalar-Tensor Theory, I}",
    doi = "10.1007/BF00653332",
    journal = "Astrophys. Space Sci.",
    volume = "14",
    pages = "454--459",
    year = "1971"
}

@article{Levin:1993wr,
    author = "Levin, Janna J. and Freese, Katherine",
    title = "{Curvature and flatness in a Brans-Dicke universe}",
    eprint = "gr-qc/9312025",
    archivePrefix = "arXiv",
    doi = "10.1016/0550-3213(94)90520-7",
    journal = "Nucl. Phys. B",
    volume = "421",
    pages = "635--661",
    year = "1994"
}

@article{Mimoso:1994wn,
    author = "Mimoso, Jose P. and Wands, David",
    title = "{Massless fields in scalar - tensor cosmologies}",
    eprint = "gr-qc/9405025",
    archivePrefix = "arXiv",
    reportNumber = "SUSSEX-AST-94-4-1",
    doi = "10.1103/PhysRevD.51.477",
    journal = "Phys. Rev. D",
    volume = "51",
    pages = "477--489",
    year = "1995"
}

@article{SupernovaSearchTeam:2003cyd,
    author = "Tonry, John L. and others",
    collaboration = "Supernova Search Team",
    title = "{Cosmological results from high-z supernovae}",
    eprint = "astro-ph/0305008",
    archivePrefix = "arXiv",
    doi = "10.1086/376865",
    journal = "Astrophys. J.",
    volume = "594",
    pages = "1--24",
    year = "2003"
}

@article{SupernovaCosmologyProject:2011ycw,
    author = "Suzuki, N. and others",
    collaboration = "Supernova Cosmology Project",
    title = "{The Hubble Space Telescope Cluster Supernova Survey: V. Improving the Dark Energy Constraints Above z>1 and Building an Early-Type-Hosted Supernova Sample}",
    eprint = "1105.3470",
    archivePrefix = "arXiv",
    primaryClass = "astro-ph.CO",
    doi = "10.1088/0004-637X/746/1/85",
    journal = "Astrophys. J.",
    volume = "746",
    pages = "85",
    year = "2012"
}

@article{SDSS:2014iwm,
    author = "Betoule, M. and others",
    collaboration = "SDSS",
    title = "{Improved cosmological constraints from a joint analysis of the SDSS-II and SNLS supernova samples}",
    eprint = "1401.4064",
    archivePrefix = "arXiv",
    primaryClass = "astro-ph.CO",
    reportNumber = "FERMILAB-PUB-14-013-A-AE",
    doi = "10.1051/0004-6361/201423413",
    journal = "Astron. Astrophys.",
    volume = "568",
    pages = "A22",
    year = "2014"
}

@article{SDSS:2005xqv,
    author = "Eisenstein, Daniel J. and others",
    collaboration = "SDSS",
    title = "{Detection of the Baryon Acoustic Peak in the Large-Scale Correlation Function of SDSS Luminous Red Galaxies}",
    eprint = "astro-ph/0501171",
    archivePrefix = "arXiv",
    reportNumber = "FERMILAB-PUB-05-057-A-CD",
    doi = "10.1086/466512",
    journal = "Astrophys. J.",
    volume = "633",
    pages = "560--574",
    year = "2005"
}

@article{Eisenstein:2005sbt,
    author = "Eisenstein, D. J.",
    title = "{Dark energy and cosmic sound}",
    doi = "10.1016/j.newar.2005.08.005",
    journal = "New Astron. Rev.",
    volume = "49",
    number = "7-9",
    pages = "360--365",
    year = "2005"
}

@article{Barris:2003dq,
    author = "Barris, Brian J. and others",
    title = "{23 High redshift supernovae from the IFA Deep Survey: Doubling the SN sample at z \ensuremath{>} 0.7}",
    eprint = "astro-ph/0310843",
    archivePrefix = "arXiv",
    doi = "10.1086/381122",
    journal = "Astrophys. J.",
    volume = "602",
    pages = "571--594",
    year = "2004"
}

@article{Planck:2015fie,
    author = "Ade, P. A. R. and others",
    collaboration = "Planck",
    title = "{Planck 2015 results. XIII. Cosmological parameters}",
    eprint = "1502.01589",
    archivePrefix = "arXiv",
    primaryClass = "astro-ph.CO",
    doi = "10.1051/0004-6361/201525830",
    journal = "Astron. Astrophys.",
    volume = "594",
    pages = "A13",
    year = "2016"
}

@article{Planck:2018vyg,
    author = "Aghanim, N. and others",
    collaboration = "Planck",
    title = "{Planck 2018 results. VI. Cosmological parameters}",
    eprint = "1807.06209",
    archivePrefix = "arXiv",
    primaryClass = "astro-ph.CO",
    doi = "10.1051/0004-6361/201833910",
    journal = "Astron. Astrophys.",
    volume = "641",
    pages = "A6",
    year = "2020",
    note = "[Erratum: Astron.Astrophys. 652, C4 (2021)]"
}

@article{Allen:2007ue,
    author = "Allen, S. W. and Rapetti, D. A. and Schmidt, R. W. and Ebeling, H. and Morris, G. and Fabian, A. C.",
    title = "{Improved constraints on dark energy from Chandra X-ray observations of the largest relaxed galaxy clusters}",
    eprint = "0706.0033",
    archivePrefix = "arXiv",
    primaryClass = "astro-ph",
    reportNumber = "SLAC-PUB-12542",
    doi = "10.1111/j.1365-2966.2007.12610.x",
    journal = "Mon. Not. Roy. Astron. Soc.",
    volume = "383",
    pages = "879--896",
    year = "2008"
}

@article{Hicken:2009dk,
    author = "Hicken, Malcolm and Wood-Vasey, W. Michael and Blondin, Stephane and Challis, Peter and Jha, Saurabh and Kelly, Patrick L. and Rest, Armin and Kirshner, Robert P.",
    title = "{Improved Dark Energy Constraints from 100 New CfA Supernova Type Ia Light Curves}",
    eprint = "0901.4804",
    archivePrefix = "arXiv",
    primaryClass = "astro-ph.CO",
    reportNumber = "SLAC-PUB-14938",
    doi = "10.1088/0004-637X/700/2/1097",
    journal = "Astrophys. J.",
    volume = "700",
    pages = "1097--1140",
    year = "2009"
}

@article{DES:2018paw,
    author = "Abbott, T. M. C. and others",
    collaboration = "DES",
    title = "{First Cosmology Results using Type Ia Supernovae from the Dark Energy Survey: Constraints on Cosmological Parameters}",
    eprint = "1811.02374",
    archivePrefix = "arXiv",
    primaryClass = "astro-ph.CO",
    reportNumber = "FERMILAB-PUB-18-590-AE",
    doi = "10.3847/2041-8213/ab04fa",
    journal = "Astrophys. J. Lett.",
    volume = "872",
    number = "2",
    pages = "L30",
    year = "2019"
}

@article{SupernovaSearchTeam:2004lze,
    author = "Riess, Adam G. and others",
    collaboration = "Supernova Search Team",
    title = "{Type Ia supernova discoveries at z \ensuremath{>} 1 from the Hubble Space Telescope: Evidence for past deceleration and constraints on dark energy evolution}",
    eprint = "astro-ph/0402512",
    archivePrefix = "arXiv",
    doi = "10.1086/383612",
    journal = "Astrophys. J.",
    volume = "607",
    pages = "665--687",
    year = "2004"
}

@article{Riess:2006fw,
    author = "Riess, Adam G. and others",
    title = "{New Hubble Space Telescope Discoveries of Type Ia Supernovae at $z \geq 1$: Narrowing Constraints on the Early Behavior of Dark Energy}",
    eprint = "astro-ph/0611572",
    archivePrefix = "arXiv",
    reportNumber = "46455850950",
    doi = "10.1086/510378",
    journal = "Astrophys. J.",
    volume = "659",
    pages = "98--121",
    year = "2007"
}

@article{Turner_2002,
doi = {10.1086/338580},
url = {https://dx.doi.org/10.1086/338580},
year = {2002},
month = {apr},
publisher = {},
volume = {569},
number = {1},
pages = {18},
author = {Michael S. Turner and Adam G. Riess},
title = {Do Type Ia Supernovae Provide Direct Evidence for Past Deceleration of the Universe?},
journal = {Astrophys. J.},
}

@article{Cunha:2008ja,
    author = "Cunha, J. V. and Lima, J. A. S.",
    title = "{Transition Redshift: New Kinematic Constraints from Supernovae}",
    eprint = "0805.1261",
    archivePrefix = "arXiv",
    primaryClass = "astro-ph",
    doi = "10.1111/j.1365-2966.2008.13640.x",
    journal = "Mon. Not. Roy. Astron. Soc.",
    volume = "390",
    pages = "210--217",
    year = "2008"
}

@article{Cunha:2008mt,
    author = "Cunha, J. V.",
    title = "{Kinematic Constraints to the Transition Redshift from SNe Ia Union Data}",
    eprint = "0811.2379",
    archivePrefix = "arXiv",
    primaryClass = "astro-ph",
    doi = "10.1103/PhysRevD.79.047301",
    journal = "Phys. Rev. D",
    volume = "79",
    pages = "047301",
    year = "2009"
}

@article{A.C.C.Guimaraes_2009,
    author = "Guimaraes, A. C. C. and Cunha, J. V. and Lima, J. A. S.",
    title = "{Bayesian Analysis and Constraints on Kinematic Models from Union SNIa}",
    eprint = "0904.3550",
    archivePrefix = "arXiv",
    primaryClass = "astro-ph.CO",
    doi = "10.1088/1475-7516/2009/10/010",
    journal = "JCAP",
    volume = "10",
    pages = "010",
    year = "2009"
}

@article{Lu:2011ue,
    author = "Lu, Jianbo and Xu, Lixin and Liu, Molin",
    title = "{Constraints on kinematic models from the latest observational data}",
    eprint = "1105.1871",
    archivePrefix = "arXiv",
    primaryClass = "astro-ph.CO",
    doi = "10.1016/j.physletb.2011.04.022",
    journal = "Phys. Lett. B",
    volume = "699",
    pages = "246--250",
    year = "2011"
}

@article{Cattoen:2007sk,
    author = "Cattoen, Celine and Visser, Matt",
    title = "{The Hubble series: Convergence properties and redshift variables}",
    eprint = "0710.1887",
    archivePrefix = "arXiv",
    primaryClass = "gr-qc",
    doi = "10.1088/0264-9381/24/23/018",
    journal = "Class. Quant. Grav.",
    volume = "24",
    pages = "5985--5998",
    year = "2007"
}

@article{Guth:1980zm,
    author = "Guth, Alan H.",
    editor = "Fang, Li-Zhi and Ruffini, R.",
    title = "{The Inflationary Universe: A Possible Solution to the Horizon and Flatness Problems}",
    reportNumber = "SLAC-PUB-2576",
    doi = "10.1103/PhysRevD.23.347",
    journal = "Phys. Rev. D",
    volume = "23",
    pages = "347--356",
    year = "1981"
}

@article{Linde:1981mu,
    author = "Linde, Andrei D.",
    editor = "Fang, Li-Zhi and Ruffini, R.",
    title = "{A New Inflationary Universe Scenario: A Possible Solution of the Horizon, Flatness, Homogeneity, Isotropy and Primordial Monopole Problems}",
    reportNumber = "LEBEDEV-81-229",
    doi = "10.1016/0370-2693(82)91219-9",
    journal = "Phys. Lett. B",
    volume = "108",
    pages = "389--393",
    year = "1982"
}

@book{Ryden:1970vsj,
    author = "Ryden, B.",
    title = "{Introduction to cosmology}",
    doi = "10.1017/9781316651087",
    isbn = "978-1-107-15483-4, 978-1-316-88984-8, 978-1-316-65108-7",
    publisher = "Cambridge University Press",
    year = "1970"
}

@book{Plebanski:2006sd,
    author = "Plebanski, J. and Krasinski, Andrzej",
    title = "{An introduction to general relativity and cosmology}",
    year = "2006"
}

@article{SupernovaCosmologyProject:1998vns,
    author = "Perlmutter, S. and others",
    collaboration = "Supernova Cosmology Project",
    title = "{Measurements of $\Omega$ and $\Lambda$ from 42 high redshift supernovae}",
    eprint = "astro-ph/9812133",
    archivePrefix = "arXiv",
    reportNumber = "LBNL-41801, LBL-41801",
    doi = "10.1086/307221",
    journal = "Astrophys. J.",
    volume = "517",
    pages = "565--586",
    year = "1999"
}

@article{SupernovaSearchTeam:1998fmf,
    author = "Riess, Adam G. and others",
    collaboration = "Supernova Search Team",
    title = "{Observational evidence from supernovae for an accelerating universe and a cosmological constant}",
    eprint = "astro-ph/9805201",
    archivePrefix = "arXiv",
    doi = "10.1086/300499",
    journal = "Astron. J.",
    volume = "116",
    pages = "1009--1038",
    year = "1998"
}

@article{SupernovaSearchTeam:1998bnz,
    author = "Schmidt, Brian P. and others",
    collaboration = "Supernova Search Team",
    title = "{The High Z supernova search: Measuring cosmic deceleration and global curvature of the universe using type Ia supernovae}",
    eprint = "astro-ph/9805200",
    archivePrefix = "arXiv",
    doi = "10.1086/306308",
    journal = "Astrophys. J.",
    volume = "507",
    pages = "46--63",
    year = "1998"
}

@article{Riess:1998dv,
    author = "Riess, Adam G. and others",
    title = "{BV RI light curves for 22 type Ia supernovae}",
    eprint = "astro-ph/9810291",
    archivePrefix = "arXiv",
    doi = "10.1086/300738",
    journal = "Astron. J.",
    volume = "117",
    pages = "707--724",
    year = "1999"
}

@article{Jordan:1949zz,
    author = "Jordan, Pascual",
    title = "{Formation of the Stars and Development of the Universe}",
    doi = "10.1038/164637a0",
    journal = "Nature",
    volume = "164",
    pages = "637--640",
    year = "1949"
}

@article{Brans:1961sx,
    author = "Brans, C. and Dicke, R. H.",
    title = "{Mach's principle and a relativistic theory of gravitation}",
    doi = "10.1103/PhysRev.124.925",
    journal = "Phys. Rev.",
    volume = "124",
    pages = "925--935",
    year = "1961"
}

@article{Nordtvedt:1970uv,
    author = "Nordtvedt, Jr., Kenneth",
    title = "{PostNewtonian metric for a general class of scalar tensor gravitational theories and observational consequences}",
    doi = "10.1086/150607",
    journal = "Astrophys. J.",
    volume = "161",
    pages = "1059--1067",
    year = "1970"
}

@article{Wagoner:1970vr,
    author = "Wagoner, Robert V.",
    title = "{Scalar tensor theory and gravitational waves}",
    doi = "10.1103/PhysRevD.1.3209",
    journal = "Phys. Rev. D",
    volume = "1",
    pages = "3209--3216",
    year = "1970"
}

@book{Fujii:2003pa,
    author = "Fujii, Y. and Maeda, K.",
    title = "{The scalar-tensor theory of gravitation}",
    doi = "10.1017/CBO9780511535093",
    isbn = "978-0-521-03752-5, 978-0-521-81159-0, 978-0-511-02988-2",
    publisher = "Cambridge University Press",
    series = "Cambridge Monographs on Mathematical Physics",
    month = "7",
    year = "2007"
}

@book{Faraoni:2004pi,
    author = "Faraoni, Valerio",
    title = "{Cosmology in scalar tensor gravity}",
    doi = "10.1007/978-1-4020-1989-0",
    isbn = "978-1-4020-1988-3",
    year = "2004"
}

@article{Quiros:2019ktw,
    author = "Quiros, Israel",
    title = "{Selected topics in scalar\textendash{}tensor theories and beyond}",
    eprint = "1901.08690",
    archivePrefix = "arXiv",
    primaryClass = "gr-qc",
    doi = "10.1142/S021827181930012X",
    journal = "Int. J. Mod. Phys. D",
    volume = "28",
    number = "07",
    pages = "1930012",
    year = "2019"
}

@article{tHooft:1993dmi,
    author = "'t Hooft, Gerard",
    title = "{Dimensional reduction in quantum gravity}",
    eprint = "gr-qc/9310026",
    archivePrefix = "arXiv",
    reportNumber = "THU-93-26",
    journal = "Conf. Proc. C",
    volume = "930308",
    pages = "284--296",
    year = "1993"
}

@article{Susskind:1994vu,
    author = "Susskind, Leonard",
    title = "{The World as a hologram}",
    eprint = "hep-th/9409089",
    archivePrefix = "arXiv",
    reportNumber = "SU-ITP-94-33",
    doi = "10.1063/1.531249",
    journal = "J. Math. Phys.",
    volume = "36",
    pages = "6377--6396",
    year = "1995"
}

@article{Bekenstein:1993dz,
    author = "Bekenstein, Jacob D.",
    title = "{Entropy bounds and black hole remnants}",
    eprint = "gr-qc/9307035",
    archivePrefix = "arXiv",
    reportNumber = "UCSBTH-93-23",
    doi = "10.1103/PhysRevD.49.1912",
    journal = "Phys. Rev. D",
    volume = "49",
    pages = "1912--1921",
    year = "1994"
}

@article{Cohen:1998zx,
    author = "Cohen, Andrew G. and Kaplan, David B. and Nelson, Ann E.",
    title = "{Effective field theory, black holes, and the cosmological constant}",
    eprint = "hep-th/9803132",
    archivePrefix = "arXiv",
    reportNumber = "BUHEP-98-7, DOE-ER-40561-358, INT-98-00-6, UW-PT-97-24",
    doi = "10.1103/PhysRevLett.82.4971",
    journal = "Phys. Rev. Lett.",
    volume = "82",
    pages = "4971--4974",
    year = "1999"
}

@article{Li:2004rb,
    author = "Li, Miao",
    title = "{A Model of holographic dark energy}",
    eprint = "hep-th/0403127",
    archivePrefix = "arXiv",
    doi = "10.1016/j.physletb.2004.10.014",
    journal = "Phys. Lett. B",
    volume = "603",
    pages = "1",
    year = "2004"
}

@article{Horava:2000tb,
    author = "Horava, Petr and Minic, Djordje",
    title = "{Probable values of the cosmological constant in a holographic theory}",
    eprint = "hep-th/0001145",
    archivePrefix = "arXiv",
    reportNumber = "CALT-68-2257, CITUSC-00-006",
    doi = "10.1103/PhysRevLett.85.1610",
    journal = "Phys. Rev. Lett.",
    volume = "85",
    pages = "1610--1613",
    year = "2000"
}

@article{Thomas:2002pq,
    author = "Thomas, Scott D.",
    title = "{Holography stabilizes the vacuum energy}",
    doi = "10.1103/PhysRevLett.89.081301",
    journal = "Phys. Rev. Lett.",
    volume = "89",
    pages = "081301",
    year = "2002"
}

@article{Fischler:1998st,
    author = "Fischler, W. and Susskind, Leonard",
    title = "{Holography and cosmology}",
    eprint = "hep-th/9806039",
    archivePrefix = "arXiv",
    reportNumber = "SU-ITP-98-39A, UTTG-06-98",
    month = "6",
    year = "1998"
}

@article{Cataldo:2001bn,
    author = "Cataldo, Mauricio and Cruz, Norman and del Campo, Sergio and Lepe, Samuel",
    title = "{Holographic principle and the dominant energy condition for Kasner type metrics}",
    eprint = "gr-qc/0104028",
    archivePrefix = "arXiv",
    doi = "10.1016/S0370-2693(01)00490-7",
    journal = "Phys. Lett. B",
    volume = "509",
    pages = "138--142",
    year = "2001"
}

@article{Guberina:2005mp,
    author = "Guberina, B. and Horvat, R. and Nikolic, Hrvoje",
    title = "{Generalized holographic dark energy and the IR cutoff problem}",
    eprint = "astro-ph/0507666",
    archivePrefix = "arXiv",
    reportNumber = "IRB-TH-20-05",
    doi = "10.1103/PhysRevD.72.125011",
    journal = "Phys. Rev. D",
    volume = "72",
    pages = "125011",
    year = "2005"
}

@article{Wang:2005jx,
    author = "Wang, Bin and Gong, Yun-gui and Abdalla, Elcio",
    title = "{Transition of the dark energy equation of state in an interacting holographic dark energy model}",
    eprint = "hep-th/0506069",
    archivePrefix = "arXiv",
    doi = "10.1016/j.physletb.2005.08.008",
    journal = "Phys. Lett. B",
    volume = "624",
    pages = "141--146",
    year = "2005"
}

@article{Qing-GuoHuang_2004,
    author = "Huang, Qing-Guo and Li, Miao",
    title = "{The Holographic dark energy in a non-flat universe}",
    eprint = "astro-ph/0404229",
    archivePrefix = "arXiv",
    doi = "10.1088/1475-7516/2004/08/013",
    journal = "JCAP",
    volume = "08",
    pages = "013",
    year = "2004"
}

@article{Gong:2004cb,
    author = "Gong, Yun-gui and Wang, Bin and Zhang, Yuan-Zhong",
    title = "{The Holographic dark energy revisited}",
    eprint = "hep-th/0412218",
    archivePrefix = "arXiv",
    doi = "10.1103/PhysRevD.72.043510",
    journal = "Phys. Rev. D",
    volume = "72",
    pages = "043510",
    year = "2005"
}

@article{Nojiri:2005pu,
    author = "Nojiri, Shin'ichi and Odintsov, Sergei D.",
    title = "{Unifying phantom inflation with late-time acceleration: Scalar phantom-non-phantom transition model and generalized holographic dark energy}",
    eprint = "hep-th/0506212",
    archivePrefix = "arXiv",
    doi = "10.1007/s10714-006-0301-6",
    journal = "Gen. Rel. Grav.",
    volume = "38",
    pages = "1285--1304",
    year = "2006"
}

@article{Wang:2005ph,
    author = "Wang, Bin and Lin, Chi-Yong and Abdalla, Elcio",
    title = "{Constraints on the interacting holographic dark energy model}",
    eprint = "hep-th/0509107",
    archivePrefix = "arXiv",
    doi = "10.1016/j.physletb.2006.04.009",
    journal = "Phys. Lett. B",
    volume = "637",
    pages = "357--361",
    year = "2006"
}

@article{Pavon:2005yx,
    author = "Pavon, Diego and Zimdahl, Winfried",
    title = "{Holographic dark energy and cosmic coincidence}",
    eprint = "gr-qc/0505020",
    archivePrefix = "arXiv",
    doi = "10.1016/j.physletb.2005.08.134",
    journal = "Phys. Lett. B",
    volume = "628",
    pages = "206--210",
    year = "2005"
}

@article{Zimdahl:2007zz,
    author = "Zimdahl, Winfried and Pavon, Diego",
    title = "{Interacting holographic dark energy}",
    eprint = "astro-ph/0606555",
    archivePrefix = "arXiv",
    doi = "10.1088/0264-9381/24/22/011",
    journal = "Class. Quant. Grav.",
    volume = "24",
    pages = "5461--5478",
    year = "2007"
}

@article{LixinXu_2009,
    author = "Xu, Lixin",
    title = "{Holographic Dark Energy Model with Hubble Horizon as an IR Cut-off}",
    eprint = "0907.1709",
    archivePrefix = "arXiv",
    primaryClass = "astro-ph.CO",
    reportNumber = "ITP-DUT-2009-07",
    doi = "10.1088/1475-7516/2009/09/016",
    journal = "JCAP",
    volume = "09",
    pages = "016",
    year = "2009"
}

@article{Caldwell:1999ew,
    author = "Caldwell, R. R.",
    title = "{A Phantom menace?}",
    eprint = "astro-ph/9908168",
    archivePrefix = "arXiv",
    doi = "10.1016/S0370-2693(02)02589-3",
    journal = "Phys. Lett. B",
    volume = "545",
    pages = "23--29",
    year = "2002"
}

@article{Feng:2004ad,
    author = "Feng, Bo and Wang, Xiu-Lian and Zhang, Xin-Min",
    title = "{Dark energy constraints from the cosmic age and supernova}",
    eprint = "astro-ph/0404224",
    archivePrefix = "arXiv",
    doi = "10.1016/j.physletb.2004.12.071",
    journal = "Phys. Lett. B",
    volume = "607",
    pages = "35--41",
    year = "2005"
}

@article{Cai:2006dm,
    author = "Cai, Yi-fu and Li, Hong and Piao, Yun-Song and Zhang, Xin-min",
    title = "{Cosmic Duality in Quintom Universe}",
    eprint = "gr-qc/0609039",
    archivePrefix = "arXiv",
    doi = "10.1016/j.physletb.2007.01.027",
    journal = "Phys. Lett. B",
    volume = "646",
    pages = "141--144",
    year = "2007"
}

@article{Cai:2007gs,
    author = "Cai, Yi-fu and Li, Ming-zhe and Lu, Jian-Xin and Piao, Yun-Song and Qiu, Tao-tao and Zhang, Xin-min",
    title = "{A String-Inspired Quintom Model Of Dark Energy}",
    eprint = "hep-th/0701016",
    archivePrefix = "arXiv",
    reportNumber = "USTC-ICTS-07-03",
    doi = "10.1016/j.physletb.2007.05.056",
    journal = "Phys. Lett. B",
    volume = "651",
    pages = "1--7",
    year = "2007"
}

@article{Cai:2008gk,
    author = "Cai, Yi-Fu and Wang, Jing",
    title = "{Dark Energy Model with Spinor Matter and Its Quintom Scenario}",
    eprint = "0806.3890",
    archivePrefix = "arXiv",
    primaryClass = "hep-th",
    doi = "10.1088/0264-9381/25/16/165014",
    journal = "Class. Quant. Grav.",
    volume = "25",
    pages = "165014",
    year = "2008"
}

@article{Kamenshchik:2001cp,
    author = "Kamenshchik, Alexander Yu. and Moschella, Ugo and Pasquier, Vincent",
    title = "{An Alternative to quintessence}",
    eprint = "gr-qc/0103004",
    archivePrefix = "arXiv",
    doi = "10.1016/S0370-2693(01)00571-8",
    journal = "Phys. Lett. B",
    volume = "511",
    pages = "265--268",
    year = "2001"
}

@article{Bento:2002ps,
    author = "Bento, M. C. and Bertolami, O. and Sen, A. A.",
    title = "{Generalized Chaplygin gas, accelerated expansion and dark energy matter unification}",
    eprint = "gr-qc/0202064",
    archivePrefix = "arXiv",
    doi = "10.1103/PhysRevD.66.043507",
    journal = "Phys. Rev. D",
    volume = "66",
    pages = "043507",
    year = "2002"
}

@article{Jackiw:2000mm,
    author = "Jackiw, R.",
    title = "{A Particle field theorist's lectures on supersymmetric, nonAbelian fluid mechanics and d-branes}",
    eprint = "physics/0010042",
    archivePrefix = "arXiv",
    reportNumber = "MIT-CTP-3000",
    month = "10",
    year = "2000"
}

@article{LAmendola_2003,
    author = "Amendola, L. and Finelli, Fabio and Burigana, C. and Carturan, D.",
    title = "{WMAP and the generalized Chaplygin gas}",
    eprint = "astro-ph/0304325",
    archivePrefix = "arXiv",
    doi = "10.1088/1475-7516/2003/07/005",
    journal = "JCAP",
    volume = "07",
    pages = "005",
    year = "2003"
}

@article{Bento:2002yx,
    author = "Bento, Maria da Conceicao and Bertolami, O. and Sen, A. A.",
    title = "{Generalized Chaplygin gas and CMBR constraints}",
    eprint = "astro-ph/0210468",
    archivePrefix = "arXiv",
    reportNumber = "DF-IST-10-2002",
    doi = "10.1103/PhysRevD.67.063003",
    journal = "Phys. Rev. D",
    volume = "67",
    pages = "063003",
    year = "2003"
}

@article{Armendariz-Picon:1999hyi,
    author = "Armendariz-Picon, C. and Damour, T. and Mukhanov, Viatcheslav F.",
    title = "{k - inflation}",
    eprint = "hep-th/9904075",
    archivePrefix = "arXiv",
    doi = "10.1016/S0370-2693(99)00603-6",
    journal = "Phys. Lett. B",
    volume = "458",
    pages = "209--218",
    year = "1999"
}

@article{Garriga:1999vw,
    author = "Garriga, Jaume and Mukhanov, Viatcheslav F.",
    title = "{Perturbations in k-inflation}",
    eprint = "hep-th/9904176",
    archivePrefix = "arXiv",
    reportNumber = "UAB-FT-466",
    doi = "10.1016/S0370-2693(99)00602-4",
    journal = "Phys. Lett. B",
    volume = "458",
    pages = "219--225",
    year = "1999"
}

@article{Chiba:1999ka,
    author = "Chiba, Takeshi and Okabe, Takahiro and Yamaguchi, Masahide",
    title = "{Kinetically driven quintessence}",
    eprint = "astro-ph/9912463",
    archivePrefix = "arXiv",
    reportNumber = "UTAP-352",
    doi = "10.1103/PhysRevD.62.023511",
    journal = "Phys. Rev. D",
    volume = "62",
    pages = "023511",
    year = "2000"
}

@article{Armendariz-Picon:2000nqq,
    author = "Armendariz-Picon, C. and Mukhanov, Viatcheslav F. and Steinhardt, Paul J.",
    title = "{A Dynamical solution to the problem of a small cosmological constant and late time cosmic acceleration}",
    eprint = "astro-ph/0004134",
    archivePrefix = "arXiv",
    doi = "10.1103/PhysRevLett.85.4438",
    journal = "Phys. Rev. Lett.",
    volume = "85",
    pages = "4438--4441",
    year = "2000"
}

@article{Armendariz-Picon:2000ulo,
    author = "Armendariz-Picon, C. and Mukhanov, Viatcheslav F. and Steinhardt, Paul J.",
    title = "{Essentials of k essence}",
    eprint = "astro-ph/0006373",
    archivePrefix = "arXiv",
    doi = "10.1103/PhysRevD.63.103510",
    journal = "Phys. Rev. D",
    volume = "63",
    pages = "103510",
    year = "2001"
}

@article{NimaArkani-Hamed_2004,
    author = "Arkani-Hamed, Nima and Creminelli, Paolo and Mukohyama, Shinji and Zaldarriaga, Matias",
    title = "{Ghost inflation}",
    eprint = "hep-th/0312100",
    archivePrefix = "arXiv",
    reportNumber = "HUTP-03-A079",
    doi = "10.1088/1475-7516/2004/04/001",
    journal = "JCAP",
    volume = "04",
    pages = "001",
    year = "2004"
}

@article{Barger:2000jg,
    author = "Barger, Vernon D. and Marfatia, Danny",
    title = "{Supernova data may be unable to distinguish between quintessence and k-essence}",
    eprint = "astro-ph/0009256",
    archivePrefix = "arXiv",
    reportNumber = "MAD-PH-00-1192",
    doi = "10.1016/S0370-2693(00)01368-X",
    journal = "Phys. Lett. B",
    volume = "498",
    pages = "67--73",
    year = "2001"
}

@article{Li:2002wd,
    author = "Li, Mingzhe and Zhang, Xinmin",
    title = "{k-essential leptogenesis}",
    eprint = "hep-ph/0209093",
    archivePrefix = "arXiv",
    reportNumber = "BIHEP-TH-2002-40",
    doi = "10.1016/j.physletb.2003.08.041",
    journal = "Phys. Lett. B",
    volume = "573",
    pages = "20--26",
    year = "2003"
}

@article{Malquarti:2003nn,
    author = "Malquarti, Michael and Copeland, Edmund J and Liddle, Andrew R and Trodden, Mark",
    title = "{A New view of k-essence}",
    eprint = "astro-ph/0302279",
    archivePrefix = "arXiv",
    doi = "10.1103/PhysRevD.67.123503",
    journal = "Phys. Rev. D",
    volume = "67",
    pages = "123503",
    year = "2003"
}

@article{Malquarti:2003hn,
    author = "Malquarti, Michael and Copeland, Edmund J. and Liddle, Andrew R.",
    title = "{K-essence and the coincidence problem}",
    eprint = "astro-ph/0304277",
    archivePrefix = "arXiv",
    doi = "10.1103/PhysRevD.68.023512",
    journal = "Phys. Rev. D",
    volume = "68",
    pages = "023512",
    year = "2003"
}

@article{Sen:1999mh,
    author = "Sen, Ashoke",
    title = "{Descent relations among bosonic D-branes}",
    eprint = "hep-th/9902105",
    archivePrefix = "arXiv",
    reportNumber = "MRI-PHY-P990203",
    doi = "10.1142/S0217751X99001901",
    journal = "Int. J. Mod. Phys. A",
    volume = "14",
    pages = "4061--4078",
    year = "1999"
}

@article{Sen:2002in,
    author = "Sen, Ashoke",
    title = "{Tachyon matter}",
    eprint = "hep-th/0203265",
    archivePrefix = "arXiv",
    reportNumber = "CGPG-02-3-3",
    doi = "10.1088/1126-6708/2002/07/065",
    journal = "JHEP",
    volume = "07",
    pages = "065",
    year = "2002"
}

@article{Sen:2002nu,
    author = "Sen, Ashoke",
    title = "{Rolling tachyon}",
    eprint = "hep-th/0203211",
    archivePrefix = "arXiv",
    reportNumber = "MRI-P-020303",
    doi = "10.1088/1126-6708/2002/04/048",
    journal = "JHEP",
    volume = "04",
    pages = "048",
    year = "2002"
}

@article{Garousi:2000tr,
    author = "Garousi, Mohammad R.",
    title = "{Tachyon couplings on nonBPS D-branes and Dirac-Born-Infeld action}",
    eprint = "hep-th/0003122",
    archivePrefix = "arXiv",
    reportNumber = "IPM-P-2000-013",
    doi = "10.1016/S0550-3213(00)00361-8",
    journal = "Nucl. Phys. B",
    volume = "584",
    pages = "284--299",
    year = "2000"
}

@article{Bergshoeff:2000dq,
    author = "Bergshoeff, E. A. and de Roo, M. and de Wit, T. C. and Eyras, E. and Panda, Sudhakar",
    title = "{T duality and actions for nonBPS D-branes}",
    eprint = "hep-th/0003221",
    archivePrefix = "arXiv",
    reportNumber = "UG-00-02, DAMTP-2000-33, MRI-PHY-P20000308",
    doi = "10.1088/1126-6708/2000/05/009",
    journal = "JHEP",
    volume = "05",
    pages = "009",
    year = "2000"
}

@article{Kluson:2000iy,
    author = "Kluson, J.",
    title = "{Proposal for nonBPS D-brane action}",
    eprint = "hep-th/0004106",
    archivePrefix = "arXiv",
    reportNumber = "MU-2022",
    doi = "10.1103/PhysRevD.62.126003",
    journal = "Phys. Rev. D",
    volume = "62",
    pages = "126003",
    year = "2000"
}

@article{Garousi:2002wq,
    author = "Garousi, Mohammad R.",
    title = "{On shell S matrix and tachyonic effective actions}",
    eprint = "hep-th/0209068",
    archivePrefix = "arXiv",
    reportNumber = "IPM-P-2002-038",
    doi = "10.1016/S0550-3213(02)00903-3",
    journal = "Nucl. Phys. B",
    volume = "647",
    pages = "117--130",
    year = "2002"
}

@article{Garousi_2003,
    author = "Garousi, Mohammad R.",
    title = "{Slowly varying tachyon and tachyon potential}",
    eprint = "hep-th/0304145",
    archivePrefix = "arXiv",
    reportNumber = "IPM-P-2003-019",
    doi = "10.1088/1126-6708/2003/05/058",
    journal = "JHEP",
    volume = "05",
    pages = "058",
    year = "2003"
}

@article{Gibbons:2002md,
    author = "Gibbons, G W",
    title = "{Cosmological evolution of the rolling tachyon}",
    eprint = "hep-th/0204008",
    archivePrefix = "arXiv",
    reportNumber = "DAMTP-2002-38",
    doi = "10.1016/S0370-2693(02)01881-6",
    journal = "Phys. Lett. B",
    volume = "537",
    pages = "1--4",
    year = "2002"
}

@article{Padmanabhan:2002cp,
    author = "Padmanabhan, T.",
    title = "{Accelerated expansion of the universe driven by tachyonic matter}",
    eprint = "hep-th/0204150",
    archivePrefix = "arXiv",
    reportNumber = "IUCAA-16-2002",
    doi = "10.1103/PhysRevD.66.021301",
    journal = "Phys. Rev. D",
    volume = "66",
    pages = "021301",
    year = "2002"
}

@article{Bagla:2002yn,
    author = "Bagla, J. S. and Jassal, Harvinder Kaur and Padmanabhan, T.",
    title = "{Cosmology with tachyon field as dark energy}",
    eprint = "astro-ph/0212198",
    archivePrefix = "arXiv",
    doi = "10.1103/PhysRevD.67.063504",
    journal = "Phys. Rev. D",
    volume = "67",
    pages = "063504",
    year = "2003"
}

@article{Abramo:2003cp,
    author = "Abramo, L. Raul W. and Finelli, Fabio",
    title = "{Cosmological dynamics of the tachyon with an inverse power-law potential}",
    eprint = "astro-ph/0307208",
    archivePrefix = "arXiv",
    doi = "10.1016/j.physletb.2003.09.065",
    journal = "Phys. Lett. B",
    volume = "575",
    pages = "165--171",
    year = "2003"
}

@article{Aguirregabiria:2004xd,
    author = "Aguirregabiria, J. M. and Lazkoz, Ruth",
    title = "{Tracking solutions in tachyon cosmology}",
    eprint = "hep-th/0402190",
    archivePrefix = "arXiv",
    doi = "10.1103/PhysRevD.69.123502",
    journal = "Phys. Rev. D",
    volume = "69",
    pages = "123502",
    year = "2004"
}

@article{Zong-KuanGuo_2004,
    author = "Guo, Zong-Kuan and Zhang, Yuan-Zhong",
    title = "{Cosmological scaling solutions of the tachyon with multiple inverse square potentials}",
    eprint = "hep-th/0403151",
    archivePrefix = "arXiv",
    doi = "10.1088/1475-7516/2004/08/010",
    journal = "JCAP",
    volume = "08",
    pages = "010",
    year = "2004"
}

@article{Copeland:2004hq,
    author = "Copeland, Edmund J. and Garousi, Mohammad R. and Sami, M. and Tsujikawa, Shinji",
    title = "{What is needed of a tachyon if it is to be the dark energy?}",
    eprint = "hep-th/0411192",
    archivePrefix = "arXiv",
    doi = "10.1103/PhysRevD.71.043003",
    journal = "Phys. Rev. D",
    volume = "71",
    pages = "043003",
    year = "2005"
}

@article{Chevallier:2000qy,
    author = "Chevallier, Michel and Polarski, David",
    title = "{Accelerating universes with scaling dark matter}",
    eprint = "gr-qc/0009008",
    archivePrefix = "arXiv",
    doi = "10.1142/S0218271801000822",
    journal = "Int. J. Mod. Phys. D",
    volume = "10",
    pages = "213--224",
    year = "2001"
}

@article{Linder:2002et,
    author = "Linder, Eric V.",
    title = "{Exploring the expansion history of the universe}",
    eprint = "astro-ph/0208512",
    archivePrefix = "arXiv",
    doi = "10.1103/PhysRevLett.90.091301",
    journal = "Phys. Rev. Lett.",
    volume = "90",
    pages = "091301",
    year = "2003"
}

@article{Jassal:2004ej,
    author = "Jassal, H. K. and Bagla, J. S. and Padmanabhan, T.",
    title = "{WMAP constraints on low redshift evolution of dark energy}",
    eprint = "astro-ph/0404378",
    archivePrefix = "arXiv",
    doi = "10.1111/j.1745-3933.2005.08577.x",
    journal = "Mon. Not. Roy. Astron. Soc.",
    volume = "356",
    pages = "L11--L16",
    year = "2005"
}

@article{Jassal:2006gf,
    author = "Jassal, Harvinder K. and Bagla, J. S. and Padmanabhan, T.",
    title = "{Understanding the origin of CMB constraints on Dark Energy}",
    eprint = "astro-ph/0601389",
    archivePrefix = "arXiv",
    doi = "10.1111/j.1365-2966.2010.16647.x",
    journal = "Mon. Not. Roy. Astron. Soc.",
    volume = "405",
    pages = "2639--2650",
    year = "2010"
}

@article{Barboza:2008rh,
    author = "Barboza, Jr., E. M. and Alcaniz, J. S.",
    title = "{A parametric model for dark energy}",
    eprint = "0805.1713",
    archivePrefix = "arXiv",
    primaryClass = "astro-ph",
    doi = "10.1016/j.physletb.2008.08.012",
    journal = "Phys. Lett. B",
    volume = "666",
    pages = "415--419",
    year = "2008"
}

@article{Wetterich:2004pv,
    author = "Wetterich, Christof",
    title = "{Phenomenological parameterization of quintessence}",
    eprint = "astro-ph/0403289",
    archivePrefix = "arXiv",
    reportNumber = "HD-THEP-04-08",
    doi = "10.1016/j.physletb.2004.05.008",
    journal = "Phys. Lett. B",
    volume = "594",
    pages = "17--22",
    year = "2004"
}

@article{Ma:2011nc,
    author = "Ma, Jing-Zhe and Zhang, Xin",
    title = "{Probing the dynamics of dark energy with novel parametrizations}",
    eprint = "1102.2671",
    archivePrefix = "arXiv",
    primaryClass = "astro-ph.CO",
    doi = "10.1016/j.physletb.2011.04.013",
    journal = "Phys. Lett. B",
    volume = "699",
    pages = "233--238",
    year = "2011"
}

@article{Albrecht:2006um,
    author = "Albrecht, Andreas and others",
    title = "{Report of the Dark Energy Task Force}",
    eprint = "astro-ph/0609591",
    archivePrefix = "arXiv",
    reportNumber = "FERMILAB-FN-0793-A",
    month = "9",
    year = "2006"
}

@article{Virey:2006em,
    author = "Virey, Jean-March and Ealet, A.",
    title = "{Sensitivity and figures of merit for dark energy supernovae surveys}",
    eprint = "astro-ph/0607589",
    archivePrefix = "arXiv",
    reportNumber = "CPT-P55-2006",
    doi = "10.1051/0004-6361:20066116",
    journal = "Astron. Astrophys.",
    volume = "464",
    pages = "837",
    year = "2007"
}

@article{Linder:2007wa,
    author = "Linder, Eric V.",
    title = "{The Dynamics of Quintessence, The Quintessence of Dynamics}",
    eprint = "0704.2064",
    archivePrefix = "arXiv",
    primaryClass = "astro-ph",
    doi = "10.1007/s10714-007-0550-z",
    journal = "Gen. Rel. Grav.",
    volume = "40",
    pages = "329--356",
    year = "2008"
}

@article{Wang:2003gz,
    author = "Wang, Yun and Mukherjee, Pia",
    title = "{Model - independent constraints on dark energy density from flux - averaging analysis of type Ia supernova data}",
    eprint = "astro-ph/0312192",
    archivePrefix = "arXiv",
    doi = "10.1086/383196",
    journal = "Astrophys. J.",
    volume = "606",
    pages = "654--663",
    year = "2004"
}

@article{Wang:2004py,
    author = "Wang, Yun and Tegmark, Max",
    title = "{New dark energy constraints from supernovae, microwave background and galaxy clustering}",
    eprint = "astro-ph/0403292",
    archivePrefix = "arXiv",
    doi = "10.1103/PhysRevLett.92.241302",
    journal = "Phys. Rev. Lett.",
    volume = "92",
    pages = "241302",
    year = "2004"
}

@article{Wang:2004ru,
    author = "Wang, Yun and Freese, Katherine",
    title = "{Probing dark energy using its density instead of its equation of state}",
    eprint = "astro-ph/0402208",
    archivePrefix = "arXiv",
    doi = "10.1016/j.physletb.2005.10.083",
    journal = "Phys. Lett. B",
    volume = "632",
    pages = "449--452",
    year = "2006"
}

@article{Einstein:1917ce,
    author = "Einstein, Albert",
    title = "{Cosmological Considerations in the General Theory of Relativity}",
    journal = "Sitzungsber. Preuss. Akad. Wiss. Berlin (Math. Phys. )",
    volume = "1917",
    pages = "142--152",
    year = "1917"
}

@article{Padmanabhan:2002ji,
    author = "Padmanabhan, T.",
    title = "{Cosmological constant: The Weight of the vacuum}",
    eprint = "hep-th/0212290",
    archivePrefix = "arXiv",
    doi = "10.1016/S0370-1573(03)00120-0",
    journal = "Phys. Rept.",
    volume = "380",
    pages = "235--320",
    year = "2003"
}

@article{Sahni:1999gb,
    author = "Sahni, Varun and Starobinsky, Alexei A.",
    title = "{The Case for a positive cosmological Lambda term}",
    eprint = "astro-ph/9904398",
    archivePrefix = "arXiv",
    reportNumber = "IUCAA-25-2000",
    doi = "10.1142/S0218271800000542",
    journal = "Int. J. Mod. Phys. D",
    volume = "9",
    pages = "373--444",
    year = "2000"
}

@article{Carroll:2000fy,
    author = "Carroll, Sean M.",
    title = "{The Cosmological constant}",
    eprint = "astro-ph/0004075",
    archivePrefix = "arXiv",
    reportNumber = "EFI-2000-13",
    doi = "10.12942/lrr-2001-1",
    journal = "Living Rev. Rel.",
    volume = "4",
    pages = "1",
    year = "2001"
}

@article{Peebles:2002gy,
    author = "Peebles, P. J. E. and Ratra, Bharat",
    title = "{The Cosmological Constant and Dark Energy}",
    eprint = "astro-ph/0207347",
    archivePrefix = "arXiv",
    reportNumber = "KSUPT-02-3",
    doi = "10.1103/RevModPhys.75.559",
    journal = "Rev. Mod. Phys.",
    volume = "75",
    pages = "559--606",
    year = "2003"
}

@article{Frieman:2008sn,
    author = "Frieman, Joshua and Turner, Michael and Huterer, Dragan",
    title = "{Dark Energy and the Accelerating Universe}",
    eprint = "0803.0982",
    archivePrefix = "arXiv",
    primaryClass = "astro-ph",
    reportNumber = "FERMILAB-PUB-08-613-A",
    doi = "10.1146/annurev.astro.46.060407.145243",
    journal = "Ann. Rev. Astron. Astrophys.",
    volume = "46",
    pages = "385--432",
    year = "2008"
}

@article{Amendola:2010ub,
    author = "Amendola, Luca and Kainulainen, Kimmo and Marra, Valerio and Quartin, Miguel",
    title = "{Large-scale inhomogeneities may improve the cosmic concordance of supernovae}",
    eprint = "1002.1232",
    archivePrefix = "arXiv",
    primaryClass = "astro-ph.CO",
    doi = "10.1103/PhysRevLett.105.121302",
    journal = "Phys. Rev. Lett.",
    volume = "105",
    pages = "121302",
    year = "2010"
}

@article{Mehrabi:2018dru,
    author = "Mehrabi, A.",
    title = "{Growth of perturbations in dark energy parametrization scenarios}",
    eprint = "1804.09886",
    archivePrefix = "arXiv",
    primaryClass = "astro-ph.CO",
    doi = "10.1103/PhysRevD.97.083522",
    journal = "Phys. Rev. D",
    volume = "97",
    number = "8",
    pages = "083522",
    year = "2018"
}

@book{Liddle:2000cg,
    author = "Liddle, Andrew R. and Lyth, D. H.",
    title = "{Cosmological inflation and large scale structure}",
    isbn = "978-0-521-57598-0, 978-0-521-82849-9",
    year = "2000"
}

@article{Padmanabhan:2004av,
    author = "Padmanabhan, T.",
    title = "{Dark energy: The Cosmological challenge of the millennium}",
    eprint = "astro-ph/0411044",
    archivePrefix = "arXiv",
    journal = "Curr. Sci.",
    volume = "88",
    pages = "1057",
    year = "2005"
}

@article{Copeland:2006wr,
    author = "Copeland, Edmund J. and Sami, M. and Tsujikawa, Shinji",
    title = "{Dynamics of dark energy}",
    eprint = "hep-th/0603057",
    archivePrefix = "arXiv",
    doi = "10.1142/S021827180600942X",
    journal = "Int. J. Mod. Phys. D",
    volume = "15",
    pages = "1753--1936",
    year = "2006"
}

@article{Ratra:1987rm,
    author = "Ratra, Bharat and Peebles, P. J. E.",
    title = "{Cosmological Consequences of a Rolling Homogeneous Scalar Field}",
    reportNumber = "PUPT-1072",
    doi = "10.1103/PhysRevD.37.3406",
    journal = "Phys. Rev. D",
    volume = "37",
    pages = "3406",
    year = "1988"
}

@article{Wetterich:1987fm,
    author = "Wetterich, C.",
    title = "{Cosmology and the Fate of Dilatation Symmetry}",
    eprint = "1711.03844",
    archivePrefix = "arXiv",
    primaryClass = "hep-th",
    reportNumber = "PRINT-87-0756, DESY-87-123",
    doi = "10.1016/0550-3213(88)90193-9",
    journal = "Nucl. Phys. B",
    volume = "302",
    pages = "668--696",
    year = "1988"
}

@article{Sen:2001xu,
    author = "Sen, A. A. and Sethi, S.",
    title = "{Quintessence model with double exponential potential}",
    eprint = "gr-qc/0111082",
    archivePrefix = "arXiv",
    doi = "10.1016/S0370-2693(02)01547-2",
    journal = "Phys. Lett. B",
    volume = "532",
    pages = "159--165",
    year = "2002"
}

@article{Zlatev:1998tr,
    author = "Zlatev, Ivaylo and Wang, Li-Min and Steinhardt, Paul J.",
    title = "{Quintessence, cosmic coincidence, and the cosmological constant}",
    eprint = "astro-ph/9807002",
    archivePrefix = "arXiv",
    doi = "10.1103/PhysRevLett.82.896",
    journal = "Phys. Rev. Lett.",
    volume = "82",
    pages = "896--899",
    year = "1999"
}

@article{Zlatev:1998yg,
    author = "Zlatev, Ivaylo and Steinhardt, Paul J.",
    title = "{A Tracker solution to the cold dark matter cosmic coincidence problem}",
    eprint = "astro-ph/9906481",
    archivePrefix = "arXiv",
    doi = "10.1016/S0370-2693(99)00707-8",
    journal = "Phys. Lett. B",
    volume = "459",
    pages = "570--574",
    year = "1999"
}

@article{Steinhardt:1999nw,
    author = "Steinhardt, Paul J. and Wang, Li-Min and Zlatev, Ivaylo",
    title = "{Cosmological tracking solutions}",
    eprint = "astro-ph/9812313",
    archivePrefix = "arXiv",
    doi = "10.1103/PhysRevD.59.123504",
    journal = "Phys. Rev. D",
    volume = "59",
    pages = "123504",
    year = "1999"
}

@article{Wang:1999fa,
    author = "Wang, Li-Min and Caldwell, R. R. and Ostriker, J. P. and Steinhardt, Paul J.",
    title = "{Cosmic concordance and quintessence}",
    eprint = "astro-ph/9901388",
    archivePrefix = "arXiv",
    doi = "10.1086/308331",
    journal = "Astrophys. J.",
    volume = "530",
    pages = "17--35",
    year = "2000"
}

@article{Urena-Lopez:2000ewq,
    author = "Urena-Lopez, L. Arturo and Matos, Tonatiuh",
    title = "{A New cosmological tracker solution for quintessence}",
    eprint = "astro-ph/0003364",
    archivePrefix = "arXiv",
    doi = "10.1103/PhysRevD.62.081302",
    journal = "Phys. Rev. D",
    volume = "62",
    pages = "081302",
    year = "2000"
}

@article{Sahlen:2006dn,
    author = "Sahlen, Martin and Liddle, Andrew R and Parkinson, David",
    title = "{Quintessence reconstructed: New constraints and tracker viability}",
    eprint = "astro-ph/0610812",
    archivePrefix = "arXiv",
    doi = "10.1103/PhysRevD.75.023502",
    journal = "Phys. Rev. D",
    volume = "75",
    pages = "023502",
    year = "2007"
}

@article{Scherrer:2007pu,
    author = "Scherrer, Robert J. and Sen, A. A.",
    title = "{Thawing quintessence with a nearly flat potential}",
    eprint = "0712.3450",
    archivePrefix = "arXiv",
    primaryClass = "astro-ph",
    doi = "10.1103/PhysRevD.77.083515",
    journal = "Phys. Rev. D",
    volume = "77",
    pages = "083515",
    year = "2008"
}

@article{Scherrer:2008be,
    author = "Scherrer, Robert J. and Sen, A. A.",
    title = "{Phantom Dark Energy Models with a Nearly Flat Potential}",
    eprint = "0808.1880",
    archivePrefix = "arXiv",
    primaryClass = "astro-ph",
    doi = "10.1103/PhysRevD.78.067303",
    journal = "Phys. Rev. D",
    volume = "78",
    pages = "067303",
    year = "2008"
}

@article{Dalmazi:2006ws,
    author = "Dalmazi, D. and de Souza Dutra, A. and Abreu, E. M. C.",
    title = "{Generalizing the Soldering procedure}",
    eprint = "hep-th/0607102",
    archivePrefix = "arXiv",
    doi = "10.1103/PhysRevD.79.109902",
    journal = "Phys. Rev. D",
    volume = "74",
    pages = "025015",
    year = "2006",
    note = "[Erratum: Phys.Rev.D 79, 109902 (2009)]"
}

@article{Gupta:2014uea,
    author = "Gupta, Gaveshna and Rangarajan, Raghavan and Sen, Anjan A.",
    title = "{Thawing quintessence from the inflationary epoch to today}",
    eprint = "1412.6915",
    archivePrefix = "arXiv",
    primaryClass = "astro-ph.CO",
    doi = "10.1103/PhysRevD.92.123003",
    journal = "Phys. Rev. D",
    volume = "92",
    number = "12",
    pages = "123003",
    year = "2015"
}

@article{Chiba:2012cb,
    author = "Chiba, Takeshi and De Felice, Antonio and Tsujikawa, Shinji",
    title = "{Observational constraints on quintessence: thawing, tracker, and scaling models}",
    eprint = "1210.3859",
    archivePrefix = "arXiv",
    primaryClass = "astro-ph.CO",
    doi = "10.1103/PhysRevD.87.083505",
    journal = "Phys. Rev. D",
    volume = "87",
    number = "8",
    pages = "083505",
    year = "2013"
}

@article{Pantazis:2016nky,
    author = "Pantazis, G. and Nesseris, S. and Perivolaropoulos, L.",
    title = "{Comparison of thawing and freezing dark energy parametrizations}",
    eprint = "1603.02164",
    archivePrefix = "arXiv",
    primaryClass = "astro-ph.CO",
    reportNumber = "IFT-UAM-CSIC-16-023",
    doi = "10.1103/PhysRevD.93.103503",
    journal = "Phys. Rev. D",
    volume = "93",
    number = "10",
    pages = "103503",
    year = "2016"
}

@article{Roy:2013wqa,
    author = "Roy, Nandan and Banerjee, Narayan",
    title = "{Tracking quintessence: a dynamical systems study}",
    eprint = "1312.2670",
    archivePrefix = "arXiv",
    primaryClass = "gr-qc",
    doi = "10.1007/s10714-013-1651-5",
    journal = "Gen. Rel. Grav.",
    volume = "46",
    pages = "1651",
    year = "2014"
}

@article{Roy:2014yta,
    author = "Roy, Nandan and Banerjee, Narayan",
    title = "{Quintessence Scalar Field: A Dynamical Systems Study}",
    eprint = "1402.6821",
    archivePrefix = "arXiv",
    primaryClass = "gr-qc",
    doi = "10.1140/epjp/i2014-14162-7",
    journal = "Eur. Phys. J. Plus",
    volume = "129",
    pages = "162",
    year = "2014"
}

@article{Sahni:1999qe,
    author = "Sahni, Varun and Wang, Li-Min",
    title = "{A New cosmological model of quintessence and dark matter}",
    eprint = "astro-ph/9910097",
    archivePrefix = "arXiv",
    reportNumber = "IUCAA-46-2000",
    doi = "10.1103/PhysRevD.62.103517",
    journal = "Phys. Rev. D",
    volume = "62",
    pages = "103517",
    year = "2000"
}

@article{Barreiro:1999zs,
    author = "Barreiro, T. and Copeland, Edmund J. and Nunes, N. J.",
    title = "{Quintessence arising from exponential potentials}",
    eprint = "astro-ph/9910214",
    archivePrefix = "arXiv",
    reportNumber = "SUSX-TH-016",
    doi = "10.1103/PhysRevD.61.127301",
    journal = "Phys. Rev. D",
    volume = "61",
    pages = "127301",
    year = "2000"
}

@article{Kim:1998kx,
    author = "Kim, Jihn E.",
    title = "{Axion and almost massless quark as ingredients of quintessence}",
    eprint = "hep-ph/9811509",
    archivePrefix = "arXiv",
    reportNumber = "SNUTP-98-137, KIAS-P98042",
    doi = "10.1088/1126-6708/1999/05/022",
    journal = "JHEP",
    volume = "05",
    pages = "022",
    year = "1999"
}

@article{Frieman:1995pm,
    author = "Frieman, Joshua A. and Hill, Christopher T. and Stebbins, Albert and Waga, Ioav",
    title = "{Cosmology with ultralight pseudo Nambu-Goldstone bosons}",
    eprint = "astro-ph/9505060",
    archivePrefix = "arXiv",
    reportNumber = "FERMILAB-PUB-95-066-A",
    doi = "10.1103/PhysRevLett.75.2077",
    journal = "Phys. Rev. Lett.",
    volume = "75",
    pages = "2077--2080",
    year = "1995"
}

@article{Brax:1999yv,
    author = "Brax, Philippe and Martin, Jerome",
    title = "{The Robustness of quintessence}",
    eprint = "astro-ph/9912046",
    archivePrefix = "arXiv",
    reportNumber = "SACLAY-SPH-T-99-075",
    doi = "10.1103/PhysRevD.61.103502",
    journal = "Phys. Rev. D",
    volume = "61",
    pages = "103502",
    year = "2000"
}

@article{Brax:1999gp,
    author = "Brax, Philippe and Martin, Jerome",
    title = "{Quintessence and supergravity}",
    eprint = "astro-ph/9905040",
    archivePrefix = "arXiv",
    doi = "10.1016/S0370-2693(99)01209-5",
    journal = "Phys. Lett. B",
    volume = "468",
    pages = "40--45",
    year = "1999"
}

@article{Albrecht:1999rm,
    author = "Albrecht, Andreas and Skordis, Constantinos",
    title = "{Phenomenology of a realistic accelerating universe using only Planck scale physics}",
    eprint = "astro-ph/9908085",
    archivePrefix = "arXiv",
    doi = "10.1103/PhysRevLett.84.2076",
    journal = "Phys. Rev. Lett.",
    volume = "84",
    pages = "2076--2079",
    year = "2000"
}

@article{Dodelson:2000jtt,
    author = "Dodelson, Scott and Kaplinghat, Manoj and Stewart, Ewan",
    title = "{Solving the Coincidence Problem : Tracking Oscillating Energy}",
    eprint = "astro-ph/0002360",
    archivePrefix = "arXiv",
    reportNumber = "FERMILAB-PUB-01-279-A, FERMILAB-PUB-00-070-T",
    doi = "10.1103/PhysRevLett.85.5276",
    journal = "Phys. Rev. Lett.",
    volume = "85",
    pages = "5276--5279",
    year = "2000"
}

@article{endo1977cosmological,
  title={The cosmological term and a modified Brans-Dicke cosmology},
  author={End{\=o}, Makoto and Fukui, Takao},
  journal={General Relativity and Gravitation},
  volume={8},
  pages={833--839},
  year={1977},
  publisher={Springer}
}

@article{Canuto:1977zz,
    author = "Canuto, V. and Hsieh, S. H. and Adams, P. J.",
    title = "{Scale-Covariant Theory of Gravitation and Astrophysical Applications}",
    doi = "10.1103/PhysRevLett.39.429",
    journal = "Phys. Rev. Lett.",
    volume = "39",
    pages = "429--432",
    year = "1977"
}

@article{Bertolami:1986bg,
    author = "Bertolami, O.",
    title = "{Time Dependent Cosmological Term}",
    doi = "10.1007/BF02728301",
    journal = "Nuovo Cim. B",
    volume = "93",
    pages = "36--42",
    year = "1986"
}

@article{berman1990brans,
  title={Brans-Dicke models with time-dependent cosmological term},
  author={Berman, Marcelo Samuel and Som, MM},
  journal={Int.J.Theor.Phys.},
  volume={29},
  pages={1411--1414},
  year={1990},
  publisher={Springer}
}

@article{Beesham:1994ni,
    author = "Beesham, A.",
    title = "{Bianchi type I cosmological models with variable G and Lambda}",
    doi = "10.1007/BF02105151",
    journal = "Gen. Rel. Grav.",
    volume = "26",
    pages = "159--165",
    year = "1994"
}

@article{Lopez:1995eb,
    author = "Lopez, Jorge L. and Nanopoulos, Dimitri V.",
    title = "{A New cosmological constant model}",
    eprint = "hep-ph/9501293",
    archivePrefix = "arXiv",
    reportNumber = "CERN-TH-95-6, CERN-TH-95-06, CERN-TH-95-006, CTP-TAMU-69-94, ACT-25-94",
    doi = "10.1142/S0217732396000023",
    journal = "Mod. Phys. Lett. A",
    volume = "11",
    pages = "1--7",
    year = "1996"
}

@article{OZER1986363,
title = {A possible solution to the main cosmological problems},
journal = {Physics Letters B},
volume = {171},
number = {4},
pages = {363-365},
year = {1986},
issn = {0370-2693},
doi = {https://doi.org/10.1016/0370-2693(86)91421-8},
url = {https://www.sciencedirect.com/science/article/pii/0370269386914218},
author = {Murat Özer and M.O. Taha},
}

@article{kalligas1992flat,
  title={Flat FRW models with variable G and $\Lambda$},
  author={Kalligas, D and Wesson, P and Everitt, CWF},
  journal={Gen. Rel. Grav.},
  volume={24},
  pages={351--357},
  year={1992},
  publisher={Springer}
}

@article{Kalligas:1995qh,
    author = "Kalligas, D. and Wesson, P. S. and Everitt, C. W. F.",
    title = "{Bianchi type I cosmological models with variable G and Lambda: A Comment}",
    doi = "10.1007/BF02108066",
    journal = "Gen. Rel. Grav.",
    volume = "27",
    pages = "645--650",
    year = "1995"
}

@article{Beesham:1993bh,
    author = "Beesham, Aroonkumar",
    title = "{Cosmological models with a variable cosmological term and bulk viscous models}",
    doi = "10.1103/PhysRevD.48.3539",
    journal = "Phys. Rev. D",
    volume = "48",
    pages = "3539--3543",
    year = "1993"
}

@article{Spindel:1993bz,
    author = "Spindel, P. and Brout, R.",
    title = "{Entropy production from vacuum decay}",
    eprint = "gr-qc/9310023",
    archivePrefix = "arXiv",
    reportNumber = "UMH-ULB-TH-05-93",
    doi = "10.1016/0370-2693(94)90651-3",
    journal = "Phys. Lett. B",
    volume = "320",
    pages = "241--244",
    year = "1994"
}

@article{Overduin:1998zv,
    author = "Overduin, J. M. and Cooperstock, F. I.",
    title = "{Evolution of the scale factor with a variable cosmological term}",
    eprint = "astro-ph/9805260",
    archivePrefix = "arXiv",
    doi = "10.1103/PhysRevD.58.043506",
    journal = "Phys. Rev. D",
    volume = "58",
    pages = "043506",
    year = "1998"
}

@article{OZER1987776,
title = {A model of the universe free of cosmological problems},
journal = {Nuclear Physics B},
volume = {287},
pages = {776-796},
year = {1987},
issn = {0550-3213},
doi = {https://doi.org/10.1016/0550-3213(87)90128-3},
url = {https://www.sciencedirect.com/science/article/pii/0550321387901283},
author = {Murat Özer and M.O. Taha},
}

@article{Abdel-Rahman:1992msa,
    author = "Abdel-Rahman, A. M. M.",
    title = "{Singularity - free decaying vacuum cosmologies}",
    doi = "10.1103/PhysRevD.45.3497",
    journal = "Phys. Rev. D",
    volume = "45",
    pages = "3497--3511",
    year = "1992"
}

@article{Chen:1990jw,
    author = "Chen, W. and Wu, Y. S.",
    title = "{Implications of a cosmological constant varying as R**(-2)}",
    doi = "10.1103/PhysRevD.41.695",
    journal = "Phys. Rev. D",
    volume = "41",
    pages = "695--698",
    year = "1990",
    note = "[Erratum: Phys.Rev.D 45, 4728 (1992)]"
}

@article{Gott227,
    author = {Gott, J. Richard, III and Rees, Martin J.},
    title = "{Astronomical constraints on a string-dominated universe}",
    journal = {Mon. Not. Roy. Astron. Soc.},
    volume = {227},
    number = {2},
    pages = {453-459},
    year = {1987},
    month = {07},
issn = {0035-8711},
    doi = {10.1093/mnras/227.2.453},
    url = {https://doi.org/10.1093/mnras/227.2.453},
    eprint = {https://academic.oup.com/mnras/article-pdf/227/2/453/2877800/mnras227-0453.pdf},
}

@article{Abdussattar:1997aa,
    author = "Abdussattar and Vishwakarma, R. G.",
    title = "{Some FRW models with variable G and Lambda}",
    doi = "10.1088/0264-9381/14/4/011",
    journal = "Class. Quant. Grav.",
    volume = "14",
    pages = "945--953",
    year = "1997"
}

@article{Rajeev:1983pr,
    author = "Rajeev, S. G.",
    title = "{Why Is the Cosmological Constant Small?}",
    reportNumber = "SU-4217-246, COO-3533-246",
    doi = "10.1016/0370-2693(83)91255-8",
    journal = "Phys. Lett. B",
    volume = "125",
    pages = "144--146",
    year = "1983"
}

@article{Kazanas:1980tx,
    author = "Kazanas, D.",
    title = "{Dynamics of the Universe and Spontaneous Symmetry Breaking}",
    doi = "10.1086/183361",
    journal = "Astrophys. J. Lett.",
    volume = "241",
    pages = "L59--L63",
    year = "1980"
}

@article{Copeland:1997et,
    author = "Copeland, Edmund J. and Liddle, Andrew R and Wands, David",
    title = "{Exponential potentials and cosmological scaling solutions}",
    eprint = "gr-qc/9711068",
    archivePrefix = "arXiv",
    reportNumber = "SUSX-TH-97-022, SUSSEX-AST-97-11-1, PU-RCG-97-20",
    doi = "10.1103/PhysRevD.57.4686",
    journal = "Phys. Rev. D",
    volume = "57",
    pages = "4686--4690",
    year = "1998"
}

@article{Ferreira:1997au,
    author = "Ferreira, Pedro G. and Joyce, Michael",
    title = "{Structure formation with a selftuning scalar field}",
    eprint = "astro-ph/9707286",
    archivePrefix = "arXiv",
    reportNumber = "CFPA-97-TH-07",
    doi = "10.1103/PhysRevLett.79.4740",
    journal = "Phys. Rev. Lett.",
    volume = "79",
    pages = "4740--4743",
    year = "1997"
}

@article{Lima:1994ni,
    author = "Lima, J. A. S. and Carvalho, J. C.",
    title = "{Dirac's cosmology with varying cosmological constant}",
    doi = "10.1007/BF02107147",
    journal = "Gen. Rel. Grav.",
    volume = "26",
    pages = "909--916",
    year = "1994"
}

@article{Wetterich:1994bg,
    author = "Wetterich, Christof",
    title = "{The Cosmon model for an asymptotically vanishing time dependent cosmological 'constant'}",
    eprint = "hep-th/9408025",
    archivePrefix = "arXiv",
    reportNumber = "HD-THEP-94-16",
    journal = "Astron. Astrophys.",
    volume = "301",
    pages = "321--328",
    year = "1995"
}

@article{Waga:1992hj,
    author = "Waga, I.",
    title = "{Decaying vacuum flat cosmological models: Expressions for some observable quantities and their properties}",
    reportNumber = "IF-UFRJ-92-30",
    doi = "10.1086/173090",
    journal = "Astrophys. J.",
    volume = "414",
    pages = "436--448",
    year = "1993"
}

@article{Salim:1992mx,
    author = "Salim, J. M. and Waga, I.",
    title = "{Thermodynamic constraints on a time dependent Lambda model}",
    doi = "10.1088/0264-9381/10/9/018",
    journal = "Class. Quant. Grav.",
    volume = "10",
    pages = "1767--1774",
    year = "1993"
}

@article{Carvalho:1991ut,
    author = "Carvalho, J. C. and Lima, J. A. S. and Waga, I.",
    title = "{On the cosmological consequences of a time dependent lambda term}",
    reportNumber = "IF-UFRJ-91-36",
    doi = "10.1103/PhysRevD.46.2404",
    journal = "Phys. Rev. D",
    volume = "46",
    pages = "2404--2407",
    year = "1992"
}

@article{Arbab:1994db,
    author = "Arbab, Arbab I. and Abdel-Rahman, A. M. M.",
    title = "{Nonsingular cosmology with a time dependent cosmological term}",
    doi = "10.1103/PhysRevD.50.7725",
    journal = "Phys. Rev. D",
    volume = "50",
    pages = "7725--7728",
    year = "1994"
}

@article{Lima:1994gi,
    author = "Lima, J. A. S. and Maia, J. M. F.",
    title = "{Deflationary cosmology with decaying vacuum energy density}",
    doi = "10.1103/PhysRevD.49.5597",
    journal = "Phys. Rev. D",
    volume = "49",
    pages = "5597--5600",
    year = "1994"
}

@article{Lima:1995ea,
    author = "Lima, J. A. S. and Trodden, M.",
    title = "{Decaying vacuum energy and deflationary cosmology in open and closed universes}",
    eprint = "astro-ph/9508049",
    archivePrefix = "arXiv",
    reportNumber = "BROWN-HET-994",
    doi = "10.1103/PhysRevD.53.4280",
    journal = "Phys. Rev. D",
    volume = "53",
    pages = "4280--4286",
    year = "1996"
}

@article{Hiscock:1986ec,
    author = "Hiscock, W. A.",
    title = "{Quantum Instabilities and the Cosmological Constant}",
    doi = "10.1016/0370-2693(86)90800-2",
    journal = "Phys. Lett. B",
    volume = "166",
    pages = "285--288",
    year = "1986"
}

@article{Reuter:1986wm,
    author = "Reuter, M. and Wetterich, C.",
    title = "{Time Evolution of the Cosmological 'Constant'}",
    reportNumber = "DESY-86-146",
    doi = "10.1016/0370-2693(87)90702-7",
    journal = "Phys. Lett. B",
    volume = "188",
    pages = "38--43",
    year = "1987"
}

@article{Caldwell:1997mh,
    author = "Caldwell, R. R. and Steinhardt, Paul J.",
    title = "{The Imprint of gravitational waves in models dominated by a dynamical cosmic scalar field}",
    eprint = "astro-ph/9710062",
    archivePrefix = "arXiv",
    doi = "10.1103/PhysRevD.57.6057",
    journal = "Phys. Rev. D",
    volume = "57",
    pages = "6057--6064",
    year = "1998"
}

@article{Caldwell:1997ii,
    author = "Caldwell, R. R. and Dave, Rahul and Steinhardt, Paul J.",
    title = "{Cosmological imprint of an energy component with general equation of state}",
    eprint = "astro-ph/9708069",
    archivePrefix = "arXiv",
    doi = "10.1103/PhysRevLett.80.1582",
    journal = "Phys. Rev. Lett.",
    volume = "80",
    pages = "1582--1585",
    year = "1998"
}

@article{hoyle1997hubble,
  title={On the Hubble constant and the cosmological constant},
  author={Hoyle, F and Burbidge, G and Narlikar, JV},
  journal={Mon. Not. Roy. Astron. Soc.},
  volume={286},
  number={1},
  pages={173--182},
  year={1997},
  publisher={Blackwell Science Ltd Oxford, UK}
}

@article{Hu:1998tk,
    author = "Hu, Wayne and Eisenstein, Daniel J. and Tegmark, Max and White, Martin J.",
    title = "{Observationally determining the properties of dark matter}",
    eprint = "astro-ph/9806362",
    archivePrefix = "arXiv",
    reportNumber = "IASSNS-AST-98-30",
    doi = "10.1103/PhysRevD.59.023512",
    journal = "Phys. Rev. D",
    volume = "59",
    pages = "023512",
    year = "1999"
}

@article{Olson:1987gy,
    author = "Olson, T. S. and Jordan, T. F.",
    title = "{Ages of the Universe for Decreasing Cosmological Constants}",
    doi = "10.1103/PhysRevD.35.3258",
    journal = "Phys. Rev. D",
    volume = "35",
    pages = "3258--3260",
    year = "1987"
}

@article{Pavon:1991uc,
    author = "Pavon, D.",
    title = "{Nonequilibrium fluctuations in cosmic vacuum decay}",
    doi = "10.1103/PhysRevD.43.375",
    journal = "Phys. Rev. D",
    volume = "43",
    pages = "375--378",
    year = "1991"
}

@article{Sahni:1991ks,
    author = "Sahni, Varun and Feldman, Hume and Stebbins, Albert",
    title = "{Loitering universe}",
    reportNumber = "PRINT-91-0070 (CITA,TORONTO)",
    doi = "10.1086/170910",
    journal = "Astrophys. J.",
    volume = "385",
    pages = "1--8",
    year = "1992"
}

@article{Maia:1994vz,
    author = "Maia, Marcos Duarte and Silva, G. S.",
    title = "{Geometrical constraints on the cosmological constant}",
    eprint = "gr-qc/9401005",
    archivePrefix = "arXiv",
    reportNumber = "FERMILAB-PUB-94-002-A",
    doi = "10.1103/PhysRevD.50.7233",
    journal = "Phys. Rev. D",
    volume = "50",
    pages = "7233--7238",
    year = "1994"
}

@article{Matyjasek:1994vp,
    author = "Matyjasek, Jerzy",
    title = "{Cosmological models with a time dependent Lambda term}",
    doi = "10.1103/PhysRevD.51.4154",
    journal = "Phys. Rev. D",
    volume = "51",
    pages = "4154--4159",
    year = "1995"
}

@article{Silveira:1994yq,
    author = "Silveira, V. and Waga, I.",
    title = "{Decaying Lambda cosmologies and power spectrum}",
    reportNumber = "FERMILAB-PUB-94-082-A",
    doi = "10.1103/PhysRevD.50.4890",
    journal = "Phys. Rev. D",
    volume = "50",
    pages = "4890--4894",
    year = "1994"
}

@article{Silveira:1997fp,
    author = "Silveira, V. and Waga, I.",
    title = "{Cosmological properties of a class of Lambda decaying cosmologies}",
    eprint = "astro-ph/9703185",
    archivePrefix = "arXiv",
    doi = "10.1103/PhysRevD.56.4625",
    journal = "Phys. Rev. D",
    volume = "56",
    pages = "4625--4632",
    year = "1997"
}

@article{John:1997ery,
    author = "John, Moncy V. and Joseph, K. Babu",
    title = "{A Low matter density decaying vacuum cosmology from complex metric}",
    eprint = "gr-qc/0007052",
    archivePrefix = "arXiv",
    doi = "10.1088/0264-9381/14/5/016",
    journal = "Class. Quant. Grav.",
    volume = "14",
    pages = "1115",
    year = "1997"
}

@article{Turner:1997npq,
    author = "Turner, Michael S. and White, Martin J.",
    title = "{CDM models with a smooth component}",
    eprint = "astro-ph/9701138",
    archivePrefix = "arXiv",
    reportNumber = "FERMILAB-PUB-97-002-A",
    doi = "10.1103/PhysRevD.56.R4439",
    journal = "Phys. Rev. D",
    volume = "56",
    number = "8",
    pages = "R4439",
    year = "1997"
}

@article{Wang:1998gt,
    author = "Wang, Li-Min and Steinhardt, Paul J.",
    title = "{Cluster abundance constraints on quintessence models}",
    eprint = "astro-ph/9804015",
    archivePrefix = "arXiv",
    doi = "10.1086/306436",
    journal = "Astrophys. J.",
    volume = "508",
    pages = "483--490",
    year = "1998"
}

@article{Gibbons:1977mu,
    author = "Gibbons, G. W. and Hawking, S. W.",
    title = "{Cosmological Event Horizons, Thermodynamics, and Particle Creation}",
    doi = "10.1103/PhysRevD.15.2738",
    journal = "Phys. Rev. D",
    volume = "15",
    pages = "2738--2751",
    year = "1977"
}

@article{Cai:2001tv,
    author = "Cai, Rong-Gen",
    title = "{Cardy-Verlinde formula and thermodynamics of black holes in de Sitter spaces}",
    eprint = "hep-th/0112253",
    archivePrefix = "arXiv",
    doi = "10.1016/S0550-3213(02)00064-0",
    journal = "Nucl. Phys. B",
    volume = "628",
    pages = "375--386",
    year = "2002"
}

@article{Frolov:2002va,
    author = "Frolov, Andrei V. and Kofman, Lev",
    title = "{Inflation and de Sitter thermodynamics}",
    eprint = "hep-th/0212327",
    archivePrefix = "arXiv",
    reportNumber = "CITA-2002-46",
    doi = "10.1088/1475-7516/2003/05/009",
    journal = "JCAP",
    volume = "05",
    pages = "009",
    year = "2003"
}

@article{Gibbons:1976ue,
    author = "Gibbons, G. W. and Hawking, S. W.",
    title = "{Action Integrals and Partition Functions in Quantum Gravity}",
    reportNumber = "PRINT-76-0995 (CAMBRIDGE)",
    doi = "10.1103/PhysRevD.15.2752",
    journal = "Phys. Rev. D",
    volume = "15",
    pages = "2752--2756",
    year = "1977"
}

@article{Redmount:1988pg,
    author = "Redmount, Ian H. and Ruiz Ruiz, Fernando",
    title = "{Thermal Equilibrium in de Sitter Space}",
    reportNumber = "DAMTP-R-88-24",
    doi = "10.1103/PhysRevD.39.2289",
    journal = "Phys. Rev. D",
    volume = "39",
    pages = "2289",
    year = "1989"
}

@article{Davies:1987ti,
    author = "Davies, P. C. W.",
    title = "{Cosmological Horizons and the Generalized Second Law of Thermodynamics}",
    reportNumber = "NCL-87-TP7",
    doi = "10.1088/0264-9381/4/6/006",
    journal = "Class. Quant. Grav.",
    volume = "4",
    pages = "L225",
    year = "1987"
}

@article{Davies:1986wx,
    author = "Davies, P. C. W. and Ford, L. H. and Page, Don N.",
    title = "{Gravitational entropy: Beyond the black hole}",
    doi = "10.1103/PhysRevD.34.1700",
    journal = "Phys. Rev. D",
    volume = "34",
    pages = "1700--1707",
    year = "1986"
}

@article{GianlucaCalcagni_2005,
    author = "Calcagni, Gianluca",
    title = "{de Sitter thermodynamics and the braneworld}",
    eprint = "hep-th/0507125",
    archivePrefix = "arXiv",
    doi = "10.1088/1126-6708/2005/09/060",
    journal = "JHEP",
    volume = "09",
    pages = "060",
    year = "2005"
}

@article{Bousso:2004tv,
    author = "Bousso, Raphael",
    title = "{Cosmology and the S-matrix}",
    eprint = "hep-th/0412197",
    archivePrefix = "arXiv",
    reportNumber = "UCB-PTH-04-36",
    doi = "10.1103/PhysRevD.71.064024",
    journal = "Phys. Rev. D",
    volume = "71",
    pages = "064024",
    year = "2005"
}

@article{Collins:1992eca,
    author = "Collins, W.",
    title = "{Mechanics of apparent horizons}",
    doi = "10.1103/PhysRevD.45.495",
    journal = "Phys. Rev. D",
    volume = "45",
    number = "2",
    pages = "495",
    year = "1992"
}

@article{JoseTomasGalvezGhersi_2011,
doi = {10.1088/1475-7516/2011/06/005},
url = {https://dx.doi.org/10.1088/1475-7516/2011/06/005},
year = {2011},
month = {jun},
publisher = {},
volume = {2011},
number = {06},
pages = {005},
author = {José Tomás Gálvez Ghersi and  Ghazal Geshnizjani and  Federico Piazza and  Sarah Shandera},
title = {Eternal inflation and a thermodynamic treatment of Einstein's equations},
journal = {JCAP},
}

@article{Hayward:1998ee,
    author = "Hayward, Sean A. and Mukohyama, Shinji and Ashworth, M. C.",
    title = "{Dynamic black hole entropy}",
    eprint = "gr-qc/9810006",
    archivePrefix = "arXiv",
    doi = "10.1016/S0375-9601(99)00225-X",
    journal = "Phys. Lett. A",
    volume = "256",
    pages = "347--350",
    year = "1999"
}

@article{Nielsen:2008kd,
    author = "Nielsen, Alex B. and Yeom, Dong-han",
    title = "{Spherically symmetric trapping horizons, the Misner-Sharp mass and black hole evaporation}",
    eprint = "0804.4435",
    archivePrefix = "arXiv",
    primaryClass = "gr-qc",
    doi = "10.1142/S0217751X09045984",
    journal = "Int. J. Mod. Phys. A",
    volume = "24",
    pages = "5261--5285",
    year = "2009"
}

@article{Davies:1988dk,
    author = "Davies, P. C. W.",
    title = "{Cosmological Horizons and Entropy}",
    reportNumber = "Print-88-0461 (NEWCASTLE)",
    doi = "10.1088/0264-9381/5/10/013",
    journal = "Class. Quant. Grav.",
    volume = "5",
    pages = "1349",
    year = "1988"
}

@article{AndreiVFrolov_2003,
    author = "Frolov, Andrei V. and Kofman, Lev",
    title = "{Inflation and de Sitter thermodynamics}",
    eprint = "hep-th/0212327",
    archivePrefix = "arXiv",
    reportNumber = "CITA-2002-46",
    doi = "10.1088/1475-7516/2003/05/009",
    journal = "JCAP",
    volume = "05",
    pages = "009",
    year = "2003"
}

@article{Wang:2005pk,
    author = "Wang, Bin and Gong, Yungui and Abdalla, Elcio",
    title = "{Thermodynamics of an accelerated expanding universe}",
    eprint = "gr-qc/0511051",
    archivePrefix = "arXiv",
    doi = "10.1103/PhysRevD.74.083520",
    journal = "Phys. Rev. D",
    volume = "74",
    pages = "083520",
    year = "2006"
}

@article{Jiang:2009kzr,
    author = "Jiang, Ke-Xia and Feng, Tsun and Peng, Dan-Tao",
    title = "{Hawking radiation of apparent horizon in a FRW universe as tunneling beyond semiclassical approximation}",
    doi = "10.1007/s10773-009-9988-y",
    journal = "Int. J. Theor. Phys.",
    volume = "48",
    pages = "2112--2121",
    year = "2009"
}

@article{Zhu:2008hn,
    author = "Zhu, Tao and Ren, Ji-Rong",
    title = "{Corrections to Hawking-like Radiation for a Friedmann-Robertson-Walker Universe}",
    eprint = "0811.4074",
    archivePrefix = "arXiv",
    primaryClass = "hep-th",
    doi = "10.1140/epjc/s10052-009-1044-9",
    journal = "Eur. Phys. J. C",
    volume = "62",
    pages = "413--418",
    year = "2009"
}

@article{Medved:2002zj,
    author = "Medved, A. J. M.",
    title = "{Radiation via tunneling from a de Sitter cosmological horizon}",
    eprint = "hep-th/0207247",
    archivePrefix = "arXiv",
    doi = "10.1103/PhysRevD.66.124009",
    journal = "Phys. Rev. D",
    volume = "66",
    pages = "124009",
    year = "2002"
}

@article{Cai:2008gw,
    author = "Cai, Rong-Gen and Cao, Li-Ming and Hu, Ya-Peng",
    title = "{Hawking Radiation of Apparent Horizon in a FRW Universe}",
    eprint = "0809.1554",
    archivePrefix = "arXiv",
    primaryClass = "hep-th",
    doi = "10.1088/0264-9381/26/15/155018",
    journal = "Class. Quant. Grav.",
    volume = "26",
    pages = "155018",
    year = "2009"
}

@article{MarcoAngheben_2005,
    author = "Angheben, Marco and Nadalini, Mario and Vanzo, Luciano and Zerbini, Sergio",
    title = "{Hawking radiation as tunneling for extremal and rotating black holes}",
    eprint = "hep-th/0503081",
    archivePrefix = "arXiv",
    doi = "10.1088/1126-6708/2005/05/014",
    journal = "JHEP",
    volume = "05",
    pages = "014",
    year = "2005"
}

@article{Nielsen:2005af,
    author = "Nielsen, Alex B. and Visser, Matt",
    title = "{Production and decay of evolving horizons}",
    eprint = "gr-qc/0510083",
    archivePrefix = "arXiv",
    doi = "10.1088/0264-9381/23/14/006",
    journal = "Class. Quant. Grav.",
    volume = "23",
    pages = "4637--4658",
    year = "2006"
}
